\documentstyle[epsfig,a4]{elsart}
\renewcommand{\i}{{\rm i}}

\newcommand{\barr}{\begin{array}}
\newcommand{\bea}{\begin{eqnarray}}
\newcommand{\be}{\begin{eqnarray}}
\newcommand{\beq}{\begin{equation}}
\newcommand{\earr}{\end{array}}
\newcommand{\eea}{\end{eqnarray}}
\newcommand{\ee}{\end{eqnarray}}

\newcommand{\eeq}{\end{equation}}
\newcommand{\nno}{\nonumber}
\newcommand{\nn}{\nonumber}

\newcommand\third{{1\over 3}}
\newcommand\twothird{{2\over 3}}
\newcommand\fourth{{1\over 4}}

\newcommand\fivehalf{{5\over 2}}
\newcommand\sevenhalf{{7\over 2}}

\newcommand{\tr}[1]{{\rm tr}_{{#1}}}
\newcommand{\Tr}[1]{{\rm Tr}_{{#1}}}
\renewcommand{\det}[1]{{\rm det}_{{#1}}}
\newcommand{\Det}[1]{{\rm Det}_{{#1}}}

\newcommand{\Ho}[2]{H^{(1)}_{#1} (k#2)}
\newcommand{\Ht}[2]{H^{(2)}_{#1} (k#2)}
\newcommand{\Jb}[2]{J_{#1} (k#2)}

\newcommand{\equa}[1]{(\ref{#1})}

\newcommand{\order}[1]{{\cal O} \left( #1 \right )}

\newcommand{\lesim}{\mbox{\raisebox{-.6ex}{$\,{\stackrel{<}{\sim}}\,$}}}
\newcommand{\gesim}{\mbox{\raisebox{-.6ex}{$\,{\stackrel{>}{\sim}}\,$}}}
\newcommand{\semiclass}{\ \stackrel{\rm s.c.}{\longrightarrow}\ }
\renewcommand{\thefootnote}{\#\arabic{footnote}}

\begin{document}
\begin{frontmatter}
\title{\bf Quantum Mechanics and Semiclassics of Hyperbolic n-Disk 
Scattering Systems
}

\author{Andreas Wirzba}
\address{Institut f\"{u}r Kernphysik, 
Technische Hochschule Darmstadt, 
Schlo{\ss}gartenstra{\ss}e~9, 
D-64289~Darmstadt, 
Germany\\
}

\date{\today}

\begin{abstract}
The scattering problems of a scalar point particle from an assembly of
$1< n <\infty$ non-overlapping and disconnected hard disks, fixed in
the two-dimensional plane, belong to the simplest realizations of
classically hyperbolic scattering systems. Their simplicity allows for
a detailed study of the quantum mechanics, semiclassics and classics
of the scattering. Here, we investigate the connection between the
spectral properties of the quantum-mechanical scattering matrix and
its semiclassical equivalent based on the semiclassical zeta-function
of Gutzwiller and Voros.  We construct the scattering matrix and its
determinant for any non-overlapping $n$-disk system (with $n<\infty$)
and rewrite the determinant in such a way that it separates into the
product over $n$ determinants of 1-disk scattering matrices --
representing the incoherent part of the scattering from the $n$-disk
system -- and the ratio of two mutually complex conjugate determinants
of the genuine multiscattering matrix ${\bf M}$ which is of 
Korringa-Kohn-Rostoker-type
and which represents the coherent multidisk aspect of the $n$-disk
scattering.  Our quantum-mechanical calculation is well-defined at
every step, as the on-shell {\bf T}--matrix and the
multiscattering kernel ${\bf M}\mbox{$-$}{\bf 1}$ are shown to be
trace-class.  The multiscattering determinant can be organized in
terms of the cumulant expansion which is the defining prescription
for the determinant over an infinite, but trace-class matrix.  The
quantum cumulants are then expanded by traces which, in turn, split into
quantum itineraries or cycles. These can be organized by a simple
symbolic dynamics.  The semiclassical reduction of the coherent
multiscattering part takes place on the level of the quantum cycles.
We show that the semiclassical analog of the $m$th quantum cumulant is
the $m$th curvature term of the semiclassical zeta function. In this
way quantum mechanics naturally imposes the curvature regularization
structured by the topological (not the geometrical) length of the
pertinent periodic orbits onto the semiclassical zeta function.
However, since the cumulant limit $m\to \infty$ and the semiclassical
limit, $\hbar \to 0$ or (wave number) $k\to \infty$, do not commute in
general, the semiclassical analog of the quantum multiscattering
determinant is a curvature expanded semiclassical zeta function which
is truncated in the curvature order.  We relate the order of this
truncation to the topological entropy of the corresponding classical
system.
%
%
We show this explicitly for the 3-disk scattering system and discuss
the consequences of this truncation for the semiclassical predictions
of the scattering resonances.
%
%
We show that, under the above mentioned truncations in the curvature
order, unitarity in $n$-disk scattering problems is preserved even at
the semiclassical level.  Finally, with the help of cluster phase
shifts, it is shown that the semiclassical zeta function of Gutzwiller and
Voros has the correct stability structure and is superior to all the
competitor zeta functions studied in the literature.  
\\[2mm]
\end{abstract}
\end{frontmatter}
\newpage
\tableofcontents
\newpage
\section{Introduction\label{chap:preintro}}
\setcounter{equation}{0}
\setcounter{figure}{0}
\setcounter{table}{0}
\setcounter{footnote}{0}
\renewcommand{\thefootnote}{\#\arabic{footnote}}

The main focus of this manuscript is on the 
transition from quantum mechanics to semiclassics in classically hyperbolic
scattering systems, and  in particular, 
on the convergence problems of periodic orbit expansions of $n$-disk
repellers.

\subsection{Motivation and historic perspective}
Why more than 70 years after the birth of textbook quantum mechanics and
in the age of supercomputers
is there still interest in semiclassical methods?
First of all, there remains the intellectual challenge to  derive classical
mechanics from quantum mechanics, especially for classically 
non-separable chaotic problems. Pure quantum mechanics is linear and of
power-law complexity, whereas classical mechanics is generically of
exponential complexity. How does the latter emerge from the former?
Secondly, in many fields (atomic physics, 
molecular physics and quantum chemistry, 
but also optics and acoustics  which are not quantum systems but 
are also characterized by the transition from wave dynamics to
ray dynamics)
semiclassical 
methods have been very powerful in the past and are still useful today
for practical calculations, from the detection of elementary
particles to the (radar)-detection of airplanes or submarines.  Third,
the numerical methods for
solving 
multidimensional, non-integrable quantum systems
are generically of ``black-box'' type, e.g.\ the diagonalization of
a large, but truncated hamiltonian matrix
in a suitably chosen basis. They 
are
computationally intense and provide little opportunity for
learning how
the  underlying dynamics organizes itself.
In contrast, semiclassical methods have a better 
chance to provide an intuitive understanding
which may even be utilized as  a vehicle for the interpretation 
of numerically calculated quantum-mechanical data.

In the days of ``old'' quantum mechanics 
semiclassical techniques provided of course the
only quantization techniques.
Because of the failure, at that time, to describe more complicated systems
such as  the helium atom (see, however, the resolution of Wintgen and 
collaborators~\cite{wintgen}; \cite{QCcourse} and also
\cite{steiner} provide  for a nice account
of the history), 
they were replaced by modern quantum mechanics based on wave mechanics. 
Here,
through WKB methods, they reappeared as approximation techniques for
1-dimensional systems and, in the generalization to the
Einstein-Brillouin-Keller (EBK) quantization, for separable 
problems~\cite{steiner,Keller_Siam,gutzwiller} where an 
$n$-degree-of-freedom
system reduces to $n$ one-degree-of-freedom systems.
Thus semiclassical methods had been limited to such systems
which  are classically nearly integrable.

It was Gutzwiller who in the late 60s and 
early 70s (see e.g.~\cite{gutzwiller} and
\cite{gut71})  (re-)introduced semiclassical methods to deal with 
multidimensional,
non-integrable quantum problems: 
with the help of Feynman path integral techniques 
the exact time-dependent propagator (heat kernel) is approximated,
in stationary phase, by the semiclassical Van-Vleck propagator. 
After a Laplace
transformation and under
a further stationary phase approximation
the energy-dependent semiclassical Green's function emerges.  
The trace of the latter is calculated and reduces 
under a third stationary
phase transformation to a  smooth Weyl term (which parametrizes the
global geometrical features) {\em and} an oscillating sum
over all periodic
orbits of the corresponding classical problem. 
Since the imaginary part of the trace of the exact Green's function
is proportional to the spectral density, the Gutzwiller trace formula
links the spectrum of eigen-energies, or at least the modulations in 
this spectrum, to the Weyl term and the 
sum over all periodic orbits.
Around the same time, Balian and Bloch
obtained similar results with the help of multiple-expansion techniques
for Green's functions, especially in billiard cavities, 
see e.g.~\cite{bb_2}.

For more than one degree of freedom, classical systems can
exhibit chaos. Generically these are, however, 
non-hyperbolic classically mixed systems with elliptic islands embedded
in chaotic zones 
and marginally stable orbits for which neither  
the Gutzwiller trace
formula nor the EBK-techniques apply, see Berry and 
Tabor~\cite{berry_tabor}.
Purely hyperbolic systems with only isolated unstable periodic orbits
are the exceptions. But in contrast to integrable systems, they are generically
stable against small perturbations~\cite{gutzwiller}. Moreover, they
allow the semiclassical periodic orbit quantization which can even be 
exact as for the case of the Selberg trace formula which relates the
spectrum of the Laplace-Beltrami operator to geodesic motion on 
surfaces of constant negative curvature~\cite{selberg}.
The Gutzwiller trace formula for generic hyperbolic systems is, however,
only an approximation, 
since its derivation is based on several semiclassical saddle-point
methods as mentioned above.

In recent years, mostly driven by the uprise of classical chaos, there has
been a resurgence of semiclassical ideas and concepts.
Considerable progress
has been made by applying semiclassical periodic orbit formulae in
the calculation of energy levels for bound-state systems or resonance poles
for scattering systems, e.g., the anisotropic Kepler
problem~\cite{gutzwiller}, the scattering problem on hard
disks~\cite{Eck_org,gr_cl,gr_sc,gr,Cvi_Eck_89,pinball}, 
the helium atom~\cite{wintgen} etc. (See
Ref.\cite{others} for a recent collection 
about periodic orbit theory.)\ 
It is well known that the periodic orbit sum for chaotic systems is divergent
in the physical region
of interest. This is the case on the real 
energy axis for bounded problems and
in the region of resonances for scattering problems, because of
the exponentially proliferating number of periodic orbits, see 
\cite{gutzwiller,scherer}. Hence refinements have been introduced
in order to transform the periodic orbit sum in the physical domain
of interest to a still conditionally convergent sum by using 
symbolic dynamics and the cycle expansion~\cite{cvi88,artuso,Cvi_Eck_89},
Riemann-Siegel lookalike formal and pseudo-orbit
expansions~\cite{berry_keats_90,berry_keats}, 
surface of section techniques~\cite{bogo,bogo92}, inside-outside 
duality~\cite{insideout}, heat-kernel regularization~\cite{steiner,steinsieb} 
etc. 
These methods tend to be motivated from other areas
in physics and mathematics~\cite{pisani} such as
topology of flows in classical chaos, the theory of the 
Riemann
zeta functions, the boundary integral method for partial differential
equations, Fredholm theory (see also \cite{prange}), quantum field theory etc.

In addition to the convergence problem, there exists the further
complication for 
{\em bounded} smooth potential and billiard problems 
that the corresponding periodic orbit sums predict in general non-hermitean
spectra. This problem is addressed by the Berry-Keating resummation 
techniques~\cite{berry_keats_90,berry_keats} -- however, in an ad-hoc fashion.
In contrast, scattering
problems circumvent this difficulty altogether 
since their corresponding resonances
are complex to start with. Moreover, the scattering resonances follow directly
from the periodic orbit sum, as the Weyl term is absent for scattering
problems. In fact, it is more correct to state that the Weyl term does not
interfer with the periodic sum, as a negative Weyl term might  
still be present, see e.g.\cite{scherer}. 
Furthermore, scattering systems
allow for a nice
interpretation of classical periodic sums in terms of survival 
probabilities~\cite{QCcourse,predchaos}. In this respect, it is an interesting
open problem why these
classical calculations do not seem to generate a Weyl term, 
whether applied for
bounded or scattering systems.
For these reasons, the study of periodic sums for 
scattering systems should be simpler than the corresponding study for
bound-state problems, as only the convergence problem is the issue.

\subsection{The $n$-disk repeller: a model for hyperbolic scattering}
\begin{figure}[htb]
 \centerline 
{\epsfig{file=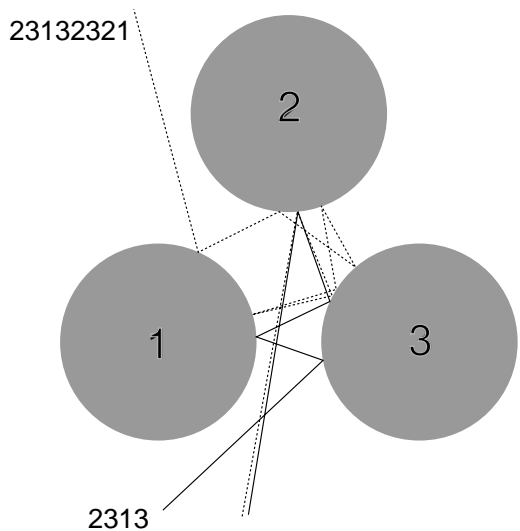,height=7cm,angle=0}}
\caption[fig:3-disk-repeller]{\small 
The three-disk repeller with the symbolic dynamics of the full domain. 
The figure is from Ref.\cite{QCcourse}.
\label{fig:3-disk-repeller}}
\end{figure}

Hence, one should look for a simple classically 
hyperbolic scattering system which can
be used to address the convergence problem. It should not be too 
special, as for example the
motion
on a surface of constant curvature, but reasonably realistic and instructive. 
Eckhardt~\cite{Eck_org} suggested such a system, the 
``classical pinball machine''. It consists of a point particle
 and a finite number (in his case three) 
identical non-overlapping disconnected
circular disks in the
plane which are centered at the corners 
of a regular polygon (in his case an equilateral
triangle). 
The point particle scatters elastically from the disks and
moves freely in between collisions. The classical mechanics, semiclassics
and quantum mechanics of this so-called three-disk system
was investigated in a series of papers by Gaspard and Rice, 
\cite{gr_cl,gr_sc,gr}, and, independently, by
Cvitanovi\'{c} and  Eckhardt~\cite{Cvi_Eck_89}, see also 
Scherer~\cite{scherer} and Ref.\cite{pinball}. 
It belongs to a class of mechanical systems
which are everywhere defocusing, hence no stable periodic orbit can exist
(see Fig.\ref{fig:3-disk-repeller}). 

The classical 
dynamics with one or two 
disks is simple, there is either no or one trapped trajectory.
The latter is obviously unstable, since a small displacement leads to
a defocusing after the reflection from the curved surface of 
disk~\cite{gr_cl}. The two-disk system is therefore one of the simplest
hyperbolic scattering systems, but it is non-chaotic.
However, with three or more disks 
there are infinitely many trapped trajectories forming a 
repeller~\cite{pinball}.
The periodic orbits corresponding to these trapped trajectories are all
isolated and
unstable because of the defocusing nature of the reflections.
Note that the one-disk and two-disk systems, although classically simple,
are nonetheless interesting.
The quantum-mechanical one-disk scattering system (since it is separable)
has been
one of
the key models for building up the semiclassical theory of 
diffraction~\cite{franz,keller,Nussenzveig}. Similarly, the two-disk 
system became the toy ground for the periodic-orbit theory of 
diffraction~\cite{vwr_prl,vwr_japan}. In fact, the two-disk system 
has infinitely many
{\em diffractive} creeping periodic orbits which can be classified by symbolic
dynamics similarly  to the infinitely many geometrical 
orbits of the three-disk
system. 
The symbolic dynamics of a general $n$-disk system is very simple, see
e.g.\ \cite{QCcourse}:
periodic orbits can be classified by a series of 
``house numbers'' of the disks which are visited by the point particle which
follows the corresponding trajectory. Not all sequences are allowed:
after each reflection from one disk, the point particle has to
proceed to a different disk, since the evolution between the disk is the
free one. Furthermore, for general geometries there may exist sequences which
correspond to trajectories which would directly pass through a disk. The
sequences corresponding to these
so-called ``ghost orbits'' have to be excluded from the classical 
consideration.  
In summary, the  geometrical periodic orbits (including ghost orbits) 
are labelled in the full domain of the $n$-disk repeller 
by itineraries (= periodic words) with $n$ different symbols (=letters)
with the trivial ``pruning'' rule that successive letters in the itinerary
must be different. The itineraries corresponding to ghost orbits have
to be removed or ``pruned'' with all their sub-branches from the symbol
tree.  Periodic trajectories which have reflections  from 
inside of a disk (i.e. the  point particle traverses first through a disk
and is then reflected from the other side of the disk) can be excluded 
from the very beginning. In fact, in our semiclassical reduction of
Sec.\ref{chap:semiclass} we will show for all repeller geometries with $n$
non-overlapping disks  that, to each specified itinerary, there belongs
uniquely one standard  periodic orbit  which might contain ghost passages
but which cannot be reflected from the inside. There is only one caveat:
our method cannot decide whether grazing trajectories (which 
are tangential to a disk surface)
belong to the
class of ghost trajectories 
or to the class of reflected trajectories. For simplicity,
we just exclude all geometries which
allow for grazing periodic orbits from our proof. Alternatively, one
might treat these grazing trajectories separately with the help of the
diffractional methods of Refs.\cite{Nussenzveig,penumbra}.

\begin{figure}[htb]
 \centerline {\epsfig{file=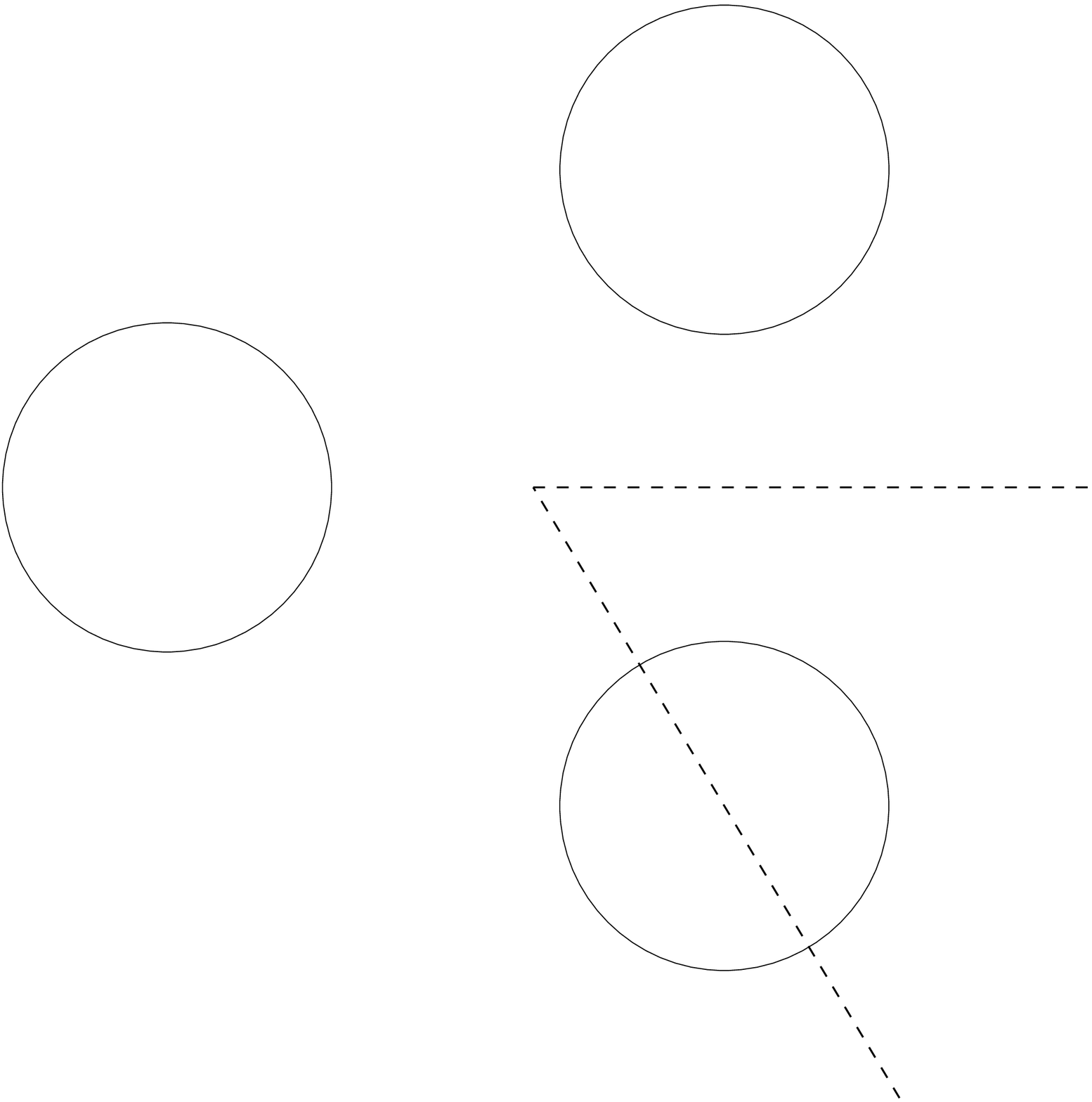,height=5cm,angle=+60}}
\caption[fig:fund_domain]{\small 
Equilateral 3-disk system and its fundamental domain.
\label{fig:fund_domain}}
\end{figure}

The symbolic dynamics described above in the full domain 
applies of course to the equilateral
three-disk system.
However, because of the discrete ${\rm C}_{\rm 3v}$ 
symmetry of that system, the dynamics can be mapped into the {\em fundamental}
domain (any one of the  1/6-th slices of the {\em full} domain 
which are  centered at the
symmetry-point of the system and which exactly cut through one half of
each disk, see Fig.\ref{fig:fund_domain}). 

In this fundamental domain the three-letter symbolic dynamics
of the full domain reduces a two-letter symbolic dynamics. The
symbol `0', say, 
labels all encounters of a periodic orbit with a disk in the
fundamental domain where the point particle in the corresponding full domain
is reflected to the disk where it was just coming from, whereas the
symbol `1', say, is reserved for encounters where the point particle
is reflected to the other disk.

Whether the full or the fundamental domain is used, the 3-disk system allows
for a  unique symbolic labelling (if
the disk separation is large enough
even without non-trivial pruning).
If a symbolic dynamics exist, the periodic orbits can
be classified by their topological length which is defined as the
length of the corresponding symbol sequence.
In this case 
the various classical and semiclassical zeta functions  
are resummable in terms of the cycle
expansion~\cite{cvi88,artuso} which can be cast to a
sum over a few fundamental cycles (or primary periodic orbits)
$t_f$ and higher
curvature corrections $C_m$ of increasing topological order $m$:
\beq
   \frac{1}{\zeta}= 1 - \sum_f t_f -\sum_m C_m \;.
 \label{zeta_gen}
\eeq
The curvature $C_m$  in \equa{zeta_gen} 
contains all allowed periodic orbits of 
topological length $m$ for a specified symbolic dynamics and 
suitable ``shadow-corrections'' of 
combinations (pseudo-orbits) of shorter
periodic orbits  with a combined topological length $m$.

Common to most studies of the semiclassics of 
the $n$-disk repellers is that they
are ``bottom-up'' approaches. 
Whether they use the
Gutzwiller trace formula~\cite{gutzwiller}, 
the Ruelle or dynamical
zeta function~\cite{ruelle}, the Gutzwiller-Voros zeta 
function~\cite{voros88}, their starting
point is the cycle expansion~\cite{cvi88,artuso,predchaos}.  
The periodic orbits are motivated from a semiclassical saddle-point
approximation.
The rest is classical in the sense that
all quantities which enter the periodic orbit calculation as {\it
e.g.} actions, stabilities and geometrical phases are determined at
the classical level (see however 
Refs.\cite{alonso,gasp_hbar,vattay_hbar} where leading
$\hbar$-corrections to the dynamical zeta function as well as the 
Gutzwiller-Voros zeta function have been calculated).
For instance, the dynamical zeta function has typically the form
\beq
  \zeta_0^{-1}(E) = \prod_p (1-t_p), \quad
     t_p = \frac{1}{\sqrt{\Lambda_p}} 
 \e^{(\i/\hbar)S_p(E) -\i\frac{\pi}{2} \nu_p}.
    \label{zeta}
\eeq
The product is over all prime cycles (prime periodic orbits) $p$.  The
quantity $\Lambda_p$ is the stability factor of the $p$-th cycle,
i.e., the expanding eigenvalue of the $p$-cycle Jacobian, $S_p$ 
is the action  and $\nu_p$ is (the sum of) the Maslov index (and the
group theoretical weight for a given representation) of the $p$-th cycle.
For $n$-disk repellers, the action is simply 
$S_p=k L_p$, the product  of the geometrical length $L_p$ of the periodic orbit
and the wave number $k=\sqrt{2m E}$ in terms of the energy $E$ and
mass $m$ of the point particle.
The semiclassical predictions for the scattering resonances 
are then extracted from the zeros of the cycle-expanded semiclassical
zeta-function. In this way one derives predictions of the 
dynamical zeta function 
for the leading resonances 
(which are the resonances closest to real $k$-axis). 
In the case of the Gutzwiller-Voros zeta function also subleading
resonances result, however, only if the resonances lie 
above a line  defined by the poles of the dynamical zeta 
function~\cite{ER92,fredh,fredh2,pinball}.
The quasiclassical zeta function of Vattay and Cvitanovi\'{c}
is entire and gives predictions for subleading resonances for
the entire lower half of the complex plane~\cite{fredh2}.

\subsection{Objective}
As the $n$-disk scattering systems are generically hyperbolic, but 
still simple enough to allow for a closed-form quantum-mechanical 
setup~\cite{gr} and detailed quantum-mechanical 
investigations~\cite{aw_chaos,aw_nucl}, we want to study
the structure of the semiclassical periodic sum for a hyperbolic scattering
system
in a ``top-down'' approach, i.e.\ in a direct derivation from the 
exact quantum  mechanics of the $n$-disk repeller. This is in contrast
to the usual semiclassical ``bottom-up'' studies of the $n$-disk 
repellers which can be  affected by uncontrolled operations during
the long and mostly formal derivation from the Gutzwiller trace formula.
Especially regularization prescriptions, like the cycle expansion, have
to be added from the outside in order to get converging 
semiclassical predictions.

Hence, for any $n$-disk scattering problem  
with a finite number of non-overlapping
disconnected disks 
we want to construct a direct link from the defining exact ${\bf S}$-matrix 
to the pertinent semiclassics (in terms of a suitable periodic-orbit expansion)
with the following qualifications:
\begin{enumerate}
\item
 The derivation should lead to a unique specification of the
periodic orbits for a given $n$-disk geometry.
The method should be able to handle $n$-disk geometries which allow
for ghost orbits, i.e., periodic orbits existing in any of the pertinent
``parent'' disk-systems (defined by the removal of one or more disks) which
are blocked by the return of at least one of the removed disks.
\item
Since disk-systems are known where the semiclassics is  strongly 
governed by diffractive orbits (see \cite{vwr_japan} and especially
\cite{WR96} for the 2-dimensional scattering analog of the two-well-potential
problem), diffractive periodic orbits should emerge together with
their standard  partners.
\item
The subleading stability structure of the standard periodic orbits
should follow from this derivation  in order to discriminate between
the Gutzwiller-Voros zeta function
and other competitors, e.g., the dynamical zeta function of 
Ruelle~\cite{ruelle} or the quasiclassical zeta function of 
Vattay and Cvitanovi\'{c}~\cite{fredh2}; in other words, we want to
derive {\em the} semiclassical spectral function.
\item
The setup of the starting-point, the  quantum-mechanical side, 
should not be plagued by formal or uncontrolled manipulations or assumptions.
Especially, if the quantum-mechanical side does not exist 
without a suitable regularization prescription, the latter should be  
provided {\em before} the semiclassical reduction is performed. This 
should
exclude that the semiclassical sums encounter {\em hidden} problems  which
are already present at the quantum-mechanical level.
\item
The link between the exact quantum mechanics and semiclassics 
should not only allow for the computation 
of scattering resonance, but should be valid for all 
values of the wave number, also away from the resonances and from the
real axis, modulo the boundary of semiclassical convergence, as this 
issue can only be addressed during the link-procedure.
Branch cuts and singularities on the quantum-mechanical
side have to be taken into account of course.
\item
The spectral function should not only result in a formal sense, but, if
necessary,
with a pertinent regularization and summation prescription that
should not be imposed from the outside.
\item 
Most importantly, the derivation should be well-defined and allow 
for a test of the summation prescription of
the period-orbit expansion. If potential problems occur, 
they should be pinpointed  in
the derivation.
\end{enumerate}
 
\subsection{Outline}
The manuscript is organized as follows.
We begin in Sec.\ref{chap:intro} with the standard approach relating
quantum-mechanical and semiclassical resonances for $n$-disk repellers.

Generalizing the
work of Gaspard and Rice~\cite{gr} to non-overlapping $n$-disk problems
of arbitrary geometry and disk sizes
we construct in 
Sec.\ref{chap:S-matrix} the ${\bf S}$-matrix
from stationary scattering theory. Details of this calculation are
relegated to App.\ref{app:construction}. Utilizing the machinery
of trace-class operators which are summarized in App.\ref{app:trace}
we construct the determinant of the $n$-disk ${\bf S}$-matrix as the
product of $n$ one-disk determinants and the ratio of the determinant
and its complex conjugate of the genuine multiscattering matrix.
It is shown how the latter determinants split under symmetry operations.
The proofs for the existence of the determinants are
relegated to App.\ref{app:suppl} and the comparison to alternative
formulations of the multiscattering kernel can be found in  
App.\ref{app:Lloyd}.

In Sec.\ref{chap:link} we state the link between the exact determinant
of the $n$-disk ${\bf S}$-matrix and the Gutzwiller-Voros curvature expansion.
We discuss the semiclassical limit of the incoherent part, whereas the
actual calculation is reported in App.\ref{app:semi1disk}.
It is shown that, under the semiclassical reduction of the quantum traces,
the Plemelj-Smithies recursion relation for the quantum cumulants 
transforms into 
the recursion relation for the semiclassical curvatures 
which are known from the
cycle expansion.

The actual semiclassical reduction is worked out in Sec.\ref{chap:semiclass}.
We start with the construction 
of the quantum cycles or itineraries built from the convolution of a finite
number of multiscattering kernels and show that they have 
the same symbolic dynamics in the full domain of an 
arbitrary 
$n$-disk system
as their semiclassical
counterparts, the geometrical periodic orbits.
We discuss  the case that the quantum-mechanical cumulant sum incorporates
geometries  which classically allow for non-trivial pruning and hence
for periodic orbits which pass undisturbed straight through a disk, 
see Refs.\cite{bb_1,Berry_KKR}. We show how these ghost orbits cancel 
against their ``parent'' periodic orbits resulting from itineraries without
the disk which is affected by the ghost passage.
For the general case of an
arbitrary quantum cycle 
Sec.\ref{chap:sc_appr_po} mirrors  the semiclassical reduction of the
convolution of two multiscattering kernels studied in detail in 
App.\ref{app:convol} with the help of the Watson contour integration and
suitable deformations of the paths in the complex angular-momentum plane.
In Sec.\ref{chap:sc_itin_geom} the geometrical limit of a quantum
cycle is studied, which is generalized to the case of $r$ times 
repeated cycles in the following section. 
In Sec.\ref{chap:sc_ghost_rule}
the ghost cancellation rule for arbitrary cycles is derived.
The semiclassical diffractive creeping contributions are constructed
and studied in Secs.\ref{chap:sc_itin_creep} and 
\ref{chap:sc_more_creep}. 
Sec.\ref{chap:semiclass} ends with the proof that 
an arbitrary
quantum itinerary reduces semiclassically to a periodic orbit of 
Gutzwiller-Voros stability, such that the link between the exact
determinant of the $n$-disk ${\bf S}$-matrix and the Gutzwiller-Voros
curvature sum is established.

Numerical tests of the semiclassical curvature expansion can be found
in Sec.\ref{chap:numerical} for the example of 
the three-disk system in the ${\rm A}_{\rm 1}$ symmetry-class
representation. First, the exact quantum-mechanical data are compared
to the  
semiclassical predictions of the Gutzwiller-Voros zeta function,
the dynamical zeta function~\cite{ruelle} and the quasiclassical zeta
function suggested in \cite{fredh2}, where all three semiclassical
zeta function are expanded in curvatures which are truncated at finite order.
Secondly, the exact cluster phase shift (defined by the phase
of the determinant of the multiscattering matrix) is compared with the
semiclassical predictions of the three zeta functions. 
Although all three zeta functions seem at first sight empirically equivalent,
as they all predict the same leading resonances closest to the
real $k$-axis, this comparison shows clearly which of the three 
is superior
and is hence the candidate for -- at least -- {\em the} FAPP 
(``for all practical purposes'') zeta function.
Sec.\ref{chap:numerical} ends with an order-by-order 
comparison of the exact cumulants with their semiclassical counterpart,
the curvatures for the Gutzwiller-Voros zeta function. 
From these numerical data we extract  
an empirical truncation rule for the curvature expansion as a function of
the wave number. We relate this rule
to the uncertainty bound resulting from finite quantum-mechanical
resolution of the exponentially proliferating details of the classically
repelling set.

Sec.\ref{chap:end} concludes with a summary.
Here we emphasis 
the preservation of unitarity under the semiclassical reduction,
the decoupling of the incoherent one-disk from the coherent $n$-disk
determinants, and
the particularities, when bounded domains are formed in the case
of (nearly) touching disks.
Furthermore, the resonance data are correlated with the truncation from
the uncertainty bound.
We discuss 
the relevance of those periodic orbits whose topological order exceeds 
the uncertainty bound. Arguments are presented 
that the Gutzwiller-Voros zeta function
ought to be interpreted 
in the asymptotic sense as an truncated
sum, whether it converges or not.
The conclusions end with a discussion on $\hbar$ corrections. 

Note that the contents of Apps.\ref{app:construction} and \ref{app:suppl} are
based on M.~Henseler's diploma thesis~\cite{mh}, while 
Sec.\ref{chap:numerical} as well as
Secs.\ref{chap:S-matrix},\ref{chap:link} and Apps.\ref{app:trace}.1-2
have partial overlap with Refs.\cite{cvw96} and \cite{wh97}, respectively.
\newpage
\section{Semiclassical resonances of the $n$-disk system\label{chap:intro}}
\setcounter{equation}{0}
\setcounter{figure}{0}
\setcounter{table}{0}
\setcounter{footnote}{0}
\renewcommand{\thefootnote}{\#\arabic{footnote}}

The connection between exact quantum mechanics, on the one side, and
semiclassics, on the other, for the $n$-disk repellers
in the standard ``bottom-up'' approach,
is rather indirect.
It has been mainly  based on a comparison of the exact
and semiclassical predictions for resonance data. In the exact
quantum-mechanical calculations
the resonance poles are extracted from the zeros of a characteristic scattering
determinant (see ref.~\cite{gr} and below), 
whereas the semiclassical predictions follow from the zeros
(poles)
of one of the semiclassical zeta functions.
These semiclassical quantities have either formally
been taken over from bounded problems (where  the semiclassical reduction
is performed via the spectral
density)~\cite{scherer,pinball} or they have just 
been extrapolated from the
corresponding classical scattering determinant~\cite{fredh,fredh2}.
Our aim is to construct a {\em direct} link between 
the quantum-mechanical and semiclassical treatment of hyperbolic scattering
in a concrete context, the $n$-disk repellers. 

Following the methods of Gaspard and Rice~\cite{gr} 
we will construct in Sec.\ref{chap:S-matrix} and App.\ref{app:construction}  
the pertinent 
on-shell {\bf T}--matrix which splits into the product of three matrices,
namely
${\bf C}(k) {\bf M}^{-1}(k) {\bf D}(k)$. The matrices
${\bf C}(k)$ and ${\bf D}(k)$ couple the incoming and 
outgoing scattering
wave   (of wave number $k$), respectively, to {\em one} of the disks, 
whereas the matrix ${\bf M}(k)$ parametrizes
the scattering interior, i.e., the  multiscattering evolution in the
multidisk geometry.
The  understanding is that
the resonance poles of the $n>1$ disk problem 
can only result from the zeros of the
characteristic  
determinant ${\rm det}{\bf M}(k)$; 
see the quantum-mechanical construction of
Gaspard and Rice~\cite{gr} for the three-disk
scattering system~\cite{Eck_org,gr_cl,Cvi_Eck_89}.
Their work refers back to Berry's 
application~\cite{Berry_KKR,Berry_LH} of the
Korringa-Kohn-Rostoker (KKR) method~\cite{KKR}
to the (infinite) 
two-dimensional
Sinai-billiard problem which, in turn, is 
based on Lloyd's multiple scattering 
method~\cite{Lloyd,Lloyd_smith}
for a finite cluster of  
non-overlapping muffin-tin potentials in three dimensions. 

The resonance poles are calculated numerically by solving
${\rm det} {\bf M}(k) =0 $ 
in a finite, but large basis, such that the
result is insensitive to an enlargement of the basis (see,
e.g., \cite{aw_chaos}).  On the semiclassical side, the geometrical
primitive periodic orbits (labelled by $p$) 
have been summed up -- including repeats (labelled by $r$) -- in the
Gutzwiller-Voros zeta function~\cite{gutzwiller,voros88} 
\be
  Z_{\rm GV} (z;k) &=& \exp\left\{ - \sum_{p} \sum_{r=1}^{\infty} \frac{1}{r}
 \frac{\left( z^{n_p}\, t_p(k) \right)^r }{1-\frac{1}{\Lambda_p^r}} \right\} 
 \nonumber \\
 &=& \prod_{p} \prod_{j=0}^{\infty}\left (1 - \frac{z^{n_p} t_p(k)}
{{\Lambda_p}^j}\right) 
 \label{GV_zeta_app}  
\ee
the dynamical zeta function of Ruelle~\cite{ruelle} 
\be
   \zeta_0^{-1} (z;k) = \exp\left\{ - \sum_{p} \sum_{r=1}^{\infty} \frac{1}{r}
  z^{r n_p}\, t_p(k)^r \right\} 
   = \prod_p \left(1 -z^{n_p} t_p\right)
 \label{dyn_zeta_app}
\ee
(which is the $j=0$ part of the Gutzwiller-Voros zeta function)
or the quasiclassical zeta function of Vattay and Cvitanovi\'{c}~\cite{fredh2} 
\be
  Z_{\rm qcl} (z;k) &=& \exp\left\{ - \sum_{p} \sum_{r=1}^{\infty} \frac{1}{r}
 \frac{\left( z^{n_p}\, t_p(k) \right)^r }{
 \left ( 1-\frac{1}{\Lambda_p^r} \right )^2
 \left ( 1-\frac{1}{\Lambda_p^{2r}}\right) } \right\}  \nonumber \\
 &=& \prod_p \prod_{j=0}^\infty \prod_{l=0}^\infty
 \left ( 1 -\frac{z^{n_p} t_p(k)}{\Lambda_p^{\,j+2l}} \right )^{j+1} \; 
 \label{qcl_zeta}
\ee
which is an entire function.
In all cases
$t_p(k)=\e^{\i  k L_p - \i  
\nu_p \pi/2}/\sqrt{|\Lambda_p|}$ is the so-called
$p^{\,\rm th}$ cycle, $n_p$ is its topological length and $z$ is a 
book-keeping variable for keeping track of the topological order.   
The input is purely geometrical, i.e., the
lengths $L_p$, the Maslov indices $\nu_p$, and the stabilities (the
leading eigenvalues of the stability matrices) $\Lambda_p$ of the
$p^{\,\rm th}$ primitive periodic orbits. Note that both expressions 
for the three zeta functions, either the exponential one or 
the reformulation in terms of infinite product(s),
are purely formal. In the physical region of interest, they may not even 
exist without regularization. (An exception is 
the non-chaotic 2-disk system, as it has only one periodic orbit, 
$t_0(k)$.)\
Therefore,
the semiclassical resonance
poles are computed from these zeta functions
in the curvature
expansion~\cite{fredh,artuso,pinball} up to a given topological length
$m$.
This procedure corresponds to a Taylor expansion of, e.g.,  $Z_{\rm GV}(z;k)$
in $z$ around $z=0$ up to order $z^m$ (with $z$ set to unity in
the end), e.g.,
\be
  Z_{\rm GV} (z;k) &=& z^0 - z\sum_{n_p=1} \frac{t_p}{1-\frac{1}{\Lambda_p}}
 \label{gutzcurv} 
 \\ 
 && \mbox{} - \frac{z^2}{2}\left\{ \sum_{n_p=2}\frac{2
 t_p}{1-\frac{1}{\Lambda_p}} \mbox{}+ \sum_{n_p=1}
 \frac{(t_p)^2}{1-\left (\frac{1}{\Lambda_p}\right )^2} -
 \sum_{n_p=1}\sum_{n_{p'}=1} \frac{t_p}{1-\frac{1}{\Lambda_p}}
 \frac{t_{p'}}{1-\frac{1}{\Lambda_{p'}}}\right\} +\cdots\; . \nonumber
\ee 
The hope is that the limit $m\to\infty$ exists ---
at least in the semiclassical regime ${\rm Re}\, k \gg 1/a$ where $a$ is
the characteristic length of the scattering potential.
We will show below that in the quantum-mechanical analog --- 
the cumulant expansion -- this limit can be taken, but that there are further
complications in
in the semiclassical case.

The cycle expansion is one way of regularizing the formal expression of
the Gutzwiller-Voros zeta function
\equa{GV_zeta_app}. Another way would be the multiplication with
a smooth cutoff function, as it is customary in quantum field theories, see 
e.g.~\cite{steiner}. This is, in principle, allowed. In order to
be able to compare quantum mechanics with semiclassics, however, the very same
cutoff function has to be introduced already on the quantum level. Candidates
for such cutoff functions which work on the quantum side {\em and} on the
semiclassical side are not so obvious, see 
e.g.\ App.\ref{app:trace_regularization}. They have to be formulated in
terms of the ${\bf T}$-matrix or the multiscattering kernel
and would introduce further
complications. Fortunately, the quantum-mechanical side of the present problem
exists
without further regularization.
Thus there is no need for an extra cutoff function.

\begin{figure}[htb]
 \centerline 
{\epsfig{file=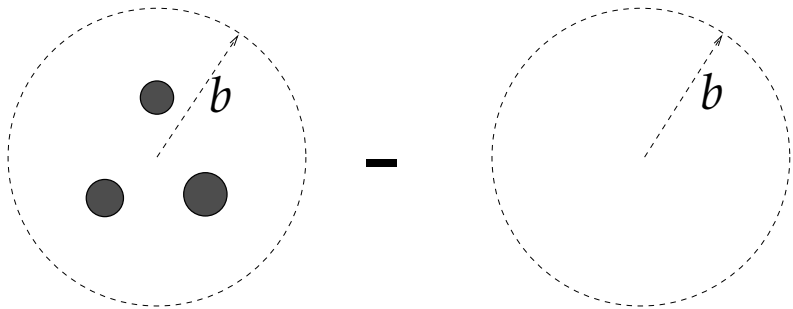,height=4cm,angle=-90}}
\caption[fig:krein]{\small 
The ``difference'' of two bounded reference systems, where one includes
the scattering system.
\label{fig:krein}}
\end{figure}

As mentioned above, the connection between quantum
mechanics and semiclassics for these scattering problems has been
the comparison of the corresponding resonance poles, the
zeros of the characteristic determinant on the one side and the zeros
of the Gutzwiller-Voros zeta function or its competitors -- 
in general in the curvature 
expansion -- on the other side. 
In the literature (see, e.g., Refs.\cite{gr_sc,scherer,pinball}
based on Ref.\cite{bb_1} or \cite{thirring}) the
link is motivated by 
the semiclassical limit of the left hand sides  
of the Krein-Friedel-Lloyd sum for the 
(integrated) spectral density~\cite{Krein,Friedel} and \cite{Lloyd,Lloyd_smith}
\be
 \lim_{\epsilon\to +0}\lim_{b\to \infty} \left ( 
 N^{(n)}(k+\i\epsilon;b)-N^{(0)}(k+\i\epsilon;b) \right )
 &=&\frac{1}{2\pi} {\rm Im} {\rm Tr}\ln {\bf S}(k)\ ,
 \label{friedel_sum_N} \\
 \lim_{\epsilon\to +0}\lim_{b\to \infty} \left ( 
 \rho^{(n)}(k+\i\epsilon;b)- \rho^{(0)}(k+i\epsilon;b) \right )
 &=& \frac{1}{2\pi} {\rm Im} 
{\rm Tr}\frac{d}{dk} \ln {\bf S}(k) \label{friedel_sum_rho}
 \; .
\ee 
See also Ref.\cite{faulkner}
for a modern discussion of the Krein-Friedel-Lloyd formula  
and Refs.\cite{thirring,gasp_varena} for the connection of 
\equa{friedel_sum_rho} to the
the Wigner time delay.
In this way the scattering problem is replaced by the difference of two
bounded  reference  billiards (e.g.\ large circular domains) 
of the same radius $b$ 
which finally will be taken
to infinity, where the first contains the
scattering region or potentials, whereas the other does not 
(see Fig.\ref{fig:krein}).
Here $\rho^{(n)}(k;b)$ ($N^{(n)}(k;b)$) and $\rho^{(0)}(k;b)$ 
($N^{(0)}(k;b)$)
are the 
spectral densities (integrated spectral densities) in the
presence or absence of the scatterers, respectively. 
In the semiclassical approximation, they will be replaced
by  a Weyl term and
an oscillating sum over periodic orbits~\cite{gutzwiller}.

Note that this expression makes only sense for wave numbers above the
real $k$-axis. Especially, if $k$ is chosen to be real, $\epsilon$ must
be greater than zero.  Otherwise, the exact left hand sides 
\equa{friedel_sum_N} and \equa{friedel_sum_rho} would give discontinuous 
staircase
or even delta function sums, respectively, 
whereas the right hand sides are continuous to start with, 
since they can be expressed
by continuous phase shifts.
Thus, the order of the two limits in \equa{friedel_sum_N} and 
\equa{friedel_sum_rho} is essential, see e.g.  
Balian
and Bloch~\cite{bb_1} who stress that smoothed level densities should be 
inserted into the Friedel sums. 
In Ref.\cite{gr_sc}, chapter IV, 
Eqs. (4.1--4), the order is, however, erroneously inverted.

Our point is that
this link between semiclassics and quantum mechanics  is 
still of very indirect nature,  as  the procedure seems to 
use the Gutzwiller-Voros zeta function for {\em bounded} systems and not
for scattering systems and as it does not
specify whether and which regularization has to be used
for the semiclassical 
Gutzwiller trace formula. Neither the
curvature regularization scheme nor other constraints on the periodic orbit
sum follow naturally in this way. For instance, as the link is made with
the help of bounded systems, the question might arise 
whether even in scattering systems 
the Gutzwiller-Voros zeta function should be resummed \`{a} 
la Berry and
Keating~\cite{berry_keats} or not.
This question is answered by the presence of the $\i\epsilon$ term and
the second limit. 
The wave number is shifted from the
real axis into the positive imaginary $k$-plane. This corresponds to
a ``de-hermitezation'' of the underlying exact hamiltonian -- the  
Berry-Keating resummation should therefore not apply, as it is concerned
with hermitean problems.
The necessity of the $+\i\epsilon$ in the semiclassical calculation 
can be understood by purely
phenomenological considerations: Without the $\i\epsilon$ term there is no
reason why one should be able to neglect spurious
periodic orbits which solely are there because of the introduction of 
the confining boundary.
The subtraction of the second (empty) reference system helps just in
the removal of 
those spurious periodic orbits which never encounter the scattering region. 
The ones that do would  still survive the first limit $b\to \infty$, if
they were not damped out by the $+\i\epsilon$ term.

Below, we will construct explicitly a {\em direct} link between  the  full 
quantum-mechanical
 {\bf S}--matrix and the Gutzwiller-Voros zeta function. 
It will be shown that
all steps in the quantum-mechanical description
are well defined, as the {\bf T}--matrix and 
the matrix
${\bf A}\equiv {\bf M}-{\bf 1}$ are trace-class matrices (i.e., the
sum of the diagonal matrix elements is 
absolutely converging in any orthonormal basis). Thus
the corresponding determinants of the {\bf S}-matrix and the
characteristic matrix {\bf M} are guaranteed to exist, 
although they are infinite matrices. 
\newpage
\section{The $n$-disk {\bf S}-matrix and its determinant\label{chap:S-matrix}}
\setcounter{equation}{0}
\setcounter{figure}{0}
\setcounter{table}{0}
%

Following the methods of Berry~\cite{Berry_KKR} and Gaspard and Rice~\cite{gr}
we here describe
the elastic 
scattering of a point particle
from $n$ hard disks
in terms of stationary scattering theory.
Because of the {\em hard-core potential} on the disk surfaces
it turns into 
a boundary value problem.
Let  ${\psi}(\pol{r}\,)$ be a solution
of the pertinent stationary Schr\"{o}dinger equation at spatial
position $\pol{r}$:  
\begin{eqnarray*}
\left({\pol{\nabla}_{{r}}}^2 + {\pol{k}}^2\right) 
    {\psi}(\pol{r}\,)  &=&  0
      \; , \qquad \pol{r} \; \mbox{outside the $n$ disks,} \\
      \psi(\pol{r}\,)  &=&  0 \; ,  \qquad \pol{r}
                 \;  \mbox{on the surfaces of the disks,} 
\end{eqnarray*}
where 
$E = {\hbar^2} {\pol{k}^2}/2m$ 
is the energy of the point-particle written
in terms of its mass $m$  and 
the wave vector $\pol{k}$ of the
incident wave.
After the wave function $\psi(\pol{r}\,)$ is expanded in a basis of
angular momentum eigenfunctions in two dimensions, it reads
\begin{eqnarray*}
 \psi(\pol{r}\,) = \sum_{m=-\infty}^{\infty} \psi_m^{k}(\pol{r}\,)
 \e^{\i m\frac{\pi}{2}} \e^{-\i m\Phi_k} \; , 
\end{eqnarray*}
where $k$ and $\Phi_k$ are the length and angle of the wave vector,
respectively. 
The scattering problem in this basis reduces to
\begin{eqnarray*}
\left({\pol{\nabla}_{{r}}}^2 + {\pol{k}}^2\right) 
    {\psi}^k_m(\pol{r}\,)  &=&  0
      \; , \qquad \pol{r} \; \mbox{outside the disks;} \\
      \psi^{k}_m(\pol{r}\,)  &=&  0 \; ,  \qquad \pol{r}
                 \;  \mbox{on the disk surfaces.} 
\end{eqnarray*}
For large distances from the scatterers ($k r \to \infty$)
the spherical components $\psi_m^k$ can be written as a superposition 
of in-coming and out-going spherical waves,
\be
     \label{psi_asymptotic}
   \psi^k_m(\pol{r}\,)  \sim      
\frac{1}{\sqrt{2\pi kr}} \sum_{m'=-\infty}^\infty
     \left[ \delta_{mm'} \e^{-\i (kr - \frac{\pi}{2}m' 
- \frac{\pi}{4})} 
+ {\bf S}_{mm'} \e^{\i  (kr - \frac{\pi}{2}m' 
 - \frac{\pi}{4})} \right]
         \e^{\i  m'\Phi_r} \; , 
\ee
where $r$ and $\Phi_r$ denote the distance and angle 
of the spatial vector $\pol{r}$ as measured
in the global two-dimensional coordinate system.
Eq.\equa{psi_asymptotic} defines the scattering matrix ${\bf S}$
which is unitary  because of probability 
conservation. In the angular-momentum basis
its
matrix elements ${\bf S}_{mm'}$ describe
the scattering of an in-coming wave with angular momentum $m$ in an
out-going wave with angular momentum $m'$. If there are no scatterers,
then ${\bf S}= {\bf 1}$ and the asymptotic expression of the plane
wave $\e^{\i  \pol{k}\cdot \pol {r}}$ in two dimensions is recovered
from $ \psi(\pol{r}\,)$.
All scattering effects are incorporated in the deviation of
 ${\bf S}$ from the unit matrix, i.e., in the {\bf T}-matrix defined
as ${\bf S}(k) = {\bf 1} - \i  {\bf T}(k)$.
In general, {\bf S} is non-diagonal and therefore
non-separable. An exception is the one-disk problem (see below).

\begin{figure}[htb]
 \centerline {\epsfig{file=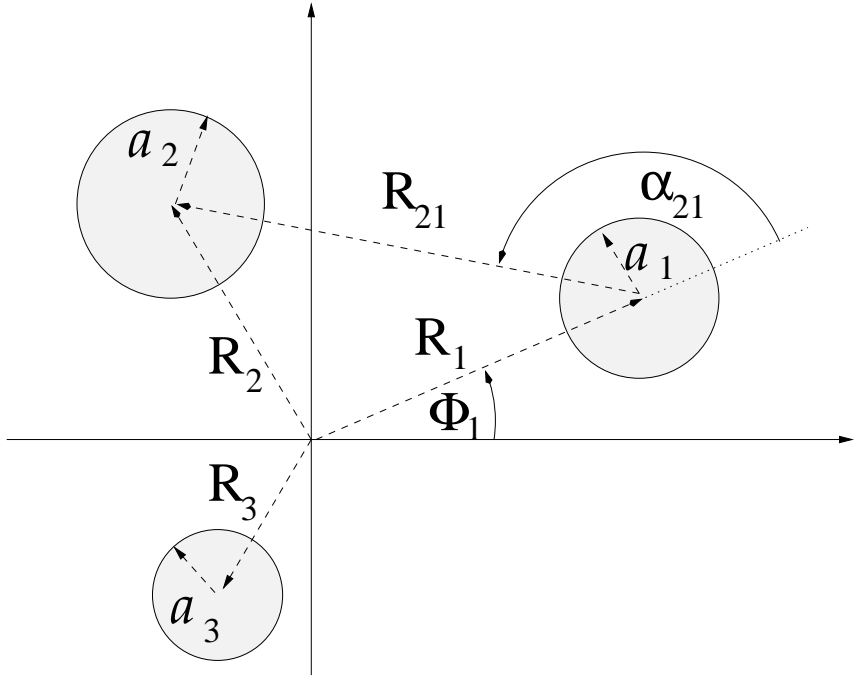,height=9cm,angle=-90}}
\caption[fig:coordinates]{\small 
Global and local coordinates for a general $3$-disk problem.
\label{fig:coordinates}}
\end{figure}

For any non-overlapping system of $n$ disks (of, in general, different
disk-radii $a_j$, $j=1,\dots,n$)
the $\bf S$-matrix can be further split up. Using 
the methods and notation of Gaspard and Rice \cite{gr} this is achieved
in the following way (see
also Ref.\cite{Lloyd_smith} and App.~\ref{app:construction} 
for a derivation of this result): 
\be 
 {\bf S}_{m m'}^{(n)}(k) 
 &=&  \delta_{mm'} - \i  {\bf T}_{mm'}^{(n)}(k) \nonumber \\
&=&
\delta_{m m'} - \i  
{\bf C}_{m l}^{\ \ j}(k) { \left ( {\bf M}^{-1}(k)\right
 ) }_{l l'}^{j j'} {\bf D}_{l'm'}^{j'}(k) \; .
\label{Smatrix} 
\ee 
Here the upper indices
$j,\,j'=1,\dots ,n <\infty$ label the $n$ different disks,
whereas the lower indices are the angular momentum quantum numbers. 
Repeated indices are
summed over.
The matrices ${\bf  C}^j$ and ${\bf D}^{j}$ 
depend on the origin and orientation of the global coordinate system
of the two-dimensional plane and are separable in the disk index $j$:
\be
 {\bf C}_{m l}^{\ \ j} &=&\frac{2\i }{\pi a_j}
    \frac{ J_{m-l}(k R_j)}{\Ho{l}{a_j}} 
   \e^{i m \Phi_{R_j}}\; , \label{Cmatrix} \\
 {\bf D}_{l'm'}^{j'} &=&- \pi a_{j'} J_{m'-l'}(k R_{j'}) J_{l'} (ka_{j'}) 
     \e^{-\i  m' \Phi_{R_{j'}}}\; , \label{Dmatrix} 
\ee
where $R_j$ and $\Phi_{R_j}$ are the distance and angle, respectively,
of the ray from the origin in the 2-dimensional plane to the center of
disk $j$ as measured in the global coordinate system 
(see Fig.\ref{fig:coordinates}).
$\Ho{l}{r}$  is the ordinary Hankel function of first kind and 
$\Jb{l}{r}$ the corresponding ordinary Bessel function. 
The  matrices ${\bf  C}^j$ and ${\bf D}^{j}$
parameterize the coupling of the incoming and outgoing scattering
waves, respectively, to the scattering interior {\em at} the
$j^{\,\rm th}$ disk.  Thus they describe only the single-disk aspects
of the scattering of a point particle from the $n$ disks.
The matrix ${\bf M}^{jj'}$ has the structure of a Kohn-Korringa-Rostoker 
(KKR)-matrix, see 
Refs.\cite{Berry_KKR,Berry_LH,Lloyd_smith}, 
\be                                         
    {\bf M}_{l l'}^{j j'} = \delta_{jj'} \delta_{l l'} +
     (1-\delta_{jj'})\, \frac{a_j}{a_{j'}}\, 
           \frac{ \Jb{l}{a_j} }{\Ho{l'}{a_{j'}} }\,
                      \Ho{l-l'}{R_{jj'}} \,
                        \Gamma_{jj'}(l,l') \; .
 \label{Mmatrix} 
\ee
without Ewald resummation~\cite{Berry_KKR},  as the number of disks is
finite.
%
Here $R_{jj'}$ is the separation between the centers of
the $j$th and $j'$th disk and $R_{jj'} = R_{j'j}$, of course. The auxiliary 
matrix
$\Gamma_{jj'}(l,l')=   \e^{\i ( l\alpha_{j' j}
          -l'(\alpha_{j j'} -\pi))}$ contains -- aside from a 
phase factor --
the angle 
$\alpha_{j'j}$ of the ray from the center of disk $j$ to
the center of disk $j'$ as measured in the local (body-fixed) coordinate
system of disk $j$ (see Fig.\ref{fig:coordinates}).

Note that $ \Gamma_{jj'}(l,l')  
= (-1)^{l-l'} (\, \Gamma_{j'j}(l',l) \,)^{\ast}$.
The ``Gaspard and Rice prefactors'' of $\bf M$, 
i.e., $(\pi a / 2 i)$ in \cite{gr},  are rescaled 
into $\bf C$ and $\bf D$.
The matrix ${\bf A}\equiv {\bf M} -{\bf 1}$ 
contains the genuine multidisk ``scattering'' 
aspects of the the $n$-disk problem,
e.g., in the pure
1-disk scattering case, $\bf A$ vanishes.
When  $({\bf M}^{-1})^{jj'}$ is expanded as a geometrical series 
about the unit matrix  $\{\delta^{jj'}\}$, a multiscattering series in
``powers''
of the matrix ${\bf   A}$ is created.

The product ${\bf C} {\bf M}^{-1} {\bf D}$ is the on-shell ${\bf T}$-matrix
of the $n$-disk system. It it the two-dimensional analog of the 
three-dimensional result of Lloyd and Smith 
for a 
finite cluster of
non-overlapping muffin-tin potentials. 
At first sight the expressions of Lloyd and
Smith (see
Eq.(98) of \cite{Lloyd_smith} and also Berry's 
form~\cite{Berry_KKR} for the infinite Sinai cluster) seem
to look simpler than ours and the original ones of 
Ref.\cite{gr}, 
as, e.g., in ${\bf M}$ the asymmetric term
$ a_j \Jb{l}{a_j}/ a_{j'} \Ho{l'}{a_{j'}}$ is replaced by a symmetric 
combination, $\Jb{l}{a_j}/\Ho{l}{a_j}$. 
Under a {\em formal} manipulation of
our matrices we can derive the same result (see App.~\ref{app:Lloyd}).
In fact, it can be checked that the (formal) 
cumulant expansion of Lloyd's and our
${\bf M}$-matrix are identical and that also numerically the determinants give
the same result. 
Note, however, that in Lloyd's case the trace-class property of ${\bf M}$ is
lost, 
such that the infinite determinant and the corresponding cumulant 
expansion converge only conditionally, and not absolutely as in our  
case.
The latter fact is based on the trace-class properties of the underlying
matrices and is an essential precondition for all further simplifications,
as e.g.\ unitary transformations, diagonalization of the matrices, etc.

A matrix is called ``trace-class'', if and only if, 
for any choice of the
orthonormal basis, the
sum of the diagonal matrix elements 
converges absolutely; it is called ``Hilbert-Schmidt'', if the
sum of the absolute squared diagonal matrix elements converges
(see M.~Reed and B.~Simon, Vol.1 and 4 
\cite{rs1,rs4} and App.~\ref{app:trace} 
for the definitions and properties of trace-class and Hilbert-Schmidt 
matrices). Here, we will only list the most important properties:
\begin{enumerate}
\item
Any trace-class matrix can be represented as the product of two
Hilbert-Schmidt matrices and any such product is again trace-class.
\item 
A matrix ${\bf B}$ is already 
Hilbert-Schmidt,
if the trace of ${\bf B}^\dagger {\bf B}$ is absolutely convergent in just 
one orthonormal basis.
\item
The  linear combination of a finite number of trace-class matrices is again
trace-class. 
\item
The hermitean-conjugate 
of a trace-class matrix is again trace-class.
\item
The product of two Hilbert-Schmidt
matrices or of a trace-class and a bounded matrix is trace-class and
commutes under the trace. 
\item
If a matrix ${\bf B}$ is trace-class, the trace
$\tr{}( {\bf B})$ is finite and independent of the basis.
\item 
If ${\bf B}$ is trace-class, the determinant
$\det{}({\bf 1}+z {\bf B})$ exists and is an entire function of $z$. 
\item 
If ${\bf B}$ is 
trace-class,
the determinant $\det{}({\bf 1}+z {\bf B})$ 
is invariant under any unitary transformation.
\end{enumerate}
In App.~\ref{app:suppl} we show explicitly that
the $l$-labelled matrices ${\bf S}^{(n)}(k)\mbox{$-$}{\bf 1}$, 
${\bf C}^{j}(k)$ and
${\bf D}^{j}(k)$ as well as the $ \{l,j\}$-labelled  matrix 
${\bf A}(k)={\bf M}(k)-{\bf 1}$
are of ``trace-class'', except at the countable
isolated zeros of $\Ho{m}{a_j}$ and  of $\Det{} {\bf M}(k)$ and
at $k \leq 0$, the branch cut 
of the Hankel functions.
The ordinary Hankel functions
have a branch cut at negative real $k$, such that even the $k$-plane is 
two-sheeted. The last property is special for even dimensions and does
not
hold in the 3-dimensional $n$-ball system~\cite{mh,N-ball}.
Therefore for almost all values of the wave number  $k$ (with the
above mentioned exceptions) the determinant of the $n$-disk ${\bf S}$-matrix
exist and the
operations of \equa{recoupl}
are mathematically well defined.
We concentrate  on the determinant, $\det{} {\bf S}$, 
of the scattering
matrix, since we are only interested in {\em spectral} properties of the
$n$-disk scattering problem, i.e. 
resonances and phase shifts, and not in wave functions.
Furthermore, the determinant  
is invariant under any change of a
complete basis expanding the $\bf S$-matrix and 
therefore also independent of the coordinate system.
\be
     \det{l} {\bf S}^{(n)} &=&  \det{l}  \left( {\bf 1} 
    - \i  {\bf C} {\bf M}^{-1} {\bf D} \right )
          = \exp\tr{l} \ln \left( {\bf 1} 
    - \i  {\bf C} {\bf M}^{-1} {\bf D} \right )
     \nno \\
           &=& \exp \left(- \sum_{N=1}^\infty \frac{\i ^N}{N} 
      \tr{l} \left[ \left( {\bf C} {\bf M}^{-1} {\bf D} \right)^N \right]  
              \right ) 
     \nno \\         
           &=& \exp \left(-  \sum_{N=1}^\infty \frac{\i ^N}{N} \Tr{L} 
 \left[  
    \left ({\bf M}^{-1} {\bf D} {\bf C} \right )^N \right ]  \right) 
   \nno \\
            &=& \exp \Tr{L} 
   \ln\left ({\bf 1}-\i  {\bf M}^{-1} {\bf D} {\bf C} \right )    
           = \Det{L} \left({\bf 1} -\i  
          {\bf M}^{-1} {\bf D} {\bf C} \right )
           \nno \\                                     
          &=& \Det{L} \left ({\bf M}^{-1} ({\bf M} -\i  {\bf D C})\right)
    \nno \\
          &=& \frac{ \Det{L} ({\bf M} - \i  
 {\bf DC})} { \Det{L} ({\bf M}) } \; .
     \label{recoupl}
\ee
We use here $\exp{{\rm tr} \ln}$ notation
as a compact 
abbreviation for the defining cumulant expansion \equa{det_def},
 since
$\det{}({\bf 1}+\mu{\bf A}) = \exp(- \sum_{N=1}^{\infty} \frac{(-\mu)^N}{N}
 \rm{tr}( {\bf A}^N))$,  
is only  valid for $|\mu| {\rm max} |\lambda_i| < 1$ 
where $\lambda_i$
is the $i$-th eigenvalue of ${\bf A}$. The determinant  is directly
defined by its cumulant expansion (see Eq.(188) of Ref.\cite{rs4} and 
Eq.\equa{det_def} of App.~\ref{app:trace}.2) which
is therefore the analytical continuation of the 
$\e^{{\rm tr} \log}$-representation.

The capital index
$L$ is a multi- or ``super''-index $L=(l,j)$.
On the l.h.s.\ of Eq.\equa{recoupl} 
the determinant and traces are only taken over small $l$,
on the r.h.s.\ they are taken over the super-indices $L=(l,j)$.
In order to signal this difference 
we will use
the following notation: $\det{}\dots$ and $\tr{}\dots$ refer to the 
$|m\rangle$ space, $\Det{}\dots$ and $\Tr{}\dots$ refer to the super-space. 
The matrices in the super-space are expanded in the complete basis 
$\{ |L\rangle\}  = \{|m;j\rangle \}$
which refers for fixed index $j$ to the origin of the $j$th disk 
and not any longer to the origin of the 
2-dimensional plane.
In deriving \equa{recoupl} the following facts were used: 
\begin{description}
\item[(a)]      
${\bf D}^j, {\bf C}^j$  
are of trace-class in the $\{|l\rangle\}$ space (see App.\,\ref{app:suppl}).
\item[(b)]
As long as the number of disks is finite,
the product
${\bf D C}$  -- now evaluated in the super-space $\{ |L\rangle \}$ --  
is of trace-class as well 
(see property (iii)).
\item[(c)]
${\bf M}-{\bf 1}$ is of trace-class (see App.~\ref{app:suppl}). Thus
the determinant 
${\rm Det}\, {\bf M}(k)$ exists.
\item[(d)] Furthermore,
${\bf M}$ is bounded (since it is the sum of a bounded 
and a trace-class matrix).
\item[(e)]
${\bf M}$ is invertible everywhere where $\Det{L} {\bf M}(k)$ is defined 
(which excludes a countable number of zeros of the Hankel 
functions $\Ho{m}{a_j}$ and the
negative real $k$-axis as there is a branch cut) 
and nonzero (which excludes a countable number
of isolated points in the lower 
$k$-plane) -- see property {(e)} of
App.~\ref{app:trace}.2. 
Therefore and because of {\bf (d)} the matrix
${\bf M}^{-1}$ is bounded.
\item[(f)\ ]
The matrices ${\bf C} {\bf M}^{-1} {\bf D}$, ${\bf M}^{-1} {\bf D C}$, 
are all of trace-class as they are the product of bounded times 
trace-class matrices and $\tr{m}[({\bf C} {\bf M}^{-1} {\bf D})^N]=
\Tr{M}[( {\bf M}^{-1} {\bf D C})^N ]$, because such products have
the cyclic permutation property under the trace  (see properties (iii)  and
(v)).
\item[(g)\ ]
 ${\bf M}-\i {\bf DC}-{\bf 1}$ is of  trace-class because of the rule 
that the sum of two trace-class matrices is
again trace-class (see property (iii)).
\end{description}
Thus all traces and determinants appearing in Eq.\equa{recoupl} 
are well-defined, except at the above mentioned isolated $k$ singularities
and branch cuts. 
In the $\{|m;j\rangle\}$ basis the trace of ${\bf
M}-{\bf 1}$ vanishes trivially because of the $\delta_{jj'}$ terms in
\equa{Mmatrix}. One should not infer from this that the trace-class property
of ${\bf M}-{\bf 1}$ is established by this fact, since
the finiteness (here vanishing) of ${\rm Tr} ({\bf
M}-{\bf 1} )$ has to be shown for every complete orthonormal
basis. After symmetry reduction (see below) ${\rm Tr} ({\bf
M}-{\bf 1} )$, calculated for each irreducible
representation separately, does not vanish any longer. However, the sum of the
traces of all irreducible representations weighted with their pertinent
degeneracies still vanishes of course. Semiclassically, this corresponds
to the fact that only in the fundamental domain there can exist 
one-letter ``symbolic words''.

After these manipulations,
the computation of the determinant of the {\bf S}-matrix is  very much
simplified
in comparison to the original formulation,
since the last term of Eq.\equa{recoupl} is completely written 
in terms of closed form
expressions and since the matrix ${\bf M}$ does not have to be inverted 
any longer.
Furthermore, as shown in App.~\ref{app:construction}.3, one can easily
construct
\be 
 {\bf M}_{l l'}^{j j'} -\i  {\bf D}_{l m'}^{j} {\bf C}_{m' l'}^{\ \ \ j'} 
 &=& \delta_{jj'} \delta_{ll'} \left (-
 \frac{\Ht{l'}{a_{j'}}} {\Ho{l'}{a_{j'}}} \right ) \nno \\ && \mbox{}
 -(1-\delta_{jj'})\, \frac{a_j}{a_{j'}}\,
 \frac{\Jb{l}{a_j}}{\Ho{l'}{a_{j'}}} \, \Ht{l-l'}{R_{jj'}} \,
 \Gamma_{jj'}(l,l') \; ,
\label{rewri} 
\ee 
where $\Ht{m}{r}$ is the
Hankel function of second kind.  
The first term on the r.h.s\ is just the {\bf S}-matrix 
for the separable scattering problem 
from a single disk, if the
origin of the coordinate system is at the center of the disk (see
App.~\ref{app:construction}.2):
\be
 {\bf S}^{(1)}_{ll'} (ka) = -\frac{\Ht{l'}{a}} {\Ho{l'}{a}} \, 
 \delta_{ll'} \; .\label{S1disk} 
\ee 
After \equa{rewri} is inserted into \equa{recoupl} and 
\equa{S1disk}  is factorized out,
the r.h.s.\ of \equa{recoupl} can be rewritten as
\beq 
 \det{l} {\bf S}^{(n)}(k) 
 = \frac{ \Det{L} [{\bf M}(k) - \i  {\bf
 D}(k){\bf C(k)}] } {\Det{L} {\bf M}(k)}  
 = \left \{
 \prod_{j=1}^{n} \left( \det{l} {\bf S}^{(1)}(k a_j) \right)
 \right \}\, \frac{\Det{L}{ {\bf M}(k^\ast) }^\dagger}{\Det{L} {\bf
 M}(k)} \label{qm}
\eeq 
where $\{ H^{(2)}_m(z)\}^{\ast} =H^{(1)}_m(z^\ast)$ has been used in the end.
All these operations are allowed, since
${\bf M}(k)-{\bf 1}$, ${\bf M}(k) - \i  {\bf
D}(k){\bf C}(k)-{\bf 1}$ and ${\bf S}^{(1)}(k) -{\bf 1}$ are trace-class
for almost every $k$ with the above mentioned exceptions. In addition,   
the zeros of the Hankel functions $H^{(2)}_m (ka_j)$ now have to
be excluded as well.
In general, the single disks
have different sizes and the corresponding 1-disk {\bf S}-matrices
should be distinguished by the index $j$. At the level of the 
determinants this labelling is taken
care of by the choice of the argument $k a_j$.
Note
that the analogous formula for the three-dimensional scattering 
of a point particle from $n$
non-overlapping  spheres (of in general different sizes) is structurally
completely the same~\cite{mh,N-ball}, except that there is no need to
exclude the negative $k$-axis any longer, since 
the {\em spherical} Hankel
functions do not posses a  branch cut.
In the above calculation it was used that
$\Gamma_{jj'}^\ast(l,l') = \Gamma_{jj'}(-l,-l')$ in general \cite{mh} and
that for symmetric systems (equilateral 3-disk-system with identical
disks, 2-disk system with identical disks): $\Gamma_{jj'}^\ast (l,l') =
\Gamma_{j'j}(l,l')$ (see \cite{gr}).  Eq.\equa{qm} is compatible with
Lloyd's {\em formal} separation of the single scattering properties  
from the multiple-scattering effects in the Krein-Friedel-Lloyd sum, see,
e.g., p.102 of Ref.\cite{Lloyd_smith} (modulo the above-mentioned conditional
convergence problems of the Lloyd formulation). 
Eq.\equa{qm} has the following properties:\\
(i)
Under the determinant of the n-disk ${\bf S}^{(n)}$-matrix, 
the 1-disk aspects separate from the multiscattering aspects, since
the determinants of the 1-disk ${\bf S}^{(1)}$ matrices factorize from
the determinants of the multiscattering matrices. 
Thus the product over the $n$ 1-disk determinants in \equa{qm} 
parametrizes 
the incoherent part of the scattering, as if the $n$-disk problem just
consisted of $n$ separate single-disk problems.\\ 
(ii)
The whole expression \equa{qm} respects unitarity as ${\bf
S}^{(1)}$ is unitary by itself, because of   $ ( H^{(2)}_{m} (z) )^\ast =
H^{(1)}_{m}(z^\ast)$ and as the quotient of the determinants of the
multiscattering matrices on 
the r.h.s.\ of \equa{qm} is manifestly unitary.\\
(iii)
The determinants over the multiscattering matrices
run over the 
super-index $L$ of the super-space.
This is the proper form for the symmetry reduction (in the super-space),
e.g., for the equilateral 3-disk system (with disks of the same size) we have
\be
    \Det{L} {\bf M}_{\rm 3\mbox{-}disk} 
    = \det{l_{{\rm A}_{\rm 1}}} {\bf M}_{{\rm A}_{\rm 1}}\, \det{l_{{\rm A}_{\rm 2}}} {\bf M}_{{\rm A}_{\rm 2}} 
          \,\left ( \det{l_{\rm E}} {\bf M}_{{\rm E}} \right )^2 \ , 
\label{3disk}
\ee
and    for the 2-disk system (with disks of the same size)
\be
    \Det{L} {\bf M}_{\rm 2\mbox{-}disk} = 
   \det{l_{{\rm A}_{\rm 1}}} {\bf M}_{{\rm A}_{\rm 1}}\, \det{l_{{\rm A}_{\rm 2}}} {\bf M}_{{\rm A}_{\rm 2}} 
     \, \det{l_{{\rm B}_{\rm 1}}} {\bf M}_{{\rm B}_{\rm 1}} \,\det{l_{{\rm B}_{\rm 2}}} {\bf M}_{{\rm B}_{\rm 2}} \; ,
 \label{2disk}
\ee  etc.
In general, if the disk configuration is characterized by a finite 
point-symmetry group ${\cal G}$, we have
\be
  \Det{L} {\bf M}_{n{\rm\mbox{-}disk}}
 = \prod_c \left ( \det{l_c} {\bf M}_{{\rm D}_c}(k) \right )^{d_c} \; ,
  \label{n_rep_disk} 
\ee
where the index $c$ runs over all conjugate classes of the symmetry group
${\cal G}$ and ${\rm D}_c$ is the $c^{\,\rm th}$ representation of dimension 
$d_c$~\cite{mh}. For the symmetric 2-disk system, these representations
are the totally symmetric ${\rm A}_{\rm 1}$, 
the totally anti-symmetric ${\rm A}_{\rm 2}$, and the
two mixed representations ${\rm B}_{\rm 1}$ and 
${\rm B}_{\rm 2}$ which are all one-dimensional.
For the symmetric equi\-triangular 3-disk system, there exist two 
one-dimensional
representations (the totally symmetric ${\rm A}_{\rm 1}$ 
and the totally anti-symmetric
${\rm A}_{\rm 2}$) and one two-dimensional representation labelled by 
${\rm E}$.
A simple check that $\Det{}{\bf M}(k)$ has been split up 
correctly is the following: the power of $\Ho{m}{a_j}$ Hankel functions 
(for fixed $m$ with 
$-\infty <m <+\infty$) in the
denominator of  $\prod_c \left [ \det{l_c} {\bf M}_{{\rm D}_c}(k) 
\right ]^{d_c}$ 
has to agree with the power of the same functions 
in $\Det{} {\bf M}(k)$ which in turn has to
be the same as in $\prod_{j=1}^n \left (\det{} {\bf S}^{(1)}(ka_j)\right)$.
Note that 
on the l.h.s.\ the determinants are calculated in the super-space $\{L\}$,
whereas on the r.h.s.\ the reduced determinants are calculated, if 
none of the disks are special in size and position,
in the normal (desymmetrized) space $\{l\}$ (however, now 
with respect to the origin of the disk in
the fundamental domain and with ranges given by the corresponding
irreducible representations). If the $n$-disk 
system has a point-symmetry where still some disks 
are special in size or position
(e.g., three equal disks in a row~\cite{WR96}), 
the determinants on the r.h.s.\
refer to a correspondingly symmetry-reduced super-space.
This summarizes the symmetry reduction on the exact quantum-mechanical level.
It can be derived from
\be
   \Det{L} {\bf M} 
 &=& \exp\left ( - \sum_{N=1}^\infty \frac{(-1)^N}{N} 
  \Tr{L}\left[ {\bf A}^N \right ] 
           \right ) \nno \\
 &=& \exp\left ( - \sum_{N=1}^\infty \frac{(-1)^N}{N} \Tr{L}
      \left[{\bf U} {\bf A}^N {\bf U}^\dagger \right ] 
           \right ) \nno \\ 
 &=& \exp\left ( - \sum_{N=1}^\infty \frac{(-1)^N}{N} \Tr{L}
      \left[\left ( {\bf U} {\bf A} {\bf U}^\dagger \right )^N\right ] 
           \right ) \nno \\
 &=& \exp\left ( - \sum_{N=1}^\infty \frac{(-1)^N}{N} \Tr{L}
      \left[{\bf A}^N_{\rm block}\right ] 
           \right ) \; ,
\ee
where ${\bf U}$ is unitary transformation which makes ${\bf A}$ 
block-diagonal in a
suitable transformed basis of the original complete set $\{|m;j\rangle\}$.
These operations are allowed because of the trace-class-property of 
${\bf A}$ and the boundedness of the unitary matrix ${\bf U}$
(see also property {(d)} of App.~\ref{app:trace}.2). 
\newpage
\section{The link between the determinant of the {\bf S}-matrix
 and the semiclassical zeta function\label{chap:link}}
\setcounter{equation}{0}
\setcounter{figure}{0}
\setcounter{table}{0}

In this chapter we will specify the semiclassical equivalent of the
determinant of the $n$-disk {\bf S}-matrix. As $\det{} {\bf S}^{(n)}$
in \equa{qm}
factorizes into a product of the 1-disk determinants and the ratio 
of the determinants
of the multiscattering matrix,
$\Det{} {\bf M}(k^\ast)^\dagger/ \Det{} {\bf M}(k)$,
the semiclassical reduction will factorize
as well into incoherent one-disk parts and an coherent multiscattering
part. Note, however, that there is an implicit connection between these
parts via the removable one-disk poles and zeros. This will be discussed
in  the conclusion section~\ref{chap:end}.

\begin{figure}[htb]
 \centerline 
{\epsfig{file=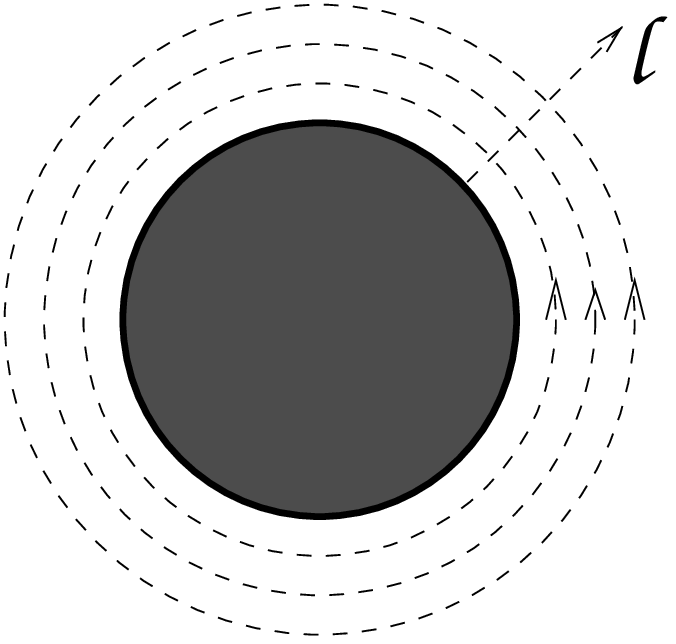,height=4cm,angle=-90}
\hspace{3cm}
\epsfig{file=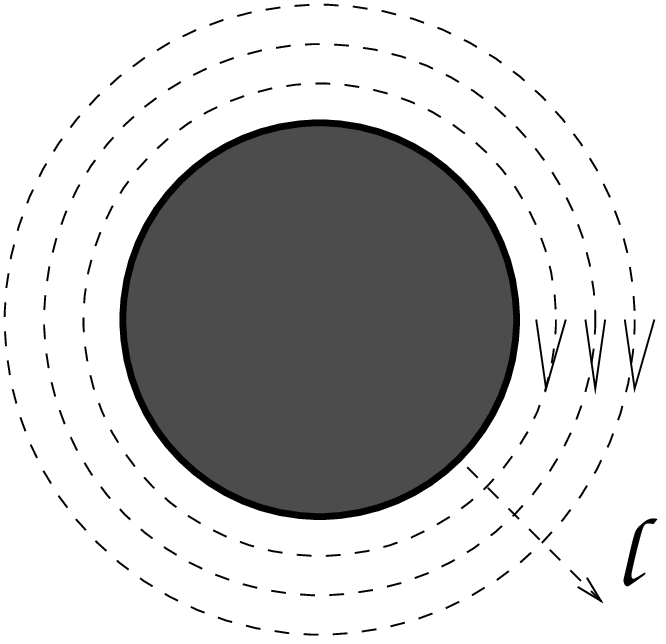,height=4cm,angle=-90}}
\caption[fig:1-disk-turns]{\small 
Left- and right-handed diffractive creeping paths of increasing
mode number $\ell$ for a single disk.
\label{fig:1-disk-turns}}
\end{figure}

In App.~\ref{app:semi1disk}, 
the  semiclassical expression for
the determinant of the 1-disk {\bf S}-matrix is constructed in analogous 
fashion to the semiclassical constructions of Ref.\cite{aw_chaos} 
which in turn is based on the work of Ref.\cite{franz}:
\be
  \det{l} {\bf S}^{(1)}(ka) \approx  
               \left( \e^{-\i  \pi N(ka)} \right)^2 
         \frac
         { \left ( \prod_{\ell=1}^{\infty} 
          \left[1 - \e^{-\i  
            2\pi {\bar \nu}_\ell(ka)}\right] \right)^2 } 
         { \left ( \prod_{\ell=1}^{\infty} 
          \left[1 - \e^{+\i  2\pi       
           \nu_\ell(ka)}\right] \right)^2 }
  \label{sc1disk} 
\ee
with the creeping exponential (for more details, see 
App.~\ref{app:semi1disk} and the definitions of App.~\ref{app:convol_residua})
\be
    \nu_\ell (ka) &=& ka + \e^{+\i \pi/3} (ka/6)^{1/3}q_\ell+\cdots 
  = ka +\i  \alpha_\ell (ka) +\cdots\; ,
  \label{nu_k} \\
  {\bar \nu}_\ell (ka) &=& ka + \e^{-\i \pi/3} (ka/6)^{1/3}q_\ell 
     +\cdots 
 = ka -\i  
( \alpha_\ell (k^\ast a))^\ast +\cdots \nno \\
 &=& \left(\nu_\ell(k^\ast a)\right )^\ast \; , 
  \label{bar_nu_k}
\ee
and $N(ka)= (\pi a^2 k^2)/4\pi + \cdots$ 
the leading term in the Weyl approximation for the staircase function of the 
wave-number eigenvalues in the disk interior. From the point of view of
the scattering particle
the interior domains of the disks are excluded relatively to the
free evolution without scattering obstacles (see, e.g., \cite{scherer}).
Therefore the negative sign in front of the Weyl term. For the same reason,  
the subleading boundary term  has here a Neumann structure, 
although the disks have Dirichlet boundary conditions. 
Lets us abbreviate the r.h.s.\ of \equa{sc1disk} for a specified disk $j$ as
\be
 \det{l}{\bf S}^{(1)}(ka_j) \semiclass \left ( \e^{-\i 
 \pi N(ka_j)} \right )^2
      \, \frac{ { {\widetilde Z}_{\rm 1\mbox{-}disk(l)} (k^\ast a_j) }^\ast}
            {{\widetilde Z}_{\rm 1\mbox{-}disk(l)} (ka_j)}\, 
\frac{ { {\widetilde Z}_{\rm 1\mbox{-}disk(r)} (k^\ast a_j) }^\ast}
            {{\widetilde Z}_{\rm 1\mbox{-}disk(r)} (k a_j)}\ ,
\ee
where ${\widetilde Z}_{\rm 1\mbox{-}disk(l)}(ka_j)$ and $
{\widetilde Z}_{\rm 1\mbox{-}disk(r)}(k a_j)$
are the {\em diffractional} zeta functions (here and in the following
we will label semiclassical zeta-functions {\em with} diffractive corrections
by a tilde)
for creeping
orbits around the $j$th disk  in the left-handed 
sense and the right-handed sense,
respectively (see Fig.\ref{fig:1-disk-turns}). 
The two orientations of the creeping orbits are the reason
for the exponents two in \equa{sc1disk}.
Eq.\equa{sc1disk} describes the semiclassical approximation to the
incoherent part (= the curly bracket on the r.h.s.) of the exact expression
\equa{qm}. 

We now turn  to
the semiclassical approximation of the coherent part of \equa{qm}, namely
the ratio of the determinants 
of the multiscattering
matrix {\bf M}.
Because of the trace-class property of ${\bf A} = {\bf M}-{\bf 1}$, the
determinants in the numerator and denominator of this ratio exist 
individually and their semiclassical approximations can be studied separately. 
In fact, because of $\Det{} {\bf M}(k^\ast)^\dagger=
( \Det{} {\bf M}(k^\ast) )^\ast$, the semiclassical reduction of
 $\Det{} {\bf M}(k^\ast)^\dagger$
follows
directly from the corresponding result  of $\Det{} {\bf M}(k)$ under
complex conjugation. 
The semiclassical  reduction of  $\Det{} {\bf M}(k)$
will be done in the cumulant
expansion, since the latter is the defining prescription for the computation
of an infinite matrix that is of the form ${\bf 1}+ {\bf A}$ where ${\bf A}$
is trace-class: 
\be
 \Det{}\left[ {\bf 1}+z {\bf A}(k)\right ] &=& 1 - (-z) \Tr{}[ {\bf A}(k)]
 - \frac{z^2}{2} \left\{ \Tr{}\left[ {\bf A}^2(k)\right]-\left[\Tr{} 
{\bf A}(k)\right ]^2 \right\} + \cdots \nonumber \\
&=& \sum_{n_c=0}^{\infty} z^{n_c} Q_{n_c} ({\bf A}) \qquad 
\mbox{with}\ Q_0({\bf A})\equiv 1
\ee
where we have introduced here a
book-keeping variable $z$ which we will finally set to one.
This allows us to express the determinant of the 
multiscattering matrix solely by
the traces of the matrix ${\bf A}$, 
$\Tr{}[ {\bf A}^m (k)]$ with $m=1,2,3,\cdots$. The cumulants and traces
satisfy the 
(Plemelj-Smithies) recursion relations~\equa{app:ps_recursion}
\be
Q_{n_c}({\bf A}) = 
 \frac{1}{n_c} \sum_{m=1}^{n_c} (-1)^{m+1} Q_{n_c-m}({\bf A}) 
               \Tr{}\left[{\bf A}^m\right] \qquad \mbox{for}\ n_c \geq 1 
 \label{PS-recursion}
\ee
in terms of the traces.
In the next section we will utilize Watson resummation
techniques~\cite{Watson,franz} which help to replace the {\em angular momentum
sums} of the traces by continuous integrals which, in turn, allow for
semiclassical saddle-point approximations. With these techniques 
and under 
complete induction
we will show  that for any geometry of $n$ disks, as long as the number
of disks is finite, the disks do not overlap and grazing or penumbra
situations~\cite{Nussenzveig,penumbra} 
are excluded (in order to guarantee unique isolated saddles),
the semiclassical reduction reads as follows:
\beq
  \Tr{}\left[ {\bf A}^m(k) \right ] \semiclass (-1)^m \sum_p \sum_{r>0}
 \delta_{m,r n_p} n_p \frac{ {t_p(k)}^r}{1 
 -\left (\frac{1}{\Lambda_p}\right)^r} + \mbox{creeping p.o.'s} \; 
  \label{sc-reduction-step}
\eeq
with inputs as defined below \equa{qcl_zeta}.
The reduction is of course only valid, if 
${\rm Re}\, k$ is sufficiently  large compared 
to the inverse of the smallest
length scale of the problem. 
The right hand side of 
Eq.\equa{sc-reduction-step} can  be inserted into the recursion relation
\equa{PS-recursion} which then reduces to 
a recursion relation for the semiclassical approximations of the
quantum cumulants 
\beq
C_{n_c}({\rm s.c.}) = -\frac{1}{n_c} \sum_{m=1}^{n_c} 
C_{n_c-m}({\rm s.c.}) \sum_p \sum_{r>0}
 \delta_{m,r n_p} n_p \frac{ {t_p(k)}^r}{1 
 -\left (\frac{1}{\Lambda_p}\right)^r} \quad \mbox{for}\ n_c \geq 1
 \label{appr-PS-recursion}
\eeq
where we have neglected the creeping orbits for the time being.
Under the assumption 
that the semiclassical limit ${\rm Re}\, k \to \infty$
and the cumulant limit $n_c \to \infty$ commute (which might be
problematic as we will discuss later), the approximate cumulants 
$C_{n_c}({\rm s.c.})$
can be summed to infinity, $\sum_{n_c=0}^{\infty} z^{n_c}C_{n_c}({\rm s.c.})$,
in analogy to the exact cumulant sum. The latter 
exists since ${\bf A}$ is trace-class.
The infinite ``approximate cumulant sum'',  however,
is nothing but the curvature expansion of the Gutzwiller-Voros zeta
function, i.e., 
\beq
 Z_{\rm GV}(z;k)|_{\rm curv.\,reg.}  
 = \sum_{n_c=0}^{\infty} z^{n_c} C_{n_c}({\rm s.c.}) \; ,
      \label{zgv_z_expansion}
\eeq
since Eq.\equa{appr-PS-recursion} 
is exactly the recursion relation of the semiclassical
curvature terms~\cite{QCcourse}.

If, in addition, the creeping periodic orbits  are summed as well, 
the standard Gutzwiller-Voros zeta function generalizes to the diffractive 
one discussed in Refs.\cite{vwr_prl,vwr_japan,vwr_stat} which we will denote
here by a tilde.
In summary, we have
\be
    \Det{}{\bf M}(k) \semiclass  
 {\widetilde Z}_{\rm GV} (k)|_{\rm curv.\,reg.} 
\ee
for a general geometry
and
\be
    \det{} {\bf M}_{{\rm D}_c}(k) \semiclass  
{\widetilde Z}_{{\rm D}_c} (k)|_{\rm curv.\,reg.} 
\ee      
for the case that there is a finite point-symmetry and the determinant
of the multiscattering matrix splits into the product of 
determinants of matrices belonging to the pertinent representations 
${\rm D}_c$,
see Eq.\equa{n_rep_disk}.
Thus the semiclassical limit of the r.h.s.\ of Eq.\equa{qm}
is 
\be
  \det{l} {\bf S}^{(n)}(k) 
       &=& \left\{ \prod_{j=1}^{n} 
 \det{l} {\bf S}^{(1)}(ka_j) \right \} \,
            \frac{\Det{L} { {\bf M}(k^\ast)}^\dagger}
{\Det{L} {\bf M}(k)} 
           \nno \\
  &\semiclass &
       \left\{ \prod_{j=1}^n \left ( \e^{-\i  
\pi N(ka_j )} \right )^{2}
      \, \frac{ {  {\widetilde Z}_{ {\rm 1\mbox{-}disk(l)} }
      (k^\ast a_j) }^\ast }
            {{\widetilde Z}_{ {\rm 1\mbox{-}disk(l)}} (k a_j)}\, 
      \, \frac{ {  {\widetilde Z}_{ {\rm 1\mbox{-}disk(r)}}
             (k^\ast a_j)  }^\ast }
            {{\widetilde Z}_{ {\rm 1\mbox{-}disk(r)}} (k a_j)}
        \, \right \}\,
            \frac{ { {\widetilde Z}_{\rm GV}(k^\ast) }^\ast}
  {{\widetilde Z}_{\rm GV}(k)}\; ,  \nno \\
    \label{gen}
\ee
where, from now on, we will suppress the qualifier 
$\cdots|_{\rm curv.\,reg.}$.
For systems which allow for complete symmetry 
reductions (i.e., equivalent disks under a finite point-symmetry 
with $a_j =a\ \forall j$) the link reads
\be
  \det{l} {\bf S}^{(n)}(k) 
   &=& \left (  \det{l}{\bf S}^{(1)} (ka)\right )^n 
     \frac{ \prod_c \left ( \det{l_c} { {\bf M}_{{\rm D}_c} (k^\ast) }^\dagger
 \right )^{d_c}}
 { \prod_c \left ( \det{l_c} { {\bf M}_{{\rm D}_c} (k) }
 \right )^{d_c}}
   \nno \\
  &\semiclass &
         \left ( \e^{-\i  \pi N(ka)} \right )^{2n}
      \, \left (
\frac{  {{\widetilde Z}_{\rm 1\mbox{-}disk(l) }(k^\ast a)}^\ast}
            {{{\widetilde Z}_{\rm 1\mbox{-}disk(l)} (ka)}}\, 
        \frac{{ {\widetilde Z}_{\rm 1\mbox{-}disk(r)}
              (k^\ast a)}^\ast}
            {{{\widetilde Z}_{\rm 1\mbox{-}disk(r)} (ka)}} \right )^n
 \times \nno \\
  && \qquad \qquad \qquad\times
            \frac{ \prod_c \left ({ {\widetilde Z}_{{\rm D}_c} (k^\ast)}^\ast 
          \right)^{d_c}} 
              {\prod_c \left ( {{\widetilde Z}_{{\rm D}_c}(k)} \right )^{d_c}}
  \label{gen_sym_full}
\ee
in obvious correspondence. 
Note that the symmetry reduction from the right hand side of \equa{gen} to
the right hand side of \equa{gen_sym_full} is compatible with the
semiclassical results of 
Refs.\cite{Lauritzen,cvi_eck_93}.

In the next section we will prove the semiclassical 
reduction step \equa{sc-reduction-step} 
for any $n$-disk scattering system under the conditions that
the number of
disks is finite, the disks do not overlap, and geometries with grazing
periodic orbits are excluded.
We will also derive the general expression for creeping periodic orbits
for $n$-disk repellers from exact quantum mechanics and show that ghost
orbits drop out of the expansion of $\Tr{}\ln({\bf 1}+{\bf A})$
and therefore out of the cumulant  expansion.

%
%
\newpage
\section{Semiclassical 
approximation and periodic orbits\label{chap:semiclass}}
\setcounter{equation}{0}
\setcounter{figure}{0}
\setcounter{table}{0}
In  this section we will work out the semiclassical reduction of
$\Tr{}[\, {\bf A}^m (k)\,]$ for non-overlapping, finite $n$-disk systems 
where
\beq
{\bf A}^{j j'}_{l l'} = (1- \delta_{jj'}) 
                  \frac{ a_j   \Jb{l }{a_j   } }
                       { a_{j'}\Ho{l'}{a_{j''}} }
                  (-1)^{l'} 
                  \e^{\i ( l \alpha_{j' j} - l' \alpha_{j j'} ) } 
                  \Ho{l-l'}{R_{jj'}} \; . 
 \label{A-kernel-ansatz}
\eeq
As usual, $a_j$, $a_{j'}$ are the radii of disk $j$ and $j'$, 
$1\leq j,j'\leq n$, 
$R_{j j'}$ is the distance between the centers of these disks, and
$  \alpha_{j' j}$ is the angle of the ray from the origin of 
disk $j$ to the one of disk $j'$ as
measured in the local coordinate system of disk $j$. The  angular momentum
quantum numbers $l$ and $l'$  can be interpreted geometrically in terms of
the positive-- or negative-valued distances 
(impact parameters) $l/k$ and $l'/k$ from the center 
of disk $j$ and disk $j'$, respectively, see \cite{Berry_KKR}.

Because of the finite set of $n$ disk-labels and because of the 
cyclic nature of the trace,  the object $\Tr{}[\, {\bf A}^m (k)\, ]$ contains
all periodic itineraries of total symbol length $m$ 
with an alphabet of $n$ symbols, i.e.\,
${\bf A}^{j_1 j_2}{\bf A}^{j_2 j_3} \cdots  {\bf A}^{j_{m-1} j_m} 
{\bf A}^{j_m j_1}$ with $j_i\in \{1,2,\dots,n\}$.
Here the disk indices are not summed over and the 
angular momentum quantum numbers are suppressed for simplicity.
The delta-function part $(1-\delta_{jj'})$ 
generates the trivial pruning rule (valid for the full $n$-disk domain) 
that successive symbols
have to be different.  We will show that these periodic 
itineraries correspond in the
semiclassical limit, $ka_{j_i}\gg 1$, to {\em geometrical} 
periodic orbits with the same symbolic dynamics. For periodic orbits with
creeping sections~\cite{aw_chaos,aw_nucl,vwr_prl,vwr_japan,vwr_stat} 
the symbolic alphabet has to be extended. 
Furthermore, depending on the geometry, there might be 
non-trivial pruning rules based on the so-called ghost 
orbits, see Refs.\cite{bb_2,Berry_KKR}. We will discuss such cases in
Sec.\ref{chap:sc_ghosts}.

\subsection{Quantum itineraries\label{chap:sc_qu_itineraries}}   
As mentioned, 
the quantum-mechanical trace can be structured
by a simple 
symbolic dynamics, where the sole (trivial)  pruning rule is automatically  
taken care of by the
$1-\delta_{jj'}$ factor appearing in ${\bf A}^{jj'}_{ll'}$.
Thus we only have to consider the semiclassical
approximation of a quantum-mechanical {\em itinerary} of length $m$:
\be
\lefteqn{ 
   {\bf A}^{j_1 j_2}
   {\bf A}^{j_2 j_3} 
   \cdots  
   {\bf A}^{j_{m-1} j_m}
   {\bf A}^{j_{m} j_1}  \nonumber} \\
&:=&
\sum_{l_1=-\infty}^{+\infty} 
 \sum_{l_2=-\infty}^{+\infty} 
 \sum_{l_3=-\infty}^{+\infty} 
    \cdots 
  \sum_{l_{m-1}=-\infty}^{+\infty}\,
  \sum_{l_m=-\infty}^{+\infty}\,
   {\bf A}^{j_1 j_2}_{l_1 l_2}
   {\bf A}^{j_2 j_3}_{l_2 l_3} 
   \cdots  
   {\bf A}^{j_{m-1} j_m}_{l_{m-1} l_m} 
   {\bf A}^{j_m j_1}_{l_m l_1} 
  \label{itinerary}
\ee
with $j_i\in \{1,2,\dots,n\}$. This is still a trace in the angular momentum
space, but not any longer with respect to the superspace.
Since 
the  trace, $\Tr{}{\bf A}^m$, itself is simply the sum of all itineraries of
length $m$, i.e.
\be
 \Tr{}{\bf A}^m = \sum_{j_1=1}^n 
                  \sum_{j_2=1}^n
                   \sum_{j_3=1}^n
                   \cdots
                  \sum_{j_{m-1}=1}^n
                  \sum_{j_{m}=1}^n \,
                 {\bf A}^{j_1 j_2}
                 {\bf A}^{j_2 j_3} 
                 \cdots  
                 {\bf A}^{j_{m-1} j_m}
                {\bf A}^{j_{m} j_1} \; , 
\ee      
its semiclassical approximation follows directly from
the semiclassical approximation of its itineraries.
Note that we here distinguish between 
a given itinerary and its cyclic permutation. All of them
give the same result, such that their contributions can finally 
be summed up
by an integer-valued factor $n_p:=m/r$, where 
the integer $r$ counts the number of repeated
periodic subitineraries.
Because of the pruning rule $1-\delta_{jj'}$, we only have to consider traces
and itineraries with $n\geq 2$ as ${\bf A}^{j j}_{l l'}=0$ implies that
$\Tr{} {\bf A} = 0$ in the full domain.

We will show in this section that, with the help of the Watson 
method~\cite{Watson,franz} 
(studied for the convolution of two ${\bf A}$ matrices 
in App.\ref{app:convol} which should be consulted
for details), 
the 
semiclassical approximation of the periodic itinerary 
\[
   {\bf A}^{j_1 j_2}
   {\bf A}^{j_2 j_3} 
   \cdots  
   {\bf A}^{j_{m-1} j_m}
   {\bf A}^{j_{m} j_1} 
\]
becomes a standard periodic orbit labelled by the
symbol sequence $j_1 j_2 \cdots  j_{m}$.
Depending on the geometry, the individual legs $j_{i-1}\to j_{i}\to j_{i+1}$ 
result either from a standard specular reflection at disk $j_i$ or from  a 
ghost path passing straight through disk $j_i$.
If furthermore creeping contributions are taken into account,
the symbolic dynamics has to be generalized from single-letter 
symbols $\{j_i\}$ to 
triple-letter symbols 
$\{j_i, s_i\times \ell_i\}$ with $\ell_i\geq 1$ integer-valued and 
$s_i=0,\pm 1$~\footnote{Actually, these are double-letter symbols as 
$s_i$ and  $l_i$ are only counted as a product.}  
By definition, the value $s_i=0$  represents the non-creeping case, such that
$\{j_i,0\times \ell_i\}=\{j_i,0\} = \{j_i\}$ reduces to the 
old single-letter symbol.
The magnitude
of a non-zero $\ell_i$ 
corresponds to  creeping sections of mode number $|\ell_i|$, whereas
the sign $s_i=\pm 1$ signals whether the creeping path turns around the
disk $j_i$ in the positive or negative sense. Additional full
creeping turns around a disk $j'$ can be summed up as a geometrical series;
therefore they do not lead to the introduction of  a further symbol.

\subsection{Ghost contributions\label{chap:sc_ghosts}}
\begin{figure}[htb]
 \centerline {\epsfig{file=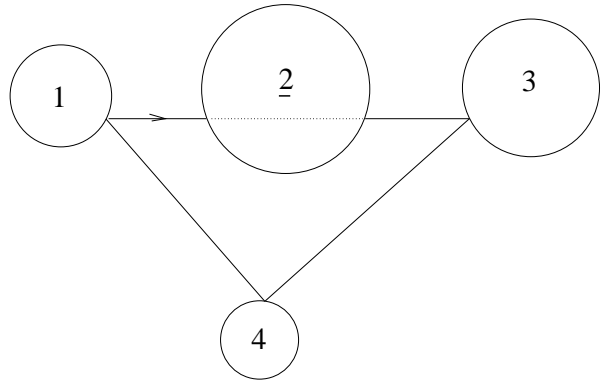,height=5cm,angle=-90}}
\caption[fig:ghost_itin]{\small 
The ghost itinerary $(1,\underline{2},3,4)$.
\label{fig:ghost_itin}}
\end{figure}
An itinerary with a semiclassical ghost section at, say, disk $j_i$ 
will be shown to have the same weight 
as the corresponding itinerary without the
$j_i$\,th symbol. Thus, semiclassically, they cancel each other  in the 
$\Tr{}\ln({\bf 1} +{\bf A})$ expansion, where they are 
multiplied by the permutation factor
$m/r$ with the integer $r$ counting 
the repeats. E.g.\ let  $(1,\underline{2},3,4)$ be
a non-repeated periodic itinerary with a ghost section at disk 2 
steming 
from the 4th-order trace $\Tr{}A^4$, where the convention is introduced 
that an underlined disk index signals a ghost passage (see 
Fig.\ref{fig:ghost_itin}).
Then its semiclassical, geometrical
contribution to $\Tr{}\ln({\bf 1} +{\bf A})$ 
cancels exactly against the one of
its ``parent'' itinerary $(1,3,4)$ (see Fig.\ref{fig:parent}) resulting from the 3rd-order trace:
\begin{figure}[htb]
 \centerline {\epsfig{file=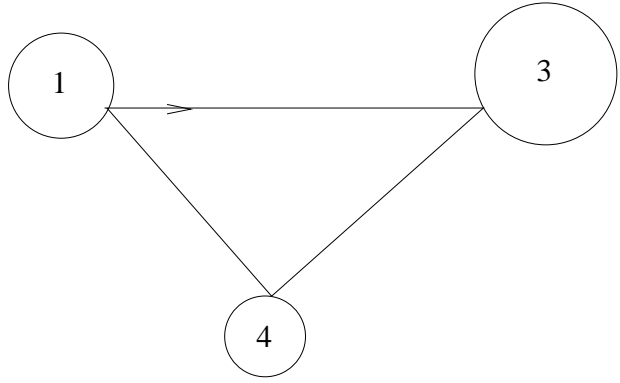,height=5cm,angle=-90}}
\caption[fig:parent]{\small 
The parent itinerary $(1,3,4)$.
\label{fig:parent}}
\end{figure}
\be 
\lefteqn{
 -\frac{1}{4} \left ( 4\, {\bf A}^{1,\underline{2}}_{\rm ghost} 
                        {\bf A}^{\underline{2},3}_{\rm ghost}  
                        {\bf A}^{3,4}_{\rm geom}  
                        {\bf A}^{4,1}_{\rm geom} \right )  
 +\frac{1}{3} \left ( 3\, {\bf A}^{1,3}_{\rm geom} 
                        {\bf A}^{3,4}_{\rm geom}  
                        {\bf A}^{4,1}_{\rm geom} \right )\nonumber} \\
             &=&(-1+1)\, {\bf A}^{1,3}_{\rm geom} 
                        {\bf A}^{3,4}_{\rm geom}  
                        {\bf A}^{4,1}_{\rm geom} =0\; .\nonumber \\
\ee
The prefactors $+1/3$ and $-1/4$ are due to the expansion of the logarithm,
the factors $3$ and $4$ inside the brackets result from the cyclic permutation
of the periodic itineraries,
and the cancellation stems from the rule 
\be
 \cdots {\bf A}^{i, \underline{i+1}}_{\rm ghost}
{\bf A}^{\underline{i+1}, i+2}_{\rm ghost}\cdots=
 \cdots{\bf A}^{i, i+2}_{\rm geom} \cdots \;.  \label{ghost-rule}
\ee
We have checked this rule in App.\ref{app:convol_ghost} 
for the convolution of two ${\bf A}$-matrices, but
in Sec.\ref{chap:sc_ghost_rule} 
 we will prove it 
to hold also inside an arbitrary (periodic) itinerary.
Of course the same cancellation holds in case that there are 
two and more ghost segments. For instance, 
consider the itinerary $(1,\underline{2},3,4,\underline{5},6)$
with ghost sections at disk $2$ and $5$ resulting
from the sixth order trace. Its geometrical contribution 
cancels in the trace-log expansion against the geometrical
reduction of 
the itineraries $(1,\underline{2},3,4,,6)$,
$(1,3,4,\underline{5},6)$ from the 5th-order trace
with ghost sections at disk 2 or 5, respectively, and against the
geometrical reduction of the itinerary
 $(1,3,4,6)$ of the 4th-order trace with
no ghost contribution:
\be
\lefteqn{
-\frac{1}{6} \left( 6\,{\bf A}^{1,\underline{2}}_{\rm ghost} 
         {\bf A}^{\underline{2},3}_{\rm ghost}
         {\bf A}^{3,4}_{\rm geom}
         {\bf A}^{4,\underline{5}}_{\rm ghost}
         {\bf A}^{\underline{5},6}_{\rm ghost}   
        {\bf A}^{6,1}_{\rm geom}  \right )} \nonumber\\
&&\mbox{} +\frac{1}{5} \left ( 5\, {\bf A}^{1,\underline{2}}_{\rm ghost} 
         {\bf A}^{\underline{2},3}_{\rm ghost} 
         {\bf A}^{3,4}_{\rm geom}
         {\bf A}^{4,6}_{\rm geom}
         {\bf A}^{6,1}_{\rm geom} 
      +5\, {\bf A}^{1,3}_{\rm geom} 
         {\bf A}^{3,4}_{\rm geom} 
         {\bf A}^{4,\underline{5}}_{\rm ghost}
         {\bf A}^{\underline{5},6}_{\rm ghost}
         {\bf A}^{6,1}_{\rm geom}\right ) \\ 
 && \mbox{}
 -\frac{1}{4}\left ( 4\, {\bf A}^{1,3}_{\rm geom} 
         {\bf A}^{3,4}_{\rm geom}
         {\bf A}^{4,6}_{\rm geom}
         {\bf A}^{6,1}_{\rm geom}\right )
  \nonumber \\                        
 &=& (-1+2-1)\,  {\bf A}^{1,3}_{\rm geom} 
         {\bf A}^{3,4}_{\rm geom}
         {\bf A}^{4,6}_{\rm geom}
         {\bf A}^{6,1}_{\rm geom} =0 \;.
\ee
Again, the prefactors $-1/4$, $+1/5$, $-1/6$ result from the trace-log
expansion, the factors 4, 5, 6 inside the brackets are due to
the cyclic 
permutations, and the rule \equa{ghost-rule} was used. If there are two
or more ghost segments adjacent to each other, 
the ghost rule \equa{ghost-rule}
has to be generalized to
\be
\lefteqn{
 \cdots
 {\bf A}^{i, \underline{i+1}}_{\rm ghost}
 {\bf A}^{\underline{i+1}, \underline{i+2}}_{\rm ghost} \cdots
 {\bf A}^{\underline{i+k}, \underline{i+k+1}}_{\rm ghosts} \cdots 
 {\bf A}^{\underline{i+n-1}, i+n}_{\rm ghosts}\cdots} \nonumber \\
&=& \cdots{\bf A}^{i, \underline{i+2}}_{\rm ghost}
 \cdots
 {\bf A}^{\underline{i+k}, \underline{i+k+1}}_{\rm ghosts} \cdots 
 {\bf A}^{\underline{i+n-1}, i+n}_{\rm ghosts}\cdots \\
&=&{
 \cdots
 {\bf A}^{i, \underline{i+3}}_{\rm ghost}
 \cdots
 {\bf A}^{\underline{i+k}, \underline{i+k+1}}_{\rm ghosts} \cdots 
 {\bf A}^{\underline{i+n-1}, i+n}_{\rm ghosts}\cdots} \\
&=& \cdots{\bf A}^{i, i+n}_{\rm geom}\cdots \; . \label{ghost-rule-extended}
\ee
Finally, let us discuss one case with a repeat, e.g.\ the itinerary
$(1,\underline{2},3,4,1,\underline{2},3,4)$ with repeated 
ghost sections at disk 2 in the semiclassical limit.
The cancellations proceed in the trace-log expansion as follows:
\be
\lefteqn{
-\frac{1}{8}\left( 4 \,{\bf A}^{1,\underline{2}}_{\rm ghost} 
         {\bf A}^{\underline{2},3}_{\rm ghost}
         {\bf A}^{3,4}_{\rm geom}
         {\bf A}^{4,1}_{\rm geom} 
         {\bf A}^{1,\underline{2}}_{\rm ghost} 
         {\bf A}^{\underline{2},3}_{\rm ghost}
         {\bf A}^{3,4}_{\rm geom}
         {\bf A}^{4,1}_{\rm geom}\right )} \nonumber \\
&& \mbox{} +\frac{1}{7}\left(7\,  {\bf A}^{1,\underline{2}}_{\rm ghost} 
         {\bf A}^{\underline{2},3}_{\rm ghost}
         {\bf A}^{3,4}_{\rm geom}
         {\bf A}^{4,1}_{\rm geom} 
         {\bf A}^{1,3}_{\rm geom} 
         {\bf A}^{3,4}_{\rm geom}
         {\bf A}^{4,1}_{\rm geom}\right )   \nonumber \\
&& \mbox{} - \frac{1}{6} \left(3\, {\bf A}^{1,3}_{\rm geom} 
         {\bf A}^{3,4}_{\rm geom}
         {\bf A}^{4,1}_{\rm geom} 
         {\bf A}^{1,3}_{\rm geom} 
         {\bf A}^{3,4}_{\rm geom}
         {\bf A}^{4,1}_{\rm geom}\right) \nonumber \\
 &=&  \left(-\frac{1}{2} + 1 -\frac{1}{2}\right )\, 
   \left[{\bf A}^{1,3}_{\rm geom} 
         {\bf A}^{3,4}_{\rm geom}
         {\bf A}^{4,1}_{\rm geom} \right]^2 =0
\ee
Note that the cyclic permutation factors of the
8th and 6th order trace are halved because of the repeat. The occurrence
of the ghost segment in the second part of the 7th order itinerary is taken
care of by the weight factor 7.

The reader might study more complicated examples and convince him- or
herself that the rule \equa{ghost-rule-extended} is sufficient to cancel
any primary or repeated periodic orbit with one or more 
ghost sections completely
out of the
expansion of $\Tr{}\ln({\bf 1}+{\bf A})$ and therefore also
out of the cumulant expansion in the semiclassical limit: 
Any periodic orbit of length $m$ 
with $n\mbox{($< m$)}$ ghost sections is cancelled by the sum of
all `parent' periodic orbits of length $m-i$ (with 
$1\leq i\leq n$  and $i$ ghost sections removed) weighted
by their cyclic permutation factor and by the prefactor resulting from
the {\em trace-log} expansion. This is the way in which
the non-trivial pruning 
for the $n$-disk billiards can be derived from the exact quantum-mechanical 
expressions in the semiclassical limit. 
Note that there must exist at least one index $i$ in any given 
{\em periodic} itinerary which
corresponds to a non-ghost section, since otherwise the itinerary in
the semiclassical limit could only 
be straight and therefore non-periodic. 
Furthermore, the series in the ghost cancelation has to
stop at the 2nd-order trace, $\Tr{} {\bf A}^2$, as $\Tr{} {\bf A}$ itself 
vanishes identically in the
full domain which is  considered here.  

\subsection{Semiclassical approximation of a periodic 
itinerary\label{chap:sc_appr_po}}
The procedure for the semiclassical approximation of a general
periodic itinerary, Eq.\equa{itinerary}, of length $m$ follows exactly
the calculation of App.\ref{app:convol} for the convolution of
two ${\bf A}$-matrices. 
The reader interested in the details of the semiclassical reduction is advised
to consult this appendix before proceeding with the remainder of
the section.
First, for any index $i$, $1\leq i \leq m$, the sum
over the integer angular momenta, $l_i$,  will be symmetrized as in
Eq.\equa{convolution} with the help
of the weight function $d(l_i)$ [$d(l_i\mbox{$\neq$}0) \equiv 1$, 
$d(l_i\mbox{=}0)=1/2$].  
\be
\lefteqn{ 
   {\bf A}^{j_1 j_2}
   \cdots
    {\bf A}^{j_{i-1} j_i} 
    {\bf A}^{j_i j_{i+1}} 
 \cdots 
   {\bf A}^{j_{m} j_1}  \nonumber} \\
&=&
\sum_{l_1=-\infty}^{+\infty} 
 \sum_{l_2=-\infty}^{+\infty} 
    \cdots 
 \sum_{l_{i-1}=-\infty}^{+\infty} 
 \sum_{l_{i}=-\infty}^{+\infty} 
 \sum_{l_{i+1}=-\infty}^{+\infty} 
 \cdots
  \sum_{l_m=-\infty}^{+\infty}
   {\bf A}^{j_1 j_2}_{l_1 l_2}
   \cdots
   {\bf A}^{j_{i-1} j_i}_{l_{i-1} l_i}
   {\bf A}^{j_i j_{i+1}}_{l_i l_{i+1}}   
   \cdots
   {\bf A}^{j_m j_1}_{l_m l_1}
  \nonumber \\
&=&
\sum_{l_1=0}^{+\infty}\, 
 \sum_{l_2=0}^{+\infty} 
    \cdots 
 \sum_{l_{i-1}=0}^{+\infty}\, 
 \sum_{l_{i}=0}^{+\infty}\, 
 \sum_{l_{i+1}=0}^{+\infty} 
 \cdots
  \sum_{l_m=0}^{+\infty}\,\,
 \sum_{s_1, \dots,  s_m=-1,1}\,
   d(l_1) \cdots d(l_{i-1}) d(l_i) \cdots d(l_m) \nonumber \\
&& \qquad\qquad\qquad \mbox{} \times
{\bf A}^{j_1 j_2}_{(s_1 l_1) (s_2 l_2)}
   \cdots
   {\bf A}^{j_{i-1} j_i}_{(s_{i-1} l_{i-1}) (s_i l_i)}
   {\bf A}^{j_i j_{i+1}}_{(s_i l_i) (s_{i+1} l_{i+1})}   
   \cdots
   {\bf A}^{j_m j_1}_{(s_m l_m) (s_1 l_1)} \; .
  \nonumber \\
\ee
Furthermore, the angles 
$\Delta \alpha_{j_i}\equiv \alpha_{j_{i+1}j_i}-\alpha_{j_{i-1}j_i}$ [the
analogs of $\alpha_{j''j'}-\alpha_{j j'}$ in Eq.\equa{convolution}]  
will be replaced
by $\widetilde \Delta\alpha_{j_i, \sigma_i}=\Delta \alpha_{j_i}-\sigma_i 2\pi$
where $\sigma_i =0,2,1$. This will be balanced by multiplying 
Eq.\equa{itinerary} with $(-1)^{\sigma_i' l_i}$ where $\sigma_i' = 
\sigma_i$ for
$\sigma_i=1$ and zero otherwise.
The three choices for $\sigma_i$ are, at this stage, equivalent, but
correspond in the semiclassical reduction to the three geometrical 
alternatives:
specular reflection at disk $j_i$ to the right, to the left or ghost
tunneling. In order not to be  bothered by borderline cases between
specular reflections and ghost tunneling, we exclude
disk configurations which allow classically 
grazing or penumbra periodic orbits~\cite{Nussenzveig,penumbra}.

Then, the sum over the integer angular momentum $l_i$ 
will be replaced by a Watson contour
integration over the complex angular momentum $\nu_i$
\beq
 \sum_{l_i=0}^{+\infty}\, (-1)^{l_i(1-\sigma'_i)} d(l_i) X_{l_i} 
 = \frac{1}{2\i} \oint_{C_+} {\rm d}{\nu_i}\,\frac{1}{\sin({\nu_i} \pi)}
\e^{-\i\nu_i\pi\sigma_i'} X_{\nu_i} \; ,
 \label{k-contour}
\eeq 
as in Eq.\equa{Watson-start}. 
The quantity $X_{l_i}$ abbreviates here 
\be
  X_{l_i}\!&\equiv&
      \frac{\Jb{l_i}{a_{j_i}   } }
                               {\Ho{l_i}{a_{j_i}   } } 
\sum_{s_i=-1,1}\,
\Ho{s_{i-1}l_{i-1}-s_i l_i}{R_{j_{i-1}j_i}}  
 \Ho{s_{i} l_i - s_{i+1}l_{i+1}}{R_{j_ij_{i+1}}} 
             \e^{\i  s_i l_i \widetilde{\Delta}\alpha_{j_i\sigma_i } } 
 \nonumber \\
 \!\!&\equiv&  \frac{\Jb{l_i}{a_{j_i}   } }
                               {\Ho{l_i}{a_{j_i}   } } Y_{l_i}
  \label{x_l_i}
\ee
where the expression has simplified because of 
$\Jb{s_i l_i}{a_{j_i}}/ \Ho{s_i l_i}{a_{j_i}}=
\Jb{l_i}{a_{j_i}}/ \Ho{l_i}{a_{j_i}}$, since $l_i$ is an integer.
The quantity $Y_{l_i}$ abbreviates the sum in \equa{x_l_i}.
The next steps are completely the same as in 
App.\ref{app:convol_Watson}--\ref{app:convol_paths}.
The paths below the real $\nu_i$ axis will be transformed above the
axis. The expressions split into a $\sin(\nu_i\pi)$-dependent contour integral
in the upper complex plane and into a $\sin(\nu_i\pi)$-independent  
straight-line integral from $\i\infty(1+\i\delta_i)$ to 
$-\i\infty(1+\i\delta_i)$.
Depending on the choice of $\sigma_i$,
the sum \equa{k-contour} becomes
exactly one of the three expression \equa{alt-1}, \equa{alt-2} or \equa{alt-3},
where the prefactor $W^{jj''}_{ll''}$ in App.\ref{app:convol_paths}
should be,  of course, replaced by all
the $l_i$-independent terms of Eq.\equa{itinerary} and where
 $j,j',j''$ are substituted by by $j_{i-1},j_i, j_{i+1}$. The angular momenta
$l$ and $l''$  are here identified with $s_{i-1}\times l_{i-1}$ and
$s_{i+1}\times l_{i+1}$, respectively. 
After the Watson resummation of the other sums,
e.g., of the  $l_{i-1}$ sum etc., $l$ has to be replaced by $\nu_{i-1}$
and $l''$ by $\nu_{i+1}$. 
If the penumbra scattering case~\cite{Nussenzveig,penumbra} is excluded,
the choice of $\sigma_i$ is, in fact, uniquely determined from the empirical
constraint that the creeping amplitude has to decrease during the creeping
process, as tangential rays are constantly emitted. In mathematical terms, it
means that the creeping angle has to be positive. As discussed at the
beginning of App.\ref{app:convol_paths}, the positivity of the
{\em two} creeping angles for the left {\em and} right turn  
uniquely specifies which of the three alternatives $\sigma_i$ is realized.
In other words, the geometry is encoded via the positivity of the two 
creeping paths into a unique choice of the $\sigma_i$.
Hence, the existence of the saddle-point \equa{i-saddle} is guaranteed.

The final step is the semiclassical approximation of the analog expressions
to Eqs.\equa{alt-1}-- \equa{alt-3} as
in App.\ref{app:convol_straightline}--\ref{app:convol_results}. 
Whereas the results
for the creeping contributions can be directly taken over from 
Eqs.\equa{res-alt-1} -- \equa{res-alt-3}, there is a subtle change in the 
semiclassical evaluation of the straight-line sections.
In the convolution problem of App.\ref{app:convol_straightline} and 
\ref{app:convol_results} we have only 
picked up second-order fluctuating terms with respect to the saddle solution
$\nu'_s$
from the $\nu'$ integration. Here, we will pick up quadratic terms 
$(\nu_i \mbox{$-$}\nu_i^s)^2$ from the 
$\nu_i$ integration {\em and} mixed terms 
$(\nu_i\mbox{$-$}\nu_i^s)(\nu_{i\pm 1}\mbox{$-$}\nu_{i\pm 1}^s)$ from the 
neighboring $\nu_{i-1}$ and $\nu_{i+1}$
integrations as well. 
Thus instead of having $m$ one-dimensional decoupled 
Gauss integrations,
we have one coupled $m$-dimensional one. Of course, also the
saddle-point equations [the analog to Eq.\equa{cond-alt-1} 
or \equa{cond-alt-3}] are now
coupled:
\be
(1-\delta_{\sigma'_i,1})
2\arccos[\nu_{i}^{s}/k a_{j_i}] 
&=&
\arccos[(\nu_{i}^{s}-\nu_{i-1}^{s})/kR_{j_{i-1} j_i}]
+\arccos[(\nu_{i}^{s}-\nu_{i+1}^{s})/kR_{j_{i} j_{i+1}}] \nonumber \\
&& \mbox{} -(\alpha_{j_{i+1} j_{i}}-\alpha_{j_{i-1} j_{i}}-\sigma_i 2\pi) 
\label{i-saddle}
\ee
where the saddle  $\nu_i^{s}$ of the $i$th integration depends on the values of
the 
saddles of the $(i-1)$th and $(i+1)$th integration and so on. 
Indeed, all $m$ saddle-point equations are coupled. 
This corresponds
to the fact that the starting- {\em and} end-point of a period orbit is not
fixed from the outside, but has  to be determined self-consistently, namely 
on the same 
footing as all the intermediate points.
         
In order to keep the resulting expressions simple we will discuss in
the following subsection just
the geometrical contributions, and leave the discussion of the ghost 
and creeping contributions for later sections.

\subsection{Itineraries in the geometrical limit\label{chap:sc_itin_geom}}
We will prove that 
the itinerary
$(-1)^m{\bf A}^{j_1 j_2}
   {\bf A}^{j_2 j_3} 
   \cdots  
   {\bf A}^{j_{m-1} j_m}
   {\bf A}^{j_{m} j_1}
$  
leads, in the semiclassical reduction, 
to the following geometrical contribution:
\be
(-1)^m
{\bf A}^{j_1 j_2}
   {\bf A}^{j_2 j_3} 
   \cdots  
   {\bf A}^{j_{m-1} j_m}
   {\bf A}^{j_{m} j_1}|_{\rm geom} = \frac{\e^{\i k L_{1\to m} -\i2m (\pi/2)}}
                {|\Lambda_{1\to m}|^{1/2}\left (1 -\frac{1}
 {\Lambda_{1\to m}}\right )}
  \; ,
 \label{itinerary-to-cycle}
\ee
where the factor $(-1)^m$ results from 
the trace-log expansion
$\Tr{}ln({\bf 1} -[-{\bf A}])$, as the periodic orbit expansion corresponds
to this choice of sign.  
The quantity 
$L_{1\to m}$ is the length of the periodic orbit with this itinerary.
$\Lambda_{1\to m}>1$ is the expanding  eigenvalue of the corresponding monodromy
matrix and
$\mu_{1\to m}=2m$
is the corresponding Maslov index indicating that the orbit 
is reflected from $m$ disks (all with Dirichlet boundary conditions).
Thus, for $n$-disk Dirichlet 
problems, the Maslov indices come out automatically.
[Under Neumann boundary conditions, 
there arises an additional minus sign per disk label
$j_i$, since
$\{\frac{d}{dk} \Ht{\nu_i}{a_{j_i}}\}/\{\frac{d}{dk}
\Ho{\nu_i}{a_{j_i}}\}\simeq
- \Ht{\nu_i}{a_{j_i}}/\Ho{\nu_i}{a_{j_i}}$
in the Debye approximation. 
The minus sign on the
right-hand side cancels the original minus sign from the trace-log
expansion such that the total Maslov index
becomes trivial. Otherwise, the Neumann case is exactly the same.]\
If the itinerary is the $r$th repeat of a primary itinerary of 
topological length $p$, the length, Maslov index and stability eigenvalue
will be shown to 
satisfy the relations: $L_{1\to m} =r L_{1\to p}$, $\mu_{1\to m}=
r\mu_{1\to p}$ and 
$\Lambda_{1\to m} = (\Lambda_{1\to p})^r$. 
  
Let us define the abbreviations
\be
 d_{i-1,i}&\equiv& \sqrt{R_{j_{i-1} j_i}^2-[(\nu_{i-1}^s-\nu_{i}^s)/k]^2}
 =d_{i,i-1} \label{d-new-def}\\
 \rho_{i}&\equiv& \sqrt{a_{j_i}^2-(\nu_i^s/k)^2} \label{rho-i-def}\\
 L_{i-1,i}&\equiv& d_{i-1,i}-\rho_{i-1}-\rho_{i}=L_{i,i-1} \label{L_ij} \\
 \delta \nu_i &\equiv& \nu_i - \nu_i^s \nonumber \\ 
 \widetilde {\delta  \nu}_i &\equiv& \frac{\delta \nu_i}{d_{i-1,i}^{1/2}}
 \nonumber
\ee
with $i$ evaluated modulo $m$, especially $i=0$ is identified with $i=m$ and
$i=m+1$ with $i=1$. The quantity $d_{i-1,i}$ is the geometrical 
length of the straight line between the impact parameter  $\nu_{i-1}^s/k$ at
disk $j_{i-1}$ and the impact parameter $\nu_i^s/k$ at disk $j_i$ in terms
of the saddle points 
$\nu_{i-1}^s$  and $\nu_i^s$. 
The latter  are determined by the saddle-point
condition \equa{i-saddle} which can be re-written for  
non-ghost scattering ($\sigma_i\neq 1$) as a condition on
the reflection angle at disk $j_i$:
\be
 \theta_{j_i}&\equiv&\arcsin[\nu_i^s/ka_{j_i}] \nonumber \\
 &=&\arcsin\left[\frac{\nu_{i}^{s}\mbox{$-$}\nu_{i-1}^{s}}
{kR_{j_{i-1} j_i}}\right]
+\arcsin\left[\frac{\nu_{i}^{s}\mbox{$-$}\nu_{i+1}^{s}}{kR_{j_{i} j_{i+1}}}
\right] 
 +(\alpha_{j_{i+1} j_{i}}\mbox{$-$}\alpha_{j_{i-1} j_{i}}\mbox{$-$}
\sigma_i 2\pi) \; . 
  \label{i-refl-angle}
\ee
Thus, $\rho_i$ is the  radius $a_{j_i}$ of the disk $j_i$ 
{\em times}  the cosine of the
reflection angle and 
$L_{i-1,i}$ is the geometrical length of the straight-line segment between 
the $(i\mbox{$-$}1)$th and
$i$th point of reflection.
Under the condition that the disks do not overlap,
the inequalities 
$L_{i-1,i} < d_{i-1,i}< R_{j_{i-1}j_i}$ hold and exclude
the possibility that the reflection points are in the mutual shadow region of
disks. For each itinerary there is at most one reflection per disk-label $j_i$
modulo repeats, of course.

Then in analogy to App.\ref{app:convol_results} the geometric limit
of the itinerary \equa{itinerary} becomes
\be
\lefteqn{(-1)^m
{\bf A}^{j_1 j_2}
   {\bf A}^{j_2 j_3} 
   \cdots  
   {\bf A}^{j_{m-1} j_m}
   {\bf A}^{j_{m} j_1}|_{\rm geom} \nonumber }\\
&=&(-1)^m\left \{
\prod_{j=1}^m\int_{-\e^{-\i\pi/4}\infty}^{\e^{-\i\pi/4}\infty} 
{\rm d} \delta \nu_j \right \}
\prod_{i=1}^m \half \e^{\i \pi/4} \sqrt{\frac{2}{\pi}}
\frac{\e^{\i k (d_{i-1,i}- 2\rho_i) } } { k^{1/2} d_{i-1,i}^{1/2} }\nonumber \\
&&\qquad\qquad\qquad \mbox{}\times
 \e^{-\i\frac{1}{2k} (\delta \nu_i)^2\left (\frac{2}{\rho_i} 
 -\frac{1}{d_{i-1,i}}
 -\frac{1}{d_{i,i+1}} \right )}
\e^{-\i\frac{1}{2k}\delta\nu_i \delta\nu_{i+1}\frac{1}{d_{i,i+1}}}
\e^{-\i\frac{1}{2k}\delta\nu_{i-1} \delta\nu_{i}\frac{1}{d_{i-1,i}}}\nonumber\\
&=& \left\{\prod_{i=1}^m  \frac{\e^{\i \pi/4}} {\sqrt{2\pi k}}
\e^{\i k (d_{i-1,i}- 2\rho_i) -\i \pi} \right\}\nonumber \\
&&\qquad \times\int_{-\e^{-\i\pi/4}\infty}^{\e^{-\i\pi/4}\infty} 
{\rm d} \widetilde{\delta \nu}_1
\cdots 
\int_{-\e^{-\i\pi/4}\infty}^{\e^{-\i\pi/4}\infty} 
{\rm d} \widetilde{\delta \nu}_m\, 
\e^{-\i\frac{1}{2k}(\widetilde{\delta \nu}_1 \cdots 
\widetilde{\delta\nu}_m) 
{\bf F}_{1\to m}
(\widetilde{\delta \nu}_1 \cdots 
 \widetilde{\delta\nu}_m)^T} \nonumber \\
&=&
\frac{\e^{\i k\sum_{i=1}^{m}\,L_{i-1,i}-\i m\pi}}{|D_{1\to m}|^{1/2}}
  \label{geometric-fluctuations}
\ee 
where we have 
used that $\sum_{i=1}^{m}\,2\rho_i=\sum_{i}^{m}(\rho_{i-1}+\rho_{i})$
since $\rho_0=\rho_m$. $L_{1\to m}\equiv \sum_{i=1}^m L_{i-1,i}$ is the total 
geometrical length of the geometrical path around the itinerary, see 
App.~G of Ref.\cite{Berry_KKR}. Note that we used the saddle-point condition
\equa{i-saddle} in order to remove not only the linear fluctuations, but 
all terms of 
linear order in the $\nu_i^s$'s from the exponents. 
Only the zeroth-order terms and the quadratic
fluctuations remain.
$D_{1\to m}$ is the determinant of the $m\times m$ matrix 
${\bf F}_{1\to m}$ ($\mbox{$\equiv$}{\bf F}$) with 
\beq \begin{array}{lclcl}

  {\bf F}_{1,1} 
          &=&\frac{2d_{m,1}}{\rho_1}- 1-\frac{d_{m,1}}{d_{1,2}} 
         &\equiv& a_1 \\
  {\bf F}_{i,i} 
          &=&\frac{2d_{i-1,i}}{\rho_i}- 1-\frac{d_{i-1,i}}{d_{i,i+1}} 
         &\equiv& a_i \quad {\rm for\ } 2\leq i \leq m-1\\
 {\bf F}_{m,m}
         &=&\frac{2d_{m-1,m}}{\rho_m}- 1-\frac{d_{m-1,m}}{d_{m,1}} &\equiv&
       a_m \\
  {\bf F}_{i,i+1}
          &=& \frac{d_{i-1,i}^{1/2}}{d_{i,i+1}^{1/2}} 
  = {\bf F}_{i+1,i}           &\equiv& b_{i,i+1} \quad {\rm for}\ 1\leq i 
  \leq m-1\\
  {\bf F}_{m,1}
          &=& \frac{d_{m-1,m}^{1/2}}{d_{m,1}^{1/2}} = 
{\bf F}_{1,m} &\equiv& b_{m,1} \\
  {\bf F}_{i,j}&=& 0 \quad {\rm otherwise} && \end{array}  
\label{G-definition}
\eeq
for $m\geq 3$. [For $m=2$ the off-diagonal matrix elements 
read instead 
\beq
  {\bf F}_{1,2}={\bf F}_{2,1}=\frac{d_{2,1}^{1/2}}{d_{1,2}^{1/2}} +
   \frac{d_{1,2}^{1/2}}{d_{2,1}^{1/2}}=2\; .
\eeq 
The corresponding diagonal matrix elements are given as above, but 
simplify  because of $d_{2,1}=d_{1,2}$.]\ 
Thus in general, the determinant reads 
\beq
 D_{1\to m}= \det{} \left(\begin{array}{ccccccc}
    a_1   & b_{1,2} & 0       & 0       &\cdots &  0        &  b_{m,1} \\
  b_{1,2} &     a_2 & b_{2,3} & 0       &\cdots &  0        &    0     \\
    0     & b_{2,3} & a_3     & b_{3,4} &\ddots & \cdots    &   \vdots  \\
    0     & 0       & b_{3,4} & a_4     &\ddots & \ddots    &  \vdots   \\ 
   \vdots & \cdots  & \ddots  & \ddots  &\ddots &b_{m-2,m-1}&    0      \\
    0     & 0       & \cdots & \ddots   & \ddots&a_{m-1}   &b_{m-1,m}   \\
  b_{m,1} & 0       &  \cdots & \cdots  &   0   &b_{m-1,m}  &a_m 
\end{array} \right ) \; .
\eeq
Note that determinants of this structure can also be found
in Balian and Bloch~\cite{bb_2} and Berry~\cite{Berry_KKR}.
Our task, however, is to simplify this expression, such that the stability
structure of an isolated unstable periodic orbit emerges in the end.
In order to derive a simpler expression for $D_{1\to m}$,
let us consider the determinant  $D_{1\to m}^{(0)}$
of the auxiliary $m\times m$ matrix ${\bf F}^{(0)}_{1\to m}$  
($\mbox{$\equiv$} {\bf F}^{(0)}$) 
which has the
same matrix elements as ${\bf F}$ with the exception that 
$b_{m,1}=0$.
The original determinant $D_{1\to m}$ can now be expressed as
\beq
 D_{1\to m}=D_{1\to m}^{(0)}-\frac{d_{m-1,m}}{d_{m,1}} 
  D_{2\to m-1}^{(0)}+2 (-1)^{m+1}
 \label{D-recursion-start}
\eeq
where the last term follows from $\prod_{i=1}^{m} b_{i,i+1} =1$.
Here and in the following  $D_{l\to k}^{(0)}$ is defined as 
the determinant of the auxiliary
$(k\mbox{$-$}l\mbox{+}1)
\times(k\mbox{$-$}l\mbox{+}1)$ matrix  
${\bf F}^{(0)}_{l\to k}$ with matrix elements
${\bf F}^{(0)}_{l\to k}|_{ij}={\bf F}_{ij}$ for $l\leq i,j\leq k(\leq m)$.
Furthermore we define $D_{k+1\to k}^{(0)}\equiv {\bf F}_{k+1\to k}^{(0)}
\equiv D_{1\to 0}^{(0)}\equiv 1$.
The $D_{l\to k}^{(0)}$ determinants fulfill the following recursion relations
\be
D_{l\to k}^{(0)}&=&\left(\frac{2 d_{k-1,k}}{\rho_k} -1 
 -\frac{d_{k-1,k}}{d_{k,k+1}}\right )D_{l\to k-1}^{(0)}
 - \frac{d_{k-2,k-1}}{d_{k-1,k}} D_{l\to k-2}^{(0)} \; , 
 \label{D0-recursion-1} \\
D_{l\to k}^{(0)}&=&\left(\frac{2 d_{l-1,l}}{\rho_l} -1 
 -\frac{d_{l-1,l}}{d_{l,l+1}}\right )D_{l+1\to k}^{(0)}
 - \frac{d_{l-1,l}}{d_{l,l+1}} D_{l+2\to k}^{(0)} \; ,
 \label{D0-recursion-2}
\ee
such that  $D_{l\to k}^{(0)}$ can be constructed from all the lower
determinants $D_{l\to j}^{(0)}$ and $D_{i\to j}^{(0)}$ with $l\leq i<j< k$. 
For example,
\be
D_{1\to k}^{(0)}&=& -\frac{d_{k-1,k}}{d_{k,k+1}}  D_{1\to k-1}^{(0)}
 +(-1)^k\left( 1 -  \frac{2d_{m,1}}{\rho_1} D_{1\to 0}^{(0)}
                + \frac{2d_{1,2}}{\rho_2} D_{1\to 1}^{(0)}
                - \cdots 
\right. \nonumber \\
&& \qquad\qquad\qquad\qquad\qquad\quad  \mbox{}+\left.
(-1)^i \frac{2d_{i-1,i}}{\rho_i}D_{1\to i-1}^{(0)}
                  \cdots + (-1)^k \frac{2d_{k-1,k}}{\rho_k}D_{1\to k-1}^{(0)}
                     \right ) \nonumber \\
 \label{D0-recursion-full}
\ee
and
\be
D_{l\to k}^{(0)}&=& -\frac{d_{k-1,k}}{d_{k,k+1}}  D_{l\to k-1}^{(0)}
 +(-1)^{k-l+1}
\left( 1 -  \frac{2d_{l-1,l}}{\rho_l} D_{l\to l-1}^{(0)}
                + \frac{2d_{l,l+1}}{\rho_{l+1}} D_{l\to l}^{(0)}
\right. \nonumber \\
&& \qquad\qquad\qquad\qquad\qquad\qquad  \mbox{}+\left.
                  \cdots + (-1)^{k-l+1} \frac{2d_{k-1,k}}{\rho_k}
D_{l\to k-1}^{(0)}
                     \right )
 \label{D0-recursion-full-l}
\ee
as  can be shown by complete induction.
Note that the product $d_{k,k+1} D_{l\to k}$ is a multinomial in 
$d_{j,j+1}/\rho_i$ where, for each index $j$, 
the $d_{j,j+1}/\rho_i$ factors appear  at most
once.

Replacing the $D_{1\to m}^{(0)}$ term 
in \equa{D-recursion-start} by the r.h.s.\ of Eq.\equa{D0-recursion-full} 
and using the relation \equa{D0-recursion-2} in order to simplify 
the expression
\be
 -\frac{d_{m-1,m}}{d_{m,1}}
 (D_{1\to m-1}^{(0)}-D_{2\to m-1}^{(0)} )\nonumber
\ee
recursively,
we finally
find after some algebra that
\be
D_{1\to m} &=&  \frac{2 d_{m-1,m}    }{\rho_m    } D_{1\to m-1}^{(0)}
               -\frac{2 d_{m-2,m-1}}{\rho_{m-1}} D_{1\to m-2}^{(0)}
               +\cdots
         +(-1)^{i-1}\frac{2 d_{m-i-1,m-i}}{\rho_{m-i}} D_{1\to m-i}^{(0)}
 \nonumber \\
          && \mbox{}     +\cdots
     +(-1)^{m-1}\frac{2 d_{m,1}   }{\rho_1    } 
 D_{1\to 0  }^{(0)} \nonumber \\
     && \!\!\!\!\!\!\!\!\mbox{}
              -\frac{2 d_{m-1,m} }{\rho_1    } D_{2\to m-1}^{(0)}
              +\frac{2 d_{m-1,m} }{\rho_2    } D_{3\to m-1}^{(0)}
              -\cdots
       +(-1)^{i-1} 
    \frac{2 d_{m-1,m} }{\rho_i    } D_{i\to m-1}^{(0)}\nonumber \\
&& \mbox{}
       +\cdots
    +(-1)^{m-1}\frac{2 d_{m-1,m} }{\rho_{m-1}} D_{m\to m-1}^{(0)} \; .
 \label{D-full}
\ee
By complete induction it can be shown that $D_{1\to m}$ is a multinomial
in $2d_{i-1,i}/\rho_j$ of order $m$ where the single factors appear at most
once and the highest term has
the structure $\prod_{i=1}^m\, 2d_{i-1,i}/\rho_i$. Thus, 
all the $d_{i-1,i}$'s  
are in
the numerators, whereas all the $\rho_i$'s appear the denominators of this
multinomial.
We will show in Sec.\ref{chap:sc_geom_stab} that 
\be
 D_{1\to m}= (-1)^m \left ( \Lambda_{1\to m}+ \frac{1}{\Lambda_{1\to m}} 
-2 \right )
 \label{D-Lambda}
\ee
where $\Lambda_{1\to m}$ is the expanding eigenvalue 
of the monodromy matrix which belongs to that period orbit which
is given by the
geometric path of the periodic itinerary.
If the result of Eq.\equa{D-Lambda} is inserted into 
Eq.\equa{geometric-fluctuations}
the semiclassical reduction \equa{itinerary-to-cycle} is proven.

\subsection{Itineraries with repeats\label{chap:sc_itin_repeats}}
In the following we will discuss modifications, if the periodic 
itinerary is
repeated $r$ times, i.e., let $m=p\times r$
still be the total topological length
of the itinerary, whereas $p$ is the length of the prime periodic unit which
is repeated $r$ times:
\be
{\bf A}^{j_1 j_2}\cdots {\bf A}^{j_{i-1} j_i}{\bf A}^{j_i j_{i+1}}\cdots
 {\bf A}^{j_m j_1}=
 \left[{\bf A}^{j_1 j_2}\cdots {\bf A}^{j_{i-1} j_i}{\bf A}^{j_i j_{i+1}}\cdots
 {\bf A}^{j_p j_1}\right]^r \; .
 \label{itinerary-repeated}
\ee
The length and Maslov index of the itinerary are of course $r$ times the
length and Maslov index of the primary itinerary 
${\bf A}^{j_1 j_2}\cdots{\bf A}^{j_p j_1}$, e.g., 
$L_{1\to m}= r L_{1\to p}$. The non-trivial point is the structure of the
stability determinant $D_{1\to m}$. Here we can use that the matrix 
${\bf F}_{1\to m}$ has exactly the structure of the matrices considered
by Balian and Bloch in Ref.\cite{bb_2}, Sec.\,6\,D. Let ${\bf F}_{1\to p}$ be the 
corresponding matrix of the primary  itinerary with matrix elements as in 
Eqs.\equa{G-definition} [where $m$ is replaced by $p$ of course].
Following ref.\cite{bb_2} we furthermore define a new matrix 
$\widetilde{\bf F}_p(\chi)$ with matrix elements
\be
  \widetilde{\bf F}_p(\chi)|_{1,p}&=& \e^{-\i\chi}{\bf F}_{1\to p}|_{1, p} 
  \nonumber \\
  \widetilde{\bf F}_p(\chi)|_{p,1}&=& \e^{\i\chi}{\bf F}_{1\to p}|_{p,1} 
  \nonumber\\
  \widetilde{\bf F}_p(\chi)|_{i,j}&=& {\bf F}_{1\to p}|_{i,j}  
  \quad {\rm otherwise} \; .\nonumber 
\ee
The determinant of
the total itinerary is then \cite{bb_2}
\be
 D_{1\to m}= \prod_{n=1}^r \, 
 \widetilde D_{1\to p}(\e^{\i 2\pi n/r}) \;  
\ee
in terms of the $r$th roots of unity, since after $r$ repeats, the prefactor in
front of the $(1,m)$ and $(m,1)$ matrix elements must be unity in order to
agree with the original expression  \equa{G-definition}.
Let us furthermore define
\be
 \alpha_p &\equiv& \half \left ( \sqrt{ \widetilde D_{1\to p}(0)}
                           +\sqrt{ \widetilde D_{1\to p}(\pi)} \right )
 \nonumber \\
 \beta_p &\equiv& \half \left ( \sqrt{ \widetilde D_{1\to p}(0)}
                           -\sqrt{ \widetilde D_{1\to p}(\pi)} \right )
\ee
then, according to Balian and Bloch~\cite{bb_2}, Sec.\,6\,D,
\be
  D_{1\to m} &=& (\alpha_p^r - (-\beta_p)^r)^2 \nonumber \\
             &=& \alpha_p^{2r} + \beta_p^{2r} -2(-\alpha_p \beta_p)^{r}
 \; . \nonumber
\ee
In our case we have the further simplification (in analogy to 
Eq.\equa{D-recursion-start})
\be
-\alpha_p \beta_p &=& -\frac{1}{4} 
 \left (\widetilde D_{1\to p}(0)-\widetilde D_{1\to p}(\pi) \right ) 
\nonumber\\
 &=& -\frac{1}{4} \left(2(-1)^{p+1} - \{-2(-1)^{p+1}\}\right) = (-1)^{p} \; ,
 \nonumber
\ee
as the corresponding two matrices differ only in the sign of the
their $(1,p)$ and $(p,1)$ elements. Especially we now have 
$\beta_p =(-1)^{p+1}/\alpha_p$, 
such that
\be
  D_{1\to m}= (\alpha_p^2)^r + \frac{1}{(\alpha_p^2)^r}-2(-1)^{pr}
 \nonumber
\ee
which corresponds to the usual form
\be
 D_{1\to m}= (-1)^{pr} \left (\Lambda_{1\to p}^r + \frac{1}
{\Lambda_{1\to p}^r}-2\right)
 \label{D-Lambda-repeat}
\ee
if $\Lambda_{1\to p}$ is identified with $(-1)^p\alpha_p^2$.
Note that from this the structure of Eq.\equa{D-Lambda} follows for the
special case $r=1$. Thus we have achieved so far two things: we have proven
that the determinant $D_{1\to m}$ organizes itself in the same way
as a monodromy matrix does and, in fact, that it can be written in terms
of a monodromy matrix ${\bf M}_{1\to p}$ with eigenvalues 
$\Lambda_{1\to p}$, $1/\Lambda_{1\to p}$
as follows 
\be
 D_{1\to m} = (-1)^{pr+1}\det{}( {\bf M}_{1\to p}^r - {\bf 1} ) \; .
 \label{D-monodromy-matrix}
\ee
What is left to show is that ${\bf M}_{1\to p}$  
is the very monodromy matrix belonging
to the periodic orbit with the itinerary as in Eq.\equa{itinerary-repeated}.
This will be done in Sec.\ref{chap:sc_geom_stab}. But first, we will complete
the study of the geometrical sector by deriving 
the ghost subtraction rules, and furthermore discuss periodic orbits with
creeping contributions.

\subsection{Ghost rule\label{chap:sc_ghost_rule}}
Let us now imagine that the itinerary \equa{itinerary} has, at the
disk position $j_i$, an angular domain  that
corresponds to a ghost section, i.e.\ $\sigma_i=1$, $i$ fixed:
\be
{\bf A}^{j_1 j_2}\cdots 
{\bf A}^{j_{i-1} \underline{j_i}}
 {\bf A}^{\underline{j_i} j_{i+1}}
\cdots
 {\bf A}^{j_m j_1} \; .
 \label{ghost-itinerary}
\ee
Because of the cyclic nature of the itinerary we can always choose the
label $j_i$ away from the first and end position [remember that  
at least two
disk positions of any periodic orbit must be of non-ghost nature]. 
In this case there are four changes relative to the calculation in 
Eq.\equa{geometric-fluctuations}, see also App.\ref{app:convol_ghost}: 
first, the
path of the $\nu_i$ integration 
is changed, second, there is a minus sign, third, the
saddle-point condition at disk $j_i$ is given by Eq.\equa{i-saddle}
with $\sigma'=1$ and
not by \equa{i-refl-angle}, 
fourth, the $\rho_i$ terms are absent. 
As in App.\ref{app:convol_ghost}, the saddle condition \equa{i-saddle}
at the $j_i$th disk implies that 
$d_{i-1,i}+d_{i,i+1}=d_{i-1,i+1}$. We can use this in order to 
express the length of
the ghost segment $L_{i-1,i+1}$ between the reflection point 
at disk $j_{i-1}$
and the next reflection point at disk $j_{i+1}$ in terms of the quantities
defined in Eq.\equa{L_ij}:
\beq
 L_{i-1,i+1}=d_{i-1,i+1}-\rho_{i-1}-\rho_{i+1}= 
 L_{i-1,i}+2\rho_{i}+L_{i,i+1}-\rho_{i-1}-\rho_{i+1} \; .
 \label{L-ghost-segment}
\eeq 
Thus, by adding and subtracting the $\rho_i$ contributions
we get 
\be
\lefteqn{
{\bf A}^{j_1 j_2}
   \cdots  
{\bf A}^{j_{i-1} \underline{j_i}}
 {\bf A}^{\underline{j_i} j_{i+1}}
   \cdots
   {\bf A}^{j_{m} j_1}|_{\rm geom} \nonumber }\\
&=&  \left \{\prod_{j=1,j\neq i}^m\, 
\int_{-\e^{-\i\pi/4}\infty}^{\e^{-\i\pi/4}\infty} 
{\rm d} \delta \nu_j\ \right \}
\int_{\e^{+\i\pi/4}\infty}^{-\e^{+\i\pi/4}\infty}{\rm d} \delta \nu_i\,
(-1)\e^{-\i \pi/2} \e^{\i k 2\rho_i +\i\frac{1}{2k}(\delta \nu_i)^2
\frac{2}{\rho_i}} \nonumber \\
&&\mbox{}\times
\prod_{l=1}^m \half \e^{\i \pi/4} \sqrt{\frac{2}{\pi}}
\frac{\e^{\i k (d_{l-1,l}- 2\rho_l) } } { k^{1/2} d_{l-1,l}^{1/2} }
 \e^{-\i\frac{1}{2k} (\delta \nu_l)^2\left (\frac{2}{\rho_l} 
 -\frac{1}{d_{l-1,l}}
 -\frac{1}{d_{l,l+1}} \right )}
\e^{-\i\frac{1}{2k}\delta\nu_l \delta\nu_{l+1}\frac{1}{d_{l+1}}}
\e^{-\i\frac{1}{2k}\delta\nu_{l-1} \delta\nu_{l}\frac{1}{d_{l}}}
\nonumber \\
&=& +\e^{\i k 2\rho_i}
\left\{\prod_{l=1}^m  \frac{\e^{\i \pi/4}} {\sqrt{2\pi k}}
\e^{\i k (d_{l-1,l}- 2\rho_l)}\right \} 
\prod_{j=1}^{m}
\int_{-\e^{-\i\pi/4}\infty}^{\e^{-\i\pi/4}\infty} 
{\rm d} \widetilde{\delta \nu}_j\ \, 
\e^{-\i\frac{1}{2k}(\widetilde{\delta \nu}_1 \cdots 
\widetilde{\delta\nu}_m) 
{\bf F}_{1\to m}^{g_i}
(\widetilde{\delta \nu}_1 \cdots 
 \widetilde{\delta\nu}_m)^T} \nonumber \\
&=& \frac{\e^{\i k \left (\sum_{l=1}^m L_{l-1,l} +2\rho_i\right)} }
 {|D^{g_i}_{1\to m}|^{1/2}} 
=\frac{\e^{\i k \left (\sum_{l=1}^{i-1} L_{l-1,l}+L_{i-1,i+1} 
+ \sum_{l=i+2}^{m} L_{l-1,l}\right)}} 
 {|D^{g_i}_{1\to m}|^{1/2}}  
 \label{ghost-calculation}
\; .
\ee
[Note that the 
exponent of the ghost itinerary is exactly the same
as of the one of its parent, the same itinerary without the disk $j_i$,
whose geometrical path  has the length 
$ \sum_{l=1}^{i-1} L_{l-1,l}+L_{i-1,i+1} 
+ \sum_{l=i+2}^{m} L_{l-1,l}$.]\
In writing down the last-but-one 
line we have cancelled the overall minus sign by exchanging
the upper and lower limit of the $\delta \nu_i$ integration. In addition, the  
following substitutions were applied:
\beq
  \begin{array}{lcl} 
\delta \nu_l &=:& d_{l-1,l}^{1/2}\, \widetilde{\delta \nu}_l\quad {\rm for} \ \
 l\neq i, i+1 \; , \\
\delta \nu_i &=:& \e^{\i\pi/2} \left(\frac{ d_{i-1,i} d_{i,i+1}}
                                       { d_{i-1,i+1} }\right )^{1/2}
                        \widetilde{\delta \nu}_i \; ,\\
\delta \nu_{i+1} &=:& d_{i-1,i+1}^{1/2}\, \widetilde{\delta \nu}_{i+1} \; .
\end{array}
 \label{ghost-substitution}
\eeq
In this way, the integration path and phase of the $i$th term agree with the
ones of the other terms. $D^{g_i}_{1\to m}$ is the determinant of 
the $m\times m$ matrix ${\bf F}^{g_i}_{1\to m}$ 
($\mbox{$\equiv$} {\bf F}^{g_i}$) which is affected by the substitutions in
the following way:
\beq
\begin{array}{lcl}
 {\bf F}_{l,k}^{g_i} &=& {\bf F}_{l,k} \quad {\rm for}\ l,k \neq i,i+1 \\
 {\bf F}_{i,i}^{g_i}&=& 1 \\
 {\bf F}_{i-1,i}^{g_i}&=& \i \left (\frac{d_{i-2,i-1} d_{i,i+1}}
                            {d_{i-1,i} d_{i-1,i+1}} \right )^{1/2} 
               ={\bf F}_{i,i-1}^{g_i} \\
 {\bf F}_{i,i+1}^{g_i}&=& \i \left(\frac{d_{i-1,i}}{d_{i,i+1}} \right)^{1/2}
    = {\bf F}_{i+1,i}^{g_i}=\i {\bf F}_{i,i+1}=\i{\bf F}_{i+1,i} \\
{\bf F}_{i+1,i+1}^{g_i}&=& \frac{2 d_{i-1,i+1}}{\rho_{i+1}}
 -\frac{d_{i-1,i+1}}{d_{i,i+1}}-\frac{d_{i-1,i+1}}{d_{i+1,i+2}}\\
{\bf F}_{i+1,i+2}^{g_i}&=& \left(\frac{d_{i-1,i+1}}{d_{i+1,i+2}}\right)^{1/2}
 = {\bf F}_{i+2,i+1}^{g_i}
\end{array}
\label{F-g-def}
\eeq
where ${\bf F}_{l,k}$ 
are the matrix elements as defined Eqs.\equa{G-definition}; i.e.,
\beq
D_{1\to m}^{g_i}=\det{}
\left( {\bf F}^{g_i}\right) =
\det{}\left( \begin{array}{cccccc}
 \ddots  & \ddots      & \ddots &  \vdots & \vdots &\cdots \\
 \ddots  & {\bf F}_{i-1,i-1} & {\bf F}_{i-1,i}^{g_i} & 0   & 0 &\cdots \\
 \ddots  & {\bf F}_{i,i-1}^{g_i} & 1 & {\bf F}_{i,i+1}^{g_i} & 0& \cdots \\
 \cdots  &  0              & {\bf F}_{i+1,i}^{g_i} & {\bf F}_{i+1,i+1}^{g_i}&
                             {\bf F}_{i+1,i+2}^{g_i} & \ddots \\
 \cdots & 0  & 0 &{\bf F}_{i+2,i+1}^{g_i}& {\bf F}_{i+2,i+2} &\ddots \\
 \vdots & \vdots & \vdots & \ddots & \ddots & \ddots \\
\end{array} \right )
 \label{D-g-start}
\eeq
We now  subtract the $i$th row times 
${\bf F}^{g_i}_{i,i-1}$ from the $(i\mbox{$-$}1)$th and the $i$th row times
${\bf F}_{i,i+1}^{g_i}$  form the $(i\mbox{+}1)$th, 
as both operations leave the 
determinant $D^{g_i}_{1\to m}$ unaffected. Using that the ghost segments
add, i.e.,
$d_{i-1,i+1}=d_{i-1,i}+d_{i,i+1}$, the numerators of the terms in the
$(i,i)$ and
$(i\mbox{+}1,i\mbox{+}1)$ matrix elements can be simplified. The determinant 
$D_{1\to m}^{g_i}$, expressed via the transformed
$m\times m$ matrix 
$\widetilde {\bf F}^{g_i}$,
reads
\beq
D_{1\to m}^{g_i}=\det{}\left(\widetilde{\bf F}^{g_i}\right) =
\det{}\left( \begin{array}{cccccc}
 \ddots  & \ddots      & \ddots &  \vdots & \vdots &\cdots \\
 \ddots  & \widetilde{\bf F}_{i-1,i-1} & 
\widetilde{\bf F}_{i-1,i}^{g_i} & \widetilde{\bf F}_{i-1,i+1}^{g_i} 
& 0 &\cdots \\
 \ddots  & 0  & 1 & 0 & 0& \cdots \\
 \cdots  &  \widetilde{\bf F}_{i+1,i-1}^{g_i} 
 &  \widetilde {\bf F}_{i+1,i}^{g_i} & \widetilde{\bf F}_{i+1,i+1}^{g_i}&
                             {\bf F}_{i+1,i+2}^{g_i} & \ddots \\
 \cdots & 0  & 0 &{\bf F}_{i+2,i+1}^{g_i}& {\bf F}_{i+2,i+2} &\ddots \\
 \vdots & \vdots & \vdots & \ddots & \ddots & \ddots \\
\end{array} \right ) \; , \label{D-g-shifted}
\eeq
where
\beq
 \begin{array}{lcl}
   \widetilde {\bf F}_{i-1,i-1}^{g_i} &=& 
  \frac{2d_{i-2,i-1}}{\rho_{i-1}}-1 -\frac{d_{i-2,i-1}}{d_{i-1,i+1}} \\
   \widetilde {\bf F}_{i-1,i}^{g_i} &=& \cdots \\
   \widetilde {\bf F}_{i-1,i+1}^{g_i}&=&
 \left (\frac{d_{i-2,i-1}} {d_{i-1,i+1}} \right )^{1,2}\\ 
 \widetilde {\bf F}_{i+1,i-1}^{g_i} &=& 
\left (\frac{d_{i-2,i-1}} {d_{i-1,i+1}} \right )^{1,2} \\
   \widetilde {\bf F}_{i+1,i}^{g_i} &=& \cdots \\
 \widetilde {\bf F}_{i+1,i+1}^{g_i} &=&
  \frac{2d_{i-1,i+1}}{\rho_{i+1}}-1 -\frac{d_{i-1,i+1}}{d_{i+1,i+2}} \; .
 \end{array}
\eeq  
Note that we do not have to specify the elements on the $i$th row
explicitly, as
the ones on the $i$th line satisfy  
$\widetilde{\bf F}^{g_i}_{i,l}=\delta_{i,l}$.
For the same reason we can remove the $i$th line and row altogether without
affecting the result for the determinant.
In doing so, we exactly 
recover the determinant $D_{1\to i-1,i+2\to m}$ and matrix 
${\bf F}_{ 1\to i-1,i+2\to m}$ 
of the  parent itinerary of the considered ``ghost''. 
[The parent  itinerary has the same sequel of disk indices except that 
the disk $j_i$ is missing.]\
\be
D_{1\to m}^{g_i}&=&
\det{}\left( \begin{array}{ccccc}
 \ddots  & \ddots        &  \ddots & \vdots &\cdots \\
 \ddots  & {\bf F}_{i-1,i-1}  & {\bf F}_{i-1,i+1} 
& 0 &\cdots \\
 \cdots  &  {\bf F}_{i+1,i-1} 
 & {\bf F}_{i+1,i+1}&
                             {\bf F}_{i+1,i+2} & \ddots \\
 \cdots & 0 &{\bf F}_{i+2,i+1}& {\bf F}_{i+2,i+2} &\ddots \\
 \vdots & \vdots  & \ddots & \ddots & \ddots \\
\end{array} \right )\nonumber \\
& =&\det{}{\bf F}_{ 1\to i-1,i+2\to m} 
= 
D_{1\to i-1,i+2\to m}\; .
\ee
The contribution of the ghost segment itself to the total 
``stability'' of the
itinerary in the geometric limit, i.e. to the stability factor of the
corresponding periodic orbit, is just
trivially one.
As also the geometrical lengths and signs 
of both itineraries are the same,
we  have finally found that
\beq
{\bf A}^{j_1 j_2}\cdots
 {\bf A}^{j_{i-1}\underline{j_{i}}} {\bf A}^{\underline{j_{i}} j_{i+1}}  
\cdots
 {\bf A}^{j_m j_1}|_{\rm geom}
=
{\bf A}^{j_1 j_2}\cdots
 {\bf A}^{j_{i-1}j_{i+1}} \cdots
 {\bf A}^{j_m j_1}|_{\rm geom} \; ,
\eeq 
i.e., the ghost cancellation rule \equa{ghost-rule}.
Of course, the calculation of this section can trivially be extended to 
itineraries 
with more than one ghost (with and without repeats) as the operations
in Eqs.\equa{ghost-calculation}, \equa{ghost-substitution} and
\equa{D-g-shifted} are local operations involving just the segments 
with disk labels $j_i$, $j_{i-1}$ and $j_{i+1}$. Thus they can be 
performed
successively without any interference.
Furthermore, as the
transformations of the pairs $(\delta \nu_k,\delta \nu_{k+1})$ in 
\equa{ghost-substitution} can be done iteratively (and in any order) 
for $k=i,i+1,\cdots$,
the generalization to the extended ghost cancellation rule
\equa{ghost-rule-extended} is trivial as well:
\be
 &&{ {\bf A}^{j_1 j_2}
   \cdots
   {\bf A}^{j_{i-1}\underline{j_{i}}} 
   {\bf A}^{\underline{j_{i}} \underline{j_{i+1}}}
   {\bf A}^{\underline{j_{i+1}}{j_{i+2}}}
   \cdots
   {\bf A}^{j_m j_1}|_{\rm geom} }
  \nonumber\\
&=& {
   {\bf A}^{j_1 j_2}
   \cdots
   {\bf A}^{j_{i-1}\underline{j_{i}}} 
   {\bf A}^{\underline{j_{i}} {j_{i+2}}}
   \cdots
   {\bf A}^{j_m j_1}|_{\rm geom}} 
 \nonumber\\
&=& {
   {\bf A}^{j_1 j_2}
   \cdots
   {\bf A}^{j_{i-1}\underline{j_{i+1}}} 
   {\bf A}^{\underline{j_{i+1}}{j_{i+2}}}
   \cdots
   {\bf A}^{j_m j_1}|_{\rm geom} }
 \nonumber\\
&=& {
   {\bf A}^{j_1 j_2}
   \cdots
   {\bf A}^{j_{i-1}{j_{i+2}} }
   \cdots
   {\bf A}^{j_m j_1}|_{\rm geom} }
\ee
etc.
\subsection{Itineraries with creeping terms\label{chap:sc_itin_creep}}
Let us now study  an itinerary of topological length  
$m$ which has, in the semiclassical limit, $m-1$ specular reflections 
{\em and}
a left-handed or right-handed creeping contact with one disk 
(which we can put without lost of generality at the end position), i.e.
\beq
 {\bf A}^{j_1 j_2} \cdots {\bf A}^{j_{m-2}j_{m-1}} {\bf A}^{j_{m-1}
\widetilde{j_m}}
{\bf A}^{\widetilde{j_m}  j_1}|_{\rm sc+creep} \; .
 \label{creeping-itinerary}
\eeq
We mark those disk positions with creeping contributions   by a tilde. 
Using the results and methods of App.\ref{app:convol_results} and
Sec.\ref{chap:sc_appr_po}, we find the following result for the itinerary
\equa{creeping-itinerary} 
\be
\lefteqn{
 {\bf A}^{j_1 j_2} \cdots {\bf A}^{j_{m-2}j_{m-1}} {\bf A}^{j_{m-1}
\widetilde{j_m}}
{\bf A}^{\widetilde{j_m}  j_1}|_{\rm sc+creep}} \nonumber \\
&=& -\sum_{\ell_m =1}^{\infty} \sum_{s_m=\pm 1}
   \frac{\e^{\i\pi/12} C_{\ell_m}  }
              {(k a_{j_m})^{1/6}} 
  \left( \frac{a_{j_m}}{\widetilde d_{m-1,s_m \ell_m} 
 |\widetilde D^{(0)}_{1\to m-1}|} \right )^{1/2}
\frac{\e^{ \i\pi\widetilde\nu_{\ell_m}(1-\sigma_m')}}
 {1-\e^{\i 2\pi \widetilde \nu_{\ell_m}}} \nonumber \\
&& \qquad\mbox{}\times
 \e^{\i s_m \widetilde\nu_{\ell_m}\left( 
 \Delta \alpha_{j_m}-\pi \sigma_m
 +\arccos\left[ \frac{\nu_{m-1}^s-s_m\widetilde \nu_{\ell_m}}
                {kR_{j_{m-1}j_m}} \right]
 -\arccos\left[ \frac{s_m\widetilde \nu_{\ell_m}-\nu_1^s}
                 {kR_{j_m j_1}} \right] \right)} \nonumber \\
&& \qquad\mbox{}\times
 \e^{\i k[ \widetilde L_{s_m \ell_m,1}+\sum_{i=2}^{m-1} L_{i-1,i}
   +\widetilde L_{m-1,s_m\ell_m}]} \; .
 \label{result-itinerary-creep}
\ee
Here,  $\widetilde\nu_\ell\equiv\widetilde\nu_\ell(ka_{j_m})$
is the $\ell_m$th  
zero of the Hankel function $\Ho{\nu_m}{a_{j_m}}$ in the upper complex
$\nu_m$ plane [and $C_{\ell_m}\equiv C_{\ell_m}(ka_{j_m})$ is 
the creeping coefficient
as given by Eq.\equa{C_ell}, see
App.\ref{app:convol_residua}],
$\Delta\alpha_{j_m}$,
$\sigma_m$ and $\sigma_m'$ are defined in Sec.\ref{chap:sc_appr_po},
$d_{i-1,i}$, $\rho_i$ and $L_{i-1,i}$ are defined in Eqs.\equa{d-new-def} --
\equa{L_ij} and
$\nu_i^s$ is the solution of the saddle-point equation \equa{i-saddle}
[where in the cases $i$=1 and $i$=$m\mbox{$-$}1$, the
respective saddles $\nu_{i-1}^s$ 
and $\nu_{i+1}^s$  have to be replaced by
$s_m\widetilde \nu_{\ell_m}$]. 
Furthermore, the following additional definitions have been introduced
\be
\widetilde d_{m-1,s_m\ell_m} &\equiv& 
  \left\{R_{j_{m-1}j_m}^2-\left(\frac{\nu_{m-1}^s- s_m \widetilde \nu_{\ell_m}}
                           {k} \right)^2\right \}^{1/2}
 \equiv \widetilde L_{m-1,s_m\ell_m}+\rho_{m-1} \\
 \widetilde d_{s_m\ell_m,1} &\equiv& 
  \left\{ R_{j_m j_1}^2-\left(\frac{s_m \widetilde \nu_{\ell_m}-\nu_1^s}
                       {k}\right)^2\right\}^{1/2}
 \equiv \widetilde L_{s_m\ell_m,1}+\rho_1  \; 
\ee
which correspond to the geometrical lengths to the surface of disk $j_m$ 
if $\widetilde \nu_\ell$ is
approximated by $k a_{j_m}$ (see below).
Finally, 
$\widetilde D^{(0)}_{1\to m-1}$ is the determinant of the matrix
$\widetilde{\bf F}^{(0)}_{1\to m-1}$ ($\mbox{$\equiv$}\widetilde{\bf F}^{(0)}$)
with the matrix elements
\beq \begin{array}{lclcl}
  \widetilde{\bf F}_{i,i} 
          &=&\frac{2d_{i-1,i}}{\rho_i}- 1-\frac{d_{i-1,i}}{d_{i,i+1}} 
         &\equiv& a_i  \quad {\rm for}\ 2\leq i\leq m-2\\
 \widetilde{\bf F}_{1,1}
         &=&\frac{2\widetilde d_{s_m\ell_m,1}}{\rho_1}- 1
 -\frac{\widetilde d_{s_m\ell_m,1}}{d_{1,2}} &\equiv&
       \widetilde a_1 \\
 \widetilde{\bf F}_{m-1,m-1}
         &=&\frac{2 d_{m-2,m-1}}{\rho_{m-1}}- 1
 -\frac{d_{m-2,m-1}}{ d_{m-1,s_m\ell_m}} &\equiv&
       \widetilde a_{m-1} \\
  \widetilde{\bf F}_{i,i+1} 
          &=& \left(\frac{d_{i-1,i}}{d_{i,i+1}}\right)^{1/2} 
 = \widetilde{\bf F}_{i+1,i} 
 &\equiv& b_{i,i+1}  \quad {\rm for}\ 2\leq i\leq m-2 \\
  {\bf F}_{1,2}
          &=& 
 \left(\frac{\widetilde d_{s_m\ell_m,1}}{d_{m,1}^{1/2}} \right)^{1/2} = 
 \widetilde{\bf F}_{2,1} &\equiv& \widetilde b_{1,2} \\
 \widetilde {\bf F}_{i,j}&=& 0 \quad {\rm otherwise} && \end{array}  \; .
\label{F-tilde-definition}
\eeq
Note that $\widetilde D^{(0)}_{1\to m-1}$ and 
$\widetilde {\bf F}^{(0)}_{1\to m-1}$ have exactly the form of the 
determinant $D^{(0)}_{1\to m-1}$ and the matrix ${\bf F}^{(0)}_{1\to m-1}$
defined in Sec.\ref{chap:sc_itin_geom} if the tilded lengths 
$\widetilde d_{s_m\ell,1}$ and $\widetilde d_{m-1,s_m\ell_m}$ are replaced
by the ``normal'' geometrical lengths $d_{s_m m,1}$ and $d_{m-1,s_m m}$:
\be
\widetilde d_{m-1,s_m\ell_m}&\approx&
 \left\{R_{j_{m-1}j_m}^2-(s_m a_{j_m} -\nu^s_{m-1}/k)^2\right\}\equiv 
 d_{m-1,s_m m} \\
 \widetilde d_{s_m\ell_m ,1}&\approx&
 \left\{R_{j_m j_1}^2-(s_m a_{j_m} -\nu^s_{1}/k)^2\right\}\equiv 
 d_{s_m m,1} 
\ee
As discussed in App.\ref{app:convol_residua}, 
this approximation is justified in the leading Airy approximation, 
where terms of order $\hbar^{1/3}$ or
higher are anyhow neglected. 
To this order we can approximate $\widetilde\nu_{\ell_m}$ 
everywhere by 
$k a_{j_m}$, except in the ``creeping'' exponential,  since there
the error would be of order $\hbar^{-1/3}$. 
Note that, in order to be consistent, we have to approximate
$\widetilde\nu_{\ell_m}\approx k a_{j_m}$ in the saddle-point conditions
for $\nu_{m-1}^{s}$ and $\nu_1^{s}$ as well. Thus, in this approximation, 
the saddles are manifestly real. Hence only in the overall factors in the
exponents we keep the ${\cal O}(\hbar^{-1/3})$ term of
\be
 \widetilde \nu_{\ell_m} &=& k a_{j_m} +\e^{\i\frac{\pi}{3}} 
 \left(\frac{k a_{j_m}}{6} \right )^\third q_{\ell_m} \nonumber \\
 &\equiv&   k a_{j_m}+\delta\widetilde \nu_{\ell_m} \; .  
 \label{vellm}
\ee
For all the other terms the errors from neglecting 
$\delta\widetilde \nu_{\ell_m}$
are, at least, of order 
${\cal O}(\{\delta\widetilde\nu_{\ell_m}\}^2/k) ={\cal O}(\hbar^{1/3})$
or even of 
${\cal O}(\delta\widetilde\nu_{\ell_m}/k)={\cal O}(\hbar^{2/3})$.
The expansion of  the products
$k\widetilde d_{m-1,s_m\ell_m}$ and $k\widetilde d_{s_m \ell_m, 1}$
in the exponents leads to 
potentially dangerous  linear terms of order $\delta\widetilde \nu_{\ell_m}$.
However, they cancel exactly 
against the terms in the expansion of the arccosines 
combined with 
those contributions which result if $\widetilde \nu_{\ell_m}\approx ka_{j_m}$ 
is inserted into the saddle-point relations for 
$\nu_{m-1}^{s}$ and $\nu_1^{s}$.

We will show below that in the leading Airy approximation, 
$\widetilde d_{m-1,s_m\l_m}\,|\widetilde D^{(0)}_{1\to m-1}|$ 
corresponds exactly to the effective radius of the creeping periodic orbit
$R^{\rm eff}_{m\hookrightarrow  m}$
defined in Ref.\cite{vwr_prl}. The latter quantity  is constructed,
as in Eq.\equa{Reff-conv},
in terms of the length segments 
$l_{i-1,i}=L_{i-1,i}$  
between the 
$(i\mbox{$-$}1)$th and $i$th point of reflection
(if $2\leq i \leq m\mbox{$-$}1$), the length segments
$l_{m,1}=d_{m,1}\mbox{$-$}\rho_{1}$ and 
$l_{m-1,m}=d_{m-1,m}\mbox{$-$}\rho_{m-1}$
between the (creeping) impact parameter at disk $j_m$ and the first or
last point of reflection, respectively, 
and $\rho_i$ [= the radius $a_{j_i}$ of the disk $j_i$ 
{\em times}  the cosine of the
reflection angle]:
\be
 R^{\rm eff}_{m\hookrightarrow m} 
  &=& l_{m,1} \prod_{i=1}^{m-1}\left(1+\kappa_i l_{i,i+1} \right)
                          \nonumber \\
  &=& R^{\rm eff}_{m\hookrightarrow m-1} (1 +\kappa_{m-1} l_{m-1,m})\; , 
  \label{R-eff-creep}
\ee
where the curvature $\kappa_i$ is given by the recursion relation
\beq
 \kappa_i = \frac{1}{\kappa_{i-1}^{\ -1}+l_{i-1,i}} +\frac{2}{\rho_i}
 \label{kappa_i}
\eeq
with $1/\kappa_0 \equiv 0$. The proof of the equivalence
of $\widetilde d_{m-1,s_m\l_m}\,|\widetilde D^{(0)}_{1\to m-1}|$ and
$R^{\rm eff}_{m\hookrightarrow m}$ uses the following relations,
which can be derived from Eqs.\equa{R-eff-creep} and \equa{kappa_i} 
by complete induction:
\be
 R^{\rm eff}_{m\hookrightarrow j} &=& R^{\rm eff}_{m\hookrightarrow j-1}
  +\left (1 +\frac{2 R^{\rm eff}_{m\hookrightarrow 1}}{\rho_1} 
  +\frac{2 R^{\rm eff}_{m\hookrightarrow 2}}{\rho_2}
  +\cdots +\frac{2 R^{\rm eff}_{m\hookrightarrow j-1}}{\rho_{j-1}} \right ) l_{j-1,j}
 \label{R-eff-induction}\\
 \kappa_{j-1} &=& \frac{1}{R^{\rm eff}_{m\hookrightarrow j-1}}
  \left(1 +\frac{2 R^{\rm eff}_{m\hookrightarrow 1}}{\rho_1} 
  +\frac{2 R^{\rm eff}_{m\hookrightarrow 2}}{\rho_2}
  + \cdots + \frac{2 R^{\rm eff}_{m\hookrightarrow j-1}}{\rho_{j-1}} \right ) \; ,
\ee
where $R^{\rm eff}_{m \hookrightarrow 0}\equiv 0$ and $R^{\rm eff}_{m\hookrightarrow 1}=l_{m,1}$.
For the right hand side of the proof,
the recursion relations \equa{D0-recursion-full} are applied to the
combinations
$ |\widetilde D^{(0)}_{1\to j-1}| \widetilde d_{j-1,j}
  \approx D^{(0)}_{1\to j-1}d_{j-1,j}\equiv \widetilde{R}_{m\hookrightarrow j}
$:
\be
 \widetilde{R}_{m\hookrightarrow j}&=&
 -\widetilde{R}_{m\hookrightarrow j-1}+(-1)^{j-1}
\left( 1-\frac{2 \widetilde{R}_{m\hookrightarrow 1}}{\rho_1}
 +\frac{2\widetilde{R}_{m\hookrightarrow 2}}{\rho_2}-\cdots \right. 
 \nonumber \\ 
 && \qquad\qquad\qquad\qquad\qquad \left.\mbox{}
  +(-1)^{j-1}\frac{2 \widetilde{R}_{m\hookrightarrow j-1}}
{\rho_{j-1}}\right)d_{j-1,j}\; 
 \label{R-bar-recursion}
\ee
with $\widetilde{R}_{m\hookrightarrow 0}\equiv 0$.
%
In the induction assumption one can use that, 
for $1\leq j<m$. the quantities
$\widetilde{R}_{m\hookrightarrow j}$ and $R^{\rm eff}_{m\hookrightarrow j}$ are related as
\be
 \widetilde{R}_{m\hookrightarrow j}=R^{\rm eff}_{m\hookrightarrow j} 
 +\frac{R^{\rm eff}_{m\hookrightarrow j} -R^{\rm eff}_{m\hookrightarrow j-1}}{L_{j-1,j}}\,\rho_j \; .
 \label{R-bar-eff-diff}
\ee
This follows from the difference between 
$d_{j-1,j}$
and $l_{j-1,j}$.
Note that $\widetilde{R}_{m\hookrightarrow 1}=d_{m,1}\mbox{=}
L_{m,1}\mbox{+}\rho_1$ and 
$R^{\rm eff}_{m\hookrightarrow 1}=L_{m,1}$ satisfy trivially this induction ansatz.
By complete induction it can now be shown that the recursion relation
\equa{R-bar-recursion}, applied to
$\widetilde{R}_{m\hookrightarrow m}\equiv D^{(0)}_{1\to m-1}d_{m-1,m}$, 
can be rewritten as 
\be
 \widetilde{R}_{m\hookrightarrow m}&=&R^{\rm eff}_{m\hookrightarrow m-1}
+l_{m-1,m}\left \{1+\frac{2R^{\rm eff}_{m\hookrightarrow 1}}{\rho_1}+\cdots
  \frac{2R^{\rm eff}_{m\hookrightarrow i}}{\rho_i}+\cdots +
   \frac{2R^{\rm eff}_{m\hookrightarrow m-1}}{\rho_{m-1}} \right \} \nonumber\\
&& \mbox{}+
2l_{m-1,m}\left\{ \frac{(-1)^{m-1}\mbox{$-$}1}{2} 
 +\frac{(-1)^{m-2}\widetilde{R}_{m\hookrightarrow 1}\mbox{$-$}
 R^{\rm eff}_{m\hookrightarrow 1}}{\rho_1}
 +\cdots
  \right . \nonumber \\
&& \qquad\qquad\quad \mbox{} \left .
 +\frac{(-1)^{m-i-1}\widetilde{R}_{m\hookrightarrow i}\mbox{$-$}R^{\rm eff}_{m\hookrightarrow i}}
  {\rho_i}
 +\cdots
  +\frac{\widetilde{R}_{m\hookrightarrow m-1}-R^{\rm eff}_{m\hookrightarrow m-1}}{\rho_{m-1}} 
 \right \},\nonumber
\ee
where the last bracket vanishes identically because of
\be
 \frac{\widetilde{R}_{m\hookrightarrow j}-R^{\rm eff}_{m\hookrightarrow j}}{\rho_j}
=(1-\delta_{j,1})\,\frac{\widetilde{R}_{m\hookrightarrow j-1}-R^{\rm eff}_{m\hookrightarrow j-1}
+2R^{\rm eff}_{m\hookrightarrow j-1}}
{\rho_{j-1}} +\delta_{j,1} \nonumber
\ee
for $1\leq j<m$.
This equality
can be derived from the induction ansatz Eq.\equa{R-bar-eff-diff}, 
if  Eq.\equa{R-eff-induction} 
is inserted for the remaining $R^{\rm eff}_{m\hookrightarrow j}$ on the 
left hand side and  for the remaining
$R^{\rm eff}_{m\hookrightarrow j-1}$ on the right hand side.
Thus,
the equivalence
$
 \widetilde{R}_{m\hookrightarrow m}\equiv D^{(0)}_{1\to m-1}\, 
 d_{m-1,m}=R^{\rm eff}_{m\hookrightarrow m}
$
is established in the leading Airy approximation. 

We therefore get
\be
\lefteqn{
 {\bf A}^{j_1 j_2} \cdots {\bf A}^{j_{m-2}j_{m-1}} {\bf A}^{j_{m-1}
\widetilde{j_m}}
{\bf A}^{\widetilde{j_m}  j_1}|_{\rm sc+creep}} \nonumber \\
&\approx& -\sum_{\ell_m =1}^{\infty} \sum_{s_m=\pm 1}
   \frac{\e^{\i\pi/12} C_{\ell_m}  }
              {(k a_{j_m})^{1/6}} 
  \left( \frac{a_{j_m}}{R^{\rm eff}_{m\hookrightarrow m}} \right )^{1/2}
 \e^{\i k L_{m\hookrightarrow m}(s_m)} 
 \nonumber \\
&& \qquad\mbox{}\times
 \e^{\i (k a_{j_m}+\delta\widetilde\nu_{\ell_m})\left( 
 (2-\sigma_m'-s_m \sigma_m)\pi+ s_m\Delta \alpha_{j_m}
 -\arccos\left[ \frac{a_{j_m}{-}s_m{\nu_{m-1}^s}/{k}}
                {R_{j_{m-1}j_m}} \right]
 -\arccos\left[ \frac{a_{j_m}{-}s_m{\nu_1^s}/{k}}
                 {R_{j_m j_1}} \right] \right)} \nonumber \\
&& \qquad\mbox{}\times
\frac{1}
 {1-\e^{\i 2\pi(k a_{j_m}+\delta \widetilde \nu_{\ell_m})}}
\; ,
 \label{result-itinerary-creep-first-airy}
\ee
where $L_{m \hookrightarrow m}\equiv L_{m,1}+\sum_{i=2}^{m-1}L_{i-1,i}+L_{m-1,m}$ is
the total length of the straight geometrical sections. 
The impact parameter $\nu_i^s/k$ is given by the solution of the saddle-point 
equation \equa{i-saddle}
where, in the cases $i$=1 and $i$=$m\mbox{$-$}1$, the 
respective saddles $\nu_{i-1}^s/k$ 
and $\nu_{i+1}^s/k$, 
have to be replaced by
$s_m  a_{j_m}$.

In summary, all the quantities entering the semiclassical-creeping limit
of the itinerary 
\equa{result-itinerary-creep-first-airy}
with one creeping section
 have geometrical
interpretations: 
\begin{enumerate} 
\item
The integer index $\ell\geq 1$ enumerates the creeping modes around the
boundary of disk $j_m$. With increasing $\ell$, the
impact parameter (or distance of the creeping path from the surface of
the disks) and the ``tunneling''
suppression factor increases.
\item
The index $s_m=\pm 1$ distinguishes 
between creeping paths of
positive sense or negative sense around a surface section of disk $j_m$.
\item
The coefficient $\e^{\i\pi/12} C_{\ell_m}/(ka_{j_m})^{1/6}$ is proportional to
the product of the two creeping diffraction constants at the beginning and
end of the creeping segment along the boundary of disk $j_m$
which parameterize the transition from a straight section to a creeping section
and vice versa, see
\cite{franz,keller,vwr_prl}.
\item
The second prefactor is the inverse  square root of the
effective radius $R^{\rm eff}_{m\hookrightarrow m}$, in units of disk radius $a_{j_m}$.
It is the geometrical amplitude, i.e., the geometrical stability factor.
\item
The $\widetilde \nu_{\ell}$ independent terms in the exponents are
just $\i k$ {\em times} the sum of all  lengths of the straight
geometrical segments of the periodic itinerary, i.e.,
$L_{m\hookrightarrow m}(s_m)=L_{m,1}+\sum_{i=2}^{m-1}L_{i-1,i}
+L_{m-1,m}$. 
\item
The 
geometrical length along the creeping section times $\i k$ is given by 
the sum of all exponential terms that are proportional to $a_{j_m}$.
\item
The creeping ``tunneling'' suppression factor is given by the imaginary
part of $\widetilde \nu_{\ell_m}$ or $\delta\widetilde \nu_{\ell_m}$. 
\item
The denominator 
$1-\e^{\i 2\pi \widetilde \nu_{\ell_m}}$ results from the summation of
all further complete creeping turns around the disk $j_m$, in terms 
of a
geometrical series~\cite{vwr_prl}. Note that the apparent poles at 
$1-\e^{\i 2\pi \widetilde \nu_{\ell_m}}=0$ cancel against  the corresponding
semiclassical poles of one-disk S-matrix, 
${\bf S}^{1}(k a_{j_m})$. In fact, the zeros of 
$1-\e^{\i 2\pi \widetilde \nu_{\ell_m}}$ are given by  
$\widetilde\nu_{\ell_m} = l (\rm integer)$ and 
are nothing but 
the zeros of the Hankel function $\Ho{l}{a_{j_m}}$  in the Airy approximation.
\end{enumerate}

\subsection{More than one creeping section\label{chap:sc_more_creep}}
The $r$th repeat of the itinerary \equa{creeping-itinerary} follows simply
as 
the sum,  
$\sum_{\ell}\sum_{s_m}$, over the $r$th power of  the summands
on the r.h.s.\ of Eq.\equa{result-itinerary-creep}. 
As in the case of geometrical itineraries, this rule 
is trivial for the occurring prefactors, signs, phases and 
exponential terms. The non-trivial point is the behavior of the
determinant  $\widetilde D^{(0)}_{1\to m-1}$ under the $r$th repeat.
However, as the corresponding  matrix $\widetilde {\bf F}^{(0)}_{1\to m-1}$
has zero $(1,m\mbox{$-$}1)$ and $(m\mbox{$-$}1,1)$ matrix elements, such that
repeats cannot couple here,  the determinant of the $r$th repeat corresponds
exactly to the $r$th power of the determinant of the primary itinerary. 
For the same reason, also
the determinants and corresponding effective radii of itineraries, with
more than one creeping contact (i.e., with at least two disks $j_i$ and
$j_m$ with creeping contacts), decouple from each other. The corresponding
semiclassical result for such an itinerary is thus the multiple sum,
$\sum_{\ell_i}\sum_{s_i}\sum_{\ell_m}\sum_{s_m}$ over the products of the
corresponding itinerary from disk $j_i$ to disk $j_m$ and the
itinerary from disk $j_m$ to disk $j_i$, each individually 
given by the suitably adjusted summand on 
the r.h.s. of Eq.\equa{result-itinerary-creep}, e.g.:
\be
\lefteqn{
 {\bf A}^{j_1 j_2} 
  \cdots 
 {\bf A}^{j_{i-1} \widetilde{j_i}}
{\bf A}^{\widetilde{j_i} j_{i+1}}
\cdots
 {\bf A}^{j_{m-1}
\widetilde{j_m}}
{\bf A}^{\widetilde{j_m}  j_1}|_{\rm sc+creep}} \nonumber \\
&\approx& 
\sum_{\ell_i =1}^{\infty} \sum_{s_i=\pm 1}
\sum_{\ell_m =1}^{\infty} \sum_{s_m=\pm 1}\nonumber \\
&& \qquad\mbox{}\times (-1)
   \frac{\e^{\i\pi/12} C_{\ell_i}  }
              {(k a_{j_i})^{1/6}} 
  \left( \frac{a_{j_i}}{R^{\rm eff}_{i\to m}} \right )^{1/2}
 \e^{\i k L_{i\to m}(s_i,s_m)} \,
\frac{1}
 {1-\e^{\i 2\pi(k a_{j_i}+\delta \widetilde \nu_{\ell_i})}}
 \nonumber \\
&& \qquad\mbox{}\times
 \e^{\i (k a_{j_i}+\delta\widetilde\nu_{\ell_i})\left( 
(2-\sigma_i'-s_i\sigma_i)\pi
 +s_i\Delta \alpha_{j_i}
 -\arccos\left[ \frac{a_{j_i} -s_i\nu_{i-1}^s/k}
                {R_{j_{i-1}j_i}} \right]
 -\arccos\left[ \frac{a_{j_i}-s_i\nu_{i+1}^s/k}
                 {R_{j_i j_{i+1}}} \right] \right)} \nonumber \\
&& \qquad\mbox{}\times (-1)
   \frac{\e^{\i\pi/12} C_{\ell_m}  }
              {(k a_{j_m})^{1/6}} 
  \left( \frac{a_{j_m}}{R^{\rm eff}_{m\hookrightarrow i}} \right )^{1/2}
 \e^{\i k L_{m\hookrightarrow i}(s_m,s_i)} 
\,\frac{1}
 {1-\e^{\i 2\pi (ka_{j_m}+\delta\widetilde \nu_{\ell_m})}}
 \nonumber \\
&& \qquad\mbox{}\times
 \e^{\i (ka_{j_m}+\delta\widetilde\nu_{\ell_m})\left(
(2-\sigma_m'-s_m\sigma_m)\pi 
 +s_m\Delta \alpha_{j_m}
 -\arccos\left[ \frac{a_{j_m} -s_m\nu_{m-1}^s/k}
                {R_{j_{m-1}j_m}} \right]
 -\arccos\left[ \frac{a_{j_m}-s_m\nu_1^s/k}
                 {R_{j_m j_1}} \right] \right)}\; , 
 \nonumber
\ee
etc. [If there are two (or more) 
creeping contacts next to each other, e.g., $j_i=
j_{m-1}$, then, in the above formula, 
the corresponding impact parameters $\nu_{i+1}^s/k$ and $\nu_{m-1}^s/k$ 
have to be replaced by $a_{j_{m}}$ and $a_{j_{m-1}}$, respectively.]

The physical reason for the simple rule of piecing together creeping paths,
is the 
point-like contact at e.g.\ disk $j_i$ 
between the
creeping sections on the one hand and the geometrical sections on the other
hand which is mediated by the diffraction constants $C_{\ell_j}$.
[Mathematically, this corresponds to the fact that $\widetilde \nu_{\ell_j}$ 
is uniquely determined as
the $\ell_j$th zero of the Airy integral and {\em not}
by a semiclassical saddle-point equation that
would couple with the saddle-point equations at the disks $j_{i-1}$ and 
$j_{i+1}$.]\
Because of this point-like contact [the independent determination 
of $\widetilde\nu_\ell$] the semiclassical  itineraries multiply for fixed
value of the mode numbers $\ell_j$ and creeping orientation $s_j$.
Especially, if we limit the mode number to $\ell=1$, periodic orbits
with common creeping sections can exactly be split up into their primary
periodic orbits, see Ref.\cite{vwr_prl}.

Finally, the ghost cancellation works for itineraries with creeping sections
in the same way 
as for itineraries which, semiclassically, are purely geometrical.
The reason is two-fold: First, by construction 
(see App.\ref{app:convol_results}), ghost segments 
can only occur in  the 
geometrical part of the creeping itinerary. Second, the ghost cancellation
rules of 
Sec.\ref{chap:sc_ghost_rule} are based
on the local properties of the segments $i\mbox{$-$}1\to i$ and 
$i\to i\mbox{+}1$. Let us now assume, for simplicity, 
that the disk $j_i$ is cut by the ghost
section. 
If there is no creeping at the neighboring disks $j_{i-1}$ and 
$j_{i+1}$, the reduction of the stability matrix $\widetilde{\bf F}^{(0)}$
and
of the 
phases and lengths of the segments is precisely the same as
in the purely geometrical case (see the substitutions 
\equa{ghost-substitution} and the analogous steps of 
Sec.\ref{chap:sc_ghost_rule}). If there is a creeping 
contact at disk  $j_{i-1}$
or/and disk $j_{i+1}$, the substitutions \equa{ghost-substitution} simplify,
as $\delta \nu_{i-1}$ or/and $\delta \nu_{i+1}$ do not exist. 
Thus, the elements ${\bf F}_{i-1,i}^{g_i}$ or/and
${\bf F}_{i+1,i+1}^{g_i}$ of Eq.\equa{F-g-def} are zero and the 
$i$th row of the determinant \equa{D-g-start} has only to be subtracted from 
the $(i+1)$th or the $(i-1)$th row or from none, 
in order that Eq.\equa{D-g-start} becomes
Eq.\equa{D-g-shifted}. The reduction in the lengths and phases hold in
these cases as before.

In summary, the ghost cancellation works 
for geometrical orbits with creeping sections
as well as for purely geometrical orbits, studied in 
Sec.\ref{chap:sc_ghost_rule}.
Semiclassically, neither the ghost itineraries nor their parent itineraries
[which have the same symbol sequence except that the  ghost labels are
removed subsequently]
contribute to the semiclassical trace-log expansion and to 
the cumulant/curvature expansion.
Thus, one can omit these ``ghost-affected'' periodic orbits altogether from
the curvature expansion.

Deep inside the negative complex $k$-plane the limitations of the first
Airy correction introduce rather big errors, see Ref.\cite{FranzGalle}. 
In this case it is advisable
to use to the original expression  \equa{result-itinerary-creep}
for semiclassical ``creeping'' itineraries 
with $\widetilde \nu_{\ell_m}$ and $C_{\ell_m}$, as given in 
Eqs.\equa{w:nuzero} and \equa{C_ell}, instead of
Eq.\equa{result-itinerary-creep-first-airy}.

To summarize,  for the special case of $n$-disk repellers,
the creeping periodic orbits of Ref.\cite{vwr_prl}
have been recovered directly from quantum mechanics, whereas the
construction of Ref.\cite{vwr_prl} has relied on Keller's semiclassical
theory of diffraction~\cite{keller}. Furthermore, 
the symbol dynamics has to be generalized from the
single-letter labelling $\{j_i\}$ to the two-letter labelling
$\{j_i, s_i \times \ell_i\}$ with $s_i=0,\pm 1$ and $\ell_i=1,2,3,\cdots$.

\subsection{Geometrical stabilities\label{chap:sc_geom_stab}}
In this subsection we will return to purely geometrical periodic orbits and
show that 
 Eqs.\equa{D-Lambda-repeat} and
\equa{D-monodromy-matrix} are correct, i.e. 
that the determinant $D_{1\to m}$ satisfies in fact
\be
 D_{1\to m} &=& (-1)^{m+1}\det{} ( {\bf M}_{1\to m} - {\bf 1} )\nonumber \\
            &=& (-1)^m \left ( \Lambda_{1\to m}+ \frac{1}{\Lambda_{1\to m}}
                   -2 \right ) \; 
 \label{D-compare-monodromy}
\ee
irrespective, whether there are repeats or not.
Here ${\bf M}_{1\to m}$ is the 2$\times$2 dimensional real monodromy matrix of
the purely geometrical periodic orbit of total topological length $m$ ($m=pr$
if there are $r$ repeats of a primary orbit of topological length $p$), 
that is, the  semiclassical
limit of the itinerary $A^{j_m j_1} \cdots A^{j_m j_1}$. Because of 
phase-space
conservation, the determinant of ${\bf M}_{1\to m}$ is unity. For this
reason and as the
matrix elements of ${\bf M}_{1\to m}$ are real (see below), 
the two eigenvalues of the
matrix are related as
$ \Lambda_{1\to m}$ and $1/ \Lambda_{1\to m}$. 
We do not have to treat repeated orbits explicitly here, as this case was
already studied in  Sec.\ref{chap:sc_itin_repeats}.

In Ref.\cite{pinball} it was shown that, for  any two-dimensional 
scalar billiard problem (whether a bound state problem 
or a scattering problem),
the monodromy matrix 
${\bf M}_{1\to m}$ of a periodic orbit 
with $m$ collisions with the billiard walls is
given by the 2$\times$2 dimensional 
Jacobian belonging
to the infinitesimal 
evolution of the vector $( \delta p_\perp,\delta x_\perp)^T$
perpendicular to this classical trajectory in phase space, i.e., by the
product
\be
 {\bf M}_{1\to m} = \prod_{i=1}^{m} {\bf T}_{i-1,i} {\bf R}_i \; .
 \label{Jacobian}
\ee
Here the matrix 
\beq
 {\bf T}_{i-1,i} =\left(  \begin{array}{c c} 1 & 0 \\
                                              L_{i-1,i} & 1 \end{array} 
        \right)
 \label{J-translation}
\eeq
parametrizes the  translational (straight ray) evolution
of the vector $( \delta p_\perp, \delta x_\perp)^T$  
[or rather  $(\delta \theta_p,\delta x_\perp)^T$ with $\theta_p$ 
being the
angle of the momentum $p$, since  the modulus of $p$ is conserved anyhow]
between the $(i\mbox{$-$}1)$th and $i$th collision where $L_{i-1,i}$ is the
corresponding length segment. As usual $i\mbox{=}0$ should be identified with 
$i\mbox{=}m$. 
The matrix 
\beq
 {\bf R}_{i} =\left(  \begin{array}{c c}
                        -1 & -2/\rho_i \\
                        0  & -1 \end{array} 
        \right)
 \label{J-reflection}
\eeq
parametrizes the evolution of the vector $(\delta \theta_p,\delta x_\perp)^T$
from immediately before to immediately after the $i$th
collision. The quantity $\rho_i\mbox{=}a_i \cos \theta_i$ is, in general, 
the product of the local radius of curvature $a_i$ and the cosine of
the reflection angle $\theta_i$ at the $i$th collision with the billiard
walls. Especially for our  
$n$-disk scattering problems,
$a_i$ is of course nothing but $a_{j_i}$, the radius of the disk $j_i$,
whereas $\theta_i$ should be identified with the scattering angle 
$\theta_{j_i}$ of Eq.\equa{i-refl-angle}, the solution of the saddle-point
equation.  Since the determinants $\det{}{\bf T}_{i-1,i}$ and 
$\det{}{\bf R}_{i}$ are trivially unity, the determinant of the product 
${\bf M}_{1\to m}$ is
unity as well, as it should because of Liouville's theorem. Furthermore,
the matrix elements
of ${\bf M}_{1\to m}$ have to be real, since the matrices 
$\det{}{\bf T}_{i-1,i}$ and  $\det{}{\bf R}_{i}$ are real, by definition. 
Thus the two  eigenvalues of   ${\bf M}_{1\to m}$ have the structure
$\Lambda_{1\to m}$ and $1/\Lambda_{1\to m}$.

\subsubsection{Monodromy matrix in closed form}
In the following we will construct 
a closed-form expression for the matrix elements of the 
matrix  ${\bf M}_{1\to n}$,
$1\leq n$,  by complete induction. 
Let us denote these matrix elements as
\beq
 {\bf M}_{1\to n} \equiv (-1)^n\left ( \begin{array}{cc} A_n & B_n \\
                                                     C_n & D_n \end{array}
                             \right ) \; . 
\eeq 
By inserting Eqs. \equa{J-translation} and \equa{J-reflection} into
Eq.\equa{Jacobian}, one can show that
\beq
 \begin{array}{lcl}
 A_1 &=& 1\\
 B_1 &=& \frac{2}{\rho_1}\\
 C_1 &=& L_{0,1}\\
 D_1 &=& 1 + \frac{2 L_{0,1}}{\rho_1} \end{array}
 \label{m-e-induction-start}
\eeq
and that 
the matrix elements of  ${\bf M}_{1\to n+1}$ and ${\bf M}_{1\to n}$ are
related as follows:
\be
    A_{n+1}&=& A_n + B_n L_{n,n+1} \label{A-induction}\\
    B_{n+1}&=& B_n \left(1 + \frac{2 L_{n,n+1}}{\rho_{n+1}} \right ) 
               +A_n \frac{2}{\rho_{n+1}} \label{B-induction}\\
    C_{n+1}&=& C_n + D_n L_{n,n+1} \label{C-induction}\\
    D_{n+1}&=& D_n \left(1 + \frac{2 L_{n,n+1}}{\rho_{n+1}} \right ) 
               +C_n \frac{2}{\rho_{n+1}} \; . \label{m-e-induction-step}
\ee
In order to be able to perform the induction
step, we do not make use of the cyclic permutation, i.e., in the
following we do not 
replace   
$L_{0,1}$ with $L_{n,1}$ or $L_{n+1,1}$, respectively, and 
$L_{n,n+1}$ or $L_{n+1,n+2}$ with $L_{n,1}$, 
but keep the original labelling.
From Eqs.\equa{m-e-induction-start} and \equa{m-e-induction-step} it 
follows, by
complete induction, that
\be
 A_n &=& 1 +\sum_{i=1}^{n-1} \frac{2 R^{\rm eff}_{i \to n}}{\rho_i}
 \label{A-result} \\
 B_n &=& \sum_{i=1}^{n}\left( 1 +\sum_{j=i+1}^n 
\frac{2 R^{\rm eff}_{i\to j}}{\rho_j}
  \right )\frac{2}{\rho_i} = \sum_{i=1}^n 
 \frac{ R^{\rm eff}_{i\to n+1}-R^{\rm eff}_{i\to n}}{L_{n,n+1}}\,
   \frac{2}{\rho_i } \label{B-result}
 \\
 C_n &=& R^{\rm eff}_{0\to n}  \label{C-result}\\
 D_n &=& 1 + \sum_{i=1}^n \frac{2 R^{\rm eff}_{0\to i}}{\rho_i}
 \label{D-result}
\ee
with $R^{\rm eff}_{j\to n}$ given as in Eq.\equa{R-eff-creep} where we 
identify
$l_{i-1,i}$ with $L_{j+i-1,j+i}$. In analogy to Eq.\equa{R-eff-induction}, 
we can derive the recursion relation 
\be
R^{\rm eff}_{j\to n} &=& R^{\rm eff}_{j\to n-1}
+\left (1 +\sum_{i=1}^{n-1-j} R^{\rm eff}_{j\to j+i} \frac{2}{\rho_{j+i}}
 \right ) L_{n-1,n}
 \label{R-eff-recursion} \; ,
\ee
where $R^{\rm eff}_{j\to n}|_{j\geq n} \equiv 0$. 
Thus $R^{\rm eff}_{n\to n}$ should not
be mixed up with the quantity $R^{\rm eff}_{n\hookrightarrow n}$ of
Sec.\ref{chap:sc_itin_creep} that rather corresponds here to 
$R^{\rm eff}_{0\to n}$ with $l_{0,1}\equiv L_{0,1}$, of course.
Note that the first iteration of Eq.\equa{R-eff-recursion} leads to 
$R^{\rm eff}_{j\to j+1}\equiv L_{j,j+1}$.
For later purposes we also define here the effective radius 
$R^{\rm eff}_{j\leftarrow n}$ which is, of course, equal to 
$R^{\rm eff}_{j\to n}$ and which satisfies the recursion relation
\be
R^{\rm eff}_{j\leftarrow n} &=& R^{\rm eff}_{j+1\leftarrow n}
+L_{j,j+1}\left (1 +\sum_{i=1}^{n-1-j}
 \frac{2}{\rho_{n-i}} R^{\rm eff}_{n-i\leftarrow n} \right ) 
 \label{R-eff-recursion-left} \; ,
\ee
where again $R^{\rm eff}_{j\leftarrow n}|_{j\geq n}\equiv 0$, such that
$ R^{\rm eff}_{n-1\leftarrow n}= L_{n-1,n}$.

The second equation of Eq.\equa{B-result} follows trivially from
\equa{R-eff-recursion}.
By inserting the Ans\"{a}tze \equa{C-result} and \equa{D-result} into
the induction step \equa{C-induction}, one can easily show, with the
help of the recursion relation \equa{R-eff-recursion} (for the case $j=0$), 
that the result for
$C_{n+1}$ is given by 
Eq.\equa{C-result},   with $n$ replaced by $n+1$.  
Similarly, by inserting the  Ans\"{a}tze \equa{A-result} and the
last identity of \equa{B-result} into the induction step \equa{A-induction},
one finds that $A_{n+1}$ is compatible with  \equa{A-result}. Here we
used the identity
$R^{\rm eff}_{n\to n} \equiv 0$. Applying the recursion relation 
\equa{R-eff-recursion} to $R^{\rm eff}_{0\to n+1}$, it easy to show that
$D_{n+1}$ is compatible with Eq.\equa{C-result} as well. Finally, for proving
that  $B_{n+1}$ is compatible with Eq.\equa{B-result}, one inserts the
first equation of \equa{B-result}  and \equa{A-result} 
into \equa{B-induction}, uses Eq.\equa{R-eff-recursion}
for re-expressing $R^{\rm eff}_{i\to n+1}$ and the fact that 
$L_{n,n+1}=R^{\rm eff}_{n\to n+1}$.

Having a closed form expression for the matrix elements of ${\bf M}_{1\to m}$
we could now construct the corresponding eigenvalues 
$\Lambda_{1\to m}^{\pm 1}$. But, in fact, we only need the linear combination
$(-1)^m(\Lambda_{1\to m}+1/\Lambda_{1\to m})$ which is equal to 
the sum $A_m+D_m$.

In summary, we have now a closed form expression for the right hand sides
of Eq.\equa{D-compare-monodromy} 
\be
(-1)^{m+1} \det{} ({\bf M}_{1\to m} - {\bf 1} ) &=& 
 (-1)^m \left (\Lambda_{1\to m}+\frac{1}{\Lambda_{1\to m}} -2 \right ) 
 \nonumber\\
 &=& 2(1+(-1)^{m+1}) +\sum_{i=1}^{m} R^{\rm eff}_{0\to i} \frac{2}{\rho_i}
     +\sum_{i=1}^{m-1}\frac{2 }{\rho_i} R^{\rm eff}_{i\leftarrow m}
 \; ,
 \label{D-compare-monodromy-right}
\ee
where we  used the identity 
$R^{\rm eff}_{i\to m}= R^{\rm eff}_{i\leftarrow m}$ in writing down 
the last relation. 

\subsubsection{Stability determinant in closed form}
In analogy to the definitions of Sec.\ref{chap:sc_itin_creep} 
[see Eq.\equa{R-bar-recursion}] we define here
$D^{(0)}_{l+1\to k-1}d_{k-1,k} \equiv 
{\overline R}_{l\to k}\equiv {\overline R}_{l\leftarrow k}$. These quantities
satisfy,
according to  Eq.\equa{D0-recursion-1}, the recursion relations
\be
{\overline R}_{l\to k}+{\overline R}_{l\to k-1}
 &=& \left ({\overline R}_{l\to k-1}\frac{2}{\rho_{k-1}}
-\left [ {\overline R}_{l\to k-1}+{\overline R}_{l\to k-2}\right ]
  \frac{1}{d_{k-2,k-1}} \right )d_{k-1,k} 
\ee
and, according to Eq.\equa{D0-recursion-2}, the recursion relations
\be
{\overline R}_{l\leftarrow k}+{\overline R}_{l+1\leftarrow k}
 &=& d_{l,l+1}\left (\frac{2}{\rho_{l+1}}{\overline R}_{l+1\leftarrow k} 
-\frac{1}{d_{l+1,l+2}}
 \left [ {\overline R}_{l+1\leftarrow k}+{\overline R}_{l+2\leftarrow k}
\right ]
  \right ) \; .
\ee
By complete induction, these 
recursion relations can be summed up to give
\be     
{\overline R}_{l\to k}&=& -{\overline R}_{l\to k-1}
+(-1)^{k-l-1}
\left( 1+\sum_{i=1}^{k-l-1} (-1)^i \,\frac{2{\overline R}_{l\to l+i}}
                                        {\rho_{l+i}} \right ) d_{k-1,k}
 \label{R-bar-recursion-full} \\
= {\overline R}_{l\leftarrow k} 
&=& -{\overline R}_{l+1\leftarrow k}+(-1)^{k-l-1}\,d_{l,l+1}
\left( 1+\sum_{i=1}^{k-l-1} (-1)^i \,\frac{2{\overline R}_{k-i\leftarrow k}}
                                        {\rho_{k-i}} \right ) 
 \label{R-bar-recursion-full-left}
\ee
with ${\overline R}_{l\to k}|_{l\geq k}
={\overline R}_{l\leftarrow k}|_{l\geq k}=0$.

According to Eq.\equa{D-full} of Sec.\ref{chap:sc_itin_geom}, 
the stability determinant $D_{1\to m}$ can
be rewritten in terms of  the ${\overline R}_{l\to k}$'s as follows 
\beq
 D_{1\to m}= \sum_{i=1}^{m} (-1)^{m-i}\, 
           {\overline R}_{0\to i}\frac{2}{\rho_i} 
             +\sum_{i=1}^{m-1} (-1)^i\,\frac{2}{\rho_i} 
       {\overline R}_{i\leftarrow m} \; .      
 \label{D-compare-stability-left}
\eeq
Adding and subtracting Eq.\equa{D-compare-monodromy-right}
to \equa{D-compare-stability-left} we get
\be
  D_{1\to m}&=& 
 2(1\mbox{+}(-1)^{m+1}) +\sum_{i=1}^{m} R^{\rm eff}_{0\to i} \frac{2}{\rho_i}
     \mbox{+}
 \sum_{i=1}^{m-1}\frac{2 }{\rho_i} R^{\rm eff}_{i\leftarrow m} 
 \nonumber \\
&& \mbox{}+2((-1)^m\mbox{$-$}1)
\sum_{i=1}^m \left ((-1)^{m-i}\, {\overline R}_{0\to i}
 \mbox{$-$}R^{\rm eff}_{0\to i} \right )\frac{2}{\rho_i}
   + \sum_{i=1}^{m-1}\frac{2}{\rho_i} 
 \left ( (-1)^i\, {\overline R}_{i\leftarrow m}
\mbox{$-$}R^{\rm eff}_{i\leftarrow m}
 \right ) \; . \nonumber \\
 \label{D-compare-left-right}
\ee
The equality \equa{D-compare-monodromy} is established, if we can
show that the second line of \equa{D-compare-left-right} is identically zero.

Note that 
the effective radius ${\widetilde R}_{m\hookrightarrow j}$,
in the
creeping case, fulfills the recursion relations 
\equa{R-bar-recursion-full} and \equa{R-bar-recursion-full-left} as well,
see e.g., Eq.\equa{R-bar-recursion}.  However, as here 
$d_{0,1}(\equiv d_{m,1}) = \rho_{0}+L_{0,1} +\rho_{1}$ and
$d_{m-1,m} =\rho_{m-1}+L_{m-1,m}+d_{m}$, 
whereas in Sec.\ref{chap:sc_itin_creep} $d_{0,1}=L_{0,1} +\rho_{1}$ and
$d_{m-1,m}=\rho_{m-1}+L_{m-1,m}$, 
the relation between the
${\overline R}_{i \to j}$'s and the $R^{\rm eff}_{i\to j}$'s have  to be 
modified in comparison to the relation \equa{R-bar-eff-diff}
between the 
 ${\widetilde R}_{m\hookrightarrow j}$'s and the 
$R^{\rm eff}_{m\hookrightarrow j}$'s.
In fact, instead of Eq.\equa{R-bar-eff-diff},
we get
\be
{\overline R}_{0\to i} &=& 
 \left\{ 1 + \rho_0 \frac{\partial}{\partial L_{0,1}}\right \}
 \left (  R^{\rm eff}_{0\to i}+
 \frac{ R^{\rm eff}_{0\to i}- R^{\rm eff}_{0\to i-1}}{L_{i-1,i}} \rho_i \right)
 \label{R-bar-eff-to-right}
\\
{\overline R}_{i\leftarrow m} &=&
 \left\{ 1 + \rho_m \frac{\partial}{\partial L_{m-1,m}}\right \}
\left (  R^{\rm eff}_{i\leftarrow m}+
 \frac{ R^{\rm eff}_{i\leftarrow m}- R^{\rm eff}_{i+1\leftarrow m}}
 {L_{i-1,i}} \rho_i \right)\; ,
 \label{R-bar-eff-to-left}
\ee
where the differentiations with respect to $L_{0,1}$ and $L_{m-1,m}$ produce
the additional $\rho_0$ and
$\rho_m$ pieces in $d_{0,1}$
and $d_{m-1,m}$, respectively, relative to Eq.\equa{R-bar-eff-diff}.
As in Sec.\ref{chap:sc_itin_creep}, these relations can be proven by complete
induction.

Now, by solely 
inserting Eqs.\equa{R-bar-eff-to-right} and \equa{R-bar-eff-to-left}
into the second line of Eq.\equa{D-compare-left-right} and collecting terms, 
we get for this expression 
\be
\lefteqn{ 2((-1)^m\mbox{$-$}1) + \sum_{i=1}^m 
 \left ((-1)^{m-i}\, {\overline R}_{0\to i}
 \mbox{$-$} R^{\rm eff}_{0\to i} \right )\frac{2}{\rho_i}
   + \sum_{i=1}^{m-1}\frac{2}{\rho_i} 
 \left ( (-1)^i\, {\overline R}_{i\leftarrow m}\mbox{$-$} 
R^{\rm eff}_{i\leftarrow m}
 \right )}\nonumber \\
&=&
 \rho_0 \frac{\partial}{\partial L_{0,1}}
 \left ( 1\mbox{+}\!\!\sum_{i=1}^{m-1}  
 \frac{2 R^{\rm eff}_{0\to i}}{\rho_i}\right )
+\rho_0 \frac{\partial}{\partial L_{0,1}} 
                                  \frac{2R^{\rm eff}_{0\to m} }{\rho_m}
-\left\{ 2\mbox{+}\rho_m \frac{\partial}{\partial L_{m-1,m}} \right \}
\left (1\mbox{+}\!\!\sum_{i=1}^{m-1}
 \frac{2 R^{\rm eff}_{i\leftarrow m}}{\rho_i} \right )  .
\nonumber \\
 \label{to-show}
\ee 
With the help of the recursion relations \equa{R-eff-recursion} and
\equa{R-eff-recursion-left}  
this expression can be rewritten as follows
\be
\equa{to-show} &=& \rho_0 \frac{\partial}{\partial L_{0,1}}
{
\left ( 
  \frac{R^{\rm eff}_{0\to m}\mbox{$-$}R^{\rm eff}_{0\to m-1}}
 {L_{m-1,m}}\right )
 }
+ 2 \frac{\partial}{\partial L_{0,1}} 
        R^{\rm eff}_{0\to m} \nonumber \\
 && \qquad\qquad\qquad\mbox{}
-\left\{ 2\mbox{+}\rho_m\frac{\partial}{\partial L_{m-1,m}} \right \}
 { \left( 
 \frac{R^{\rm eff}_{0\leftarrow m}\mbox{$-$}R^{\rm eff}_{1\leftarrow m}}
         {L_{0,1}}\right)}  \nonumber \\
&=&\underbrace{
 \left\{ 1\mbox{+}\rho_0 \frac{\partial}{\partial L_{0,1}} \right\}
   \left( \frac{R^{\rm eff}_{0\to m}}{\rho_m}+
      \frac{R^{\rm eff}_{0\to m}\mbox{$-$}R^{\rm eff}_{0\to m-1}}{L_{m-1,m}} 
  \right )}_{ {\overline R}_{0\to m}}
-\frac{R^{\rm eff}_{0\to m}}{\rho_m}
-\frac{R^{\rm eff}_{0\to m}\mbox{$-$}R^{\rm eff}_{0\to m-1}}{L_{m-1,m}}
\nonumber \\
&&\mbox{}
-\frac{\rho_0}{\rho_m}\frac{\partial}{\partial L_{0,1}}R^{\rm eff}_{0\to m}
+2\frac{\partial}{\partial L_{0,1}}R^{\rm eff}_{0\to m}
  +\frac{\partial}{\partial L_{m-1,m}}R^{\rm eff}_{0\leftarrow m}
 -\frac{R^{\rm eff}_{0\leftarrow m}\mbox{$-$}R^{\rm eff}_{1\leftarrow m}}
   {L_{0,1}}
 \nonumber \\
&&\mbox{} 
+\frac{R^{\rm eff}_{0\leftarrow m}}{\rho_0}
-\underbrace{
 \left\{ 1+\rho_m \frac{\partial}{\partial L_{m-1,m}} \right\}
   \left( \frac{R^{\rm eff}_{0\leftarrow m}}{\rho_0}+
      \frac{R^{\rm eff}_{0\leftarrow m}-R^{\rm eff}_{0\leftarrow m-1}}
   {L_{0,1}} 
  \right )}_{ {\overline R}_{0\leftarrow m}}
 \; . \label{Next-step}
\ee
Under 
$\rho_0\mbox{$\equiv$}\rho_m$, ${\overline R}_{0\to m}\mbox{=}
{\overline R}_{0\leftarrow m}$ and 
$R^{\rm eff}_{0\to m}\mbox{=}
R^{\rm eff}_{0\leftarrow m}$, the expression simplifies
[furthermore, note that  $R^{\rm eff}_{m-i\leftarrow m}$  
and $R^{\rm eff}_{0\to i}$ are independent of
$L_{0,1}$ and  $L_{m-1,m}$, respectively, if $i\leq m-1$]:  
\be
\equa{Next-step}&=& 
\frac{\partial}{\partial L_{0,1}}R^{\rm eff}_{0\leftarrow m}
-\frac{R^{\rm eff}_{0\leftarrow m}\mbox{$-$}R^{\rm eff}_{1\leftarrow m}}
 {L_{0,1}}
+\frac{\partial}{\partial L_{m-1,m}}R^{\rm eff}_{0\to m}
-\frac{R^{\rm eff}_{0\to m}\mbox{$-$}R^{\rm eff}_{0\to m-1}}{L_{m-1,m}}
 \nonumber \\
&=& \frac{\partial}{\partial L_{0,1}}
 \left[  R^{\rm eff}_{1\leftarrow m}
\mbox{+}L_{0,1}
 \left ( 1\mbox{+}\sum_{i=1}^{m-1} \frac{2 R^{\rm eff}_{m-i\leftarrow m}}
                                    {\rho_{m-i}} \right ) 
  \right] 
- \left ( 1\mbox{+}\sum_{i=1}^{m-1} \frac{2 R^{\rm eff}_{m-i\leftarrow m}}
                                    {\rho_{m-i}} \right ) \nonumber \\
&& \mbox{}+ \frac{\partial}{\partial L_{m-1,m}}
 \left[  
R^{\rm eff}_{0\to m-1}
\mbox{+}L_{m-1,m}
 \left ( 1 \mbox{+}\sum_{i=1}^{m-1} \frac{2 R^{\rm eff}_{0\to i}}
                                    {\rho_{i}} \right ) 
  \right]
 - \left ( 1 \mbox{+}\sum_{i=1}^{m-1} \frac{2 R^{\rm eff}_{0\to i}}
                                    {\rho_{i}} \right ) 
\nonumber \\
&=& 0 \quad {\rm q.\, e.\, d.}
\ee
The identity \equa{D-compare-monodromy} is therefore established. 
In summary, we have proven that the geometrical semiclassical 
limit of a quantum 
itinerary for any non-overlapping $n$-disk system 
[see Eq.\equa{geometric-fluctuations}]
is exactly 
the corresponding periodic orbit with the Gutzwiller weight. 
Hence, the
validity of
Eq.\equa{itinerary-to-cycle} for {\em any} non-overlapping finite 
$n$-disk system (with the exclusion of the grazing geometries)
is shown in the semiclassical limit. 
Note, however, that this is no general 
proof of the convergence of that the curvature series, 
since two limits are involved: the semiclassical limit 
$p/\hbar= k\to \infty$ (or $\hbar \to 0$) and the cumulant limit $m\to \infty$.
In general, these two limits do not commute. For purely chaotic classical
$n$-disk systems with a positive value for the topological entropy, the
exponential proliferating number of orbits and, therefore of classical input,
is not compatible to the  just 
algebraically rasing number of operations, needed to solve for the zeros of
the quantum determinant of the multi-scattering kernel. In these cases,
the curvature sum of the periodic orbits has to deviate from the cumulant
sum involving the quantum itineraries. The semiclassical limit and the
cumulant limit should better not commute. 
We will study this numerically in the next section.
\newpage
\section{Numerical tests of semiclassical 
curvature expansions against exact data\label{chap:numerical}}  
\setcounter{equation}{0}
\setcounter{figure}{0}
\setcounter{table}{0}
In this section which overlaps partly with Ref.\,\cite{cvw96}
we test the predictions of the curvature expanded 
Gutzwiller-Voros zeta function, the 
dynamical zeta function~\cite{ruelle} and the quasiclassical zeta function of 
Refs.\cite{fredh,fredh2}
against
the exact quantum-mechanical data for 
the 3-disk-system
in the ${\rm A}_{\rm 1}$-representation.

\begin{figure}[htb]
 \centerline {\epsfig{file=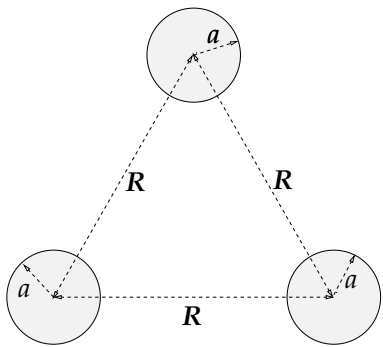,height=4.5cm,angle=-90}}
\caption[fig:3-disk-geo]{\small The three-disk system with center-to-center
separation $R=6a$.
\label{fig:3-disk-geo}}
\end{figure}

As mentioned in the introduction,
the 3-disk repeller~\cite{Eck_org,gr_cl,gr,Cvi_Eck_89,pinball,fredh}
is one of the simplest, classically completely chaotic,
scattering systems and provides a convenient numerical laboratory for 
computing exact quantum-mechanical spectra
as well as for testing the semiclassical ideas.
It consists of a free point particle which moves in the two-dimensional
plane and which scatters off 
three identical hard 
disks of radius $a$ centered at the corners of an equilateral
triangle of side length $R$,
see Fig.\ref{fig:3-disk-geo}. 
The discrete ${\rm C}_{\rm 3v}$ symmetry reduces the
dynamics to motion in the fundamental domain (which is a $1/6$th slice 
of the full
domain and which exactly contains one half of one disk), 
and the spectroscopy to irreducible subspaces
${\rm A}_{\rm 1}$, ${\rm A}_{\rm 2}$ and ${\rm E}$. 
All our calculations are performed for the fully
symmetric subspace ${\rm A}_{\rm 1}$~\cite{gr,aw_chaos}.

The genuine multiscattering data in the ${\rm A}_{\rm 1}$ subspace are computed
from
the determinant $\det{}[1 + {\bf A}_{{\rm A}_{\rm 1}}(k)]$ where the
multiscattering kernel
${\bf A}_{{\rm A}_{\rm 1}}(k)$, expressed in 
the angular momentum basis relative to the half-disk in
the fundamental regime, reads~\cite{gr}
\be
 {\bf A}_{{\rm A}_{\rm 1}}(k)_{m,m'} &=& d(m) d(m') \frac{\Jb{m}{a}}{\Ho{m'}{a}}\left\{
\cos\left(\frac{\pi}{6}(5m-m')\right) \Ho{m-m'}{R} \right.\nn \\
 &&\qquad \mbox{}\left.
+(-1)^{m'}\cos\left(\frac{\pi}{6}(5m+m')\right) \Ho{m+m'}{R} 
\right\} 
\ee
with $ 0\leq m,m' < \infty$ and 
\be
 d(m):=\left \{ \begin{array}{ccc} \sqrt{2} & {\rm for} & m>0 \\
                                   1        & {\rm for} & m=0 \ . 
              \end{array} \right.
 \nn
\ee
As ${\bf A}$ is trace-class for any $n$-disk geometry,
the determinant exists and can numerically be calculated in a truncated
Hilbert space. The Hilbert space is here the space of angular momentum
eigenfunctions $\{|m\rangle\}$
on the surface of the half-disk in the fundamental domain which can be 
truncated by an upper angular momentum $m_{\rm max}$.
From the study, in App.\ref{app:suppl}, of the asymptotic 
behaviour of ${\bf A}_{m,m'}$ with respect to the angular momentum
one can derive the following inequality for 
the truncation point $m_{\rm max}$: 
\beq
   m_{\rm max}\gesim \frac{e}{2} ka \approx 1.5 ka \; . 
  \label{m-max}
\eeq
This agrees, of course, with the numerical findings.   
The truncated matrix  ${\bf M}_{{\rm A}_{\rm 1}}(k)$ 
is then numerically transformed to an upper triangular matrix 
and the determinant is
calculated from the product of the diagonal elements.
This procedure is faster than the computation of the determinant from
the product of the eigenvalues of ${\bf M}_{{\rm A}_{\rm 1}}(k)$ 
(see \equa{det_prod}). The numerical results  for both ways
agree, of course, up to computer accuracy. The zeros of the determinant,
$\det{}{\bf M}_{{\rm A}_{\rm 1}}(k)$, in the lower complex wave-number plane determine
the scattering resonances, whereas the phase of the determinant evaluated
on the real $k$-axis gives the cluster phase shift. The cumulants can be
constructed either from the Plemelj-Smithies recursion formula 
\equa{ps_formula} or from
the multinomials of the eigenvalues~\equa{cumulant}. 
The latter procedure is numerically more
stable, especially deep inside the negative complex wave-number plane.
This concludes the numerical setup for the exact calculation.

As shown in Secs.\ref{chap:link} and \ref{chap:semiclass}
the classical analog of the characteristic determinant 
(actually of $\det{}\{{\bf 1}+z {\bf A}(k)\}$ to be precise) is the
 semiclassical zeta function of Gutzwiller~\cite{gutzwiller} 
and Voros~\cite{voros88} which, prior to a regularization, is given by
$Z_{GV} (z;k)$ (see \equa{GV_zeta_app}). 
However, in the literature there exist other competitors 
for a semiclassical zeta function,
e.g., 
the dynamical zeta function   $\zeta_0^{-1} (z;k)$
of Ruelle\cite{ruelle} (see \equa{dyn_zeta_app})
which is the $j$=0 part of the Gutzwiller-Voros zeta function as well as
the quasiclassical zeta-function $ Z_{\rm qcl} (z;k)$
of ref.\cite{fredh2} (see \equa{qcl_zeta}). 
As usual, for all three choices, 
$
 t_p(k)= { \e^{\i k L_p - \i \nu_p \pi/2}}/ {|\Lambda_p|^{\half}}$
is the
$p^{\,\rm th}$ primary cycle, $n_p$ its topological length,
$L_p$ is its geometrical length, 
$\nu_p$ its Maslov index together with the 
group theoretical weight of the studied 
${\rm C}_{\rm 3v}$-representation (in the
present case
the ${\rm A}_{\rm 1}$-representation), 
and  $\Lambda_p$  its  stability (the
expanding eigenvalues of the stability matrix) --- see 
refs.\cite{fredh,fredh2} for further details.
The variable
$z$ is a 
book-keeping device for keeping track of the topological order in the
cycle- or curvature expansion~\cite{cvi88,artuso} (see Eqs.\equa{gutzcurv} 
and \equa{zgv_z_expansion}).
In the following, the various curvature-expanded zeta functions 
are truncated at a given curvature (i.e., total topological)
order $n_c$. The semiclassical predictions for the scattering 
resonances are determined from the zeros
of these truncated zeta functions, the predictions for the cluster phase
shifts discussed in Sec.\ref{chap:num_clusterphase} from the phases on the
real $k$-axis and the curvatures from the terms of order $z^m$ in
the curvature expansion.
Input data for the lengths $L_p$ , stabilities $\Lambda_p$  and
Maslov indices $\nu_p$ of the periodic orbits of the 3-disk system in
the ${\rm A}_{\rm 1}$-representation have been  
taken from Rosenqvist~\cite{Per_thesis,Per_private},
Scherer~\cite{scherer}
and Eckhardt~\cite{Eck_private}.

\subsection{Exact versus semiclassical resonances\label{chap:num_resonances}}
In this chapter we compare the numerically computed exact quantum-mechanical
resonances 
of the 3-disk repeller 
with the corresponding semiclassical predictions of the three 
semiclassical zeta functions: the Gutzwiller-Voros
zeta function \equa{GV_zeta_app}, the dynamical zeta function 
\equa{dyn_zeta_app} and
the quasiclassical zeta function \equa{qcl_zeta}.

For the 3-disk-repeller with 
center-to-center separation $R\mbox{=}6a$,
we have computed all exact quantum-mechanical ${\rm A}_{\rm 1}$ resonances
(numerically determined from the zeros of 
$\det{} {\bf M}_{{\rm A}_{\rm 1}}(k)$)  as well as
all the corresponding approximate ones (from the zeros of the 
at finite curvature order
$n_c$ truncated  zeta functions) 
in the wave-number window: $0 \leq {\rm Re}\, k \leq 250/a$ and $0\geq
{\rm Im}\, k \geq -1.6/a$. This window  contains several hundreds  of leading
and subleading  resonances, from the lowest ones onwards.
In Figs.\,\ref{fig:e_gv1}--\ref{fig:e_gv12}, for increasing
curvature  order, the resonances are plotted
as the real part of the wave number (resonance ``energy'') versus the
imaginary part of the wave number (resonance ``width'').

Some features of the resonance spectra allow for an immediate 
interpretation~\cite{gr,scherer,pinball}: 
The mean spacing of the resonances is approximately 
$2\pi/{\bar L}$,
where ${\bar L}$ is the average of the geometrical lengths
of the shortest periodic orbits, namely the lengths $L_0$ and $L_1$ of 
the two periodic orbits of topological length one. The 
data also  exhibit various beating patterns 
resulting from the interference of the
periodic orbits of nearly equal length; e.g., 
the leading beating pattern is of order  
$2\pi/{\Delta L}$,
where $\Delta L$ is the difference of the lengths $L_1$ and $L_0$.


In Figs.\,\ref{fig:e_gv1} -- \ref{fig:e_gv4} 
a comparison is made from the first- up to the fourth
order in the curvature expansion. Already at fourth order 
the four leading
resonance bands are well approximated by the Gutzwiller-Voros
zeta-function (in fact, for ${\rm Re}\, k
\lesim 75/a$ already the second curvature order is enough to describe
the first two leading resonance bands). This is in agreement with the
rule of thumb that
any new resonance band is linked with a new  curvature or
cumulant order.
%
Neither the dynamical
zeta-function  nor the quasiclassical one perform as well to
fourth order.
The reason is that the quasiclassical as well as
the dynamical zeta-function predict extra resonances which are absent 
in the exact quantum-mechanical calculation. Thus the third and
fourth curvature order of these zeta-functions are distributed over
the average of the third and fourth resonance bands and the spurious
extra resonances.
In the window plotted one can classify the exact data into four leading
resonance bands closest to the real wave-number axis 
and 2 subleading ones shielded by the leading resonances.
Thus, 
just periodic orbits of topological
length up to four are needed 
in order to reproduce the qualitative trend of the exact data closest
to the real axis.  The
3-disk-system has  8 periodic orbits up to this topological
length.  Actually,  
the 3-disk-system with center-to-center separation $R=6a$ is not very
chaotic at these $k$ values.
All experimentally accessible spectral data in this regime (which can be 
extended
up to ${\rm Re}\, k \approx 950/a$ as only 
about there the subleading resonance
bands mix with the four leading ones) can be
parameterized by 16 real numbers, i.e., 8 periodic orbit lengths, 8
stabilities, and 8 Maslov indices. Experimentalists can stop
here. The subleading bands are completely shielded (up to
${\rm Re}\, k \approx 950/a)$ by the above mentioned four bands. The
subleading bands (below ${\rm Re}\, k \approx 950/a$)  
are only of theoretical interest, as they can be used
to test the semiclassical zeta functions.

In Figs.\,\ref{fig:e_gv5} a comparison is made up to fifth
curvature order.  The Gutzwiller-Voros zeta-function does at least as well as
in Fig.\,\ref{fig:e_gv4}a for the leading four resonance bands, but now
it also describes the peak position of the fifth resonance band for
large enough values of ${\rm Re}\,k$.    
Note the {\em diffractive} band of exact resonances
from $k\approx(0. -\i 0.5)/a$ to $k\approx (100.-\i 1.6)/a$ which our
semiclassical zeta functions fail to describe.
As shown in Refs.\cite{vwr_prl,vwr_japan,vwr_stat} the 
diffractive band of resonances can be accounted for by inclusion of
creeping periodic orbits which have been omitted from our semiclassical
calculations.
The dynamical and quasiclassical zeta functions show 
a slight improvement with respect to the four leading resonance
bands; however, no agreement with the fifth one.

In  Figs.\,\ref{fig:e_gv6}
a comparison is made up to sixth curvature order.
The Gutzwiller-Voros zeta-function fails for the third resonance band below
${\rm Re}\, k \approx 20/a$, for the fourth below ${\rm Re}\, k \approx 45/a$,
for the fifth and sixth below  ${\rm Re}\, k \approx 70/a$ and 80/$a$, 
respectively.  Below these values, the last two curvature orders try to
build up an accumulation line. 
Above these values, the qualitative agreement
with the data is rather good. 
The dynamical zeta-functions at this order
just improves
the description of the four leading resonance bands. 
Furthermore, it builds
up a sharp line of accumulation for the subleading resonances,
the border of convergence controlled by the location of the nearest
poles of the dynamical zeta function, see \cite{ER92,fredh}.
The quasiclassical zeta function also improves
the description of the four resonance bands,
although it is still not of the same
quality as the Gutzwiller-Voros  one 
even at two curvature orders lower.
Note that the quasiclassical zeta function is trying to build up
two bands of spurious resonances in agreement with our rule of thumb.

In Figs.\,\ref{fig:e_gv7} a comparison is made up to
seventh curvature order.  
The first four resonance bands of the Gutzwiller-Voros zeta function 
have converged and the accumulation  
line has moved up. Only above ${\rm Re}\, k \approx 140/a$ the
fifth and sixth resonance band emerge, now with improved 
accuracy, however. Also the seventh resonance band is approximated. 
The dynamical zeta-function now clearly produces its line of convergence
(the accumulation line of resonances). Above this line, the resonances
(except the ones very close to the accumulation) 
are approximated as well as in the
Gutzwiller-Voros case; below, no agreement is found. 
At this order the
quasiclassical zeta-function is doing as well as the Gutzwiller-Voros
zeta-function did already at curvature order four. None of the
subleading bands are described by the quasiclassical zeta-function.
Instead another band of spurious resonances emerges.

In Figs.\,\ref{fig:e_gv8} and \ref{fig:e_gv12} 
the comparison is made up to the eighth and twelfth
curvature order, respectively.  
The border of convergence of the Gutzwiller-Voros
zeta-function has now moved (in the plotted region) above the fifth
and sixth band of the exact resonances. 
It has moved also closer to the very sharp accumulation
line of resonances of the dynamical zeta function. However, these lines
are still not identical even at twelfth curvature order.  
The subleading
quasiclassical resonances have stabilized onto the spurious bands.
Furthermore, some subleading resonances move further down into the
lower complex $k$-plane~\footnote{Note that the quasiclassical results of this
figure are
directly comparable with the results of the so-called Quantum-Fredholm
determinant of Ref.\cite{fredh} (see Fig.\,4b in Ref.\cite{fredh}) 
as both calculations involve periodic orbits
of topological length up to eight. As we now know, all 
the subleading resonances of
that figure have nothing to do with quantum mechanics.}. 
Eventually (see also \cite{Per_thesis}), starting
with curvature order 10 and 12 the fifth and sixth resonance bands are
approximated --- in addition, to four or six
spurious resonance
bands, respectively. 
Thus the quasiclassical zeta-function seems to find the
subleading resonance bands, but at the cost of many extra spurious
resonances. Note that at these high curvature orders the
quasiclassical zeta-function has numerical convergence problems for large
negative imaginary $k$ values (especially for low values of 
${\rm Re}\, k$). This is in agreement with 
the expected large cancellations
in the curvature expansion at these high curvature orders.
Furthermore, periodic orbits  of larger topological order than twelve
would be needed to falsify the success of the quasiclassical zeta
function, since it barely  manages to approximate
the two bands of subleading resonances at this  curvature order.

Qualitatively, the results can be summarized as follows. The
Gutzwiller-Voros zeta-function does well above its line of convergence,
defined by the dynamical zeta-function, already at very low curvature
orders where the dynamical zeta-functions still has problems. 
Below this line we observe that the Gutzwiller-Voros
zeta-function  works only  as an asymptotic expansion. However, when it
works, it works very well and very efficiently. This implies that
the additional ($\Lambda_p$-dependent)  
terms  of the Gutzwiller-Voros zeta function, relative
to the simpler dynamical zeta function, are the correct
ones. This is of course in agreement with the findings of our semiclassical
reduction in Sec.\ref{chap:semiclass}.
Eventually,
the dynamical
zeta-function does as well for the leading resonances as
the Gutzwiller-Voros one. As experimentally these are the only
resonances accessible, one might -- for practical purposes -- limit the
calculation just to this zeta function, see, however, 
Sec.\ref{chap:num_clusterphase}.
The quasiclassical
zeta-function seems to find all subleading geometrical
resonances. Unfortunately, the highest periodic orbits at our
disposal are of topological length 12; the very length where the sixth
resonance band seems to emerge. Thus higher orbits would be needed to
confirm this behavior.  But all this comes 
at a very high price: The rate
of convergence is slowed down tremendously (in comparison with the
asymptotically working Gutzwiller-Voros zeta-function), as this
zeta-function is producing additional spurious resonance bands which do
not have quantum-mechanical counter parts, but only classical 
ones~\cite{cvw96}. Without a quantum
calculation, one could therefore not tell the spurious from the real
resonances. 

As a by-product we have a confirmation of our empirical rule of thumb that
`each new cumulant or curvature order is connected with a new line of 
subleading resonances'.
This rule therefore relates the curvature truncation limit,
$m\to\infty$, either to the limit ${\rm Im}\, k \to -\infty$, if there is
no accumulation of subleading resonances, i.e., if the zeta function is
entire~\cite{fredh,fredh2},  or to
the formation of an accumulation band of resonances. Both facts 
support our
claim that, in general, 
the curvature limit $m\to\infty$ and the semiclassical limit
${\rm Re}\, k \to \infty$ cannot and should not commute deep inside the
lower complex $k$-plane, as the subleading resonances of increasing
cumulant order are approximated
worse and worse.  Only an asymptotic expansion should be possible,
in agreement with our findings for the Gutzwiller-Voros zeta function.
%
%
\subsection
{Exact versus semiclassical cluster phase shifts\label{chap:num_clusterphase}}

In the last chapter 
the semiclassical
zeta functions were judged by  the comparison of 
their resonances predictions
with the exact resonances poles (especially
the subleading ones),
as was done in the past,
see e.g.\ Refs.\cite{gr,pinball,fredh,aw_chaos,aw_nucl,vwr_prl,vwr_stat}.
Since the deviations between the zeta functions themselves and from
the exact data are most pronounced
for the subleading resonances (which are shielded by the leading ones), one
could argue that empirically  it does not matter which of
the three zeta functions are used to describe the measured data, since all
three give the same predictions for the leading 
resonances~\cite{fredh,fredh2}.

Below, however, we will show that even experimentally 
one can tell the three semiclassical zeta functions apart and that, in fact,
the Gutzwiller-Voros one is by far the best.

\subsubsection{Cluster phase shifts}
In Sec.\ref{chap:link} the exact and semiclassical expressions for
the determinant of ${\bf S}$-matrix for non-overlapping $n$-disk systems
have been constructed. For the case of the 3-disk system they read
\be
  \det{l} {\bf S}^{(3)}(k) 
   &=& \left (  \det{l}{\bf S}^{(1)} (ka)\right )^3 
     \frac{ \det{l} { {\bf M}_{{\rm A}_{\rm 1}} (k^\ast) }^\dagger}
          { \det{l}  {\bf M}_{{\rm A}_{\rm 1}} (k) }\,
     \frac{ \det{l}  { {\bf M}_{{\rm A}_{\rm 2}} (k^\ast) }^\dagger}
          { \det{l}  {\bf M}_{{\rm A}_{\rm 2}} (k) }\,
   \frac{\left (  {\det{l} {\bf M}_{\rm E} (k^\ast)}^\dagger\right )^2 }
          { \left( \det{l}  {\bf M}_{\rm E} (k)\right )^2 }    \nno \\
  &\semiclass &
         \left ( \e^{-\i \pi N(k)} \right )^{2\times 3}
      \, \left( \frac{  {{\widetilde Z}_{\rm 1\mbox{-}disk(l) }(k^\ast)}^\ast}
            {{{\widetilde Z}_{\rm 1\mbox{-}disk(l)} (k)}}\, 
        \frac{{ {\widetilde Z}_{\rm 1\mbox{-}disk(r)}
              (k^\ast)}^\ast}
            {{{\widetilde Z}_{\rm 1\mbox{-}disk(r)} (k)}}\right )^3 
 \times \nno \\
  && \qquad \qquad \qquad\times
            \frac{{ {\widetilde Z}_{{\rm A}_{\rm 1}} (k^\ast)}^\ast  }
                     {{\widetilde Z}_{{\rm A}_{\rm 1}}(k)}\,
               \frac{ {{\widetilde Z}_{{\rm A}_{\rm 2}} (k^\ast)}^\ast }
                {{\widetilde Z}_{{\rm A}_{\rm 2}}(k)}
           \,
              \frac{ { {{\widetilde Z}_{{\rm E}} (k^\ast)}^\ast  }^2}
                      {{{\widetilde Z}_{{\rm E}}(k)}^2} \nno \\
 \label{3disk-link}
\ee 
where the tilde indicates that diffractive corrections have to be included,
in general.
Especially for the ${\rm A}_{\rm 1}$-representation of the 3-disk system we therefore have
the relation between the quantum-mechanical kernels and the Gutzwiller-Voros
zeta functions
\be
   \frac{ \det{l} { {\bf M}_{{\rm A}_{\rm 1}} (k^\ast) }^\dagger}
          { \det{l}  {\bf M}_{{\rm A}_{\rm 1}} (k) }
\semiclass 
 \frac{{ {Z}_{{\rm A}_{\rm 1}} (k^\ast)}^\ast  }
                     {{Z}_{{\rm A}_{\rm 1}}(k)} \ ,
    \label{A1-link}
\ee
where we have now neglected diffractive corrections.
As argued in the conclusion section \ref{chap:end} both sides
 of 
Eq.\equa{3disk-link} and Eq.\equa{A1-link} 
respect unitarity; 
the quantum-mechanical side exactly and for 
the semiclassically side under the
condition that the curvature expansion converges or that it is truncated.
As all the $n$-disk resonances for non-overlapping $n$-disk repellers 
are below the real $k$-axis, the border of absolute convergence, defined
by the closest resonances to the real axis~\cite{pinball,fredh} is inside
the lower complex wave-number plane and unitarity on the real axis is
guaranteed.
Thus, if
the wave number $k$ is real, 
the left hand sides and also the right hand sides
of eqs.\equa{3disk-link} and \equa{A1-link} can be written as 
$\exp\{ \i 2 \eta(k) \}$ with a real {\em phase shift} $\eta(k)$.
In fact, we can define a total phase shift for the coherent part of the 3-disk
scattering problem (here always understood in the ${\rm A}_{\rm 1}$-representation) 
for exact quantum mechanics as well as for the three
semiclassical candidates by:
\be
    \e^{2\i \eta_{\rm qm}(k)} &:=&    
\frac{ \det{l} { {\bf M} (k^\ast) }^\dagger}
          { \det{l}  {\bf M} (k) } \label{eta_qm} \\
      \e^{2\i \eta_{\rm GV}(k)}&:=&
 \frac{{ { Z}_{\rm GV} (k^\ast)}^\ast  }
                     {{ Z}_{\rm GV}(k)} \label{eta_gv}\\
      \e^{2\i \eta_{\rm dyn}(k)}&:=&
 \frac{{  \zeta_0^{-1} (k^\ast)}^\ast  }
                     { \zeta_0^{-1} (k)}   \label{eta_dy0} \\
      \e^{2\i \eta_{\rm qcl}(k)}&:=&
 \frac{{ { Z}_{qcl} (k^\ast)}^\ast  }
                     {{ Z}_{qcl}(k)} \label{eta_qcl}\ .
\ee
This phase shift definition should be compared with the cluster 
phase shift given in 
Sec.\,4 of Lloyd and Smith~\cite{Lloyd_smith}. For a separable system,
as e.g.\ the 1-disk system (in the angular momentum representation),
the cluster phase shift just corresponds to the sum 
\be
 \eta(k) = \sum_{l=-\infty}^{\infty} \eta_l (k) \ ,
\ee
as the ${\bf S}$-matrix of the one-disk system (evaluated with respect
to the center of the disk) reads 
\be
 {\bf S}_{ll'}(k) &=& \frac{-\Ht{l}{a}}{\Ho{l}{a}} \delta_{ll'} \nn \\
                  &=& \e^{2\i\eta_l(k)}  \delta_{ll'} \ ,
 \label{num:1-disk}
\ee
such that 
\be
  \det{}{\bf S}(k) = \prod_{l=-\infty}^{+\infty} \e^{2\i\eta_l(k)} \ .
 \label{num:1-disk-cluster}
\ee 
Let us once more stress: the coherent or cluster phase shift is an 
experimentally accessible quantity: from the 
measured differential cross sections 
the elastic scattering amplitudes have to be constructed. This leads to
the full phase shift of the 3-disk system including the contribution from 
the
single disks. However, the incoherent part can be subtracted by
either making reference experiments with just 
single disks at the same position
where they used to be in the 3-disk problem or by numerical subtractions
as the one-disk phase shifts are known  analytically, since the system
is separable, see \equa{num:1-disk} and \equa{num:1-disk-cluster}. 
In this way one can separate
the incoherent phase shifts  from the coherent ones. 

Thus $\eta_{\rm qm}(k)$ is ``measurable'' in principle. 
We next use these cluster phase shifts in order to
discriminate between the various zeta functions.
Below, we compare the exact quantum-mechanical cluster phase shift
$\eta_{qm}$ with
\begin{enumerate}
\item 
the semiclassical cluster phase shift $\eta_{\rm GV} (k)$ of
the Gutzwiller-Voros zeta function \equa{GV_zeta_app},
\item
with the semiclassical cluster phase shift $\eta_{\rm dyn} (k)$ of
the dynamical zeta function \equa{dyn_zeta_app},
\item
and with the semiclassical cluster phase shift $\eta_{\rm qcl} (k)$ of
the quasiclassical zeta function \equa{qcl_zeta}.
\end{enumerate}
The zeta functions in the numerator as well as in the 
denominator of $Z(k)^\ast/Z(k)$ have been expanded to curvature order 
(=topological length) 12. For the Gutzwiller-Voros zeta function this is 
an overkill as already curvature order 4 should describe the data below 
${\rm Re} k = 950/a$. In fact, 
we have not seen any difference in the Gutzwiller-Voros calculation 
between the 
curvature order 3 and 12 results for $k\leq 120/a$ and 
up  to figure accuracy. Curvature order
2, however, gives in the regime $100/a \leq k \leq 120/a$ noticeable 
deviations.
On the other hand, as mentioned in Sec.\ref{chap:num_resonances}, 
the quasiclassical zeta function has problems for lower curvature orders with
predicting the  
\mbox{(sub-)}leading resonances; therefore, these high curvature orders
are used in order 
to give the quasiclassical zeta function as fair a chance as possible.
The coherent phase shifts are compared in
the window $ 104/a \leq k \leq 109/a$, which is a typical window
narrow enough to resolve the rapid oscillations with $k$ sufficiently
large such that diffractive effects can be safely neglected.
Furthermore, although we have no physical interpretation in terms of the 
${\bf S}$-matrix, we also compare in the same window 
the exact quantum-mechanical product 
$\det{} {\bf M} (k) \det{} {\bf M}(k^\ast)^\dagger$ with the
squared modulus 
of the Gutzwiller-Voros zeta function
 $Z_{\rm GV}(k) Z_{\rm GV}(k^\ast)^\ast$, the dynamical zeta function 
 $ \zeta_0^{-1} (k) 
 \zeta_0^{-1}(k^\ast)^\ast$, and
the quasiclassical zeta function
$Z_{\rm qcl} (k) Z_{\rm qcl}(k^\ast)^\ast $.
Here
$k$  is taken to be real 
and the case of the 3-disk system in the ${\rm A}_{\rm 1}$-representation
with center-to-center separation $R=6a$ is studied.

Consider finally the general quasiclassical zeta functions of 
Ref.\cite{fredh2} and especially the
ratio
\be
  Z(k):= \frac{F_{+}(\half     ,k) F_{-}(\sevenhalf,k)}
              {F_{-}(\threehalf,k) F_{+}(\fivehalf,k)}
  \label{num:ratio}
\ee
with $F_{+}(\beta,k;z)$ and $F_{-}(\beta,k;z)$ being defined as follows
\be
   F_{+}(\beta,k;z) &=& \exp\left\{ - \sum_{p} \sum_{r=1}^{\infty} \frac{1}{r}
 \frac{\left( z^{[p]}\, t_p(k) \right)^r }{
 \left ( 1-\frac{1}{\Lambda_p^r} \right )^2
 \left ( 1-\frac{1}{\Lambda_p^{2r}}\right) } |\Lambda_p^r|^{-\beta+\half} 
\right\} 
 \nn \\
   F_{-}(\beta,k;z) &=& \exp\left\{ - \sum_{p} \sum_{r=1}^{\infty} \frac{1}{r}
 \frac{\Lambda_p^r}{|\Lambda_p^r|}\,
\frac{\left( z^{[p]}\, t_p(k) \right)^r }{
 \left ( 1-\frac{1}{\Lambda_p^r} \right )^2
 \left ( 1-\frac{1}{\Lambda_p^{2r}}\right) } |\Lambda_p^r|^{-\beta+\half} 
\right\} \; . \nn
\ee
Here the
subleading factor $(1+ |\Lambda_p^r|^{2\beta-4})$ of Eq.(11) in 
ref.\cite{fredh2}
has been removed as in Eq.(12) of ref.\cite{fredh2}.
When Eq.\equa{num:ratio} is used, 
the corresponding coherent phase shift
\be
 \e^{\i 2\eta_{\rm rat}(k)}= \frac{Z(k^\ast)^\ast}{Z(k)}
\ee
works 
on the real wave-number axis and  in the limit 
$n \to \infty$ (where $n$ is the curvature order) as well as the original 
Gutzwiller-Voros zeta function. Hence, it does not matter here whether the 
Gutzwiller-Voros zeta function is directly expanded in the curvature expansion
or whether 
the individual determinants 
$F_{+}(\half,k)$, $F_{-}(\sevenhalf,k)$, $F_{-}(\threehalf,k)$ and
$F_{+}(\fivehalf,k)$ are each expanded in separate curvature expansions up
to the {\em same} curvature order and
{\em then} inserted in the ratio \equa{num:ratio}. Note that the presence or
absence of the subleading factor $(1+ |\Lambda_p^r|^{2\beta-4})$ 
in the definitions of $F_{+}(\beta,k;z)$ and $ F_{-}(\beta,k;z)$ 
does not change the results up to figure accuracy.

\renewcommand{\baselinestretch}{0.7}
\noindent\begin{figure}[hbt]
\centerline{(a) \epsfig{file=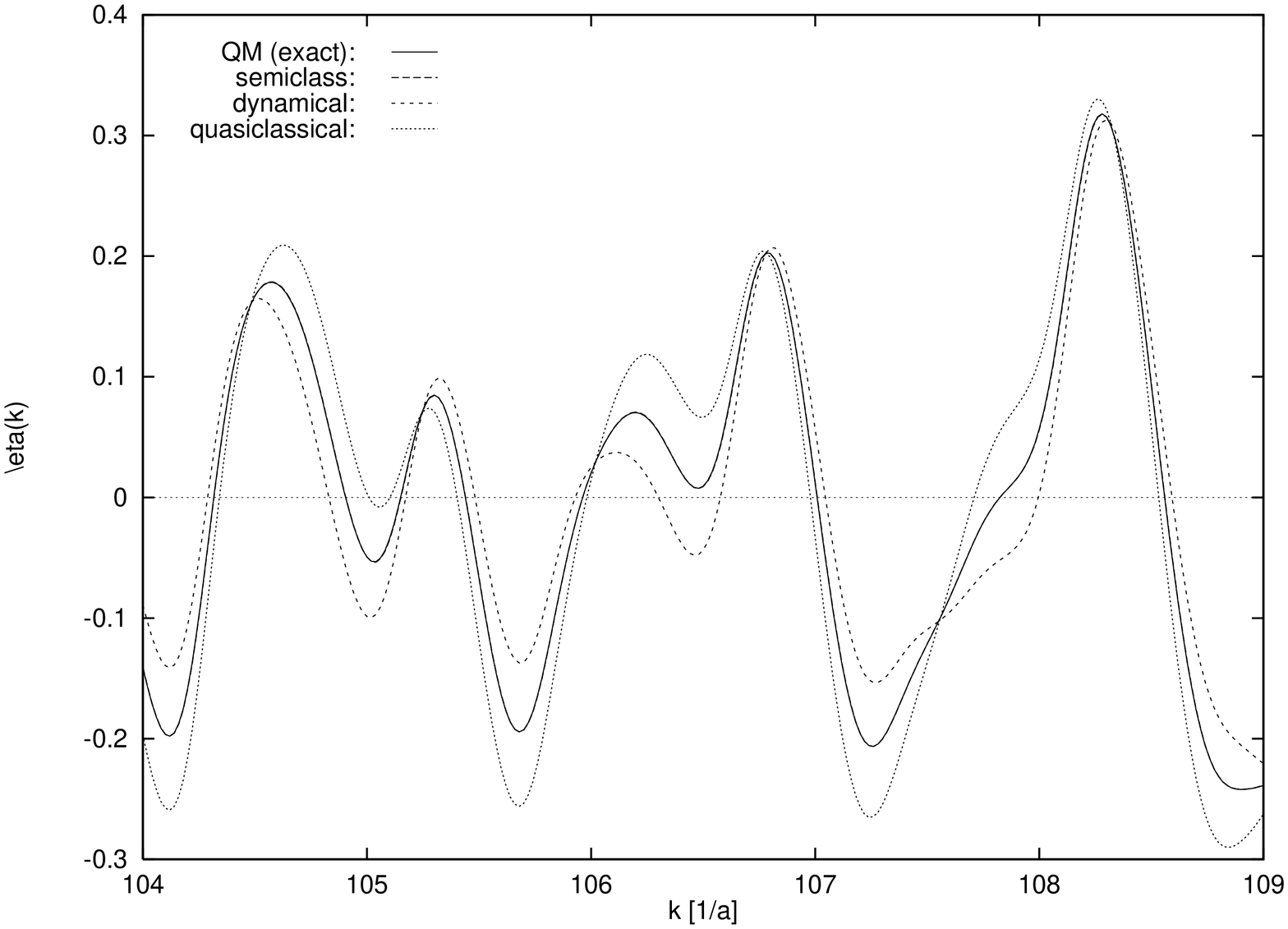,height=9cm,angle=-90}}

\centerline{(b) \epsfig{file=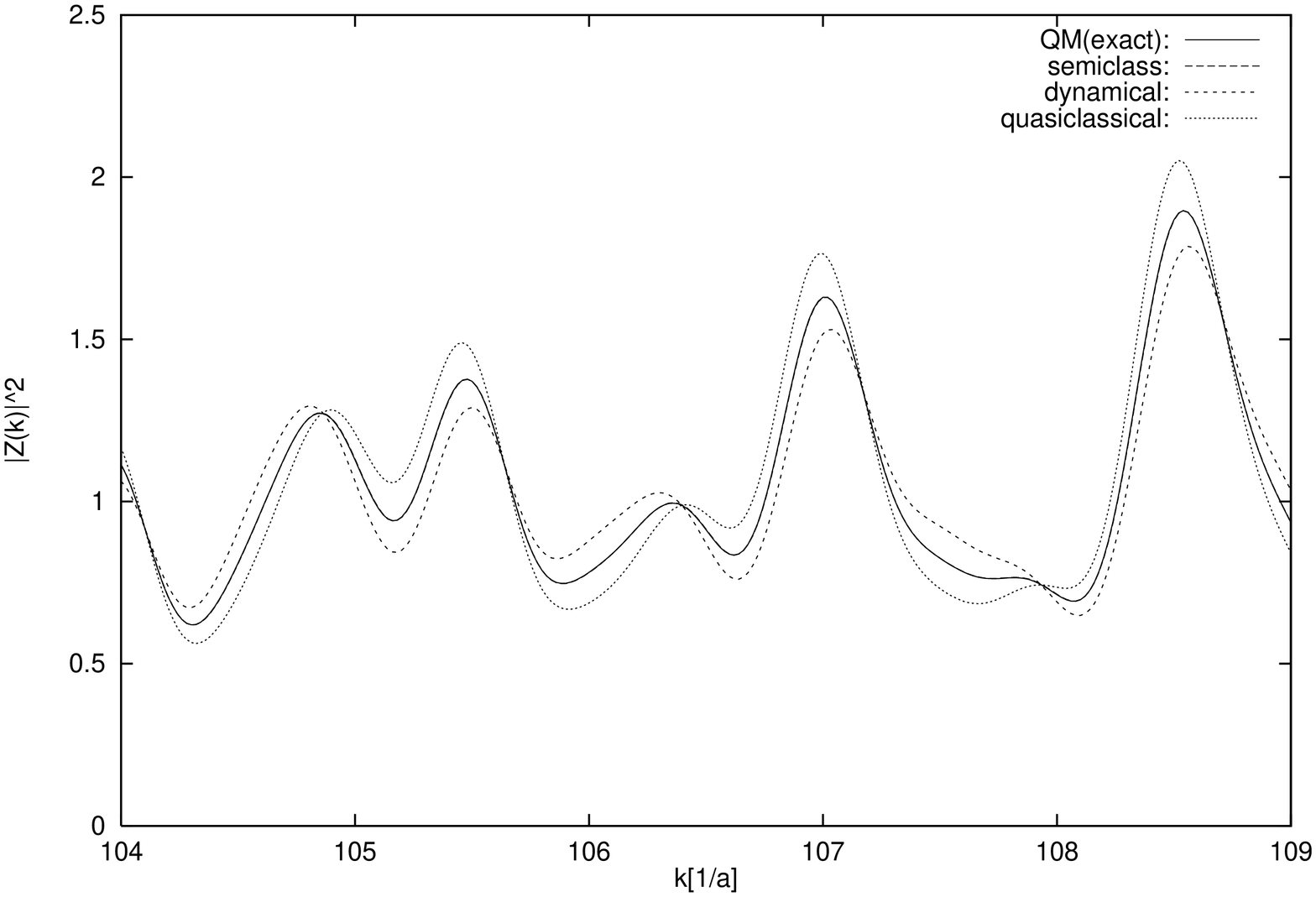,height=9cm,angle=-90}}
\caption[fig_ph_all]{\small
(a)\ 
The coherent cluster phase shifts of the 3-disk scattering system 
in the ${\rm A}_{\rm 1}$-representation  with center-to-center separation $R=6a$. The 
exact quantum-mechanical data are compared 
to the predictions of 
the Gutzwiller-Voros zeta function \equa{GV_zeta_app},
the dynamical zeta function \equa{dyn_zeta_app} and the quasiclassical
zeta function \equa{qcl_zeta} calculated up to 12th order in the
curvature expansion. 
(b)\ The same for the squared moduli of the exact spectral determinant 
and the semiclassical zeta functions.
The predictions of the Gutzwiller-Voros zeta function 
and exact
quantum mechanics coincide within the resolutions of the 
plots.
} 
\end{figure}

Let us stress that phase shifts are 
not only of theoretical interest,  
as are the subleading resonances (which
are completely shielded by the leading resonances), but {\em hard} 
data which can be extracted, in principle, from measured differential 
cross sections.

In summary, even empirically, 
one can tell the three semiclassical zeta functions apart and see which
is the best. Again 
the Gutzwiller-Voros one --- 
whether used directly
or whether defined as the ratio \equa{num:ratio} 
of four quasiclassical determinants as in ref.\cite{fredh2} --- 
is by far the best.

\subsection{The quantum-mechanical cumulant expansion versus the
semiclassical curvature expansion\label{chap:num_cumulant}}

In this subsection it will be shown that the Gutzwiller-Voros zeta function
approximates its quantum-mechanical counterpart, the characteristic 
KKR-type determinant~\cite{KKR,Lloyd_smith,Berry_KKR}, 
only in an  asymptotic sense, such that it always should be understood
as a truncated series. 

As shown in Sec.\ref{chap:link},
the characteristic determinant and the Gutzwiller-Voros zeta-function 
are related as
\beq
     \det{} {\bf M}(k)  \semiclass  Z_{\rm GV}(k) \; .
\eeq

Let $Q_m(k)$ denote the $m^{\,\rm th}$ cumulant of $\det{} {\bf M}(k)$ 
-- i.e.\ the term proportional to $z^m$ in the Taylor expansion of
$\det{}\{{\bf 1}+z {\bf A}(k)\}$ --
which
satisfies the Plemelj-Smithies recursion relation \equa{PS-recursion} 
(see also App.\ref{app:trace}).
%
Since the Plemelj-Smithies recursion formula is plagued by
cancellations of very large numbers,
we have not used the Plemelj-Smithies recursion relations for  our numerical
calculation of $Q_m(k)$,  but instead we construct this quantity 
directly from the eigenvalues 
$\{\lambda_j(k) \}$
of the trace-class matrix ${\bf A}(k)$, i.e. 
\be
  Q_m(k)= \sum_{1\leq j_1 < \cdots < j_m<\infty}
 \lambda_{j_1}(k) \cdots \lambda_{j_m}(k)   \label{product}
\ee
(see again App.\ref{app:trace}
for more details).
Unfortunately, a semiclassical analog to this exact formula 
has not been found so far. 
Thus $C_m(k)$, the corresponding 
semiclassical $m$th-order curvature term, 
of $Z_{\rm GV}(k)$, can
only be constructed from the semiclassical 
equivalent of the Plemelj-Smithies
recursion relation \equa{appr-PS-recursion} which exactly corresponds to the standard
curvature expansion of refs.\cite{artuso,fredh,pinball})
and is therefore inherently plagued by large cancellations.
The cumulant and curvature expansions, truncated 
at $n$th order, read:
\be
      \det{} {\bf M}(k)|_n &=& \sum_{m=0}^{n}  Q_m (k) \\ 
       Z_{\rm GV}(k)|_n   &=&  \sum_{m=0}^{n}  C_m(k)        \ .    
\ee

Let us recapitulate what we already know about these series.
From Sec.\ref{chap:link} together with the appendices \ref{app:trace}
and \ref{app:suppl} we deduce that the cumulant sum
\beq  
     \lim_{n\to \infty}   \det{} {\bf M}(k)|_n =   \lim_{n\to \infty} 
 \sum_{m=0}^{n} Q_m(k) = \det{}{\bf M}(k)
\eeq
is absolutely convergent, i.e.\
\beq
    \sum_{m=0}^{\infty} | Q_m(k) | < \infty \ ,
\eeq
because of the trace-class property of 
${\bf A}(k)\equiv {\bf M}(k)-{\bf 1}$ for non-overlapping, disconnected
$n$-disk systems.
On the other hand,
as discussed in Refs.\cite{ER92,fredh,fredh2},
the Gutzwiller-Voros curvature sum converges only above 
an accumulation line  (running below and approximately 
parallel to the real wave-number axis, see Sec.\ref{chap:num_resonances}) 
which is given by the
first poles of the  dynamical zeta function, 
$\zeta_0^{-1}(k)$, or the leading zeros of the subleading zeta function.
However, as shown in Sec.\ref{chap:num_resonances}, even
below this boundary of convergence 
the truncated Gutzwiller-Voros curvature sum, 
$Z_{\rm GV}(k)|_n$ approximates the
quantum-mechanical data as an asymptotic series.

\begin{figure}[hbt]
\centerline{ 
\epsfig{file=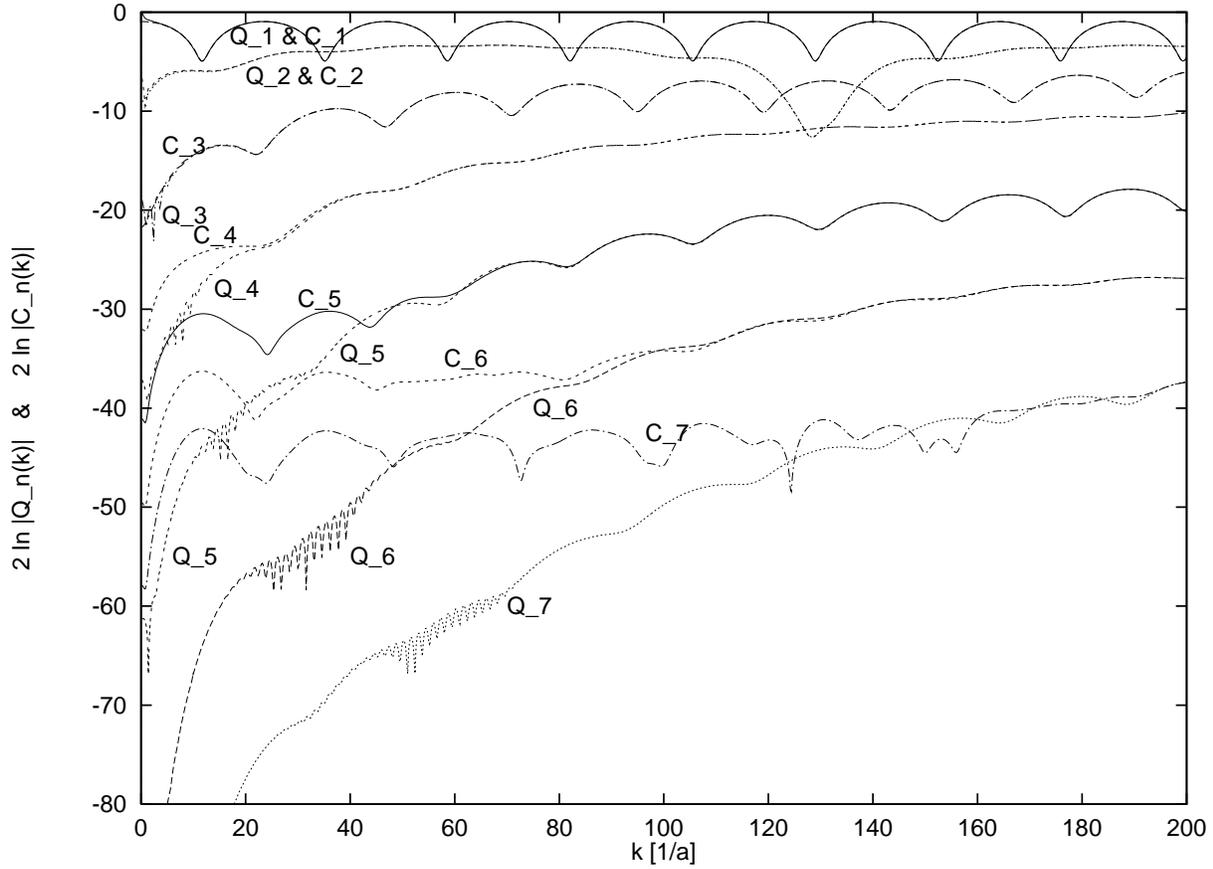,height=11.5cm,angle=-90}}

\caption[fig_cum_mod]{\small
Comparison of the squared moduli (on a logarithmic scale)
of the first seven quantum-mechanical
cumulant terms, $|Q_n(k)|^2$, 
with the corresponding semiclassical curvature terms, $|C_n(k)|^2$, of the
Gutzwiller-Voros zeta function \equa{GV_zeta_app} 
evaluated on the real wave-number axis $k$. 
The system is the ${\rm A}_{\rm 1}$ three-disk repeller with center-to-center
separation $R=6a$.
\label{fig_cum_mod}} 
\end{figure}

In addition,
a very important property for the discussion 
of the cumulant and curvature
terms is the existence of the 
scaling formulas (established by us numerically) which 
relate the $m$th 
cumulants or curvatures inside the complex wave-number plane to the
corresponding quantities on the real $k$-axis:
\be
    Q_m({\rm Re}\, k + \i {\rm Im}\, k) 
   &\sim&  Q_m({\rm Re}\, k) \e^{- m \bar{L}  {\rm Im}\, k} 
  \label{Qscale} \\
    C_m({\rm Re}\, k + \i {\rm Im}\, k) &\sim&   C_m({\rm Re}\, k) 
   \e^{ - m \bar{L}  {\rm Im}\, k} \ .
  \label{Cscale}
\ee
(For this to hold, 
diffractive effects have to be negligible, i.e.  
$-{\rm Im}\, k \ll {\rm Re}\, k$.)\  
Here $\bar{L}\approx R-2a$ is 
the average of the geometrical  lengths of the shortest periodic orbits,
the two   orbits of topological
length one. The scaling can be motivated by the approximate
relation $\Tr{}[ {\bf A}^m(k) ] \approx \{ \Tr{} {\bf A}(k)\}^m$ which,  of
course, cannot be exact, as otherwise the cumulants would be identically zero.
Nevertheless, the overall behaviour follows from this, since
\[ 
\Tr{}\left[ {\bf A}( {\rm Re}\, k + \i {\rm Im}\, k) \right]
\ \sim  \Tr{}
 \left[ {\bf A}( {\rm Re}\, k) \right ] 
\e^{ -  \bar{L}  {\rm Im}\, k}  \; .
\] 
From Fig.\ref{fig_cum_mod}
one can deduce
that the
deviations between quantum-mechanical cumulants  and semiclassical
curvatures (as evaluated on the real $k$-axis) decrease 
with increasing ${\rm Re}\, k$, but
increase with increasing curvature order $m$. 
The value of ${\rm Re}\, k$ where the quantum-mechanical and
semiclassical curves join is approximately given by 
${\rm Re}\, k a \sim 2^{m+1}$.
Approximately the same transition points can be generated from a comparison
of the phases of the cumulant and curvatures.
 
\renewcommand{\baselinestretch}{0.7}
\noindent

By varying the center-to-center distance we have numerically 
verified 
that the above limits generalize to the following relations valid
on the real wave-number axis ($k$ real and positive):
\be
   C_m(k) \approx Q_m(k) \quad {\rm with}\ \ 
1 \gg |C_m(k)| \approx |Q_m(k)|     \qquad  
  {\rm if}\  \ ka  \gesim\,   2^{m-1} \frac{\bar L}{a}  
  \label{asymp}
\ee                      
and        
\be
     1 \gg |C_m(k)|  \gg | Q_m(k) | \qquad       {\rm if}\ \  ka  \lesim\,   
   2^{m-1}\frac{\bar L}{a}  
   \label{non_asymp} \; .
\ee
What is the interpretation of \equa{asymp} and \equa{non_asymp}?  
For fixed $k$,
even in the regime, where $Z_{\rm GV}(k)|_n$ 
converges, e.g., on the real $k$-axis, the Gutzwiller-Voros zeta function
is only an {\em asymptotic} approximation
to the true quantum-mechanical
cumulant sum, since for $m>m_{\rm crit}$, defined by $ka \approx   
2^{m_{\rm crit}-1}\frac{\bar L}{a}$,
the exact quantum-mechanical cumulants $Q_m(k)$ and the semiclassical 
curvatures $C_m(k)$ are grossly different.
These deviations  can be enhanced
by  
the $m^{\,{\rm th}}$ derivative, $m>m_{\rm crit}$,
with respect to the book-keeping 
variable $z$, since this operation eliminates all approximately equal
terms, such that 
the corresponding cumulant and curvature series are transformed to 
completely different expressions.
The fact that $Z_{\rm GV}(k)|_n$   
-- even in its convergence regime -- is only an 
asymptotic expansion to the exact quantum mechanics is normally
not  visible, as the terms in \equa{asymp} are exponentially small on or close
to the
real axis and therefore sum
to a tiny quantity. In other words, close to the real axis
the absolute error $|C_n(k)-Q_n(k)|$ 
for $m> m_{\rm crit}$
is still small.  The relative error 
$|C_n(k)/Q_n(k)|$ on the other hand is tremendous 
(see Fig.\ref{fig_cum_mod}).
Deeper inside the negative complex wave-number plane, however,
under the scaling rules \equa{Qscale}
and \equa{Cscale}, the deviations \equa{non_asymp} are blown up, such that the
relative errors $|C_n(k)/Q_n(k)|$ 
eventually become visible as absolute errors $|C_n(k)-Q_n(k)|$ in the 
resonance calculation of Sec.\ref{chap:num_resonances}.
If  ${\rm Im}\, k$  is  above the boundary line of convergence, 
these errors still sum up to a finite  
quantity which might, however, not be negligible any longer, as  was
the case on or close to the real $k$-axis. Below the convergence line
these errors sum up to infinity.
Thus the Gutzwiller-Voros curvature expansion $Z_{\rm GV}(k)|_n$
does not suddenly become an asymptotic approximation to 
$\det{} {\bf M}(k)|_n$, it
always is an asymptotic approximation (as shown by the relative error
$|C_n(k)/Q_n(k)|$), {\em even} in its convergence regime
above the accumulation line {\em and even} on the real axis, 
where the zeta function itself
is in its domain of absolute convergence~\cite{scherer}.

Thus, the value of ${\rm Im}\, k$ where --- for a given $m$ --- 
the $Z_{\rm GV}(k)|_n$  sum 
deviates from $\det{} {\bf M}(k)|_n$ is governed by the real part of $k$ 
and the scaling rules  \equa{Qscale} and \equa{Cscale}. 
It has nothing  to do with the 
boundary line between the convergence region and the asymptotic region of
$Z_{\rm GV}(k)$, as the asymptotic expansion is given by a {\em finite} 
sum of all terms satisfying \equa{asymp}. Therefore, the truncated 
Gutzwiller-Voros expansion describes the quantum-mechanical resonance data
even {\em below} the line of convergence of the infinite curvature 
series, see \ref{chap:num_resonances}.
On the other hand, the boundary line of the convergence regime of the 
Gutzwiller-Voros expansion
is solely governed by those $C_m(k)$ which have nothing to do with 
the quantum analog $Q_m(k)$, i.e.\ solely by terms of character 
\equa{non_asymp}.
The reason is, of course, that the convergence property of an infinite sum
is
governed by the {\em infinite} tail   and not by the first few terms.
Whether  the Gutzwiller-Voros expansion converges or not is therefore
not  related to  whether 
the quantum-mechanical data are described
well or not.  

The {\em convergence property} of a semiclassical zeta function 
on the one hand and the {\em approximate description of quantum mechanics} 
by these zeta functions are two {\em different things}.
It could happen that a zeta function is convergent,
but not a good description of quantum mechanics (see especially 
the failure of the entire quasiclassical zeta function to approximate
the exact cluster phase shift in Sec.\ref{chap:num_clusterphase}).
On the other hand it may not converge, in general, but its finite truncations
nevertheless approximate -- at least to some order  -- 
quantum mechanics, as it is the case for the Gutzwiller-Voros 
zeta function. 

These findings hold for any re-writing  of the
Gutzwiller-Voros zeta function, as
$Z_{\rm GV}(k)|_n$ was already shown to be {\em asymptotic} in a regime where 
the curvature
sum is still {\em absolutely convergent} and the limit
$\lim_{n\to \infty} Z_{\rm GV}(k)|_n$ exists.
Therefore, 
any re-writing of $Z_{\rm GV}(k)$, especially the one of ref.\cite{fredh2}
as the ratio of four quasiclassical zeta functions \equa{num:ratio}
will {\em at best} work as an {\em asymptotic}
expansion to the exact quantum-mechanical cumulant expansion.  
Note that, for  finite curvature order $n$
\be
  \frac{F_{+}(\half     ,k)|_n F_{-}(\sevenhalf,k)|_n}
              {F_{-}(\threehalf,k)|_n F_{+}(\fivehalf,k)|_n}
 &\neq& 
   \left.\left\{  \frac{F_{+}(\half     ,k) F_{-}(\sevenhalf,k)}
              {F_{-}(\threehalf,k) F_{+}(\fivehalf,k)} \right \}\right|_n 
  \label{GVvar} \label{f-ratio} \\
 &=& Z_{\rm GV}(k)|_n \ .
 \nonumber
\ee  
If the ratio is evaluated according to the r.h.s.\ of \equa{GVvar}, 
one obtains exactly
the same result as for the original Gutzwiller-Voros expansion using formula 
\equa{GV_zeta_app}.
If, however, the ratio is evaluated
according to l.h.s.\ of \equa{GVvar}, the relation to the quantum-mechanical 
cumulant expansion is lost: the matching of the semiclassical 
coefficients of  $z^m$ with the quantum-mechanical ones is spoiled,
as the asymptotic terms, resulting from various curvature 
orders of the $Z_{\rm GV}(k)|_n$ calculation, mix. 
If $n$ is large enough, also the l.h.s.\ of \equa{GVvar}
will deviate strongly from the quantum mechanics as
the original
formulation of the Gutzwiller-Voros expansion does  --- 
the difference is that this new expression approximates quantum mechanics 
at slower rate than the original formula, as the asymptotic terms of higher and
lower curvature order are mixed. However, at high enough curvature order $n$
also the new l.h.s.\ of \equa{f-ratio} 
will encounter terms of class \equa{non_asymp} and
will therefore --- for large negative imaginary wave numbers --  
deviate strongly from the quantum-mechanical resonance data.

What is the reason for the truncation at $ka \approx 2^{m_{\rm 
crit}-1} \bar L/a$\,?
This boundary follows from a combination of the uncertainty principle
with ray optics and the exponentially increasing number of periodic orbits
of the 3-disk repeller. For fixed wave number $k$, quantum mechanics
can only resolve the classical repelling set of the periodic orbits
up to a critical topological order $m_{\rm crit}$. The quantum
wave-packet  which explores the repelling set,  has to disentangle $2^n$
different sections of size $d\sim a/2^n$ on the ``visible'' part of the
disk surface between two successive collisions with the disk. Since
these collisions are spatially separated by the mean length 
$\bar L$, the flux spreads by a factor $\bar L/a$. 
In other words, the non-vanishing value of the  topological entropy for
the 3-disk system, $h\sim \ln 2$, 
is the reason. For comparison, the uncertainty bound on the wave number
in the hyperbolic, but non-chaotic two-disk system is independent of
the curvature order (in case diffractive creeping is negligible), as
there is only one geometrical periodic orbit and therefore the repelling
set is trivial with zero topological entropy. 

\begin{figure}[htb]
\centerline{
{\epsfig{file=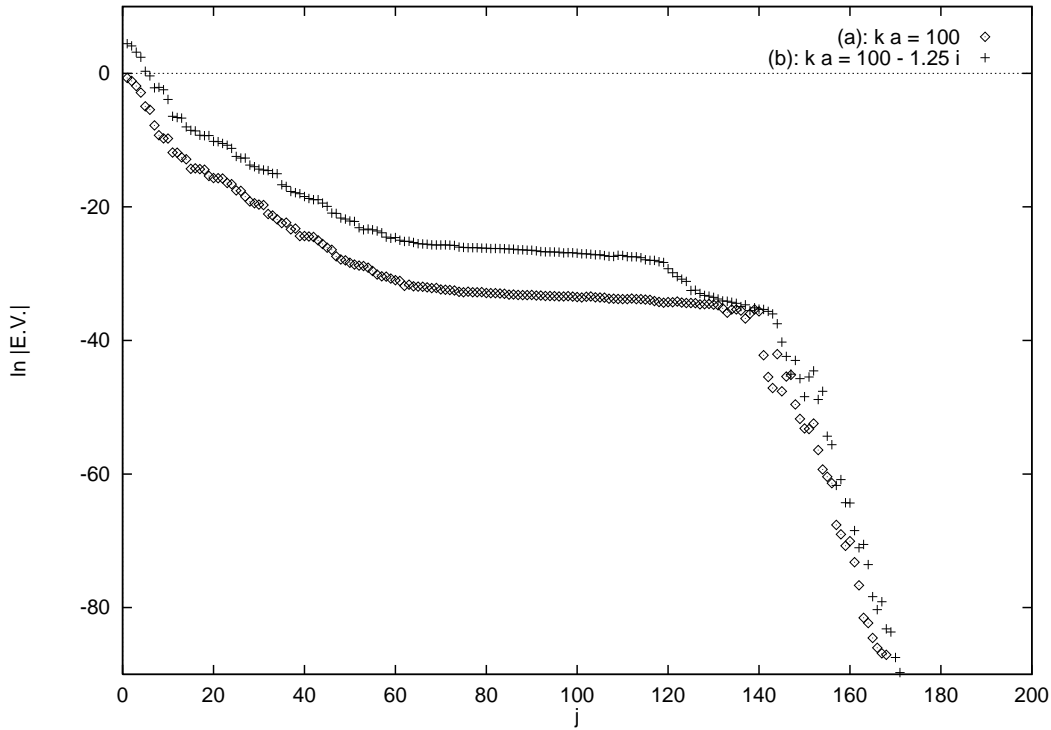,height=10cm,angle=-90}}}
\caption[fig:eigenvalues]{\small The moduli 
of the eigenvalues 
of the multiscattering   kernel 
 $A_{m,m'}(k)$ of the ${\rm A}_{\rm 1}$ 3-disk repeller with $R=6a$  
for the cases $k=100/a$ 
and $k=(100-1.25\i)/a$
on a logarithmic scale.  
The eigenvalues are displayed as function of their
index $j$ in descending order.
\label{fig:eigenvalues}}
\end{figure}

The result that the semiclassical curvature expansion has to be truncated
at finite order
for a fixed wave number $k$, is different from the fact that the (in principle
infinite) multiscattering kernel ${\bf A}_{m,m'} = 
{\bf M}_{m,m'}-\delta_{m,m'}$ can be truncated to a finite matrix.
The truncation in the curvature order is related to the 
resolution of the repelling set of periodic
orbits of the 3-disk system.
The truncation in the size of the matrix is related to the semiclassical
resolution of the single disks of the 3-disk system.
The point particle classically only scatters from the disk, if its impact
parameter is of
the size or smaller than the disk radius $a$. Note that in the
fundamental domain of the ${\rm A}_{\rm 1}$ disk system, one considers only one 
half-disk. 
Mathematically, this follows  from the asymptotic behaviour
of the ratio $\Jb{m}{a}/\Ho{m}{a}$ which governs the scaling of the  
kernel ${\bf A}_{m,m'}$ and which is 
valid for $m$ larger than $m_{\rm max}$, defined in Eq.\equa{m-max},
see App.\ref{app:suppl}.
 In order to visualize this, we have plotted
in Fig.\ref{fig:eigenvalues} the moduli of
the eigenvalues (on a logarithmic scale and in descending order) 
of the multiscattering kernel
 $A_{m,m'}(k)$ of the ${\rm A}_{\rm 1}$ 3-disk repeller with $R=6a$  
for the cases $k=100/a$ (on the real wave-number axis)
and $k=(100-1.25\i)/a$ as function of the eigenvalue index $j$.
The imaginary part of the 
latter wave number is characteristic
for a domain where the
subleading resonance bands emerge.
The one-disk resolution is clearly visible in the exponentially
decreasing tails of both curves above ${\rm Re}\,k \approx 140/a$.
In order to exhibit this feature,
the matrix itself was truncated here at a large value of $m=220$.
Furthermore, from the curves, one can read off that only the first
few eigenvalues (six for the upper curve corresponding to the case
$k=(100-1.25\i)/a$ 
 and two to four eigenvalues for the
lower curves for the case $k=100/a$) are ``essential'', i.e., are
of the order unity or bigger.
These numbers match very well the minimal topological order needed
in the semiclassical calculation to approximate the relevant resonances
at the specified $k$ values. 
Whereas inside the negative complex $k$-plane one has to go to
higher curvature orders in the truncation of
the semiclassical zeta function 
in order to find all the subleading resonances
(namely to  
order six for the specified $k$ value), on or close to the real axis
only  the leading
resonances are ``visible'', in agreement with the data for
the cluster  phase shifts which, for the specified $k$ value, 
can be well approximate by a semiclassical
calculation of order three to four. 
One can also extract from the figure what happens if the negative
imaginary part of $k$ is increased: the curve is basically 
parallelly shifted upwards. 
Thus the number of eigenvalues and the minimal curvature
order for the semiclassical description of quantum mechanics increases
the deeper one ``dives'' into the lower complex wave-number plane.
Although only a few eigenvalues are essential for the computation of 
the resonances
and phase shifts, the size of the matrix is determined by the
much bigger number $m_{\rm max}\approx 
(e/2) ka$, such that many more eigenvalues
are produced by a matrix diagonalization code.
Unfortunately, one cannot escape this mismatch, as the model space
for the matrix ${\bf M}$ has to be that large in order to guarantee stable
numerical results for the leading eigenvalues.

In summary, the minimal size of the matrix is determined by the
resolution of the single disks, whereas the maximal topological order up
to which the semiclassical curvature expansion makes sense, follows from
the Heisenberg uncertainty limit on the quantum 
resolution of the repelling set.
The topological  exponential rise of the number of periodic orbits, with
increasing curvature expansion order $n$, is the physical reason for 
the eventual 
breakdown of the curvature expansion of the semiclassical zeta function
\equa{GV_zeta_app} as compared with the exact quantum-mechanical cumulant
expansion which defines the determinant of the multiscattering matrix
in an infinite-dimensional Hilbert space.
\newpage
\section{Conclusions\label{chap:end}}
\setcounter{equation}{0}
\setcounter{figure}{0}
\setcounter{table}{0}
%

Starting from the exact quantum-mechanical ${\bf S}$-matrix we have
tried to find a direct derivation of the semiclassical spectral
function for a rather special class of classically hyperbolic
scattering systems, namely the non-overlapping disconnected finite $n$-disk
repellers in two dimensions.  We have confined our investigation to
these systems as they are on the one hand ``realistic'' enough to
capture the essence of classically hyperbolic scattering problems (or,
for certain geometries, even chaotic scattering problems) and on the
other hand simple enough to allow for a ``top-down'' approach from
exact quantum mechanics to semiclassics without, and this is the
important point, any ``formal'' step in between.  We have reason for
this ``pedantry'': It is known from the work of
Refs.\cite{ER92,fredh,fredh2} that the standard spectral function, the
cycle-expanded 
semiclassical zeta function of Gutzwiller and Voros, is not entire for
the 3-disk system and therefore fails to describe subleading
scattering resonances in the complex wavenumber plane below its 
boundary 
of convergence. The question is whether these failures are induced by
unjustified formal steps in the semiclassical reduction or whether they
are inherently
a property of the semiclassics itself. Since we expected that {\em the}
semiclassical spectral function must follow from the semiclassical
reduction of the cumulant expansion of the corresponding
multiscattering kernel~\cite{aw_chaos,aw_nucl}, we had to avoid any
bias or unjustified assumption on the exact kernel as well as on the
semiclassical spectral function, e.g., on the existence and
regularization of the quantum mechanical expression\,\footnote{In
fact, even the quantum-mechanical expression is not entire in the
whole complex $k$-plane, since it has a branch cut on the negative real
axis and poles which cancel the one-disk singularities.}, on
the structure of the period orbits, especially on the structure of
their stabilization, on the organization of the spectral function in
terms of cycles or curvatures, on the classification of these
curvatures by the topological lengths of the orbits, etc.

Our first task therefore has been to ensure
that  the quantum mechanical starting point for the
semiclassical reduction is well-defined. The ${\bf T}$-matrix of the 
$n$-disk scattering systems, derived by  the methods of stationary
scattering theory~\cite{gr}, was shown to exist on the real $k$-axis 
and to be trace-class. Therefore the actual starting point for the
{\em spectral} classification of the scattering system, 
the {\em determinant} of the ${\bf S}$-matrix, exists also  and
can be manipulated by cyclic permutations, unitary
transformations, splitting into sub-determinants, in other words, operations
which are non-trivial for matrices of infinite-dimensional Hilbert spaces.
With the  help of these (then justified) operations, 
we succeeded in transforming the determinant
of the ${\bf S}$-matrix into a form that is well suited for the
semiclassical reduction step, see Eq.\equa{qm}. 
It separates into 
the incoherent superposition of $n$ one-disk scattering determinants and into
the ratio of the determinant and the complex conjugated determinant of the
genuine multiscattering matrix, {\bf M}. 
Furthermore, the determinant of the multiscattering 
matrix can be decomposed into
sub-determinants, if the $n$-disk system has additional symmetries.
All of the above mentioned  determinants are shown to exist separately.
This is one of the key points for the semiclassical reduction, since the
existence of the ${\bf S}$-matrix alone would not guarantee that the
one-disk aspects can be separated from the multi-disk aspects  
in a well-defined manner. 
Note that the standard geometrical periodic 
orbits (without creeping) can only ``know'' about the multi-disk aspects,
and not about the single disk aspects.

 As the determinants are taken over infinite dimensional matrices, one
has to worry about their very definition. The von-Koch criterion  
(the existence of the determinant in one orthonormal basis, 
see App.\ref{app:trace_Koch}) 
is not sufficient for this task, since implicitly in the derivation
of the ${\bf S}$-matrix and explicitly under symmetry-reductions unitary
transformations are mandatory. The multiscattering matrix must be reducible
to a 
form ``unit matrix plus a trace-class matrix'' in order for 
its determinant to exist. Fortunately, we could prove
that the multiscattering
kernel ${\bf A}={\bf M}-{\bf 1}$ is trace-class for
any $n$-disk geometry as long as the disks do not overlap nor touch each other.
Furthermore, the determinant over the infinite matrix ${\bf M}$ is 
defined as a cumulant expansion which -- as shown by us -- semiclassically
reduces to the curvature expansion. Thus already quantum mechanically the
cumulant/curvature ``regularization'' emerges. Moreover, by working in
the full domain of the $n$-disk system, we could show in 
Sec.\ref{chap:semiclass} that the
cumulants split via the quantum traces 
into quantum itineraries which can be classified  by the very 
symbol dynamics of their semiclassical reductions -- the semiclassical 
periodic orbits. Thus the
cumulant/curvature ordering in terms of the {\em topological} lengths of
the quantum itineraries and hence of the 
{\em topological} lengths of the semiclassical periodic orbits is already
present on the quantum-mechanical level.
One does not need to impose it from the outside (as it would be 
the case for the
semiclassical reduction of the 
Krein-Friedel-Lloyd sums  of two bounded reference systems);
but it follows naturally
from the properties
of quantum mechanics; namely, 
from the defining cumulant expansion
of the determinant of the exact multi-scattering matrix. 

Thus the classification by the quantum itineraries is a virtue, but it 
is unfortunately also
a vice, as quantum traces and the Plemelj-Smithies recursion formulas
are involved. The latter introduce (unnecessarily) large terms  which finally
cancel in the cumulants themselves. As the cumulant sum of 
a trace-class operator converges absolutely, a direct semiclassical
reduction of a complete cumulant and not of the potentially large 
quantum itineraries or
quantum traces would be highly desirable. Unfortunately, the
direct semiclassical
reduction of a complete cumulant is not known. It might correspond to
an integration of the small {\em differences} between
the direct motion and the shadowed motion 
of the quantum wave packet. Instead,  the standard calculation 
for the complete curvatures proceeds 
through the shadowing of all full periodic orbits of a the pertinent
topological length by all 
pseudo-orbits (=products of shorter periodic orbits) 
of the same total topological length. 
Because of these large cancellations, the semiclassical reduction on the
level of the itineraries is potentially dangerous for the semiclassical
equivalent of the quantum-mechanically absolutely converging cumulant sum,
the curvature sum. There is no guarantee that it converges as well.

As mentioned, in Sec.\ref{chap:semiclass} we have managed to construct the
semiclassical equivalent for each specified quantum itinerary.
By working in the full domain and utilizing the pertinent 
simple symbol dynamics,
which is
valid under the condition that the number of disks is finite and that the
disks do not overlap nor touch, our
semiclassical reduction applies for all $n$-disk geometries, with one 
exception:  we have to veto geometries which allow for grazing
periodic orbits. In this way, we could guarantee for any specified
quantum itinerary that the
sequence of disk labels  transforms uniquely to
a sequence of non-overlapping
semiclassical saddles in the complex angular-momentum plane 
which corresponds to the standard semiclassical 
periodic orbit, specified by the same symbolic
sequence. The weight of the periodic orbits was shown to be 
identical to the one derived by  Gutzwiller~\cite{gutzwiller}. 
Furthermore, we have shown that to each itinerary that generates
a (non-creeping or creeping) periodic orbit with a ``ghost tunneling'' section 
straight through a disk there belong ``parent'' 
itineraries, such that the ghost and corresponding parent periodic orbits
cancel exactly in the semiclassical curvature sum. 
This establishes how pruning emerges from quantum
mechanics in the semiclassical reduction. We have also shown
that, to each quantum itinerary of topological length $n$, there belong
$3^n\mbox{$-$}1$ different periodic orbits which contain creeping sections and
we have specified a generalization of the symbolic labelling. By the Watson
contour method of Ref.\cite{franz} we have 
derived their structure which 
agrees with the result  of the semiclassical
construction of Refs.\cite{vwr_prl,vwr_stat} which in turn 
is based on Keller's
semiclassical theory of diffraction~\cite{keller}.
The direct  link of the determinant of the exact ${\bf S}$-matrix (via
the determinant of the multiscattering matrix, via its cumulants and 
quantum itineraries) to the periodic orbits is therefore established.
If the operations are inverted, the right hand side of Eq.\equa{gen}
emerges, modulo the caveat that the semiclassical curvature sum might
not converge, in general.

What is known about the convergence properties of the curvature
expansion of the Gutzwiller-Voros zeta function from the literature?
The Gutzwiller trace as well as the zeta-function for $n$-disk repellers 
is known to converge (even absolutely) in the complex wave-number plane above
a line specified by the resonance with largest imaginary $k$-value,
see e.g.\ Ref.\cite{scherer}. As
all resonances belong to the 
lower half of the complex wave-number plane, the zeta function converges
at least on the real $k$-axis. From Refs.\cite{ER92,fredh,fredh2} it is
known that the cycle or 
curvature expansion of the Gutzwiller-Voros sum converges
even inside the resonance region above an accumulation line 
defined by the poles of the dynamical
zeta function. Thus above this accumulation line 
(and away from the branch cut and singularities of the exact
quantum-mechanical side)  
our semiclassical limit
\equa{gen}  (or \equa{gen_sym_full}  for symmetry-reducible problems) 
is established for the full, untruncated 
Gutzwiller-Voros zeta function; see, however, below for the discussion of
the asymptotic behaviour of the curvature sum.

The relations are compatible with Berry's 
expression for the integrated spectral density in Sinai's
billiard (a {\em bounded} $n\to \infty$ disk system, 
see Eq.(6.11) of Ref.\cite{Berry_KKR}) and -- in general -- with the 
Krein-Friedel-Lloyd sums
\equa{friedel_sum_rho}.
They justify the formal manipulations of Refs.\cite{scherer,pinball,moroz}. 
Furthermore, for these scattering systems, 
unitarity is automatically 
preserved semiclassically (without reference of
any re-summation techniques 
\`{a} la Berry and Keating\cite{berry_keats} which are needed in
bounded problems). Quantummechanically, unitarity follows
from the relation
\be 
  \det{} {{\bf S}^{(n)} (k)}^\dagger = \frac{1}{\det{} {\bf S}^{(n)}(k^\ast)}
\ee 
which is manifestly the case (see  the first lines of Eq.\equa{gen} 
and   \equa{gen_sym_full}). Semiclassically, this follows from 
the second lines of \equa{gen} and \equa{gen_sym_full}, with the caveat
that curvature sums on the right hand sides must exist, i.e., they
either converge or are suitably truncated.
This is of course a very pleasant property. But,
on the other hand, unitarity can therefore not be used
in scattering problems to gain any constraints on the 
structure of
${\widetilde Z}_{\rm GV}$, as it could in bounded problems,
see \cite{berry_keats}.
Why are 
bounded problems special? In the semiclassical treatment of {\em scattering}
problems the poles of the determinant of the ${\bf S}$-matrix result
from the zeros of ${\widetilde Z}_{\rm GV}(k)$ in the lower complex 
$k$-plane (where in general -- except at the zeros --  
${\widetilde Z}_{\rm GV}(k)$ dominates 
${\widetilde Z}_{\rm GV}(k^\ast)^\ast$ which is small, but nonzero),
whereas the zeros of the determinant  of the ${\bf S}$-matrix 
are produced by  
the ones
of ${\widetilde Z}_{\rm GV}(k^\ast)^\ast$ in the upper complex $k$-plane (where
in turn  ${\widetilde Z}_{\rm GV}(k)$ is the small, but nonzero zeta-function).
For bounded problems $k$ is real and both zeta-functions become 
equally important. (A sign of this is the fact that 
the  Hankel functions of either first or second kind which appear in 
the spectral determinants  are replaced by
the corresponding Bessel functions.)\ This 
obviously calls for a fine-tuning, hence, the re-summation. Note
also the symmetry-breaking $\i\epsilon$ prescription which had to be added
to the l.h.s.\ of the 
Krein-Friedel-Lloyd sums, see 
Sec.\,\ref{chap:intro}.

As stated above, the incoherent single-disk scattering decouples from 
the genuine multi-disk scattering. 
The 1-disk poles do not influence the position of 
the {\em genuine} multi-disk poles.
However, $\Det{}{\bf M}(k)$ does not only possess zeros, but also poles. The
latter exactly cancel the poles of the product over the 1-disk
determinants, $\prod_{j=1}^{n} {\rm det} {\bf S}^{(1)}(ka_j)$, 
since both involve the same ``number'' and ``power'' 
of $\Ho{m}{a_j}$ Hankel
functions in the denominator. The same is true for the poles of 
${\Det{} { {\bf M}(k^\ast)}^\dagger}$ and the {\em zeros} of 
 $\prod_{j=1}^{n} {\rm det} {\bf S}^{(1)}(ka_j)$, 
as in this case the ``number'' of $\Ht{m}{a_j}$
Hankel functions  in the denominator of the former and the
numerator of the latter is the same --- see also Berry's 
discussion of the same cancelation in the integrated spectral density
of Sinai's billiard (Eq.(6.10) of Ref.\cite{Berry_KKR}). 
Semiclassically, this cancelation corresponds to a removal of the additional
creeping contributions of topological length zero, i.e., 
$1/(1-\exp(\i 2\pi \nu_{\ell}))$,
from ${\widetilde Z}_{GV}$ via the 1-disk diffractive
zeta functions, ${\widetilde Z}_{\rm disk({\it j}l)}$ and $
{\widetilde Z}_{\rm disk({\it j}r)}$. 
The orbits of topological length zero result
from the geometrical sums over additional creepings around the single disks,
$\sum_{n_{w}=0}^{\infty} 
(\,\exp(\i 2\pi \nu_\ell)\,)^{n_w}$ (see \cite{vwr_prl}), and multiply the
ordinary creeping paths which are classified by their topological
length.
Their cancellation
is very important for situations
where the disks nearly touch, as in such cases the full circulations by 
creeping orbits of any of
the touching disks should clearly be suppressed, as it now is.
Therefore, it is important to keep a consistent count of
the diffractive contributions in the semiclassical reduction.

What happens to the resonances, when the  spacing
between the disks becomes vanishing small such that bounded regions are
formed in the limit of $n>2$ touching disks?
Because of the ratio $\Det{}{\bf M}(k^\ast)^\dagger/\Det{}{\bf M}(k)$,
to each (quantum-mechanical or semiclassical pole) of
$\det{} {\bf S}^{(n)}(k)$ in the lower complex $k$-plane there 
belongs a zero of $\det{} {\bf S}(k)$
in the upper complex $k$-plane with the same ${\rm Re}\, k$ value, 
but opposite ${\rm Im}\, k$.
When the bounded regions are formed
some of these opposite zeros/poles move onto the real axis 
(such that their contributions cancel out of \equa{qm}).
We have convinced ourselves that for the 3-disk scattering system 
with $\epsilon>0$ separation
these
resonances approach infinitesimally the bound-state eigenvalues
of the complementary calculation of
the spectrum inside  the bounded region, see, e.g., Ref.\cite{scherer}
for the billiard  bounded by  three touching disks. 
Semiclassically,
this would be 
a non-trivial calculation as the eigen-energies have to be real which
--- without resummation \`{a} la Berry and Keating~\cite{berry_keats} --- they
are not. In this situation, one really has to think about further 
resummation techniques. 
Most
of the resonances, however,  do not move onto the real axis at all, 
as $n$-disk repellers,
{\em even} with bounded sub-domains, are still scattering systems.
The
would-be bound states, however, drop out of the exact formula
for $\det{} {\bf S}^{(n)}(k)$, as they should.


Let us come back to the numerical data of Sec.\ref{chap:numerical} and
the existence of the curvature expansion.
In this section
we have reported on numerical results for the exact quantum-mechanical
${\rm A}_{\rm 1}$ 
resonances of the three-disk system with $R=6a$ in the complex
$k$-plane in the region: $0 \leq {\rm Re}\, k \leq 250/a$ and $0\geq
{\rm Im}\, k \geq -1.6/a$.  
The first observation is that the quantum-mechanical 
resonances in this window can be grouped
into (leading and subleading) bands. In addition to the data presented in
Sec.\ref{chap:num_resonances}, 
where we have related the band structure to the semiclassical
curvature expansion, it has been 
numerically checked that the emergence of a
new band is in fact linked to a new cumulant order. 
The data of this window up to
${\rm Re}\, k =200/a$ can be fitted very well with a quantum-mechanical
cumulant expansion which is truncated at order seven.
This knowledge, together with the fact that any periodic orbit results from
the semiclassical reduction of a quantum itinerary with the same symbol
sequence (in the full domain),  tells us that 
periodic orbits of topological length eight and higher are completely 
irrelevant for the description of the presented 
quantum-mechanical data, for regions 
below ${\rm Re}\, k \approx 200/a$ and above 
${\rm Im}\, k = -1.6/a$.
Thus any deviation of semiclassical predictions from the exact data 
cannot be cured by the inclusion of higher topological orbits.
At best, they should leave the resonances untouched.

This finding seems to be at variance with the result 
of Sec.\ref{chap:num_resonances}
where the quasiclassical
zeta function of Ref.\cite{fredh2} approximates most of the exact
resonances at curvature
order twelve. However, this truncated zeta functions 
finds also six erroneous resonance bands which do not
have quantum-mechanical counter parts. This means that its topological
expansion does not match the cumulant expansion, as we know of course
from the analytical results of Sec.\ref{chap:semiclass}. The semiclassical
reduction of a cumulant sum {\em is} 
the Gutzwiller-Voros curvature sum, and not
the cycle expansion of the quasiclassical zeta function nor of the dynamical
one. In fact, as shown by the comparison of exact to semiclassical coherent
phase shifts, see Sec.\ref{chap:num_clusterphase}, the latter two zeta
functions are very inferior to the Gutzwiller-Voros zeta
function which describes the exact phase shifts 
up to the resolution of the plot.
Any competitor zeta function should do at least as well as 
the Gutzwiller-Voros one in order to be taken seriously. The question
whether it 
converges or not is not a criterion for how successfully it approximates
the  quantum-mechanical data. 

In Sec.\ref{chap:num_cumulant} we have
finally executed what was already advocated by us in 
Refs.\cite{aw_chaos,aw_nucl}. For the 3-disk example we 
have numerically compared term by
term and  order by order
the quantum-mechanical cumulants with their semiclassical counterparts,
the curvatures of the Gutzwiller-Voros zeta function. The numerical data
show
that, for a fixed value of the wave number, 
the cumulants and curvatures agree in 
magnitude and also in phase 
only up to a finite cumulant order which  is determined by the wave number
and then deviate strongly.
We have interpreted this result from the
uncertainty bound on the quantum-mechanical resolution of the details of 
the classically repelling set which exponentially grow with the topological
entropy of the system under consideration.
Close to the real axis these deviations are
hidden by the smallness in absolute value of the higher-order 
cumulants and curvatures. Therefore, the semiclassical phase shifts agree
very well with the quantum-mechanical ones, even for very high
curvature orders. However,  with increasing value for $-{\rm Im}\,k$,
inside
the lower half of the 
complex plane, the
deviations are enhanced by the scaling laws discussed in 
Sec.\ref{chap:num_cumulant} such that they eventually become noticeable.

This observation 
matches extremely well the results of the resonance comparison in
Sec.\ref{chap:num_resonances}. The resonances which are located
above  and away from the boundary of convergence are 
approximated by the Gutzwiller-Voros
curvature expansion as soon as the curvature order is sufficiently high.
The resonances at or below the boundary  of convergence, however, are 
approximated only up to the curvature order which respects the uncertainty
bound. The curvature expansion 
works there only as an asymptotic series.

Our interpretation is that eventually
quantum mechanics and (semi-)classics have to part ways, as 
the quantum-mechanical spectral data only need power-law complexity, i.e.\
$N^3$ operations if the multiscattering matrix can be truncated as
an $N\times N$ matrix, whereas the resolution of the 
classically repelling set needs exponential
complexity if the topological entropy is non-zero.
In other words, whether the curvature expansion converges or not with
respect to quantum mechanics it should be truncated at the cumulant order
specified by the uncertainty bound. All curvature terms exceeding this
order are -- from the quantum-mechanical point of view -- irrelevant.
From this perspective, the semiclassical side of the
relation \equa{gen} (and \equa{gen_sym_full})
should be interpreted to be valid just for the truncated Gutzwiller-Voros
curvature sums, where the order of the truncation increases with increasing
value of ${\rm Re}\, k$ (or, since  $k=p/\hbar$, with  decreasing $\hbar$). 
The semiclassical limit ${\rm Re}\, k\to \infty$ and the cumulant limit
$m\to\infty$ do not commute, in general, if 
the topological entropy is non-zero. These facts should be kept separated
from the  $\hbar$-effects of 
Refs.\cite{alonso,gasp_hbar,vattay_hbar} which investigate the 
${\cal O}(\hbar)$
corrections to the periodic orbits. We discuss here the $\hbar$-corrections
to the curvatures which result from the periodic orbits
via large cancellations against the pseudo-orbits.
Part of the $\hbar$-corrections of   Refs.\cite{alonso,gasp_hbar,vattay_hbar}
cancel out as well,
as can be shown from the comparison of the difference 
between the $m$-th order 
exact and
semiclassical trace which exceeds by far in magnitude the difference
between the corresponding $m$-th order cumulant and curvature.
In fact, from the discussion of the subleading
Debye corrections in App.\ref{app:convol_straightline} one can deduce
that each term $\Ht{l}{a}/\Ho{l}{a}$ 
introduces a correction
factor of order $(1+\i\hbar/4 p a)$, such that the quantum itinerary of
topological
length $m$ has at least an ${\cal O}( 1+m\i \hbar /4 pa)$ factor
relative to the corresponding periodic orbit
(assuming that all disks have the same radius $a$ for simplicity). 
However, the pseudo-itineraries of order $m$ (which are the quantum mechanical
analog of the pseudo-orbits) have the same correction factor, such that
it cancels in the corresponding cumulant. Thus, the ${\cal O}(m\hbar)$
terms cancel. But what about the ${\cal O}(m\hbar^2)$ terms which might
be of the same importance as the ${\cal O}(\hbar)$ terms, as the limits
$m\to \infty$ and $\hbar\to 0$ do not commute? Only if the cumulant sum
is truncated at a finite order, the  ${\cal O}(m\hbar^2)$ terms 
become negligible relative to the ${\cal O}(\hbar)$ terms. In principle,
the uncertainty boundary should be derivable from the semiclassical
reductions of Sec.\ref{chap:semiclass} and App.\ref{app:convol} to
the quantum itineraries, if $\hbar$-corrections are taken
into account systematically. In practice,
however, there is a very long way from the 
$\hbar$-corrections extracted from the quantum itineraries to the surviving
$\hbar$-corrections on the cumulant level
because
of the
very
large cancellations of the
quantum itineraries with the pseudo-itineraries.

\newpage
\section*{Acknowledgements}

The author would like to thank the 
Niels-Bohr-Institute and Nordita for repeated hospitality 
and
the Leon Rosenfeld Scholarship Fund 
and its board for support of his recent
visit to Copenhagen. He is indebted to Predrag
Cvitanovi\'{c}, Michael Henseler, 
Per Rosenqvist and G\'{a}bor Vattay
for many helpful contributions.
Furthermore numerous illuminating discussions with
Andy Jackson,
Debabrata Biswas, Bruno Eckhardt, Harald Friedrich,
Pierre Gaspard, Bertrand Georgeot, 
Thomas Guhr, Dieter Gr\"{a}f, Ralph Hofferbert, Ronnie Manieri,
Carmelo Pisani,
Harel Primack, Achim Richter, 
Martin Sieber, Uzy Smilansky, Frank Steiner, Gregor Tanner, 
Hans Weidenm\"{u}ller,
Niall Whelan and the late Dieter Wintgen  are gratefully acknowledged.  
He thanks  Bruno Eckhardt and Per Rosenqvist for supplying
him with numerical input for periodic orbits of the 3-disk system.
Part of the present work has overlap with the author's
publications and preprints 
\cite{cvw96,N-ball,wh97} and with Michael Henseler's diploma 
thesis~\cite{mh} which
was prepared and written 
under the author's guidance; the author would like to thank
all his collaborators and especially
Michael Henseler.

Encouraging interest of  Friedrich Beck, Gerry Brown, 
Mariana Kirchbach, Achim Richter,
Jochen Wambach and all the members of the NHC group  of the
institute of nuclear physics at Technische Hochschule
Darmstadt is gratefully
acknowledged.

The author thanks Jochen Wambach for carefully reading the manuscript and
Ralph Weichert for his help on the program {\em xfig\/}.

Numerical calculations were performed on the DEC-Alpha workstation cluster at
the Niels-Bohr-Institute, on the IBM-Risc workstation cluster at
GSI Darmstadt and the IBM-Risc and DEC-Alpha server of  the NHC group at 
the
TH-Darmstadt.
\newpage
\appendix
\section{Traces and determinants of infinite dimensional 
matrices
\label{app:trace}}
\setcounter{equation}{0}
\setcounter{figure}{0}
\setcounter{table}{0}
This Appendix summarizes the definitions and properties for trace-class
and Hilbert-Schmidt matrices and operators, the determinants over 
infinite dimensional matrices and possible regularization schemes for
matrices or operators which are not of trace-class.

\subsection{Trace class and Hilbert-Schmidt class}

This section is based on
Ref.\cite{bs} and also Refs.\cite{rs1,bs_adv,gohberg,kato} which 
should be consulted for further details and proofs.
The trace class and Hilbert-Schmidt property will be defined
here
for linear, in general nonhermitean operators ${\bf A}\in 
{\cal L}({\cal H})$: ${\cal H} 
\to {\cal H}$ (where ${\cal H}$ is a separable Hilbert space). 
The transcription to matrix elements (used in the prior chapters) is
simply $a_{ij}= \langle \phi_i, {\bf A} \phi_j\rangle$ 
where $\{ \phi_n \}$ is an
orthonormal basis of ${\cal H}$ and $\langle \ ,\ \rangle$ is the inner product
in ${\cal H}$ (see Ref.\cite{gohberg} where the theory of 
{\em von Koch matrices} of Ref.\cite{vonKoch} is discussed.). 
Thus the trace is the generalization of the usual notion of
the sum of the diagonal elements of a matrix; but because {\em infinite} 
sums are
involved, not all operators will have a trace and, if the trace exists in one
basis, it is nontrivial that it exists also in any other basis:

\begin{description}
\item[\ (A)\ ]
 An operator ${\bf A}$ is called {\bf trace-class}, ${\bf A} \in 
{\cal J}_1$, if and only if, for every orthonormal basis, $\{\phi_n\}$:
\be
 \sum_n |\langle \phi_n, {\bf A} \phi_n \rangle| < \infty\; .
 \label{tc_def}
\ee
The family of all trace-class operators is denoted by ${\cal J}_1$.
\item[\ (B)\ ] 
An operator ${\bf A}$ is called {\bf Hilbert-Schmidt},  ${\bf A} \in 
{\cal J}_2$, if and only if, for every orthonormal basis, $\{\phi_n\}$:
\be
 \sum_n \| {\bf A} \phi_n \|^2 < \infty\; .
\ee
The family of all Hilbert-Schmidt operators is denoted by ${\cal J}_2$.
\item[\ (C)\ ]
{\bf Bounded operators} ${\bf B}$ are dual to
trace-class operators. They satisfy the
the following condition: $|\langle \psi, {\bf B} \phi\rangle | 
\leq C\| \psi\| \|\phi\|$ with $C <\infty$ and $ \psi,\phi \in {\cal H}$.
If they have eigenvalues,
these are bounded as well. The family of bounded operators is denoted by 
${\cal B}({\cal H})$ with the norm $\|{\bf B}\|= {\rm sup}_{\phi \neq 0} 
\frac{\|{\bf B}\phi\|}{\|\phi\|}$
for $\phi\in {\cal H}$.
Examples for bounded operators are unitary operators and
especially the unit matrix. In fact, every bounded operator 
can be written as a linear combination of four
unitary operators~\cite{rs1}.
\item[\ (D)\ ]
An operator ${\bf A}$ is called
{\bf positive}, ${\bf A}\geq 0$, if
$\langle {\bf A}\phi,\phi \rangle \geq 0 \ \, \forall \phi\in 
{\cal H}$. Notice that ${\bf A}^\dagger {\bf A}\geq 0$. 
We define ${\bf | A|}=\sqrt{{\bf A}^\dagger {\bf A}}$.
\end{description}

The most important properties of the trace and Hilbert-Schmidt classes
can be summarized as (see Refs.\cite{rs1,bs}):
\begin{description}
\item[\ (a)\ ] 
${\cal J}_1$ and ${\cal J}_2$ are $\ast$ideals., i.e., they
are vector spaces closed under scalar multiplication, sums, adjoints, and
multiplication with bounded operators.
\item[\ (b)\ ] 
${\bf A}\in {\cal J}_1$ if and only if ${\bf A}={\bf  BC}$ with
${\bf B},{\bf C}\in {\cal J}_2$.
\item[\ (c)\ ]
${\cal J}_1 \subset{\cal J}_2$. 
%
\item[\ (d)\ ] 
For any operator ${\bf A}$, we have ${\bf A}\in {\cal J}_2$ if
$\sum_n \| {\bf A} \phi_n \|^2 < \infty$ for a single basis.\\
For any operator ${\bf A}\geq 0$, 
we have ${\bf A}\in {\cal J}_1$ if
$\sum_n |\langle \phi_n, {\bf A} \phi_n \rangle| < \infty$ for a single basis.
\item[\ (e)\ ] 
If ${\bf A} \in  {\cal J}_1$, ${\rm Tr}({\bf A})= 
\sum \langle \phi_n, {\bf A} \phi_n\rangle$ is independent of the basis used.
\item[\ (f)\ ] ${\rm Tr}$ is linear and obeys 
${\rm Tr}({\bf A}^\dagger)=\overline{{\rm Tr}(\bf A)}$; 
${\rm Tr}({\bf A B})= {\rm Tr}({\bf BA})$ if either ${\bf A}\in {\cal J}_1$ 
and  ${\bf B}$ bounded, ${\bf A}$ bounded and ${\bf B}\in {\cal J}_1$ or
both ${\bf A},{\bf B} \in {\cal J}_2$.
\end{description} 
Note that 
the most important property for proving that an operator is trace-class
is the decomposition {\bf (b)} into two Hilbert-Schmidt ones, as the 
Hilbert-Schmidt property can be easily  verified in one single orthonormal
basis (see {\bf (d)}). Property {\bf (e)} ensures then that the trace is the
same in any basis. Properties {\bf (a)} and {\bf (f)} show 
that trace-class operators behave
in complete analogy to finite-rank operators. The proof whether a matrix is
trace-class (or Hilbert-Schmidt) or not simplifies enormously 
for diagonal matrices, as then the second
part of property {\bf (d)} is directly applicable: just  the moduli
of the eigenvalues (or -- in case of Hilbert-Schmidt -- 
the absolute squares) have to be summed
in order to answer that question. A good strategy for checking 
the trace-class character of a general matrix ${\bf A}$ is therefore 
the decomposition
into two matrices ${\bf B}$ and ${\bf C}$ 
where one, say ${\bf C}$, should be chosen to be diagonal
and either just barely of Hilbert-Schmidt character leaving enough freedom
for its partner ${\bf B}$  or of trace-class character 
such that one only has to show
the boundedness for ${\bf B}$.

\subsection{Determinants det({\bf 1}+{\bf A}) of trace-class operators {\bf A}}
This section is mainly based on Refs.\cite{rs4,bs_adv} 
which should be consulted
for further details and proofs. See also Refs.\cite{gohberg,kato}.

\noindent{\bf Pre-definitions} (Alternating algebra and Fock spaces):\\
Given a Hilbert space ${\cal H}$, 
$\otimes^n {\cal H}$ is defined as the vector space of multilinear functionals
on ${\cal H}$ with $\phi_1 \otimes \cdots \otimes \phi_n \in \otimes^n 
{\cal H}$ if $\phi_1,\dots ,\phi_n \in {\cal H}$. 
$\bigwedge^n ({\cal H})$ is defined as the subspace of $\otimes^n {\cal H}$
spanned by the wedge-product
\be
 \phi_1 \wedge \cdots \wedge \phi_n = \frac{1}{\sqrt{n!}} 
     \sum_{\pi \in {\cal P}_n} \epsilon (\pi) [ \phi_{\pi(1)} \otimes \cdots
   \otimes\phi_{\pi(n)}] \; ,
\ee
where ${\cal P}_n$ is the group of all permutations of $n$ letters and 
$\epsilon(\pi) = \pm 1$ depending on whether $\pi$ is an even or odd 
permutation. The inner product in $\bigwedge^n ({\cal H})$
is given by
\be
 \left ( \phi_1 \wedge \cdots \wedge \phi_n , 
         \eta_1 \wedge \cdots \wedge \eta_n \right ) 
    = {\rm det} \left \{ (\phi_i, \eta_j) \right \} \; ,
\ee
where ${\rm det}\{ a_{ij} \} = \sum_{\pi\in {\cal P}_n} \epsilon(\pi) 
 a_{1\pi(1)}\cdots a_{n\pi(n)}$. $\bigwedge^n ({\bf A})$ is defined as 
functor (~a functor satisfies $\bigwedge^n ({\bf A}{\bf B})
=\bigwedge^n ({\bf A})
\bigwedge^n ({\bf B})$~) 
on $\bigwedge^n ({\cal H})$ with
\be
 {\bigwedge}^n({\bf A})\left( \phi_1 \wedge \cdots \wedge \phi_n \right )
 = {\bf A}\phi_1 \wedge \cdots \wedge {\bf A} \phi_n \ .
\ee

\noindent{\bf Properties:}\ 
If ${\bf A}$ trace-class, i.e., ${\bf A} \in {\cal J}_1$, then
for any positive integer $k$, 
$\bigwedge^k ( {\bf A} )$ is trace-class, and for any orthonormal
basis $\{ \phi_n \}$ the cumulant 
\be
  {\rm Tr}\left( {\bigwedge}^k ({\bf A} ) \right )
 = \sum_{i_1 < \cdots < i_k} 
\left ( (\phi_{i_1} \wedge \cdots \wedge \phi_{i_k}),
       ({\bf A}\phi_{i_1} \wedge \cdots \wedge {\bf A}\phi_{i_k}) \right )
\ee
is finite and independent of the basis.  
${\rm Tr}{\bigwedge}^0({\bf A}) \equiv 1$.

\vskip 0.3cm
{\noindent}{\bf Definition:} Let ${\bf A} \in {\cal J}_1$, then 
$\det{}({\bf 1} +{\bf A})$ is defined as
\be
 {\rm det}({\bf 1}+{\bf A})= \sum_{k=0}^{\infty} 
 {\rm Tr}\left( {\bigwedge}^k ({\bf A} ) \right ) \; .
  \label{det_def}
\ee 

\noindent{\bf Properties}:
Let ${\bf A}$ be a linear operator on a separable Hilbert space ${\cal H}$ and
$\{ \phi_j\}_1^\infty$ an orthonormal basis.
\begin{description}
\item[\ (a)\ ] 
$\sum_{k=0}^{\infty}{\rm Tr}\left( \bigwedge^k ({\bf A} ) \right )$ converges
for each ${\bf A} \in {\cal J}_1$.
\item[\ (b)\ ] 
$|{\rm det}({\bf 1}+{\bf A})| \leq \prod_{j=1}^{\infty} 
\left(1 + \mu_j({\bf A}) \right )$ 
where $\mu_j({\bf A})$ are the {\em singular} values
of ${\bf A}$, i.e., the eigenvalues of $|{\bf A}|=\sqrt{{\bf A}^\dagger 
{\bf A}}$, and  
$|{\rm det}({\bf 1}+{\bf A})| \leq \exp( {\rm Tr} |{\bf A}| )$.
\item[\ (c)\ ]
For any ${\bf A}_1, \dots, {\bf A}_n \in {\cal J}_1$,
$
 \langle  z_1, \dots, z_n \rangle \mapsto {\rm det}
 \left ( {\bf 1} + \sum_{i=1}^n z_i {\bf A}_i \right )
$
is an entire analytic function.
\item[\ (d)\ ]
If ${\bf A},{\bf B} \in {\cal J}_1$, then
\be
 {\rm det} ({\bf 1} + {\bf A}) {\rm det} ({\bf 1}+ {\bf B} )
 &=& {\rm det}\left 
 ( {\bf 1} +{\bf A}+{\bf B} + {\bf A B} \right ) \nonumber \\
 &=& {\rm det}\left( ({\bf 1} +{\bf A})({\bf 1} +{\bf B}) \right ) 
 = {\rm det}\left( ({\bf 1} +{\bf B})({\bf 1} +{\bf A}) \right )\ .
\ee
If ${\bf A}\in {\cal J}_1$ and ${\bf U}$ unitary, then
\be
{\rm det} \left ({\bf U}^{\dagger} ( {\bf 1}+{\bf A} ) {\bf U} \right )
= {\rm det} \left ({\bf 1}+{\bf U}^{\dagger}{\bf A} {\bf U} \right )
 = {\rm det} ({\bf 1} + {\bf A})\ .
\ee
\item[\ (e)\ ] 
If ${\bf A}\in {\cal J}_1$, then
$({\bf 1}+ {\bf A})$ is invertible if and only if ${\rm det}({\bf 1}+{\bf A})
\neq 0$.
\item[\ (f)\ ]
If $\lambda\neq 0$ is an $n$-times degenerate eigenvalue of ${\bf A}\in 
{\cal J}_1$, then
${\rm det}({\bf 1}+ z {\bf A})$ has a zero of order $n$ at $z=-1/\lambda$.
%
\item[\ (g)\ ]
For any ${\bf A} \in {\cal J}_1$,
\be
{\rm det} ( {\bf 1}+  {\bf A} ) = \prod_{j=1}^{ N( {\bf A} ) } \left( 1 
+ \lambda_j ( {\bf A} ) \right ) \; ,
  \label{det_prod}
\ee
where here and in the following 
$ \{ \lambda_j ( {\bf A} ) \}_{j=1}^{ N( {\bf A} )}$ are the eigenvalues of
${\bf A}$ counted with algebraic multiplicity ($N({\bf A})$ can of
course be infinite).
%
\item[\ (h)\ ] If ${\bf A}\in {\cal J}_1$, then
\be
{\rm Tr} \left ( {\bigwedge}^k ({\bf A} ) \right ) =
 \sum_{j=1}^{N \left( {\bigwedge}^k ({\bf A})\right) } 
 \lambda_j \left( {\bigwedge}^k ({\bf A})\right )
= \sum_{1\leq j_1 < \cdots < j_k \leq N( {\bf A})}
 \lambda_{j_1}({\bf  A}) \cdots \lambda_{j_k} ({\bf A} )  
< \infty \; . \nonumber
 \\
 \label{cumulant}
\ee
\item[\ (i)\ ] If ${\bf A} \in {\cal J}_1$, then
\be
{\rm det} ( {\bf} 1 + z {\bf A}) = \sum_{k=0}^{\infty}\, z^k \,
 \sum_{1\leq j_1 < \cdots < j_k \leq N( {\bf A})}
 \lambda_{j_1}({\bf  A}) \cdots \lambda_{j_k} ({\bf A} ) < \infty\; .
  \label{det_cum_red}
\ee
\item[\ (j)\ ]
If ${\bf A} \in {\cal J}_1$, then for $|z|$ small (i.e., 
$|z|\, {\rm max}| \lambda_j ({\bf A})| < 1$), the
series $\sum_{k=1}^{\infty} z^k {\rm Tr}\left( (-{\bf A})^k \right )/k$
converges and
\be
  {\rm det}( 1 + z {\bf A} ) &=& 
 \exp\left (-\sum_{k=1}^{\infty} \frac{z^k}{k} {\rm Tr}\left( (-{\bf A})^k 
\right )
  \right ) \nonumber \\
  &=& \exp\left ( {\rm Tr} \ln( {\bf 1} + z {\bf A}) \right )\ .
 \label{tracelog}
\ee 
\item[\ (k)\ ] 
{\em The Plemelj-Smithies formula:}\ Define $\alpha_m({\bf A})$
for ${\bf A}\in {\cal J}_1$ by
\be
 {\rm det} ( {\bf 1}+ z{\bf A}) = \sum_{m=0}^{\infty} z^m 
 \frac{\alpha_m ({\bf A})}{m!} \ .
 \label{ps_formula}
\ee
Then $\alpha_m({\bf A})$ is given by the $m\times m$ determinant
\be
 \alpha_m({\bf A}) = \left | \begin{array}{ccccc}
    {\rm Tr}({\bf A}) & m-1     &  0  & \cdots & 0 \\
  {\rm Tr}({\bf A}^2) & {\rm Tr}({\bf A})    &  m-2 & \cdots & 0 \\
   {\rm Tr}({\bf A}^3) &  {\rm Tr}({\bf A}^2) &  {\rm Tr}({\bf A})
    & \cdots & 0 \\
  \vdots & \vdots & \vdots & \vdots & \vdots \\
         &        &        &        &    1 \\
  {\rm Tr}({\bf A}^m) &   {\rm Tr}({\bf A}^{(m-1)}) &
    {\rm Tr}({\bf A}^{(m-2)}) & \cdots & {\rm Tr}({\bf A})
 \end{array} \right |
\ee
with the understanding that $\alpha_0({\bf A})\equiv 1$ and 
$\alpha_1({\bf A})\equiv {\rm Tr} ({\bf A})$. 
Thus the cumulants $Q_m({\bf A}) \equiv \alpha_m({\bf A})/ m!$ 
(with  $Q_0 ({\bf A}) \equiv  1 $)
satisfy
the recursion relation
\be
 Q_m({\bf A}) &=& \frac{1}{m}\sum_{k=1}^{m} (-1)^{k+1} Q_{m-k}({\bf A})\, 
   {\rm Tr}({\bf A}^k) \quad {\rm for}\ m \geq 1 \; . 
 \label{app:ps_recursion} 
\ee
\end{description}
Note that formula \equa{ps_formula} is the 
quantum analog of the curvature
expansion of the Gutzwiller-Voros zeta function with ${\rm Tr}({\bf A}^m)$
corresponding to the sum of 
all periodic orbits (primitive and also repeated ones)  of
{\em total} topological length $m$, see 
Eq.\equa{appr-PS-recursion}.
In fact, in the cumulant expansion \equa{ps_formula} (as well as in the
curvature expansion there are large 
cancellations  involved: 
Let us order -- without loss of generality  --
the eigenvalues of the operator ${\bf A}\in {\cal J}_1$ as:
\be
 |\lambda_1|\geq |\lambda_2|\geq \cdots \geq |\lambda_{i-1}|\geq 
|\lambda_i |\geq 
|\lambda_{i+1}|\geq 
 \cdots \; .\nn
\ee
This is always possible because of
$\sum_{i=1}^{N({\bf A})} |\lambda_i| < \infty$.
Then, in the standard (Plemelj-Smithies) cumulant evaluation of 
the determinant, Eq.\equa{ps_formula}, there are enormous cancellations of
large numbers, e.g., at the $k^{\,\rm th}$ cumulant order ($k>3)$, 
all the intrinsically
large ``numbers''
 $\lambda_1^k$, $\lambda_1^{k-1}\lambda_2$, $\dots$, 
$\lambda_1^{k-2}\lambda_2\lambda_3$, $\dots$  and many more have to cancel
out exactly until the r.h.s.\ of \equa{cumulant}
is finally left over.  
Algebraically, the fact that there are these large cancellations 
is of course of no importance. However, if
the determinant is calculated numerically, the large cancellations might spoil
the result or even the convergence. 

\subsection{Von Koch matrices\label{app:trace_Koch}} 
Implicitly, many of the above  properties are based
on the theory of von Koch matrices~\cite{gohberg,vonKoch,Hille}: 
An infinite matrix ${\bf 1} - {\bf A}=\| \delta_{jk}-a_{jk}\|_1^\infty$, 
consisting
of complex numbers, 
is called a matrix with an {\em absolutely convergent determinant}, if the
series $\sum| a_{j_1 k_1} a_{j_2 k_2} \cdots a_{j_n,k_n}|$  with
$n=1,2,\cdots$ converges, where
the sum extends over all pairs of systems of indices $(j_1,j_2,\cdots,j_n)$
and $(k_1,k_2,\cdots,k_n)$ which differ from each other only by a permutation,
and where $j_1 < j_2 < \cdots j_n$. Then the limit
\be
 \lim_{n\to\infty} {\rm det}\| \delta_{jk}-a_{jk}\|_1^n = 
 {\rm det}( {\bf 1} -{\bf A} ) \nn 
\ee
exists and is called the determinant of the matrix ${\bf 1}-{\bf A}$. 
The matrix ${\bf 1}-{\bf A}$ is called
{\em von Koch matrix}, if both conditions
\be
   \sum_{j=1}^{\infty} |a_{jj}| &<& \infty \; ,\label{Koch1} \\
  \sum_{j,k=1}^{\infty} |a_{jk}|^2 &<& \infty \label{Koch2}
\ee
are fulfilled. Then the following holds (see Ref.\cite{gohberg,Hille}): 
\begin{description}
\item[\ (a)\ ] Every von Koch matrix has an
absolutely convergent determinant. If the elements of a 
von Koch matrix are
functions of some parameter $\mu$ ($a_{jk}=a_{jk}(\mu)$, $j,k=1,2,\cdots$) and
both series in the defining conditions, \equa{Koch1} and \equa{Koch2},
converge uniformly in the domain of the parameter $\mu$, 
then as $n\to \infty$ the determinant 
${\rm det} \| \delta_{jk}-a_{jk}(\mu)\|_1^n$ tends to the determinant 
${\rm det}({\bf 1}+{\bf A}(\mu))$ uniformly with respect to $\mu$, over the
domain of $\mu$. 
\item[\ (b)\ ] 
If the matrices ${\bf 1}-{\bf A}$ and ${\bf 1}-{\bf B}$
are von Koch matrices, then their product ${\bf 1}-{\bf C}= 
( {\bf 1}-{\bf A}) ({\bf 1}-{\bf B})$ is a von Koch matrix, and
$
 {\rm det} ({\bf 1}-{\bf C})= 
{\rm det}( {\bf 1}-{\bf A})\, {\rm det} ({\bf 1}-{\bf B}) $.
\end{description}
Note that every trace-class matrix ${\bf A} \in {\cal J}_1$ is also a von
Koch matrix (and that any matrix satisfying condition \equa{Koch2} is
Hilbert-Schmidt and vice versa). The inverse implication, however, is not
true: von Koch matrices are not automatically trace-class. The caveat is that
the definition of von Koch matrices is basis-dependent, whereas the 
trace-class property is basis-{\em independent}. As the traces involve
infinite sums, the basis-independence is not at all trivial. An example for
an infinite matrix which is von Koch, but not trace-class is the following:
\be
      {\bf A}_{ij} = \left \{ \begin{array}{l c l c l}
        2/j & {\rm for}&  i-j=-1 &{\rm and} & j\ {\rm even}\; , \\ 
        2/i & {\rm for}&  i-j=+1 &{\rm and} & i\ {\rm even} \; , \\
         0  &  {\rm else}\; , & & &
 \end{array}
 \right . \,
\ee
i.e.,
\be
{\bf A}=
\left ( \begin{array}{ccccccc}
         0  &  1  &  0  &  0  &  0  &  0     &\cdots \\
         1  &  0  &  0  &  0  &  0  &  0     &\cdots  \\
         0  &  0  &  0  & 1/2 &  0  &  0     &\cdots  \\
         0  &  0  & 1/2 &  0  &  0  &  0     &\cdots  \\
         0  &  0  &  0  &  0  &  0  & 1/3    &\ddots   \\
         0  &  0  &  0  &  0  & 1/3 &  0     &\ddots   \\
 \vdots&\vdots&\vdots&\vdots&\ddots&\ddots    &\ddots \\ 
 \end{array} \right ) \; .\label{counterex}
\ee  
Obviously, condition \equa{Koch1} is fulfilled by definition. Secondly,
condition \equa{Koch2} is satisfied as $\sum_{n=1}^{\infty} 2/n^2 <\infty$.
However, the sum over the moduli of the eigenvalues is just twice the
harmonic series $\sum_{n=1}^{\infty} 1/n$ which does not converge. The
matrix \equa{counterex} violates the trace-class definition \equa{tc_def}, 
as in its eigenbasis the sum over
the moduli of its diagonal elements is infinite. 
Thus the {\em absolute} convergence is traded for a
{\em conditional} convergence,
since 
the sum over the eigenvalues themselves can be arranged to still be zero,
if the eigenvalues with the same modulus are summed first.
Absolute convergence  is of course essential, if sums
have to be rearranged or exchanged. Thus, 
the trace-class property is indispensable for any controlled 
unitary transformation of an infinite  determinant, as then
there will be necessarily  a change of basis and in general 
also a re-ordering of the corresponding traces.

Nevertheless, the von-Koch-criteria \equa{Koch1} and \equa{Koch2} are useful,
as any trace-class matrix has at least to meet these simple tests which can
be easily performed in any specified basis.

\subsection{Regularization \label{app:trace_regularization}}
Many interesting operators are not of trace-class (although they might be
in some ${\cal J}_p$ with \mbox{$p>1$}: an operator  $A$ is in 
${\cal J}_p$ iff
${\rm Tr} |A |^p < \infty$ in any orthonormal basis). 
In order to compute determinants of such
operators, an extension of the cumulant expansion
is needed which, in fact, corresponds to a regularization 
procedure~\cite{rs4,bs_adv}:\\
E.g., let ${\bf A} \in {\cal J}_p$ with $ p \leq n$. Define
\be
 R_n (z{\bf A}) = \left( {\bf 1} + z{\bf A} \right )\exp\left ( 
 \sum_{k=1}^{n-1}\frac{(-z)^k}{k} {\bf A}^k \right ) - {\bf 1}
\ee
as the regulated version of the operator $z{\bf A}$.
Then  the regulated operator $R_n(z{\bf A})$ is trace-class,
i.e., $R_n(z{\bf A})\in {\cal J}_1$. Define now 
$
 {\rm det}_n ({\bf 1}+z{\bf A})
   = {\rm det}({\bf 1} + R_n (z{\bf A}) )$.
Then the regulated determinant
\be
{\rm det}_n ({\bf 1}+z{\bf A})= \prod_{j=1}^{N(z{\bf A})}\left [ \left (1+ 
  z\lambda_j({\bf A})\right ) \exp \left ( \sum_{k=1}^{n-1} \frac{
\left(- z\lambda_j({\bf A})\right)^k}{k} \right ) \right ] < \infty
 \label{jp_reg}
\ee
exists and is finite. The corresponding Plemelj-Smithies formula 
for ${\rm det}_n( {\bf 1} + {\bf A})$ results from the standard 
Plemelj-Smithies formula \equa{ps_formula} by simply setting 
${\rm Tr} ({\bf A})$, ${\rm Tr} ({\bf A}^2)$, $\dots$,   
${\rm Tr} ({\bf A}^{n-1})$ to zero~\cite{bs_adv}.                

See also Ref.\cite{voros87} where 
the Fredholm determinant
\be
 \Delta(\lambda)=\prod_{k=0}^{\infty} \left(1-\frac{\lambda}{\lambda_k}\right
 )
\ee 
is regulated --- in the case $\mu \equiv d/m > 1$ ---  as 
a Weierstrass product
\be
\Delta(\lambda)=\prod_{k=0}^{\infty} \left [\left(1-\frac{\lambda}{\lambda_k}
             \right)
  \exp\left( \frac{\lambda}{\lambda_k}
            +\frac{\lambda^2}{2\lambda_k^2}
            + \cdots
            + \frac{\lambda^{[\mu]}}{[\mu] \lambda_k^{[\mu]}} \right )
               \right ] \; .
   \label{voros}                
\ee
Here
$\{\lambda_j\}$ are the eigenvalues of an  elliptic (pseudo)-differential 
operator ${\bf H}$ of order $m$ on a compact or 
bounded manifold of dimension $d$ (with 
$0 < \lambda_0 \leq \lambda_1 \leq\ \cdots$ and 
$\lambda_k \uparrow +\infty$) and 
$[\mu]$ denotes the integer part of $\mu$. Eq.\equa{voros} is
a unique entire
function of order $\mu$ with zeros at $\{\lambda_k\}$ and subject to
the normalization conditions
\be
 \ln \Delta(0)= \frac{d}{d\lambda}\ln\Delta(0) = \cdots = \frac{d^{[\mu]}}
 {d\lambda^{[\mu]}} \ln \Delta(0) =0 \ .
\ee
Clearly 
Eq.\equa{voros} is the same as \equa{jp_reg}; one just has to identify 
$z=-\lambda$, ${\bf A}=1/{\bf H}$ and $n-1=[\mu ]$.
An example is the regularization of the spectral determinant
$\Delta(E)= \det{}[(E-{\bf H})]$
which -- as it stands -- would only make sense for a finite dimensional basis
(or finite dimensional matrices). In Ref.\cite{keat_sieb}  
the regulated spectral determinant 
for the example of the hyperbola billiard in two
dimensions (thus $d=2$, $m=2$ and hence $\mu=1$) is given as
\be
  \Delta(E)= \det{}\left[(E-{\bf H})\Omega(E,{\bf H})\right] \; ,
\ee
where $\Omega(E,{\bf H}) = - {\bf H}^{-1} {\rm e}^{E {\bf H}^{-1}}$.    
Thus the spectral determinant in the eigenbasis of ${\bf H}$ (with
eigenvalues $E_n\neq 0$) reads
\be
      \Delta(E) = \prod_{n} \left(1 - \frac{E}{E_n} \right ) {\rm e}^{E/E_n}
 < \infty \ .
\ee
Note that ${\bf H}^{-1}$ is for this example of Hilbert-Schmidt character.
\newpage
\section{Exact quantization of the $n$-disk
scattering problem\label{app:construction}}
\setcounter{equation}{0}
\setcounter{figure}{0}
\setcounter{table}{0}
\newcommand{\simkr}{\stackrel{\mbox{$kr\to\infty$}{\sim}}}
In this Appendix (which
is based on M.~Henseler's diploma thesis\,\cite{mh}  
where also the corresponding formulas for
the three dimensional $n$-ball scattering problem can be found, see also
\cite{N-ball,wh97}) 
we will construct the scattering matrix for the  
scattering of a point particle from $n$ 
circular hard disks which are fixed in the two-dimensional plane.
The basic ideas go back to  Lloyd's
multiple-scattering method~\cite{Lloyd_smith}, 
an application of the KKR-method~\cite{KKR},
to three-dimensional band structure calculations as the limiting case of $n$
disjunct non-overlapping  muffin-tin potentials 
(see also Ref.\cite{Berry_KKR} for the translation of these methods to the
infinite two-dimensional Sinai-billiard) 
and to
the work of Gaspard and Rice\cite{gr}, who
introduced the techniques reported below 
to the scattering problem of a point 
particle from three equal disks in the two-dimensional plane.
Here we will present a generalization of these methods 
to the scattering from $n$ 
non-overlapping disks of -- in general -- different sizes.

\subsection{The stationary scattering problem}

As stated in Sec.\ref{chap:S-matrix},
the quantum-mechanical description of the scattering from $n$ hard disks
will be performed in the framework of the stationary scattering theory.
Let  ${\psi}^{\pol{k}}(\pol{r}\,)$ be a solution
of the scattering problem  (for a fixed incident wave vector $\pol{k}$).
The decomposition of  $\psi$ into a sum over complex exponential (angular) 
functions
\be     \label{psi_zerl_2}
   {\psi}^{\pol{k}}(\pol{r}\,) = \sum_{m=-\infty}^\infty \psi^k_m(\pol{r}\,)
          \e^{\i m(\frac{\pi}{2}-\Phi_k)}
\ee
($\Phi_k$ and $\Phi_r$ are the angles of $\pol{k}$ and $\pol{r}$, 
respectively, in the global coordinate system) leads to
$({\pol{\nabla}_{{r}}}^2\mbox{+}{\pol{k}}^2) \psi^k_m(\pol{r}\,)\mbox{=}0$.
The corresponding
separation of a plane wave in two dimensions into angular eigenfunctions 
reads:
\be  \label{eb_Welle_2}
   \e^{\i \pol{k} \cdot \pol{r}} = \e^{\i  kr \cos\Phi_r} =
      \sum_{m=-\infty}^\infty J_m(kr) \e^{\i m \Phi_r}
\e^{\i m (\frac{\pi}{2}-\Phi_k)}\; .
\ee
The ordinary Bessel and Hankel functions ($ J_l(z)  =  \frac{1}{2} (
H^{(1)}_l(z) 
+ H^{(2)}_l(z) )  $) 
of integer order 
satisfy the expressions (for $|z|  \gg  1$):
\be             \label{Bess_Hank_2}
               H_l^{(2)}(z) & \sim &
 \sqrt{\frac{2}{\pi z}}\e^{-\i (z - \frac{\pi}{2}l -
                                                \frac{\pi}{4})}
         \quad \mbox{incoming,} \\
     H_l^{(1)}(z) & \sim & \sqrt{\frac{2}{\pi z}}\e^{+i(z - \frac{\pi}{2}l -
                                                \frac{\pi}{4})}
         \quad \mbox{out-going.}  \label{H1_asymp}
\ee
The to-be-constructed solution can be written as 
a superposition of incoming
and out-going spherical waves ($ kr \gg 1 $)
\be
     \label{def_S_2}
   \psi^k_m(\pol{r}\,)  \sim       
\frac{1}{\sqrt{2\pi kr}} \sum_{l=-\infty}^\infty
     \left[ \delta_{ml} \e^{-\i (kr - \frac{\pi}{2}l 
- \frac{\pi}{4})} 
+ 
{\bf S}_{ml} \e^{\i (kr - \frac{\pi}{2}l - \frac{\pi}{4})} \right]
         \e^{\i l\Phi_r} \; , 
 \label{S-matrix}
\ee
where the  matrix {\bf S} is the scattering matrix of the two-dimensional
scattering problem.

\subsection{Calculation of the {\bf S}--matrix\label{app:construction-S}}

In order to describe a generic configuration of $n$ disks we use the following
notation (see Fig.\ref{fig:coordinates}): 
The  index $j\in \{1,\cdots,n\}$ labels the $j^{\rm th}$ disk
whose radius is $a_j$. The distance between the centers of the disks
$j$ and $j'$ is called $R_{j j'}=R_{j'j}$.
To specify the $n$ disks we introduce $n$+1
different coordinate systems.
First of all, a {\em global} coordinate system~($x,~y$) is chosen with its 
origin in the neighborhood of the $n$ disks.
In case of symmetrical systems, as, e.g., three equal disks at the corners
of an equilateral triangle, the origin is best placed in the center of 
symmetry. In order to fully use the symmetry of such configurations $n$ local
coordinate systems~($x^{(j)},~y^{(j)}$) are introduced whose origins are 
placed at
the centers of the $n$ disks, respectively. The axes of these coordinate 
systems are chosen in such a way that they fully respect the 
symmetry of the configuration. 
The spatial vector to the center of the disk $j$,
as measured in the global system, is called $\pol{R}_j$, $R_j$ is its length
and $\phi_{R_j}$ its angle. 
Vectors called $\pol{s}_j$ or $\pol{S}$ are surface vectors.
The unit vector
$\hat{R}_{j j'}^{(j)} \equiv \pol{R}_{j j'}^{(j)} / R_{j j'}$ is pointing
from the center of disk  $j$ to the center of disk $j'$,
as measured in the ($j$)-system, $\alpha_{j' j}$ is its corresponding angle. 
In general, vectors with an upper index $(j)$ are measured in the
$(j)$-system, vectors without upper index are measured in the global system.

The Green's functions satisfy the differential equation
$    ({\pol{\nabla}_{{r}}}^2+{\pol{k}}^2)G(\pol{r},\pol{r}\,') 
     = \delta^2(\pol{r}-\pol{r}\,')$.
In two dimensions the free Green's function reads\cite{gr}:
\be      \label{free_green}
  G(\pol{r},\pol{r}\,')  =  - \frac{\i }{4} 
H_0^{(1)}(k|\pol{r}-\pol{r}\,'|) \; .
\ee
For the following, we will apply the Green's formula:
\be
   \int_V \! {\rm d}^2r \, (\phi(\pol{r}) {\pol{\nabla}_{{r}}^2}
\psi(\pol{r}) -
\psi(\pol {r}) {\pol{\nabla}_{{r}}^2}\phi(\pol{r})) \! & = & \!
   \int_{\partial V} \! {\rm d}\pol{S} \cdot (\phi(\pol{S}) {\pol \nabla}_S 
        \psi(\pol{S}) -
             \psi(\pol{S}) \pol{\nabla}_S \phi(\pol{S}))  \nn 
 \label{greens_int}
\ee
where $V$ is the integration volume and $\partial V$ denotes its boundary.
After inserting the expansion coefficients 
$\psi_{m}^k(\pol{r}\,)$ from~\equa{psi_zerl_2} and the (free) 
Green's function in the last equation,
one finds:
\be    \label{Green_2}
    \int_{\partial V} \, {\rm d}\pol{S} 
 \cdot [\psi_{m}^k(\pol{S}) \pol{\nabla}_S
    G(\pol{S},\pol{r}\,')
       - G(\pol{S},\pol{r}\,') \pol{\nabla}_S\psi_{m}^k(\pol{S})] 
    =
     \cases{ 0 & $\pol{r}\,' , \notin V \; ,$\cr
             \psi_{m}^k(\pol{r}\,') & $\pol{r}\,' \in V\; .$} 
\ee
The integration volume is chosen as a  big disk whose center is
in the origin of the global coordinate system and whose radius is 
large enough 
that
the asymptotic equations \equa{Bess_Hank_2} and \equa{H1_asymp}
hold for the points
far away from the origin but inside the integration volume. 
From the large disk
the small $n$ disks (as given in the concrete disk configuration) are
excluded; however, the radii of these subtracted disks have been  
increased by a small
increment  $\epsilon > 0$ in comparison to the original disks. 
In the end, the case $\epsilon \to 0$ is considered. 
In order to construct the {\bf S}--matrix,
one  has to work out  \equa{Green_2} for two different cases~\cite{gr}.
In the first case the point $\pol{r}\,'$ is on the surface of the (original) 
scattering
disk  $j$, 
such that it is now 
outside the 
integration volume $V$. In the second case $\pol{r}\,'$ is
in the integration volume; however, so far
away from all $n$ disks that the asymptotic 
equations \equa{Bess_Hank_2} and \equa{H1_asymp} are then valid. 
The boundary of $V$ 
splits into $n$+1 disjunct regions: Into the outer layer of the large disk,
$\partial_\infty V$, and into the boundaries 
$\partial_j V$ of the $n$ subtracted disks which contain and  cover 
the scattering disks.

\subsubsection{First case: $ \pol{r}\,' = \pol{X}_j \, \in \, 
\mbox{boundary of disk\,} j$}           

Because of the Dirichlet boundary conditions, the wave function 
vanishes on the boundary of the scattering disks; however,
its gradient does not vanish there: 
\beq
     \psi^k_{m}(\pol{X}_j)  =  0  
            \quad ; \quad
      \pol{n}_j \cdot \pol{\nabla} \psi^k_{m}(\pol{X}_j^{(j)}) 
     \equiv \sum_{m'=-\infty}^{+\infty} {\bf B}^{\ \ j}_{mm'}
            \e^{\i m' \theta_j}\; . 
 \label{gradient}
\eeq
Here the unit vector $\pol{n}_j$  
is chosen to point perpendicularly to $\partial_j V$ into
the complementary region of $V$. Note that $|\pol{X}_j^{(j)}|=a_j$. 
Furthermore, 
$\theta_j$ labels the direction of 
$\pol{X}_j^{(j)}$ as measured in the local coordinate system of
the disk $j$. The coefficients
${\bf B}^{\ \ j}_{mm'}$ are unknown  so far.
Eq.\equa{Green_2} now reads:
\be    \label{Green_a_2}
     0 = I^j_\infty + \sum_{j'=1}^n I^j_{j'}  \; .
\ee
The occurring integrals are: 
\be    \label{int_def_a_2}
   I^j_\infty & = & \int_{\partial_\infty V} {\rm d}\pol{S}\, \cdot
    [\psi_{m}^k(\pol{S}) \pol{\nabla}_S
    G(\pol{S},\pol{X}_j)
       - G(\pol{S},\pol{X}_j) \pol{\nabla}_S \psi_{m}^k(\pol{S})]  \\
    I^j_{j'} & = & - \int_{\partial_{j'} V} {\rm d}\pol{s_{j'}}\, \cdot
         G(\pol{s_{j'}},\pol{X}_j^{(j')}) \pol{\nabla}_{s_{j'}} 
\psi_{m}^k(\pol{s_{j'}}) \ .
     \label{int_def_a2_last}
\ee
In the following we will repeatedly apply the 
addition theorems for
Bessel and Hankel functions~\cite{Abramowitz}:
\beq
          \label{add_Bess_Hank_2}
  C_n(w) \e^{ \pm i n \beta}  =  \sum_{l=-\infty}^\infty C_{n+l}(u)
                             J_l(v) \e^{\pm i l \alpha} \; ,   
\eeq
where $w = \sqrt{u^2+v^2-2uv\cos \alpha}$, 
 $w\cos\beta =   u-v\cos\alpha$, $ |v\e^{\i \alpha}|<|u|$,
 $w\sin\beta = v\sin\alpha$ and 
 $ C_n(z)  \in  \{J_n(z), Y_n(z) , H^{(1)}_n(z), 
         H^{(2)}_n(z) \}$.

\noindent{\bf Calculation of $I^j_\infty$}           \nopagebreak

The calculation is 
performed in the global coordinate system. The 
addition theorem \equa{add_Bess_Hank_2} is used to rewrite 
the free Green's function \equa{free_green}. 
In addition, because of the large value for  $R_S$, the  
Hankel function $H^{(1)}_l(k R_S)$ is approximated according to 
\equa{H1_asymp}.
The resulting expression and
the asymptotic expression~\equa{def_S_2} for 
$\psi_m^k(\pol{S})$ 
are inserted into $ I^j_\infty$.
The terms proportional to ${\bf S}_{ml}$ cancel out, such that
\be     \label{I_j_infty_2}
  I^j_\infty = \e^{\i  m \phi_{R_j}} \sum_{m'=-\infty}^\infty J_{m-m'}(kR_j)
          J_{m'}(ka_j) \e^{\i m' \theta_j}  \, .
\ee
Only the coordinates of the center of the disk $j$ are still 
expressed in the global system, whereas the coordinates on the disk surface
have been transferred to  the local coordinate system of disk~$j$. 

\noindent{\bf Calculation of $I^j_j$}       \nopagebreak

Here we work  relative to  the local coordinate system of disk $j$. 
Using the addition theorem
\equa{add_Bess_Hank_2} for the free Green's function \equa{free_green}
and performing the angular integration under 
the boundary condition \equa{gradient}  
we obtain
\beq    \label{I_j_j_2}
  I_j^j = -\frac{\pi a_j}{2\i } \sum_{l=-\infty}^\infty {\bf B}_{ml}^{\ \ j}
     H^{(1)}_l(ka_j) J_l(ka_j) \e^{\i l \theta_j} \;  ,
\eeq
where all quantities are 
expressed in the local
coordinate system of disk $j$.

\noindent{\bf Calculation of $I^j_{j'} \; , \; j \not= j'$}   \nopagebreak

Working relative to the local coordinate system of disk $j'$, 
we have in this case: 
\beq
  G(\pol{s}_{j'}, \pol{X}_j^{(j')})  =  \frac{1}{4\i } \sum_{l=-\infty}^\infty
    H^{(1)}_l(k X_j^{(j')}) J_l(ka_{j'}) \e^{-\i l(\phi_{s_{j'}} -(\alpha_{jj'} +
     \gamma_{j'}))}  \; . 
\eeq
In writing down the last equation 
the addition theorem for Hankel functions has been used again. 
Here $\phi_j^{(j')}=\alpha_{jj'}+\gamma_{j'}$, where $\phi_j^{(j')}$ 
is the
angle of $\pol{X}_j^{(j')}$,  $\alpha_{jj'}$ is the angle of 
the ray from the center of 
disk $j'$ to the center of disk $j$ and  $\gamma_{j'}$
is the difference angle. All three angles 
are measured relative to the local coordinate
system of disk $j'$.
%
%
After insertion into $I^j_{j'}$ and the angular integration 
we apply once more the addition theorems for
Hankel functions.
Then 
$I^j_{j'}$ reads:
\beq      \label{I_j_j'_2}
  I^j_{j'} = -\frac{\pi a_{j'}}{2\i } \sum_{l,l'=-\infty}^\infty
   {\bf B}^{\ \ j'}_{ml} J_l(ka_{j'}) J_{l'}(ka_j) H^{(1)}_{l-l'}(kR_{j'j})
   \e^{\i (l\alpha_{jj'} -l'\alpha_{j'j})} (-1)^{l'} \e^{\i l' \theta_j} \; ,
\eeq
where the entries of $I^j_{j'}$ do not depend on the global coordinate
system. The $j'$ dependent quantities are expressed in the local coordinate
system of disk $j'$, the $j$-dependent ones in that of disk $j$.

The computed integrals are now inserted into the formula \equa{Green_a_2},
written as
$  I^j_\infty  =  -\sum_{j'=1}^n I^j_{j'} $, 
        \label{green_erg_2}
which leads to
\beq 
\sum_l \widehat{\bf C}^{\ \ j}_{ml} \e^{\i l \theta_j}  =  \sum_l \sum_{j',l'}
     {\bf B}^{\ \ j'}_{ml'} \widehat{\bf M}^{j'j}_{l'l} \e^{\i l \theta_j} 
  \label{hat_relation} 
\eeq
with $\widehat{\bf C}^{\ \ j}_{ml}$ abbreviating the terms in
Eq.\equa{I_j_infty_2},
whereas $\widehat{\bf M}^{j'j}_{l'l}$ stands for the
terms in \equa{I_j_j_2} and \equa{I_j_j'_2}.
Equation \equa{hat_relation} 
 holds for all points $\pol{X}_j$ on the boundary of 
the disk $j$.
Then, the coefficients 
$\widehat{\bf C}^{\ \ j}_{ml}$ 
and $\widehat{\bf M}^{j'j}_{l'l}$ are normalized in such a way, that 
in the 1--disk case the new ${\bf M}$-matrix is just the unit matrix. This
corresponds to a division of the l.h.s.\ and r.h.s.\ of 
\equa{hat_relation} 
by the diagonal 
matrix $\{  \Ho{l}{a_j}  \Jb{l}{a_j}\pi a_j /2\i  \}$. 
Asymptotically (i.e., for $|l| \gg |k a_j |$) 
the modulus of its matrix elements behaves as 
$| \Ho{l}{a_j}  \Jb{l}{a_j} | \sim 1/(\pi |l|)$. 
Therefore, this division 
does not affect  the 
``trace-character'' of
the matrices $\widehat{\bf C}^j$ , $\widehat{\bf M}^{jj'}$ and 
${\bf B}^{j'}$ (see App.~\ref{app:suppl}). 
Thus one gets the matrix equation
\be     \label{CAM_2}
  {\bf C}^j = {\bf B}^{j'} \cdot {\bf M}^{j' j}
\ee
with
\be  \label{C_N_disk}
  {\bf C}^{\ \, j}_{lm}\! 
 & = &\! \e^{\i l \phi_{R_j}} \frac{J_{l-m}(kR_j)}{H^{(1)}_m(ka_j)}
    \frac{2\i }{\pi a_j}  \; ,   \\
  {\bf M}^{j\ \ j'}_{mm'}\! & = & \!
 \delta^{jj'} \delta_{mm'}  
     +(1\mbox{$-$}\delta^{jj'}) 
    \frac{a_j J_m(ka_j)}{ a_{j'} 
 H^{(1)}_{m'}(ka_{j'})} H^{(1)}_{m-m'}(kR_{jj'})
    \e^{\i (m\alpha_{j'j} -m'\alpha_{jj'})} (-1)^{m'}   \; ,  
      \label{M_N_disk}
\ee
where $R_j$ and $\phi_{R_j}$ are the magnitude and the angle of the ray from
the origin of the global coordinate system to the center of disk $j$, 
as measured in the global
coordinate system. The angle $\alpha_{j'j}$ is the angle of the ray from disk
$j$ to disk $j'$  as measured in the local coordinate system of disk $j$,
$R_{jj'}=R_{j'j}$ is the distance of the centers of disk $j$ and $j'$,
$a_j$, $a_{j'}$ are their radii, respectively.

\subsubsection{Second case: $ \mbox{point}\ \pol{r}\,'=\pol{r}
 \in V \quad , \quad  r \; \mbox{large}$}               

For this case we obtain from \equa{Green_2}:
\be    \label{psi_km_2}
    \psi^k_{m}(\pol{r}\,) = I_\infty^{\pol{r}} +
       \sum_{j=1}^n I_j^{\pol{r}} \; .
\ee
%
In analogy to the first case, the following abbreviations have been introduced:
\be   \label{I_def_b_2}
   I^{\pol{r}}_\infty & = & \int_{\partial_\infty V} {\rm d}\pol{S} \cdot
     (\psi^k_{m}(\pol{S}\,) \pol{\nabla}_S G(\pol{S},\pol{r}\,) -
     G(\pol{S},\pol{r}\,) \pol{\nabla}_S \psi^k_{m}(\pol{S})) \; , \\
   I^{\pol{r}}_j & = & - \int_{\partial_j V} {\rm d}\pol{s}_j \cdot
     G(\pol{s}_{j},\pol{r}^{\,(j)}) \pol{\nabla}_{s_j} \psi^k_{m}(\pol{s}_{j})
 \; .
\ee

\noindent{\bf Calculation of $I_{\infty}^{\pol{r}}$}    \nopagebreak

$I_{\infty}^{\pol{r}}$ can be calculated in close analogy to $I_{\infty}^{j}$.
%
A single application of the 
addition theorems \equa{add_Bess_Hank_2} yields
$  I_{\infty}^{\pol{r}} = J_m(kr) \e^{\i m \Phi_r}$,
where $\Phi_r$ is the angle of $\pol{r}$ in the global coordinate system.

\noindent{\bf Calculation of $I_j^{\pol{r}}$}     \nopagebreak

We have
\beq
  G(\pol{s}_{j},\pol{r}^{\,(j)}\,)  =  
  \frac{1}{4\i } \sum_{l,l'=-\infty}^\infty
     H^{(1)}_l(kr) J_{l-l'}(kR_j) J_{l'}(ka_j) \e^{\pm l' i\phi_{s_j}}
     \e^{(\mp il\Phi_r \pm il\Phi_{R_j})} \; ,  
\eeq
where the addition theorem for cylindrical functions has been applied twice.
The angle $\phi_{s_j}$ of $\pol{s}_{j}$ is measured relative
to the local coordinate system of 
disk $j$. After integration over this angle,
we get:
\be   \label{I_r_j_2}
  I^{\pol{r}}_j = -\frac{\pi a_j}{2\i } \sum_{l,l'=-\infty}^\infty
    H^{(1)}_l(kr) J_{l-l'}(kR_j) J_{l'}(ka_j) {\bf B}^{\ \ j}_{ml'}\,
    \e^{\i l(\Phi_r - \Phi_{R_j})} \; ,
\ee
where all quantities, except $a_j$, 
are now defined with respect to the global coordinate
system.

Both integrals are now inserted into Eq.\equa{psi_km_2}. Taking 
into account
\equa{Bess_Hank_2} and \equa{H1_asymp}, one gets  
Eq.\equa{S-matrix} for 
$kr \gg 1$.
The {\bf S}--matrix can now
be written as
\be      \label{S_2_disk}
    {\bf S}^{(n)} = {\bf 1} - \i {\bf B}^j \cdot {\bf D}^j\; ,
\ee
where we have introduced the superscript $(n)$ in order to 
indicate  that the
${\bf S}$-matrix refers to the $n$-disk scattering problem.
The matrix ${\bf D}^j$ in the last equation is given by
\be     \label{D_N_disk}
  {\bf D}^j_{mm'} = 
 -\pi a_j J_{m'-m}(kR_j) J_m(ka_j) \e^{-\i m'\Phi_{R_j}}\; .
\ee
Using (\ref{CAM_2}), we finally get the (formal) expression
for the {\bf S}--matrix which
will be justified in App.~\ref{app:suppl}:
\beq
                 \label{SCMD_2} 
   {\bf S}^{(n)}  =  {\bf 1} - \i {\bf C}^{j} \cdot ({\bf M}^{-1})^{jj'} 
                                  \cdot {\bf D}^{j'} \; .
 \label{s_summary}
\eeq

The {\bf S}-matrix ${\bf S}^{(1)}$ of the scattering of a point particle
from a single hard disk is given by
\be     \label{S_1disk}
   {\bf S}^{(1)}_{ml}(ka_j) = -\frac{H^{(2)}_{l}(ka_j)}{H^{(1)}_{l}(ka_j)}
         \delta_{ml} \; ,  
\ee
as can be seen by comparison of the general asymptotic expression
\equa{def_S_2} for the wavefunction with the exact solution for the 1-disk
problem.

\subsection{The determination of the product ${\bf D}\cdot{\bf C}$}
In order to rewrite the determinant of the {\bf S}--matrix (see 
Sec.\ref{chap:link}) we have to determine the product
{\bf D}$\cdot${\bf C} (see Eqs. \equa{C_N_disk} and \equa{D_N_disk}).  
We apply once again the addition theorem for Bessel 
functions~\equa{add_Bess_Hank_2} using $
   \pol{R}_{j'j} = \pol{R}_{j\,'}\mbox{$-$}\pol{R}_j$,
where $R_{j'}$ and  $R_{j}$ are the magnitudes of these vectors and 
$\Phi_{R_{j'}}$ and $\Phi_{R_j}$ the corresponding angles, as measured
in the global coordinate system.
We find the following expressions 
\be      
  \sum_{\tilde{l}=-\infty}^\infty {\bf D}^{j}_{l \tilde{l}} 
              {\bf C}^{\ j'}_{\tilde{l} l'}
   & = & -2\i \left( \frac{a_j}{a_{j'}} \right)
   \frac{J_l(ka_j)}{H^{(1)}_{l'}(ka_{j'})} J_{l-l'}(kR_{jj'}) (-1)^{l'}
   \e^{\i (l\alpha_{j'j} - l'\alpha_{jj'})}  \; ,    \nn  \\
  \sum_{\tilde{l}=-\infty}^\infty {\bf D}^{j}_{l \tilde{l}} 
              {\bf C}^{\ j}_{\tilde{l} l'}
   & = & -2\i \frac{J_l(ka_j)}{H^{(1)}_{l'}(ka_{j})} \delta_{ll'} \; .     \nn
\ee
Using the expression \equa{M_N_disk}  
for ${\bf M}^{j\ \ j'}_{mm'}$ 
we finally get for ${\bf X} \equiv {\bf M} - \i {\bf D} \cdot {\bf C}$:
\be
 {\bf X}^{jj'}_{ll'}
   = - \frac{H^{(2)}_{l'}(ka_{j'})}{H^{(1)}_{l'}(ka_{j'})} 
   \delta^{jj'} \delta_{ll'}    
  -(1\mbox{$-$}\delta^{jj'}) 
  \frac{ a_j J_l(ka_j)}{ a_{j'} H^{(1)}_{l'}(ka_{j'})} 
H^{(2)}_{l-l'}(kR_{jj'}) (-1)^{l'}
   \e^{\i (l\alpha_{j'j} - l'\alpha_{jj'})}      \; .   \nn
\ee
The r.h.s.\ of this equation can be reformulated in terms of the scattering
matrix of the single-disk problem, ${\bf S}^{(1)}(ka_{j'})$,  
and by the complex conjugate
of ${\bf M}$, namely
$  {\bf X}^{jj'}_{ll'}(k) = {\bf S}^{(1)}_{ll'}(ka_{j'}) 
\left( {\bf M}^{j\ \ j'}_{-l,-l'}(k^\ast)
     \right)^\ast $.
The matrix
{\bf X} can therefore be expressed as the product,
\[     
  {\bf X}^{jj'}_{-l,-l'}  =  \sum_{\tilde{j}} \sum_{\tilde{l}} \left(
{\bf M}^{j \tilde{j}}_{l \tilde{l}} \right)^\ast  
{\bf Y}^{\tilde{j}j'}_{\tilde{l} l'} \; , \]
where the second factor is given by
$ {\bf Y}^{\tilde{j}j'}_{\tilde{l} l'}  \equiv  \delta^{\tilde{j} j'}
   \delta_{\tilde{l} l'} {\bf S}^{(1)}_{-l',-l'}  $.
Thus we get the formal expression for the determinant of
{\bf X}:
\beq
  \Det{L} {\bf X}(k) = \left( \prod_{j=1}^n \det{l} {\bf S}^{(1)}(ka_j)
    \right) \Det{L} {\bf M}(k^\ast)^\dagger  \; .
     \label{det_X}
\eeq
The last step in the formal evaluation of the determinant of 
${\bf S}^{(n)}$ (as function of the wave number $k$) is 
the insertion of Eq.\equa{det_X}
into Eq.\equa{recoupl} giving the final result (see \equa{qm}):
\be    \label{det_S_disk}
   \det{l} {\bf S}^{(n)}(k) = \underbrace{
      \frac{\Det{L} ({\bf M}(k^\ast)^\dagger)}{\Det{L} {\bf
       M}(k)}}_       {\mbox{coherent}}  \underbrace{ \left(
  \prod_{j=1}^n \det{l} {\bf S}^{(1)}(ka_j) \right) }_{\mbox{incoherent}}
                    \; .
\ee

\newpage
\section{Existence of the {\bf S}--matrix and its determinant in  
$n$-disk systems\label{app:suppl}}
\setcounter{equation}{0}
\setcounter{figure}{0}
\setcounter{table}{0}
This appendix is based on M.~Henseler's diploma thesis\,\cite{mh} and
on Ref.\cite{wh97}.

The derivations of the expression for  
{\bf S}--matrix~\equa{SCMD_2} in App.~\ref{app:construction} and 
of its determinant (see Sec.\,\ref{chap:link}) 
are of purely formal character as all the matrices involved are
of infinite size. Here, we will show that the 
operations are all well-defined. For this purpose, the trace-class 
(${\cal J}_1$)
and Hilbert--Schmidt(${\cal J}_2$)   operators will play a 
central role. The definitions and most important properties of these
operator-classes can be found in App.~\ref{app:trace}. 
As shown in App.~\ref{app:construction}  the
${\bf S}^{(n)}$--matrix can  
be written in the following form (see \equa{S_2_disk}):
\beq
   {\bf S}^{(n)} = {\bf 1} - \i{\bf T} \quad , \quad
   {\bf T} = {\bf B}^j {\bf D}^j \; .
\eeq
The {\bf T}-matrix is  trace-class on the positive
real $k$-axis ($k>0$), 
since it is the product of the matrices ${\bf D}^j$ and ${\bf B}^j$ which
will turn out to be trace-class or, in turn, are
bounded there
(see App.~\ref{app:trace}.1 for the definitions). 
Again formally,  we derived 
in App.~\ref{app:construction} that
${\bf C}^j={\bf B}^{j'}{\bf M}^{j'j}$ implies the relation ${\bf B}^{j'}=
{\bf C}^{j}({\bf M}^{-1})^{jj'}$. 
Thus, the existence of ${\bf
M}^{-1}(k)$ has to be shown, as well -- except at isolated poles in the 
lower complex $k$-plane 
below the real $k$-axis and on the branch cut on the negative real 
$k$-axis which
results
from the branch cut of the defining Hankel functions. 
As we will prove later, ${\bf M}(k)-{\bf 1}$ is trace-class, except of course 
at the above mentioned 
points in the $k$-plane.  
Therefore, using property (e) of App.~\ref{app:trace}.2  
we only have to show that $\Det{}{\bf M}(k)\neq 0$ in order to guarantee the
existence of ${\bf M}^{-1}(k)$.  At the same time, 
${\bf M}^{-1}(k)$ will be proven to be
bounded 
as all its eigenvalues and the product of its eigenvalues are then finite.
The existence of these eigenvalues  
follows from the trace-class property of
${\bf M}(k)$ which, together with $\Det{}{\bf M}(k)\neq 0$ ,
guarantees the finiteness of the eigenvalues and their product.

We have normalized ${\bf M}$ in such a way
that for the scattering from a single disk we simply have
${\bf B}={\bf C}$. 
Thus the structure of the matrix ${\bf C}^j$ does not dependent
on whether the point particle scatters only from a single disk or
from $n$ disks. Hence the properties of this matrix can 
be
determined from the single disk scattering alone.
The functional form \equa{C_N_disk} 
shows that ${\bf C}$ cannot have poles on the real positive $k$-axis ($k>0$) 
in agreement
with the structure of the ${\bf S}^{(1)}$--matrix discussed in 
App.\,\ref{app:construction}. 
If the origin of the coordinate system is placed at the origin of the disk,
the matrix  ${\bf S}^{(1)}$ is 
diagonal. 
In the same basis ${\bf C}$ becomes diagonal. Thus one can easily see that
${\bf C}$  
has no zero eigenvalue on
the positive real $k$-axis and that it will be trace-class there.
Thus neither ${\bf C}$ nor the 1--disk (or for that purpose 
the $n$--disk ) {\bf S}-matrix can possess poles or zeros on the real positive
$k$-axis. The statement about ${\bf S}^{(n)}$ 
follows simply from the unitarity of the {\bf S}-matrix which can 
be checked easily.
Since, for real positive $k$, ${\bf S}^\dagger(k) = {\bf S}^{-1}(k)$, we have 
$|{\bf S}(k)| = {\bf 1}$ on the real  axis, such that poles (and also zeros)
of {\bf S} are excluded there. Actually, for the exclusion of poles and zeros
on the real positive $k$-axis, only the weaker condition 
that $|\det{} {\bf S}^{(n)}(k)|=1$, 
$k>0$, is  needed. That this is fulfilled for all non-overlapping $n$ disk 
systems is obvious from the final expression  \equa{qm} 
for $\det{} {\bf S}^{(n)}$ in Sec.\,\ref{chap:link}. This formula even holds
for $\Det{} {\bf M}(k) \to 0$ if $k$ approaches the real positive axis,
since then $\Det{} {\bf M}(k^\ast)^\dagger$ approaches zero as well, 
such that both
terms cancel in  formula  \equa{qm}. Thus the fact that 
$|\det{} {\bf S}^{(n)}(k)|=1$ on the positive real $k$-axis cannot be used
to disprove that $\Det{} {\bf M}(k)$ could be zero there. However, if  
$\Det{} {\bf M}(k)$ were zero there, the ``would-be'' pole must cancel
out of  ${\bf S}^{(n)}(k)$. Looking at formula \equa{s_summary}, this pole
has to cancel a zero from ${\bf C}$ or ${\bf D}$ where both
matrices are already fixed on the 1-disk level.  
Now, property (g) and
(f) of App.~\ref{app:trace}.2  leave for ${\bf M}(k)$ (provided that
${\bf M}-{\bf 1}$ has been proven  trace-class)  only one chance 
to cause trouble
on the positive real $k$-axis, namely, if at least one of its eigenvalues 
(whose existence is guaranteed) 
becomes zero. On the other hand ${\bf M}$ still has to satisfy 
${\bf C}^j = {\bf B}^{j'}{\bf  M}^{j'j}$ with 
${\bf C}$ completely determined by the 1-disk scattering alone, where 
$\Det{}{\bf M}=1$ everywhere in the $k$-plane. The fact that ${\bf C}^j(k)$
cannot have zero eigenvalue for $k>0$ can be used to show 
that the following inequality
holds for the modulus of the
diagonal matrix element, $|{\bf C}^{\ \ j}_{mm}(k)| >0 $,  
for the state $|m\rangle$
of any orthonormal basis. Now choose as the basis the eigenbasis of 
${\bf M}^{j\ \ j'}_{mm'}(k)$ 
and $|m\rangle$ as the state there ${\bf M}(k)$ has
a candidate for a zero eigenvalue in the $|m,j\rangle$ space. 
Comparing the left and the right-hand side
$
 |{\bf C}^{\ \ j}_{mm}(k)| = | {\bf B}^{\ \ j'}_{ml} {\bf M}_{lm}^{j' j}|  
$
one finds a contradiction  if the corresponding eigenvalue
of ${\bf M}(k)$ were zero, i.e., the l.h.s. would be greater than zero for 
$k>0$, whereas the r.h.s. would be zero. 
Hence, such a zero eigenvalue cannot exist for 
$k>0$, hence $\Det{}{\bf M}(k)\neq 0$ for $k>0$, hence ${\bf M}(k)$ is
invertible on the real positive $k$-axis, 
provided  ${\bf M}(k)- {\bf 1}$ is trace-class.
From the existence of the inverse relation 
${\bf B}^{j'}={\bf C}^{j} ({\bf M}^{-1})^{jj'}$, the 
trace-class property
of ${\bf C}^{j}$ to be shown 
and the boundedness of $({\bf M}^{-1})^{jj'}$ follows the
boundedness of ${\bf B}^{j}$ and therefore the trace-class property of
the n-disk {\bf T}-matrix, ${\bf T}^{(n)}(k)$, except at the above excluded
$k$-values. 

What is left to prove is
\begin{description}
\item[\ (a)\ ] ${\bf M}(k)-{\bf 1}\in {\cal J}_1$ for all $k$, except at
the poles of $\Ho{m}{a_j}$ and for $k\leq 0$,
\item[\ (b)\ ] ${\bf C}^j(k), {\bf D}^j(k) \in {\cal J}_1$ with the exception
of the $k$-values  mentioned in {\bf (a)},
\item[\ (c)\ ] ${\bf T}^{(1)}(ka_j) \in {\cal J}_1$ (with the same exceptions
as in {\bf (a)} and {\bf (b)}) 
\item[\ (d)\ ] ${\bf M}^{-1}(k)$ does not only exist, but is bounded.
\end{description} 
Under these conditions all the manipulations of Sec.\,\ref{chap:link},
 are
justified and ${\bf S}^{(n)}$, as in \equa{Smatrix}, and
$\det{} {\bf S}^{(n)}$, as in \equa{qm}, are shown to exist.

\subsection{Proof that ${\bf T}^{(1)}(ka_j)$ is trace-class}
The  {\bf S}--Matrix for the $j^{\,\rm th}$ disk is given by \equa{S_1disk}.
Thus ${\bf V}\mbox{$\equiv$}-\i{\bf T}^{(1)}(ka_j) =
{\bf S}^{(1)}(ka_j)\mbox{$-$}{\bf 1}$ is diagonal:
\be  \label{B_v_k_disk}
  {\bf V}_{ll'}
 = \delta_{ll'}  
   \frac{-2 J_l(ka_j)}{H^{(1)}_{l'}(ka_j)} \ .
\ee
Hence, we can write ${\bf V} = {\bf U} {\bf | V|}$
where ${\bf U}$ is diagonal and unitary, and therefore bounded. What is
left to show (see property (a) of \ref{app:trace}.1) is that 
${\bf |V|}\in {\cal J}_1$. This is very simple since we can now use  
the second part of property (d) of App.~\ref{app:trace}.1: we just
have to show in a special orthonormal basis (here
the eigenbasis) that 
\be
\sum_{l=-\infty}^{+\infty} {\bf |V|}_{ll} = \sum_{l=-\infty}^{+\infty}
  2 \left |\frac{J_l(ka_j)}{H^{(1)}_{l}(ka_j)}\right | < \infty
\ee
as ${\bf |V|}\geq 0$ by definition.
The ordinary  Bessel and Hankel functions of integer order satisfy
\beq     \label{bess_hank_neg_Ordng}
  J_{-n}(z)  =  (-1)^n J_n(z)   \; , \; 
  H^{(1)}_{-n}(z) = \e^{\i \pi n} H^{(1)}_n(z) \;  ,  \;
  H^{(2)}_{-n}(z) = \e^{-\i \pi n} H^{(2)}_n(z)  \; , 
\eeq
\be
  \nu & \rightarrow & \infty \quad , \quad \nu \; \mbox{real}:  \nn \\
       \label{J_infty2}
  J_\nu(z)  \sim  \frac{1}{\sqrt{2\pi\nu}}
     \left( \frac{ez}{2\nu} \right)^\nu   \; & , & \;   
       \label{H1_infty}
  H^{(1)}_\nu(z)  \sim  -\i \sqrt{\frac{2}{\pi\nu}}
     \left( \frac{ez}{2\nu} \right)^{-\nu}  \; .
\ee
Thus:
\be    \label{a_l}
  \tr{}({\bf |V|})  \lesim  4 \sum_{l=0}^\infty \frac{1}{2}
   \left( \frac{e|ka_j|}{2l} \right)^{2l} \equiv 2 \sum_{l=0}^\infty
   ({\rm a}_l)^{2l} \; .
\ee
These ${\rm a}_l$ satisfy:
$  {\rm a}_l < {\rm a}_{l_0} <1 \quad \mbox{for} \quad l > l_0 \ {\rm and} \
   l_0 > \frac{e|ka_j|}{2} \; . $
The series  $\sum_{l=0}^\infty ({\rm a}_{l_0})^{2l}$ converges, and hence 
also the sum $\sum_{l=0}^\infty ({\rm a}_{l})^{2l}$ as it is bounded from above
by the previous sum. That means that ${\bf |V|}\in {\cal J}_1$ and
(because of property (a) of App.~\ref{app:trace}.1)\   
${\bf S}^{(1)}-{\bf 1}\in {\cal J}_1$, as well.
This, in turn,  
means that  $\det{} {\bf S}^{(1)}(ka_j)$ exists (see property (i)
of App.~\ref{app:trace}.2) and  also that the product 
$\prod_{j=1}^n \det{} {\bf S}^{(1)}(ka_j) < \infty$ in the case 
where $n$ is finite
(see property (d) of
the same Appendix). The limit $\lim_{n\to \infty}$ does not exist, in general,
as the individual terms $\det{} {\bf S}^{(1)}(ka_j)$ can become 
large, of course.

\subsection{Proof that ${\bf A}(k) \equiv {\bf M}(k) - {\bf 1}$ is trace-class}

The determinant of the characteristic matrix ${\bf M}(k)$ is defined, if
${\bf A}(k) \in {\cal J}_1$. In order to show this, we split ${\bf A}$ into
the product of two operators which -- as will be shown -- 
are both Hilbert-Schmidt. Then according to property (b) of 
App.~\ref{app:trace}.1 the product is trace-class.

Let therefore ${\bf A} = {\bf E} \cdot {\bf F}$,
where {\bf A} follows from \equa{M_N_disk}.
In order to simplify the decomposition of ${\bf A}$, we choose one of the
factors, namely, 
${\bf F}$,  as a diagonal matrix:
\beq
   {\bf A}^{jj'}_{ll'}  =  {\bf E}^{jj'}_{ll'} F^{j'}_{l'} \; , \quad
   {\bf F}^{jj'}_{ll'}  =  F^j_{l} \delta^{jj'} \delta_{ll'}  
 \label{A_zerl2}
\eeq
and
\be  \label{def_F2}
  F^j_l = \frac{\sqrt{H^{(1)}_{2l}(k\alpha a_j)}}{H^{(1)}_l(ka_j)}
   \quad , \; \alpha >2 \; .
\ee
Already this form 
leads to the exclusion of the zeros of the the Hankel functions
${H^{(1)}_l(ka_j)}$  
and also the  negative real $k$-axis (the branch cut
of the Hankel functions for  $k\leq 0$) 
from our final proof of ${\bf A}(k)\in {\cal J}_1$. 
First, we have to show that $\|{\bf F}\|^2=\sum_{j,l} ({\bf F^\dagger
F})^{jj}_{ll} < \infty$. We start with 
\beq
  \| {\bf F} \|^2 
   \le  \sum_{j=1}^n 2 \sum_{l=0}^\infty
    \frac{|H^{(1)}_{2l}(k\alpha a_j)|}{|H^{(1)}_l(ka_j)|^2} \; \equiv \;
    \sum_{j=1}^n 2 \sum_{l=0}^\infty {\rm a}_l \; .
\eeq
This form restricts the proof to $n$-disk configurations with 
$n$ {\em finite}. 
Using the asymptotic expressions \equa{J_infty2} 
for the Bessel and Hankel functions of large orders, it is easy to prove 
the absolute convergence of $\sum_l {\rm a}_l$ in the case $\alpha > 2$.
%
Therefore
$\|{\bf F}\|^2 < \infty$ and because of property (d) of 
App.~\ref{app:trace}.1 we get ${\bf F} \in {\cal J}_2$.

Using the decomposition \equa{A_zerl2}) and the definition of {\bf F} 
\equa{def_F2},   the second factor {\bf E}, is constructed. 
We then have to show the absolute convergence of the expression
\be
  \|{\bf E}\|^2 = \sum_{j,j'=1 \atop j \neq j'}^n
    \left( \frac{a_j}{a_{j'}} \right)^2 \sum_{l,l'=-\infty}^\infty
    \frac{|J_l(|ka_j|)|^2 |H^{(1)}_{l-l'}(|kR_{jj'}|)|^2}
         {|H^{(1)}_{2l'}(|k\alpha a_{j'}|)|}  \
\ee
in order to prove that also ${\bf E}\in {\cal J}_2$.
%
This is fulfilled, if $\sum_{l,l'} {\rm a}_{ll'}, < \infty$,
where 
\beq       \label{def_a_l_lp2}
{\rm }a_{ll'} = \frac{|J_l(|ka_j|)|^2 |H^{(1)}_{l+l'}(|kR_{jj'}|)|^2}
         {|H^{(1)}_{2l'}(|k\alpha a_{j'}|)|} \; .
\eeq
Necessary conditions for the convergence of the double sum over ${\rm a}_{ll'}$
are: $\sum_{l'} {\rm a}_{ll'} < \infty$ as well as $\sum_l {\rm a}_{ll'} <
\infty$. For the case $l \to \infty \; , \; l'$ fixed, we obtain with
the help of the  
asymptotic formulas \equa{J_infty2} 
the expression:
\beq     \label{def_b_l(lp)}
   l  \to  \infty\,:  \quad 
   {\rm a}_{ll'}  \sim  \frac{1}{\pi^2}
    \frac{ \left( \frac{e|kR_{jj'}|}{2} \right)^{-2l'} }
         { |H^{(1)}_{2l'}(|k\alpha a_{j'}|)| }
  \frac{1}{l}  \underbrace{
  \left( \frac{l+l'}{l} \right)^{2l} (l+l')^{2l'-1}
   \left( \frac{a_j}{R_{jj'}} \right)^{2l}
               }_{\equiv b_l(l')} \; .
\eeq
For any $\epsilon >0$ this yields the
estimate:
\be
  b_l(l') 
   <  (2l)^{2l'} \left( \frac{(1+\epsilon) a_j}{R_{jj'}}
 \right)^{2l} \; , \; l>l_0 
\ {\rm with} \frac{l'}{l_0}< \epsilon \; .
\ee
For $x \equiv (1+\epsilon) a_j / R_{jj'} < 1$, the series $\sum_{l=0}^\infty
x^{2l}$ converges absolutely. 
As $  \sum_l (2l)^{2l'} x^{2l}= (x \partial / \partial x)^{2l'}  
\sum_l x^{2l} < \infty$,
the series $\sum_{l=0}^\infty\, b_l(l')$ converges absolutely, as well. 
Therefore we have the absolute convergence of 
$\sum_l {\rm a}_{ll'}$ for $a_j < R_{jj'}$ with fixed~$l'$ in the limit
$\epsilon \to 0$.
In the opposite case, $l' \to \infty \; , \; l$~fixed, 
the absolute convergence of $\sum_{l'} {\rm a}_{ll'}$ for 
$\frac{\alpha}{2} a_{j'} < R_{jj'}$ can be proven analogously.
%
%
%

We must of course show the convergence  of $\sum_{l,l'} {\rm a}_{ll'}$ 
for 
the case 
$l,l' \to \infty$. Using again the asymptotic behavior of the
Bessel and Hankel functions of large order we get the following
proportionality for $l,l' \to \infty$:
\beq        \label{def_b_ll_2}
  {\rm a}_{ll'}  \propto  \frac{\sqrt{l'}}{(l+l')l} \frac{(l+l')^{2(l+l')}}
   { l^{2l} {l'}^{2l'}}
   \left( \frac{a_j}{R_{jj'}} \right)^{2l}
   \left( \frac{\alpha}{2} \frac{a_{j'}}{R_{jj'}} \right)^{2l'} 
   =  \frac{\sqrt{l'}}{(l+l')l}  b_{ll'} \; .
\eeq
The double sum $\sum_{l,l'=0}^\infty {\rm a}_{ll'}$ is convergent, if 
 $\sum_{l,l'=0}^\infty b_{ll'}$ converges. In order to show this, 
we introduce two new summation indices
($M,\, m$) as $l+l'=2M$ and $l-l'=m$.
Hence, we have
\be  \label{def_c_Mm2}
  \sum_{l,l'=0}^\infty b_{ll'} = \sum_{M=0}^\infty\, \sum_{m=-2M}^{2M} c_{Mm}
\ee
with
\be     \label{c_Mm2}
 c_{Mm} = \frac{(2M)^{4M}}{(M+\frac{m}{2})^{2(M+\frac{m}{2})}
             (M-\frac{m}{2})^{2(M-\frac{m}{2})}}
         \left( \frac{a_j}{R_{jj'}} \right)^{2(M+\frac{m}{2})}
   \left( \frac{\alpha}{2} \frac{a_{j'}}{R_{jj'}} \right)^{2(M-\frac{m}{2})}
      \; .
\ee
For sufficiently large  $M$, 
the powers occurring in the last expression can be approximately estimated
with the help of the Stirling formula, 
$ n^n \sim n!\, \e^n / \sqrt{2\pi}$.
In this way, we get for $M \to \infty$:
\beq
  c_{Mm}  \sim  2\pi \left( \frac{(2M)!}{(M+\frac{m}{2})!
    (M-\frac{m}{2})!}
   \left( \frac{a_j}{R_{jj'}} \right)^{M+\frac{m}{2}}
   \left( \frac{\alpha}{2} \frac{a_{j'}}{R_{jj'}} \right)^{M-\frac{m}{2}}
                       \right)^2   \; . 
\eeq
Hence, the total sum reads
\beq       \label{last_sum}
  \sum_{M=0}^\infty \sum_{m=-2M}^{+2M}  c_{Mm} 
     \lesim   2\pi \sum_{M=0}^\infty \left( \left(
     \frac{a_j + \frac{\alpha}{2} a_{j'}}{R_{jj'}} \right)^{2M}
       \right)^2 \; ,
\eeq
where the sum over $m$ has been performed 
with the help of the binomial formula.
The remaining series  in \equa{last_sum}
converges for $a_j + \frac{\alpha}{2} a_{j'}
< R_{jj'}$. Therefore, under the stated conditions
$\sum_{l,l'} {\rm a}_{ll'}$ converges absolutely, as well. 
We finally get the desired result:
 \newline \hspace{2cm}
The  operator {\bf E} belongs to the class  of Hilbert--Schmidt 
operators~(${\cal J}_2$), {\em if} the conditions 
$\frac{\alpha}{2}a_{j'} + a_{j}<R_{jj'}$,
$(1+\epsilon)a_{j'} < (1+\epsilon)\frac{\alpha}{2} a_{j'} < R_{jj'}$
and $(1+\epsilon) a_{j} < R_{jj'}$
are met in the limit $\epsilon \to 0$. 
In summary, this means:
${\bf E(k) \cdot F(k) = A}(k) \in {\cal J}_1$ for such finite $n$ 
disk configurations for which the disks neither overlap nor touch and for
those values of $k$ which lie neither on the zeros of the Hankel functions
$\Ho{m}{a_j}$ nor on the negative real $k$-axis ($k\leq 0$).
The zeros of the Hankel functions 
$H^{(2)}(k^\ast a_j )$ are then automatically excluded, too.
The zeros of the
Hankel functions $\Ho{m}{\alpha a_j}$ in the definition of ${\bf E}$ are
canceled by the corresponding zeros of the same Hankel functions in
the definition of ${\bf F}$ and can therefore be removed. A slight
change in $\alpha$ readjusts the positions of the zeros in the complex 
$k$-plane such that they can always be moved  to non-dangerous places.  
For these (``true'') scattering systems the  determinants $\Det{} {\bf
M}(k)$ and $\Det{} {\bf M}(k^\ast)^\dagger$ 
%
%
are defined and can be calculated with the help of 
one of the cumulant formulas given
in App.~\ref{app:trace}.2, e.g., by the 
Plemelj-Smithies formula \equa{ps_formula}
(with $\Det{} \; = \e^{\Tr{} \log \;}$, see \equa{tracelog}, 
for small arguments) or
by  Eqs.\equa{det_prod} or \equa{det_cum_red} 
if ${\bf M}$ or ${\bf A}$ can 
be diagonalized.

\subsection{Proof that ${\bf C}^j$ and ${\bf D}^j$ are trace-class}

The expressions for ${\bf D}^j$ and
${\bf C}^j$ 
can be found in \equa{D_N_disk} and \equa{C_N_disk}. 
Both matrices contain -- for a fixed value of $j$ --  
only the information of the single-disk scattering.
As in the proof of ${\bf T}^{(1)}\in {\cal J}_1$, we will go to the eigenbasis
of ${\bf S}^{(1)}$. In that basis both matrices 
${\bf D}^j$ and ${\bf C}^j$ become diagonal:
\be
   {\bf D}^j_{mm'} &=& -\pi a_j  J_m(ka_j) 
 \e^{-\i m\Phi_{R_j}}\, \delta_{m m'} \; ,
     \label{D_N_disk_diag}  \\
    {\bf C}^{\ j}_{lm} &=& \e^{\i m \Phi_{R_j}}\frac{1}
                          {H^{(1)}_m(ka_j)}  \frac{2\i }{\pi a_j}\, 
\delta_{l m} \label{C_N_disk_diag}
    \; .
\ee   
Using the same techniques as in the proof of  ${\bf T}^{(1)}\in {\cal J}_1$,
we can show that
${\bf C}^j$ and  ${\bf D}^j$ are trace-class. 
In summary, we have ${\bf D}^j \in  {\cal J}_1$ for all $k$
since the Bessel functions  which define that matrix 
possess neither poles nor branch cuts. 
The matrix 
${\bf C}^j \in {\cal J}_1$ for almost every $k$, except at the
zeros of the Hankel functions $\Ho{m}{a_j}$ and the branch cut of these
Hankel functions on the negative real $k$-axis ($k\leq 0$).
Note that the values of $\tr{} {\bf D}^j$ or $\tr{} {\bf C}^j$, 
are finite and the same whether one uses the non-diagonal
expressions \equa{D_N_disk}/\equa{C_N_disk} or the diagonal ones 
\equa{D_N_disk_diag}/\equa{C_N_disk_diag}. This is, of course, 
in agreement with
property (e) of App.~\ref{app:trace}.1.

\subsection{Existence and boundedness  of ${\bf M}^{-1}(k)$}

As ${\bf M}(k)-{\bf 1}\in {\cal J}_1$ except at the zeros of
$\Ho{m}{a_j}$ and on the negative real $k$-axis ($k\leq 0$),
${\bf M}^{-1}(k)$ exists everywhere, except at the points mentioned above {\em
and} except at $k$-values where $\Det{} {\bf M}(k) =0$. In other 
words, except at the poles of the ${\bf S}^{(n)}(k)$ matrix, see 
Eq.\equa{qm}.
With the exception of the negative real axis and the isolated zeros of
$\Ho{m}{a_j}$, ${\bf M}(k)$ is analytic. Hence, the points of the complex
$k$-plane  with $\Det{} {\bf M}(k) =0$ are 
isolated. Hence, $\Det{} {\bf M}(k)\neq 0$ almost everywhere. Thus, almost
everywhere, ${\bf M}(k)$ can be diagonalized and the product of the eigenvalues
weighted with their degeneracies is finite, see App.~\ref{app:trace}.2
for both properties. Thus ${\bf M}^{-1}(k)$ exists
and can be diagonalized as well. Hence, all the eigenvalues 
of  ${\bf M}^{-1}(k)$ (and their product)
are finite  in
the complex $k$-plane, where $\Det{} {\bf M}(k)$ is defined and nonzero. 
Thus ${\bf M}^{-1}(k)$ is bounded (and $\Det{}{\bf M}^{-1}(k)$ exists)
almost everywhere in
the complex $k$-plane. 

In summary, the formal steps in the calculation of the $n$-disk 
 {\bf S}--matrix (see App.~\ref{app:construction}) and its determinant
(see Sec.\,\ref{chap:link}) are all allowed and well-defined, {\em if}
the disk configurations are such that the disks neither touch nor overlap.
\newpage
\section{Comparison to Lloyd's {\bf T}-matrix\label{app:Lloyd}}
\setcounter{equation}{0}
\setcounter{figure}{0}
\setcounter{table}{0}
As mentioned in Sec.\ref{chap:S-matrix}, Lloyd has 
constructed a formal expression for the
${\bf T}$-matrix of a
finite cluster of muffin-tin potentials in three dimensions, 
see Eq.(98) of 
Ref.\cite{Lloyd_smith}. Transcribed to the case of a cluster of $n$
disk-scatterers fixed in the two-dimensional plane, Lloyd's ${\bf T}$-matrix
reads as $
\widetilde {\bf T} (k) =  \widetilde{\bf C}(k) 
   \left(\widetilde {\bf M}(k)\right)^{-1}\widetilde{\bf D}(k)$
with
\be
  \widetilde{\bf C}_{ml}^{\ \ j} &=& J_{m-l}(k R_j ) \e^{\i m \Phi_{R_j}}   
\left(\frac{-2\i \Jb{l}{a_{j}}}{ 
            \Ho{l}{a_{j}}} \right ) \; ,
  \\
   \widetilde{\bf D}_{l'm'}^{j'}  &=&   J_{m'-l'}(k R_{j'})
               \e^{-\i m' \Phi_{R_{j'}} } \; , \\
    \widetilde{\bf M}_{ll'}^{jj'} &=& \delta^{jj'} \delta_{ll'}
 +(1-\delta^{jj'}) \frac{\Jb{l}{a_j}}{\Ho{l}{a_j}} 
   H^{(1)}_{l-l'}(k R_{j'j}) \Gamma_{jj'}(l,l') \; ,
\ee
where the tilde is discriminating the matrices in the Lloyd representation
from the corresponding matrices in the Gaspard-Rice representation, 
defined in \equa{Cmatrix}, \equa{Dmatrix} and \equa{Mmatrix}. The Lloyd 
representation allows for a very simple interpretation. The matrix 
$\widetilde{\bf C}^j$ 
describes the regular propagation (in terms of the homogenous part of the free 
propagator) from the origin to the point $\pol R_j$ and a one-disk scattering
from a disk centered at this point, as given by the one-disk 
${\bf T}^{(1)}$-matrix. The matrix $\widetilde{\bf D}^{j'}$ describes the
(regular) propagation back from the disk $j'$ to the origin. 
The matrix $(\widetilde{\bf M}^{jj'})^{-1}$ parametrizes the 
multiscattering chain. If it is expanded around $\delta^{jj'}$, it
describes the sum  of no propagation and no scattering {\em plus} 
the propagation from disk $j$ to disk $j'$ (in terms
of the full propagator) and a scattering from disk $j'$ and so on. The 
disadvantage of the Lloyd representation is that the trace-class character
of $\widetilde {\bf A} \equiv
\widetilde {\bf M}- {\bf 1}$ is lost, as the terms
$\Jb{m}{a_j}$ and $( \Ho{m}{a_j} )^{-1}$ ``stabilize'' only 
the asymptotic behavior of the index $l$, but not of the index $l'$ any longer,
as the asymmetric Gaspard-Rice form did.
 The infinite determinant 
$\det{} \widetilde {\bf M} $ is therefore no longer absolutely 
convergent, but only conditionally. 
Any manipulation in the Lloyd representation
of the matrix 
$\widetilde{\bf M}$ and the corresponding ${\bf S}$-matrix 
has therefore 
to be
taken with great care. Note, however, that the (formal)
cumulant expansions of 
$\widetilde {\bf M}$ and ${\bf M}$ are the same as the corresponding traces
satisfy
${\rm Tr} (\widetilde{\bf A}^n)={\rm Tr}({\bf A}^n)$. In other words, if
the cumulants of $\widetilde{\bf M}$ are
summed up according to the Plemelj-Smithies form
of ${\bf M}$, the result of $\det{} \widetilde{\bf M}(k)$ and
$\det{} {\bf M}(k)$ is the same. In fact, one can derive the Lloyd
representations $\widetilde {\bf C}^j$, $\widetilde {\bf M}^{jj'}$ and
$\widetilde {\bf D}^{j'}$ from the expressions for $\widehat{\bf C}^j$,
$\widehat{\bf M}^{j'j}$ and ${\bf D}^j$ of App.~\ref{app:construction}.1
(see Eq. \equa{hat_relation})
by
the following {\em formal} manipulations: First,
$\widehat{\bf C}^{\ \ j}_{ml}$ and $\widehat{\bf M}^{j'j}_{l'l}$ are divided by
the diagonal matrix $\{ \Ho{l}{a_j}/(-2\i )\}$. This produces already
$\widetilde{\bf C}^{j}$.
Second, ${\bf B}^{\ \ j'}_{ml'}$ in the (now changed) relation
\equa{hat_relation} and in \equa{S_2_disk} is
rescaled as 
\be
{\bf B}^{\ \ j'}_{ml'}= \widetilde{\bf B}^{\ \ j'}_{ml'} 
\frac{-1}{\pi a_{j'} \Jb{l'}{a_{j'}}} \; ,
\ee
such that $\widetilde{\bf D}^{j'}$ and $\widetilde{\bf M}^{jj'}$ emerge.
Both manipulations are only of formal nature as they {\em change} the
``trace-character'' of the corresponding matrices.
\newpage
\section{1-disk determinant in the semiclassical 
approximation\label{app:semi1disk}}
\setcounter{equation}{0}
\setcounter{figure}{0}
\setcounter{table}{0}
In App.~\ref{app:construction-S} 
we have constructed the scattering matrix
for the 1-disk system (see Eq.\equa{S1disk}):
\be
 \left ({\bf S}^{(1)}(ka_j)\right )_{m m'} = 
  - \frac{\Ht{m}{a_j}}{\Ho{m}{a_j}} \, \delta_{m m'} \; .
\ee
Instead of calculating the semiclassical approximation 
to its determinant, we instead do so for 
\be
  {\bf d}(k) \equiv \frac{1}{2\pi \i } \frac{d}{dk} 
     \ln \det{}{\bf S}^{(1)}(ka_j)\ ,
    \label{delay}
\ee
the so-called {\em time delay}. Recall that the corresponding 
${\bf T}^{(1)}$-matrix is trace-class. Thus, according to properties (j)
and (c)  of
App.~\ref{app:trace}.2 the following operations are justified:
\be
 {\bf d}(k) &=& \frac{1}{2\pi \i } \frac{d}{dk}  
      \tr{}\left (\, \ln  \det{}{\bf S}^{(1)}(ka_j)
                    \, \right ) 
            = \frac{1}{2\pi \i }\tr{} 
                      \left ( \frac{ \Ho{m}{a_j} }{ \Ht{m}{a_j} }\,
                   \frac{d}{dk} \frac{ \Ht{m}{a_j}}{ \Ho{m}{a_j} } \right )
                      \nn \\
            &=& \frac{a_j}{ 2\pi \i }\tr{}
                 \left ( \frac{ {H_m^{(2)}}'(k a_j) }  { \Ht{m}{a_j}}
                        - \frac{ {H_m^{(1)}}'(k a_j) }  { \Ho{m}{a_j}}
                   \right ) \; .
              \label{delay_start}
\ee
Here the prime denotes the derivative with respect to the argument of the
Hankel functions. Let us introduce the abbreviation
\be
     \chi_{\nu} =  \frac{ {H_{\nu}^{(2)}}'(k a_j) }  { \Ht{\nu}{a_j}}
                        - \frac{ {H_{\nu}^{(1)}}'(k a_j) }  { \Ho{\nu}{a_j}}
\; .
         \label{abbrev}
\ee
Following Ref.\cite{franz}, we apply the Watson contour method~\cite{Watson} 
to \equa{delay_start} (see also Sec.\ref{chap:semiclass} 
and App.\ref{app:convol})
\beq
  {\bf d}(k) = \frac{a_j}{ 2\pi \i } \sum_{m=-\infty}^{+\infty}\,
                   \chi_m 
             =  \frac{a_j}{ 2\pi \i } \, \frac{1}{2\i } \oint_C {\rm d}\nu\,
 \frac{\e^{-\i \nu\pi}}{\sin(\nu\pi)} \chi_{\nu} \; .
  \label{watson_1_disk}
\eeq
Here the contour $C$ encircles in a counter-clock-wise manner 
a small semi-infinite 
strip $D$ which
completely covers the real $\nu$-axis~\footnote{In 
App.\ref{app:convol}, symmetrized expressions have been Watson transformed.
Thus, the corresponding $D_+$ only has to cover  the
the real {\em positive} $\nu$-axis .},  but which only has a small finite
extend into the positive and negative imaginary $\nu$ direction. As in
Ref.\cite{aw_chaos}, the contour $C$ will be split up in the path
above and below the real $\nu$-axis such that
\be
 {\bf d}(k)&=&  \frac{a_j}{ 2\pi \i } \left \{ 
 -\frac{1}{2\i }\int_{-\infty +\i \epsilon}^{+\infty+\i \epsilon}
   {\rm d}\nu\,
 \frac{\e^{-\i \nu \pi}}{\sin(\nu \pi)} \chi_{\nu} 
 + \frac{1}{2\i }\int_{-\infty -\i \epsilon}^{+\infty-\i \epsilon}
   {\rm d}\nu\,
 \frac{\e^{-\i \nu \pi}}{\sin(\nu \pi)} \chi_{\nu} \right \} \ .
\ee
Then, we perform the substitution $\nu \to - \nu$ in the second integral 
so as to get
\be
  {\bf d}(k)&=&  \frac{a_j}{ 2\pi \i } \left \{
   -\frac{1}{2\i }\int_{-\infty +\i \epsilon}^{+\infty+\i \epsilon}
   {\rm d}\nu\,
 \frac{\e^{-\i \nu \pi}}{\sin(\nu \pi)} \chi_{\nu} 
  -  \frac{1}{2\i }\int_{-\infty +\i \epsilon}^{+\infty+\i \epsilon}
   {\rm d}\nu\,
 \frac{\e^{+\i \nu \pi}}{\sin(\nu \pi)} \chi_{-\nu} \right \} \nn \\
   &=& 
   \frac{a_j}{ 2\pi \i } \left \{ 2 
    \int_{-\infty +\i \epsilon}^{+\infty+\i \epsilon}
   {\rm d}\nu\, \frac{\e^{2 \i \nu \pi}}{1- \e^{2\i \nu \pi}} \, \chi_\nu 
      +\int_{-\infty}^{+\infty}   {\rm d}\nu\, \chi_{\nu}
       \right \} \; ,
   \label{watson_1_disk_end}
\ee
where we used the fact that $\chi_{-\nu}=\chi_{\nu}$.
The contour in the last integral could be deformed to pass 
over the real $\nu$-axis since its integrand has  no Watson denominator
any longer.
We will now approximate the last expression semiclassically, i.e., under
the assumption $ka_j \gg 1$.
As the two contributions in the last line of \equa{watson_1_disk_end} 
differ by the presence
of the Watson denominator, they will have to be handled semiclassically in 
different ways: the first will be closed in the upper complex plane and
evaluated at the poles of $\chi_\nu$, the second integral will be performed
under the Debye approximation for Hankel functions. 
We will now work out the first term. The
poles of $\chi_\nu$ in the upper complex plane 
are given by the zeros of $\Ho{\nu}{a_j}$ which will be
denoted by $\nu_{\ell}(ka_j)$ and by the zeros of $\Ht{\nu}{a_j}$ which we will
denote by $-\bar \nu_{\ell}(ka_j)$, $\ell =1,2,3,\cdots$. 
In the Airy approximation to the Hankel
functions, see \cite{Abramowitz}, they are given by  Eqs.\equa{nu_k} and
\equa{bar_nu_k}:
\be
    \nu_\ell (ka) &=& ka + \e^{+\i \pi/3} (ka/6)^{1/3}q_\ell+\cdots 
  = ka +\i \alpha_\ell (k) +\cdots \; ,
 \\
  -{\bar \nu}_\ell (ka) &=& -ka - 
 \e^{-\i \pi/3} (ka/6)^{1/3}q_\ell +\cdots 
 = -ka +\i 
( \alpha_\ell (k^\ast a))^\ast +\cdots \nno \\
 &=& -\left(\nu_\ell(k^\ast a)\right )^\ast \; ,
\ee
where $\alpha_\ell (k a_j)$ is defined in \cite{vwr_prl} and $q_l$ labels
the zeros of the Airy integral \equa{Airy-integral}, 
for details see \cite{franz,aw_chaos}.
In order to keep the notation simple, we will abbreviate 
$\nu_\ell \equiv\nu_\ell (k a_j)$ and 
$\bar \nu_\ell \equiv \bar\nu_\ell (k a_j)$.
Thus the
first term of \equa{watson_1_disk_end} becomes finally
\be
 \frac{a_j}{ 2\pi \i } \left \{ 2 \int_{-\infty +\i \epsilon}^{+\infty+\i \epsilon}
   {\rm d}\nu\, 
  \frac{\e^{2 \i \nu \pi}}{1- \e^{2\i\nu \pi}} \, \chi_\nu \right \}
 = 2a_j \sum_{\ell = 1}^{\infty} \left ( \frac{\e^{2\i\nu_\ell \pi}}
                                    {1 -\e^{2\i\nu_\ell \pi}} 
   +  \frac{\e^{-2\i\bar\nu_\ell \pi}}
                                    {1 -\e^{-2\i\bar\nu_\ell \pi}} \right).
\ee
In the second term of  \equa{watson_1_disk_end} we will insert the Debye
approximations for the Hankel functions
~\cite{Abramowitz}:
\be
 H_{\nu}^{(1/2)}(x) &\sim& \sqrt{\frac{2}{{\pi \sqrt{x^2-\nu^2}}}} 
  \exp\left 
      (\pm \i \sqrt{x^2-\nu^2} \mp \i\nu \arccos \frac{\nu}{x} \mp 
\i \frac{\pi}{4}
       \right )  \ \ \mbox{for $|x|>\nu$}\; , 
 \label{H1-Debye}\\
   H_{\nu}^{(1/2)}(x) &\sim& \mp \i\,\sqrt{\frac{2}{{\pi\sqrt{\nu^2-x^2}}}}
         \exp\left(-\sqrt{\nu^2-x^2}+\nu{\rm ArcCosh}\frac{\nu}{x}\right)
    \ \ \ \ \ \ \mbox{for $|x| < \nu$}\; .
\ee
Note that for $\nu > ka_j$ the contributions in $\chi_\nu$ cancel.
Thus the second integral of \equa{watson_1_disk_end} becomes
\be
   \frac{a_j}{ 2\pi \i }  
      \int_{-\infty}^{+\infty}   {\rm d}\nu\, \chi_{\nu}
        &=& \frac{a_j}{2\pi \i }\int_{-ka_j}^{+ka_j} {\rm d}\nu \,
 \frac{(-2\i )}{a_j}
 \frac{d}{dk} \left ( \sqrt{k^2a_j^2-\nu^2} - \nu \arccos\frac{\nu}{ka_j}
      \right ) +\cdots \nn \\
     &=& -\frac{1}{k\pi}\int_{-k a_j}^{ka_j}
      {\rm d}\nu \sqrt{k^2 a_j^2 -\nu^2}
     +\cdots  = -  \frac{a_j^2}{2} k +\cdots \; ,
        \label{nu_integ}
\ee
where $\cdots$ takes care of the polynomial corrections in the Debye 
approximation and the boundary correction terms in the $\nu$ integration.

In summary, our semiclassical approximation to ${\bf d}(k)$ reads 
\be
 {\bf d}(k) =  2a_j \sum_{\ell = 1}^{\infty} \left ( 
    \frac{\e^{2\i\nu_\ell \pi}}
                                    {1 -\e^{2\i\nu_\ell \pi}} 
   + \frac{\e^{-2\i\bar\nu_\ell \pi}}
                                    {1 -\e^{-2\i\bar\nu_\ell \pi}} \right)
 - \frac{{a_j}^2}{2} k +\cdots \ .
\ee
Using the definition of the time delay \equa{delay}, 
we get the following expression
for $\det{}{\bf S}^{(1)}(k a_j)$:
\be
   \lefteqn{\ln\det{}{\bf S}^{(1)}(k a_j)- \lim_{k_0 \to 0} 
   \ln\det{}{\bf S}^{(1)}(k_0 a_j)} \nn\\
   &=&\!
\int_0^k {\rm d}\tilde k\, \left (- \i 2\pi  \frac{a^2 \tilde k}{2} 
        + 2 (\i 2\pi  a_j)   \sum_{\ell = 1}^{\infty} 
      \left ( \frac{\e^{\i 2\pi\nu_\ell(\tilde k a_j)}}
                      {1 -\e^{\i 2\pi\nu_\ell(\tilde k a_j) }}
       + \frac{\e^{-\i 2\pi\bar\nu_\ell(\tilde k a_j) }}
                            {1 -\e^{-\i 2\pi\bar\nu_\ell(\tilde k a_j)}} 
 \right) \right) +\cdots \nn \\
   &\sim & \!
-2\pi \i N(k)\!+\!2 \sum_{\ell = 1}^{\infty}  
       \int_0^k {\rm d}\tilde k\,\frac{d}{d \tilde k}\left\{ 
      -\ln\left(1\mbox{$-$}
            \e^{\i  2\pi \nu_\ell(\tilde k a_j)}\right )
      +\ln\left(1\mbox{$-$}
            \e^{-\i 2\pi \bar\nu_\ell(\tilde k a_j)}\right ) \right\}
 + \cdots ,
      \label{k_int_1disk}
\ee  
where in the last expression it has been used that semiclassically
$\frac{d}{d  k}  \nu_\ell( k a_j)\sim\frac{d}{d  k}  
\bar\nu_\ell( k a_j) \sim a_j$ and that
the Weyl term for a single disk of radius $a_j$ goes like 
$N(k) =\pi a_j^2 k^2/({4\pi})+\cdots$ (the next terms come from
the boundary terms in the $\nu$-integration in \equa{nu_integ}).
Note that for the lower limit,  $k_0\to 0$, we have two  
simplifications:
First, 
\be
\lim_{k_0\to 0} {\bf S}^{(1)}_{mm'}(k_0 a_j) =
 \lim_{k_0\to 0}\frac{ -H^{(2)}_{m}(k_0 a_j)}{H^{(1)}_{m}(k_0 a_j)} 
\delta_{mm'}
&=& 1 \times \delta_{mm'} 
\qquad \forall m,m' \nn \\
 \leadsto \quad
 \lim_{k_0\to 0} \det{}{\bf S}^{(1)}(k_0 a_j) &=& 1\; . \nn
\ee 
Secondly, for $k_0\to 0$,
the two terms in the curly bracket of \equa{k_int_1disk} cancel.
Hence, we finally obtain the semiclassical result for the determinant of
${\bf S}^{(1)}(k a_j)$
\be
 \det{}{\bf S}^{(1)}(k a_j)\semiclass \e^{ -\i 2\pi  N(k)}
\frac{\prod_{\ell=1}^{\infty} 
     \left(1-\e^{-2\i\pi \bar\nu_\ell( k a_j)}\right )^2}
      {\prod_{\ell=1}^{\infty} 
     \left(1-\e^{2\i\pi \nu_\ell( k a_j)}\right )^2} \  ,
\ee
which should be compared with expression \equa{sc1disk} 
of Sec.\,\ref{chap:link}. For more details we refer to App.\ref{app:convol}. 
\newpage
\section{Semiclassical approximation of two convoluted ${\bf A}$-matrices
\label{app:convol}}
\setcounter{equation}{0}
\setcounter{figure}{0}
\setcounter{table}{0}

\begin{figure}[htb]
 \centerline {\epsfig{file=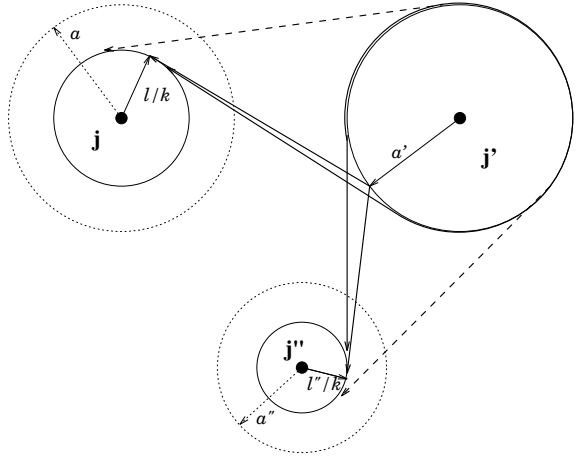,height=6cm,angle=-0}}
\caption[fig:conv1]{\small The geometry belonging to a
trajectory, $j\to j'\to j''$, specularly reflected to 
the {\em right}. Shown are the geometrical path (full line) and the
shortest allowed right handed (dashed line) and left handed (full line) 
creeping paths. All paths originate from an ``impact parameter'' circle 
of radius $|l/k|$ 
centered at disk $j$, then contact the surface of disk $j'$ (of radius $a'$) 
and end on an ``impact parameter'' circle of radius $|l''/k|$
centered at disk $j''$ . Note
that the impact radii do not have to be equal to the disk radii, $a$ 
and $a''$.
\label{fig:conv1}}
\end{figure}

\begin{figure}[htb]
 \centerline {\epsfig{file=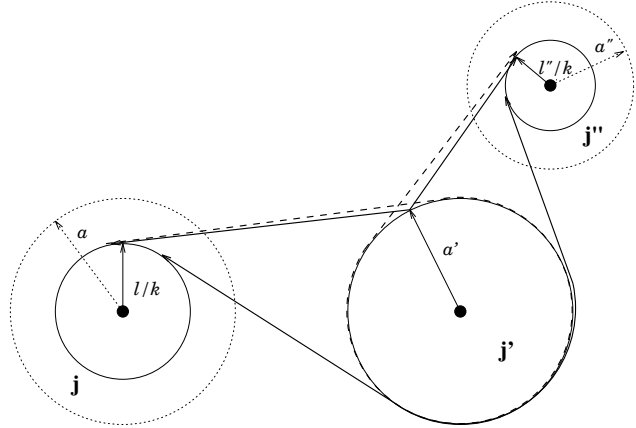,height=5.63cm,angle=0}}
\caption[fig:conv2]{\small The same as in Fig.\ref{fig:conv1} for the case of
a  specular reflection to the {\em left}.
\label{fig:conv2}}
\end{figure}

\begin{figure}[htb]
 \centerline {\epsfig{file=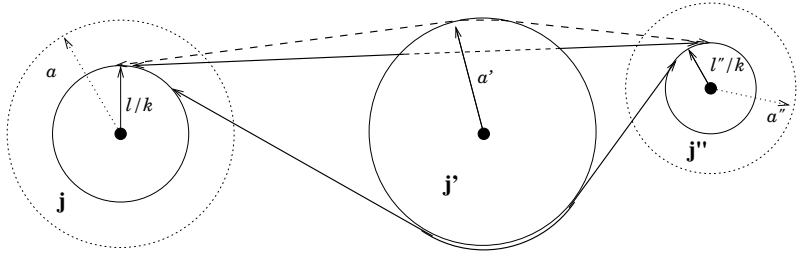,height=3.33cm,angle=-90}}
\caption[fig:conv3]{\small The geometry belonging to a {\em ghost} trajectory
$j\to j' \to j''$ which passes straight through the disk $j'$ 
(of radius $a'$). 
Shown are the
geometrical ghost path (full line and short-dashed line), and the
shortest allowed right handed (dashed line) and left handed (full line) 
creeping paths. All paths originate from an ``impact parameter'' circle 
of radius $|l/k|$ 
at disk $j$ and end on an ``impact circle'' of radius $|l''/k|$ 
centered at disk $j''$.
Note
that the impact radii do not have to be equal to the disk radii, $a$ and $a''$.
\label{fig:conv3}}
\end{figure}

In  this appendix we introduce the necessary apparatus for the
semiclassical reduction of
$\Tr{}[\, {\bf A}^m (k)\,]$ for the $n$-disk system where
\beq
{\bf A}^{j j'}_{l l'} = (1- \delta_{jj'}) 
                  \frac{ a_j   \Jb{l }{a_j   } }
                       { a_{j'}\Ho{l'}{a_{j''}} }
                  (-1)^{l'} 
                  \e^{\i ( l \alpha_{j' j} - l' \alpha_{j j'} ) } 
                  \Ho{l-l'}{R_{jj'}} \; . 
 \label{A-kernel}
\eeq
As usual, $a_j$, $a_{j'}$ are the radii of disk $j$ and $j'$, 
$1\leq j,j'\leq n$, 
$R_{j j'}$ is the distance between the centers of these disks, and
$  \alpha_{j' j}$ is the angle of the ray from the origin of 
disk $j$ to the one of disk $j'$, as
measured in the local coordinate system of disk $j$. The  angular-momentum
quantum numbers $l$ and $l'$  can be interpreted geometrically in terms of
the positive-- or negative-valued distances 
(impact parameters) $l/k$ and $l'/k$ from the center 
of disk $j$ and disk $j'$, respectively, see 
Figs.\ref{fig:conv1}--\ref{fig:conv3}.

The semiclassical approximation of the convolution of two 
kernels $\sum_{l'}\,
{\bf A}^{j j'}_{l l'}{\bf A}^{j' j''}_{l' l''}$ 
contains all (but one) essential steps
necessary for the semiclassical reduction of the quantum cycles and traces
themselves. What is
missing is the mutual interaction between successive saddles of
the quantum itinerary, including the final saddle which ``closes'' the 
semiclassical open
itinerary 
to a period orbit. This is studied in Sec.\ref{chap:semiclass}.

The idea here is to construct 
the convolution of the two kernels $\sum_{l'}\,
{\bf A}^{j j'}_{l l'}{\bf A}^{j' j''}_{l' l''}$ 
and then to compare it  
-- in the case $j\neq j''$ -- 
with the single kernel ${\bf A}^{j j''}_{l l''}$ (see
\equa{A-kernel}) in the semiclassical limit,  where
the Hankel function $\Ho{l-l''}{R_{j j''}}$ is evaluated in the 
Debye approximation \equa{H1-Debye} to leading order~\cite{Abramowitz}. 
Let us
start with
\be
\sum_{l'}\,{\bf A}^{j j'}_{l l'}{\bf A}^{j' j''}_{l' l''} \!\!&=&\!
(1- \delta_{jj'}) (1- \delta_{jj''})
                  \frac{ a_j   \Jb{l }{a_j   } }
                       { a_{j''}\Ho{l''}{a_{j''}} }
                   (-1)^{l''} 
                       \e^{\i ( l  \alpha_{j'  j  } 
                              -l''  \alpha_{j'  j''} ) } \nonumber\\
        && \quad\mbox{} \times \sum_{l'=\infty}^{+\infty}\,
          (-1)^{l'}
         \frac{\Jb{l'}{a_{j'}   } }
                               {\Ho{l'}{a_{j'}   } }
                    \Ho{l-l'}{R_{jj'}}  \Ho{l'-l''}{R_{j'j''}} 
             \e^{\i  l' ( \alpha_{j'' j' } -  \alpha_{j   j' } ) } \nonumber\\ 
       &=& \!
       W_{ll''}^{jj''}\!\sum_{l'=\infty}^{+\infty}\,
          (-1)^{l'}
         \frac{\Jb{l'}{a_{j'}   } }
                               {\Ho{l'}{a_{j'}   } }
                    \Ho{l-l'}{R_{jj'}}  \Ho{l'-l''}{R_{j'j''}} 
             \e^{\i  l' ( \alpha_{j'' j' } -  \alpha_{j   j' } ) } \\
     &=& \!
   W_{ll''}^{jj''}\!\sum_{l'=0}^{+\infty}\,
 (-1)^{l'} d(l')        \frac{\Jb{l'}{a_{j'}   } }
                               {\Ho{l'}{a_{j'}   } }
\left[     \Ho{l-l'}{R_{jj'}}  \Ho{l'-l''}{R_{j'j''}} 
             \e^{\i  l' ( \alpha_{j'' j' } -  \alpha_{j   j' } ) }
    \right.\nonumber \\ 
    &&\qquad\left.\mbox{}  +  \Ho{l+l'}{R_{jj'}}  \Ho{-l'-l''}{R_{j'j''}} 
             \e^{-\i  l' ( \alpha_{j'' j' } -  \alpha_{j   j' } ) }
 \right] \; ,
 \label{convolution}
\ee
where we have introduced the abbreviations  $W^{j j''}_{l l''}$ for the
$l'$-independent pieces and the weight factor $d(l')=1$ for $l'\neq 0$
and $d(0)=1/2$.
We  have symmetrized this expression with respect to $l'$ for simplicity using
that   
$\Jb{-l'}{a_{j'}}=(-1)^{l'}\Jb{l'}{a_{j'}}$ and
$\Ho{-l'}{a_{j'}}=(-1)^{l'}\Ho{l'}{a_{j'}}$, valid for $l'$ integer.
We will furthermore abbreviate 
$\Delta\alpha_{j'} \equiv \alpha_{j''j'}-\alpha_{j j'}$ where
$0\leq \Delta\alpha_{j'} < 2\pi$. However,  in order to be able to get three 
domains 
for this angle (which we will  later identify with the three different cases:
specular 
reflection from disk $j'$ to the right (see Fig.\ref{fig:conv1}), 
to the left (see Fig.\ref{fig:conv2}) and  the ghost 
``tunneling'' case (see Fig.\ref{fig:conv3})) 
we define
$\widetilde{\Delta}\alpha_{j'\sigma} \equiv \alpha_{j''j'}-\alpha_{j j'}
-\sigma \pi$ where $\sigma=0,2,1$, respectively, 
and balance this by multiplying accordingly the
right hand side of 
Eq.\equa{convolution} with the phase factor 
$(-1)^{l'\sigma}$ which is only nontrivial for $\sigma=1$.
We denote this nontrivial phase by  $(-1)^{-l'\sigma'}$ where $\sigma'=\sigma$
for $\sigma=1$ and zero otherwise. The three choices for the value
of $\sigma$ are still equivalent at this stage.

\subsection{The Watson contour resummation\label{app:convol_Watson}}
It will be shown that \equa{convolution} contains, in
the semiclassical limit $ka_{j'}\gg 1$
-- depending on the pertinent angles and distances -- 
a classical path or a possible
ghost path and two creeping paths, all starting under the impact parameter
$l/k$ with respect to the origin of disk $j$  and
ending at an impact parameter $l''/k$ relative to the origin of disk $j''$.
This calculation will be performed with the help of the Watson
contour method~\cite{Watson,franz}, i.e., under the replacement
\beq
 \sum_{l'=0}^{+\infty}\, (-1)^{l'(1-\sigma')} d(l') X_{l'} 
 = \frac{1}{2\i} \oint_{C_+} {\rm d}{\nu'}\,\frac{1}{\sin({\nu'} \pi)}
\e^{-\i\nu'\pi\sigma'} X_{\nu'} \; .
 \label{Watson-start}
\eeq
Here 
\be
  X_{l'}&\equiv& 
      \frac{\Jb{l'}{a_{j'}   } }
                               {\Ho{l'}{a_{j'}   } } 
\left[     \Ho{l-l'}{R_{jj'}}  \Ho{l'-l''}{R_{j'j''}} 
             \e^{\i  l' \widetilde{\Delta}\alpha_{j'\sigma } }
 \right. \nonumber  \\
 && \qquad\qquad\quad \left.
 \mbox{}+  \Ho{l+l'}{R_{jj'}}  \Ho{-l'-l''}{R_{j'j''}} 
             \e^{-\i  l' \widetilde{\Delta}\alpha_{j'\sigma }  }
 \right] \nonumber \\
 &\equiv&  \frac{\Jb{l'}{a_{j'}   } }
                               {\Ho{l'}{a_{j'}   } } Y_{l'}
\ee
stands for the 
integrand  and $Y_{l'}$ for the symmetrized square bracket in 
Eq.\equa{convolution}.
The contour $C_+$ is the boundary of a narrow semi-infinite strip 
$D_+$ which completely
covers the positive real ${\nu'}$-axis.

\begin{figure}[htb]
 \centerline {\epsfig{file=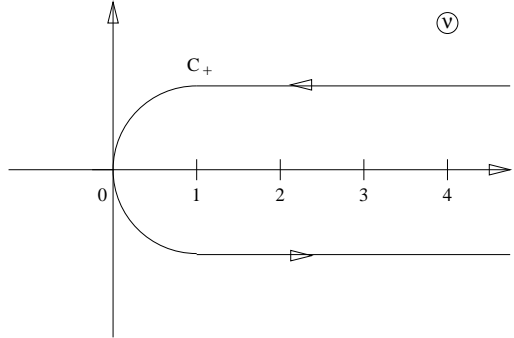,height=4.5cm,angle=-90}}
\caption[fig:Watson-contour]{\small The Contour $C_+$ in the complex $\nu$-plane.
\label{fig:Watson-contour}}
\end{figure}

$C_+$ has been chosen in such a way that it encircles in a positive sense
all poles of the Watson denominator $\sin(\nu'\pi)$ at 
$\nu'=1,2,3,\cdots$ exactly once (see Fig.\ref{fig:Watson-contour}). 
At $\nu'=0$ the weight factor
$d(0)=1/2$ is taken into account by a principle value description,
i.e., by the average of the two contour integrals whose paths cross
the real $\nu$-axis symmetrically 
just to the right and left of the point $\nu'=0$. 
A precondition on the validity of the Watson replacement
is  the analyticity of $X_{\nu'}$ in this strip $D_+$. This is the
case if $D_+$ has been chosen narrow enough in the imaginary $\nu$ direction
that the poles of $X_{\nu'}$, the zeros of the Hankel function 
$\Ho{\nu'}{a_{j'}}$ and  $\Ho{-\nu'}{a_{j'}}$   
lie either above or below the strip in the complex $\nu'$-plane [for $k$ real
and positive]. 
The contour can now be split
into four parts:
\be
\oint_{C_+} {\rm d}
 {\nu'}\,\frac{\e^{-\i\nu'\pi\sigma'}}{\sin({\nu'} \pi)} X_{\nu'}
&=& +\int_{+\infty+\i\epsilon}^{0+\i\epsilon}{\rm d}{\nu'}\,
            \frac{\e^{-\i\nu'\pi\sigma'}}{\sin({\nu'} \pi)} X_{\nu'}
 +{\rm P}\int_{0+\i\epsilon}^{0}{\rm d}{\nu'}\,
            \frac{\e^{-\i\nu'\pi\sigma'}}{\sin({\nu'} \pi)} X_{\nu'} 
 \nonumber \\
 &&  \mbox{}
 +{\rm P}\int_{0}^{0-\i\epsilon}{\rm d}{\nu'}\,
            \frac{\e^{-\i\nu'\pi\sigma'}}{\sin({\nu'} \pi)} X_{\nu'}
 +\int_{0-\i\epsilon}^{\infty-\i\epsilon}{\rm d}{\nu'}\,
            \frac{\e^{-\i\nu'\pi\sigma'}}{\sin({\nu'} \pi)} X_{\nu'}\; ,
\ee
where ${\rm P}\int\cdots$ denotes the principal value integration.
The next step in the evaluation is a shift of the contour paths below the
real ${\nu'}$-axis to  paths above this axis by the 
substitution ${\nu'} \to -{\nu'}$
in the corresponding integrals:
\be
 \oint_{C_+} {\rm d}{\nu'}\,\frac{\e^{-\i\nu'\pi\sigma'}}{\sin({\nu'} \pi)} 
                                                    X_{\nu'}
&=& -\int_{0+\i\epsilon}^{\infty+\i\epsilon}{\rm d}{\nu'}\,
            \frac{\e^{-\i\nu'\pi\sigma'}}{\sin({\nu'} \pi)} X_{\nu'}
 -\int_{-\infty+\i\epsilon}^{0+\i\epsilon}{\rm d}{\nu'}\,
            \frac{\e^{\i\nu'\pi\sigma'}}{\sin({\nu'} \pi)} X_{-\nu'}
 \nonumber \\
&&\mbox{} +{\rm P}\int_{0+\i\epsilon}^{0}{\rm d}{\nu'}\,
            \frac{1}{\sin({\nu'} \pi)} 
 \left[\e^{-\i\nu'\pi\sigma'}X_{\nu'}-\e^{\i\nu'\pi\sigma'}X_{-\nu'}\right]
\; .
\ee

We insert $X_{\nu'}= (\Jb{\nu'}{a_{j'}}/\Ho{\nu'}{a_{j'}}) Y_{\nu'}$
and use that
for general complex-valued angular momenta ${\nu'}$, the transformation laws
for the Hankel and Bessel functions
read 
\be
\Ho{-{\nu'}}{a_{j'}}  &=& \e^{\i{\nu'}\pi}\Ho{{\nu'}}{a_{j'}}\; , \label{hotrafo} \\
 \Ht{-{\nu'}}{a_{j'}} &=& \e^{-\i{\nu'}\pi}\Ht{{\nu'}}{a_{j'}}\; , 
 \label{httrafo}
\ee
such that
\be
 \frac{\Jb{-{\nu'}}{a_{j'}}}{\Ho{-\nu'}{a_{j'}}}
   &=& \frac{ \Jb{{\nu'}}{a_{j'}}}{\Ho{\nu'}{a_{j'}}}
  - \i \e^{-\i\nu'\pi}\sin(\nu'\pi)
        \frac{ \Ht{{\nu'}}{a_{j'}}}{\Ho{\nu'}{a_{j'}}} 
 \label{jminus} \; , \\
 \e^{\i\nu'\pi} \frac{\Ht{-{\nu'}}{a_{j'}}}{\Ho{-{\nu'}}{a_{j'}}}
 &=& \e^{-\i\nu'\pi}\frac{\Ht{{\nu'}}{a_{j'}}}{\Ho{{\nu'}}{a_{j'}}}
 \label{geominus} \; .
\ee
Furthermore, by definition, we have $Y_{-\nu}=Y_{+\nu}$.
Thus 
\be
\frac{1}{2\i} \oint_{C_+} {\rm d}{\nu'}\,
 \frac{\e^{-\i\nu'\pi\sigma'}}{\sin({\nu'} \pi)}
 X_{\nu'}
&=& 
-\frac{1}{2\i}\int_{-\infty+\i\epsilon}^{\infty+\i\epsilon}{\rm d}{\nu'}\,
            \frac{\e^{-\i\nu'\pi\sigma'}}{\sin({\nu'} \pi)} 
              \frac{\Jb{\nu'}{a_{j'}   } }
                               {\Ho{\nu'}{a_{j'}   } } Y_{\nu'} \nonumber \\
&&\mbox{}+\frac{1}{4}{\rm P}\int_{-\infty+\i\epsilon}^{+\infty-\i\epsilon}
      \!\!\! {\rm d}\nu'\, \e^{-\i\nu'\pi(1-\sigma')} \left[
         \frac{\Ht{\nu'}{a_{j'}}}{\Ho{\nu'}{a_{j'}}} 
         \mbox{$-$}2\delta_{\sigma',1} 
 \frac{\Jb{\nu'}{a_{j'}}}{\Ho{\nu'}{a_{j'}}}
   \right ]Y_{\nu'} \; ,\nonumber\\
\ee
where \equa{geominus} and the symmetry of $Y_{\nu'}$ 
has been used in order to reflect
the resulting $\sin(\nu'\pi)$-independent integrals at $\nu'=0$ such that 
they combine to the symmetric integral:
\beq
+{\rm P}\!\!\int_{-\infty +\i\epsilon}^{+\infty
-\i\epsilon}{\rm d}\nu'\cdots \equiv
 \int_{-\infty +\i\epsilon}^{0+\i\epsilon}{\rm d}\nu'\cdots
+{\rm P}\!\!\int_{-0 +\i\epsilon}^{0}{\rm d}\nu'\cdots
+{\rm P}\!\!\int_{0}^{0-\i\epsilon}{\rm d}\nu'\cdots
+\!\!\int_{0-\i\epsilon}^{+\infty-\i\epsilon}{\rm d}\nu'\cdots.
\eeq
Furthermore, in the case $\sigma'=1$, the identity 
$\e^{\i\nu'\pi}=\e^{-\i\nu'\pi} +2\i\sin(\nu'\pi)$ has been employed
in order to group
the terms resulting from the paths below the real $\nu'$-axis into   the
terms belonging to the paths above this axis. 
Altogether we have so far that 
\be
\sum_{l'}\,
{\bf A}^{j j'}_{l l'}{\bf A}^{j' j''}_{l' l''} &=&
W^{j j''}_{l l''} \left \{
-\frac{1}{2\i} \int_{-\infty+\i\epsilon}^{+\infty+\i\epsilon} {\rm d} {\nu'}\,
\frac{\e^{-\i\nu'\pi\sigma'}}{\sin({\nu'}\pi)}
               \frac{\Jb{{\nu'}}{a_{j'}   } }
                    {\Ho{{\nu'}}{a_{j'}   } } \right .
 \nonumber \\
&& \qquad\qquad \times
\left[
        \Ho{l-{\nu'}}{R_{jj'}}  \Ho{{\nu'}-l''}{R_{j'j''}} 
             \e^{\i  {\nu'} \widetilde{\Delta}\alpha_{j',\sigma} }
 \right. \nonumber \\
&& \qquad\qquad \qquad \left.\mbox{}
 +\Ho{l+{\nu'}}{R_{jj'}}  \Ho{-{\nu'}-l''}{R_{j'j''}} 
             \e^{-\i  {\nu'} \widetilde{\Delta}\alpha_{j',\sigma} } \right ]
     \nonumber \\
&& \mbox{} +\frac{1}{4} {\rm P}
  \int_{-\infty+\i\epsilon}^{+\infty-\i\epsilon} {\rm d} {\nu'}\,
\e^{-\i\nu'\pi(1-\sigma')}\left[
\frac{\Ht{{\nu'}}{a_{j'}   } }
                    {\Ho{{\nu'}}{a_{j'} }} 
  -2\delta_{\sigma',1}\frac{\Jb{{\nu'}}{a_{j'}   } }
                    {\Ho{{\nu'}}{a_{j'} }} 
\right ] \nonumber \\
&& \qquad\qquad \mbox{} \times
\left[
                    \Ho{l-{\nu'}}{R_{jj'}}  \Ho{{\nu'}-l''}{R_{j'j''}} 
             \e^{\i  {\nu'} \widetilde{\Delta}\alpha_{j',\sigma} } 
\right. \nonumber\\
 && \left.\left.\qquad\qquad\qquad
 \mbox{}+    \Ho{l+{\nu'}}{R_{jj'}}  \Ho{-{\nu'}-l''}{R_{j'j''}} 
             \e^{-\i  {\nu'} \widetilde{\Delta}\alpha_{j',\sigma} } \right ]
\right \} \; .
    \label{sofar}
\ee
Note that both integrals on the right hand side exist separately. The
one with the Watson ``$\sin$''-denominator is finite, 
because the zeros of the $\sin(\nu'\pi)$ function in the denominator 
are avoided by the
$+\i\epsilon$ prescription and because the rapid convergence
of the ratio $\Jb{\nu'}{a_{j'}}/\Ho{\nu'}{a_{j'}}$ counterbalances
the diverging $R_{jj'}$ and $R_{j'j''}$-dependent Hankel functions, as long
as the disks do not touch. This is basically the same argument by which
one can show the existence of the sum on the left hand side. However,
we do not
have to prove this separately, because we already 
know from App.\ref{app:suppl} that ${\bf A}$ is trace-class.
The existence of the principal value integral follows from the symmetric
nature of the path and of the integrand (see below for more details).

It will be shown that  the  term with the Watson ``$\sin$''-denominator,
$-1/[2\i\sin({\nu'} \pi)]= \sum_{n=0}^\infty \e^{\i (2n+1){\nu'} \pi}$, 
will lead in the semiclassical reduction to 
paths with left handed and right
handed  creeping sections around
the middle disk $j'$ [where the index $n$ counts further complete 
turns around this disk]. On the other hand the term without this denominator 
will give  either a semiclassical path specularly reflected from the 
disk $j'$ (to the left or right) 
or a ghost path passing undisturbed through disk $j'$.

\subsection{The integration paths\label{app:convol_paths}}
Thus the third step is to close the path of the ``$\sin$''-dependent integral
in the upper complex ${\nu'}$-plane. 
\begin{figure}[htb]
 \centerline {\epsfig{file=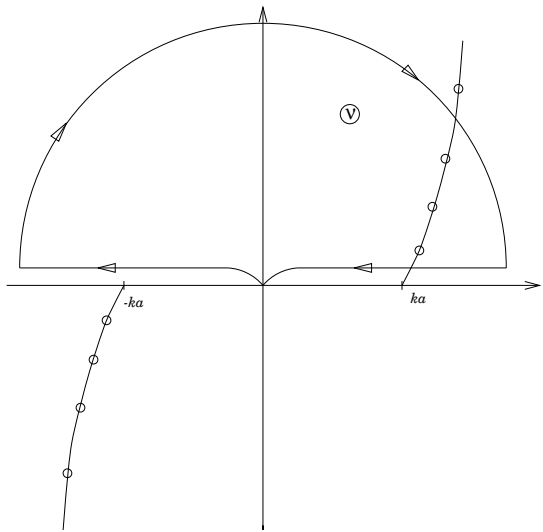,height=7cm,angle=-90}}
\caption[fig:upper_contour]
{\small The path for the ``sin''-dependent integral.
The lines denoting the zeros of $\Ho{\nu}{a}$ in
the upper
and of  $\Ht{\nu}{a}$ in the lower complex $\nu$-plane are shown as well.
\label{fig:upper_contour}}
\end{figure}

For given values of $\alpha_{jj'}$,
$\alpha_{j''j'}$, $l/k$ and $l''/k$, i.e., for a given geometry, this selects
which value of $\sigma$ has to be inserted into Eq.\equa{sofar}.
The reason is that
the closing of contour will be performed under
the condition that the corresponding semicircular integration arc 
vanishes, such that
the integral is solely given by  its residua which are here 
the zeros ${\nu'_\ell}$ ($\ell =1,2,3,\dots$) of the
Hankel function $\Ho{{\nu'}}{a_{j'}}$ in the upper complex ${\nu'}$-plane.
At ``optical boundaries'' this clear separation is not 
possible~\cite{Nussenzveig,penumbra}. 
This is the realm of ``penumbra'' scattering. 
In order not to be plagued by these difficulties, we exclude geometries
which allow for grazing classical paths from the further consideration.

In the  Airy {\em approximation} to leading order,
the zeros of these Hankel functions are given by Eq.\equa{nu_k}, 
modulo $ {\cal O}\left ([k a_{j'}]^{-1/3}\right)$ corrections.
A necessary  condition for the vanishing of the semicircular arc,
which, in turn, determines the choice among the three values for $\sigma$, 
is that the
total angle $\beta$ of the 
integrand's ``creeping
exponential'' $\exp\{\i {\nu'} \beta({\nu'})\}$ 
(including the terms resulting from
the Hankel functions) must be positive [and large enough to exclude
the penumbra region in the ``optical shadow'' and ``optically lit'' 
region] for ${\nu'}$ given by Eq.\equa{nu_k},
i.e.,
${\nu'} \approx k a_{j'}$.  
A violation of this condition would correspond semiclassically 
to a negative creeping path
which has to be excluded for physics reasons: during the creeping
the modulus of the wave has to decrease and not to increase~\cite{franz}, as
tangential rays are  continuously ejected, while the path creeps around a
convex bending. 
The positivity of the creeping exponential actually only  guarantees the
vanishing of the integrand on the arc to the left of the line of zeros
${\nu'_\ell}$ of the
Hankel function $\Ho{{\nu'}}{a_{j'}}$ and to the right of the line of zeros
${\nu'_\ell}^{(2)}$ of the Hankel function
$\Ht{{\nu'}}{a_{j'}}$ in the upper complex ${\nu'}$-plane. 
The vanishing of the remainder of the arc is a consequence
of the strongly decreasing $\Jb{{\nu'}}{a_{j'}}/\Ho{{\nu'}}{a_{j'}}$ term which
dominates the behavior of the integrand to the right of the ${\nu'_\ell}$'s 
and to the left of the ${\nu'_\ell}^{(2)}$'s. Whereas the  
 ${\nu'_\ell}^{(2)}$ line does not cause any problems, the ${\nu'_\ell}$ 
line is
potentially dangerous as the Hankel function in the denominator is vanishing.
The remedy is to put the path right in between 
two adjacent zeros~\cite{franz}.

\begin{figure}[htb]
 \centerline {\epsfig{file=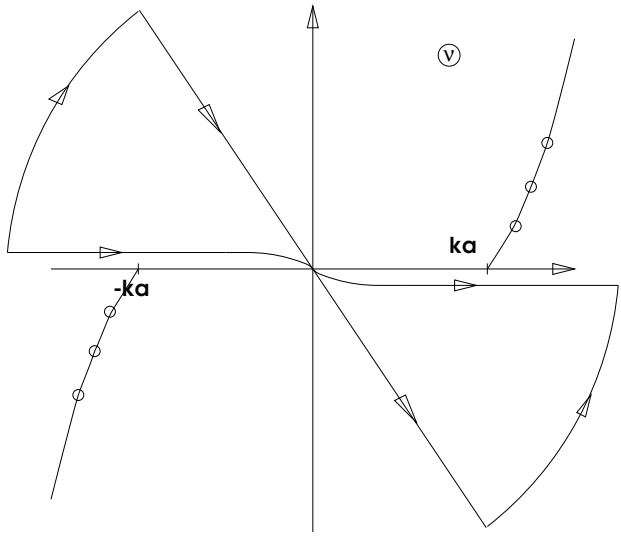,height=7cm,angle=-90}}
\caption[fig:geometric_path]{\small 
The original and the deformed contour of the ``sin''-independent integral 
for the case $\sigma'=0$. The lines of zeros are as 
in Fig.\ref{fig:upper_contour}.
\label{fig:geometric_path}}
\end{figure}

\begin{figure}[htb]
 \centerline {\epsfig{file=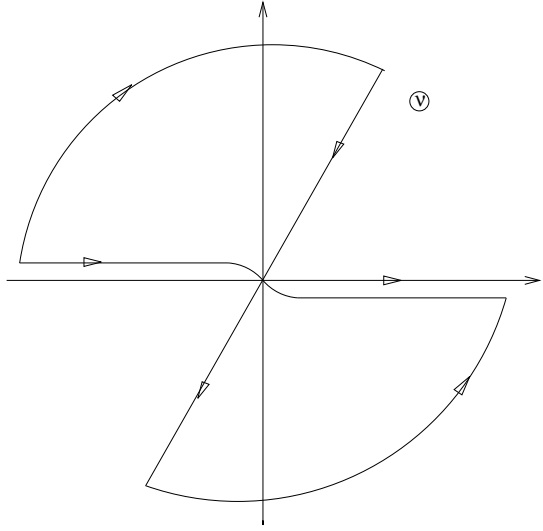,height=7cm,angle=-90}}
\caption[fig:ghost_path]{\small 
The contour of the ``sin''-independent integral 
in the case $\sigma'=1$ corresponding to a {\em ghost}. Note that 
the lines of zeros from Fig.\ref{fig:upper_contour} are absent.
\label{fig:ghost_path}}
\end{figure}

As already mentioned, the ``$\sin$''-independent integral is symmetric
in path and integrand. Because of this the path can be symmetrically deformed
as follows (the preserved symmetry takes care of 
the original principal value description):
It is replaced by an arc  from $-\infty+\i\epsilon$ to $+\i\infty(1+\i\delta)$,
a straight line from $+\i\infty(1+\i\delta)$ to $-\i\infty(1+\i\delta)$ and
finally a symmetric arc (to the first one) from $-\i\infty(1+\i\delta)$
to $+\infty-\i\epsilon$, where in the case $\sigma'=0$, the parameter 
$\delta$ is chosen positive and small enough such 
that $|{\rm Re}\, {\nu'}| \ll
ka_{j'}$. [This allows later to use the
Debye approximation of the Hankel functions,
$\Ho{{\nu'}}{a_{j'}}$ and $\Ht{{\nu'}}{a_{j'}}$.]\ 
See Fig.\ref{fig:geometric_path}. The deformation of the path
is justified as the sum of the new path and the (negatively traversed) original
one do not encircle any singularities of the integrand. Since 
the integrand is
symmetric under the exchange ${\nu'} \to -{\nu'}$, the integrals over the two
symmetric arcs completely cancel, such that only the straight
line segment from $+\i\infty(1+\i\delta)$ to $-\i\infty(1+\i\delta)$  
gives a contribution. This expression is finite since
it is symmetric under ${\nu'}\to -{\nu'}$ and since  
the integrand vanishes
rapidly
for $|\nu|\to \infty$, as long as the slope of the straight line section
is negative. In the case $\sigma'=1$ the parameter $\delta$ has to be chosen
negative since the integrand only vanishes rapidly for a straight line section
with positive slope (see Fig.\ref{fig:ghost_path}). 
The reason for this difference is the presence of
the
ratio $\Ht{{\nu'}}{a_{j'}}/\Ho{{\nu'}}{a_{j'}}$ in the first case which
is replaced by $(\Ht{{\nu'}}{a_{j'}}/\Ho{{\nu'}}{a_{j'}})
-2\Jb{{\nu'}}{a_{j'}}/\Ho{{\nu'}}{a_{j'}}= -1$ in the case $\sigma'=1$.
(See also below the discussion of the pertinent Fresnel integrals
in the semiclassical saddle-point approximation). 

As mentioned,
the actual result 
depends on the concrete geometry and on the impact parameters
$l/k$ and $l''/k$, i.e., on the value of 
$\Delta\alpha_{j'}=\alpha_{j''j'}-\alpha_{jj'}$, on the value of $\sigma$ and
the
angles $\phi_\pm(l,{\nu'}) \equiv \arccos(({\nu'}\pm l)/k R_{jj''})$ and 
 $\phi_\pm''(l'',{\nu'}) \equiv \arccos(({\nu'}\pm l'')/k R_{j'j''})$  
resulting from the asymptotic Debye approximation of the Hankel
$\Ho{l\pm{\nu'}}{kR_{jj'}}$ and $\Ho{\pm{\nu'}-l''}{k R_{j'j''}}$, 
respectively.
Since $\sigma$ can take three values 
there exist three mutually exclusive alternatives:\\ 
The first one corresponds to $\sigma=0$ and $0 < 2\pi \mp \Delta\alpha_{j'} - 
\phi_\pm(l,ka_{j'}) - \phi_\pm''(l'',ka_{j'}) 
\leq 2\pi$ 
(this geometry allows only a classical 
path from
disk $j$ [under the impact parameter $l/k$] 
to disk $j''$ [under the impact parameter $l''/k$] 
which is specularly reflected to the {\em right} 
at disk $j'$, see Fig.\ref{fig:conv1}):
\be
\sum_{l'}\,
{\bf A}^{j j'}_{l l'}{\bf A}^{j' j''}_{l' l''} \!&=&\!
W^{j j''}_{l l''} \left \{
  \sum_{n=0}^{\infty} 
\oint_{\partial C_+}\!\! {\rm d} {\nu'}
           \frac{\Jb{{\nu'_\ell}}{a_{j'}   } }
                    {
                     \Ho{{\nu}}{a_{j'}   }}  
 \left[
\Ho{l-{\nu'}}{R_{jj'}}  \Ho{{\nu'}-l''}{R_{j'j''}} 
      \e^{\i  {\nu'} (\pi+\Delta\alpha_{j'} +2n\pi) }  \right. 
\right. \nonumber \\
 &&\left.\left.\qquad\qquad\mbox{}+
  \Ho{l+{\nu'}}{R_{jj'}}  \Ho{-{\nu'}-l''}{R_{j'j''}} 
             \e^{\i  {\nu'} (\pi-\Delta\alpha_{j'} +2n\pi) }
\right ]
\right.
    \nonumber \\
 && \left. \quad
\mbox{} + \frac{1}{4}  
\int_{\i\infty(1        +\i \delta)}^{-\i\infty(1+\i \delta)}\!\!\! 
   {\rm d} {\nu'}\,
 \frac{\Ht{{\nu'}}{a_{j'}   } }
                    {\Ho{{\nu'}}{a_{j'}   } } \left[
  \Ho{l-{\nu'}}{R_{jj'}}  \Ho{{\nu'}-l''}{R_{j'j''}} 
             \e^{\i  {\nu'} (-\pi+ \Delta\alpha_{j'}) }  
  \right.\right. \nonumber \\
 &&\left.\left.\qquad\qquad\mbox{}+
  \Ho{l+{\nu'}}{R_{jj'}}  \Ho{-{\nu'}-l''}{R_{j'j''}} 
             \e^{\i  {\nu'} (-\pi-\Delta\alpha_{j'}) } \right] \right \}
 \label{alt-1}
\ee
The second case is $\sigma=2$ and
 $0< \Delta\alpha_{j'} -\phi_{-}(l,ka_{j'}) 
-\phi_{-}''(l'',ka_{j'})\leq 2\pi$ and
 $0<4\pi-\Delta\alpha_{j'} 
- \phi_{+}(l,ka_{j'}) -\phi_{+}''(l'',ka_{j'})\leq 2\pi$
(this geometry allows only a classical 
path from
disk $j$  
to disk $j''$  
which is specularly reflected to the {\em left} 
at disk $j'$, see Fig.\ref{fig:conv2}):
\be
 \sum_{l'}\,
 {\bf A}^{j j'}_{l l'}{\bf A}^{j' j''}_{l' l''} \!&=&\!
 W^{j j''}_{l l''} \left \{ \sum_{n=0}^{\infty}
  \oint_{\partial C_+}\!\! {\rm d} {\nu'}
           \frac{\Jb{{\nu'}}{a_{j'}   } }
                    {\Ho{{\nu'}}{a_{j'}   } }  
 \left[
\Ho{l-{\nu'}}{R_{jj'}}  \Ho{{\nu'}-l''}{R_{j'j''}} 
      \e^{\i  {\nu'} (\Delta\alpha_{j'} +2n\pi-\pi) }  \right. 
\right. \nonumber \\
 &&\left.\left.\qquad\qquad\mbox{}+
  \Ho{l+{\nu'}}{R_{jj'}}  \Ho{-{\nu'}-l''}{R_{j'j''}} 
             \e^{\i  {\nu'} (3\pi-\Delta\alpha_{j'} +2n\pi) }
\right ]
\right.
    \nonumber \\
 && \left. \quad
\mbox{} + \frac{1}{4}  
\int_{+\i\infty(1+\i \delta)}^{-\i\infty(1+\i \delta)}\!\!\! {\rm d} {\nu'}\,
 \frac{\Ht{{\nu'}}{a_{j'}   } }
                    {\Ho{{\nu'}}{a_{j'}   } } \left[
  \Ho{l-{\nu'}}{R_{jj'}}  \Ho{{\nu'}-l''}{R_{j'j''}} 
             \e^{\i  {\nu'} (-3\pi+ \Delta\alpha_{j'}) }  
 \right.\right. \nonumber \\
 &&\left.\left.\qquad\qquad \mbox{}+
  \Ho{l+{\nu'}}{R_{jj'}}  \Ho{-{\nu'}-l''}{R_{j'j''}} 
             \e^{\i  {\nu'} (\pi-\Delta\alpha_{j'}) } \right] \right \}
 \; .
 \label{alt-2}
\ee
The third alternative is  $\sigma=\sigma'=1$ and
$0< \Delta\alpha_{j'} -\phi_{-}(l,ka_{j'}) 
-\phi_{-}''(l'',ka_{j'})\leq 2\pi$ and
 $0<2\pi-\Delta\alpha_{j'} 
- \phi_{+}(l,ka_{j'}) -\phi_{+}''(l'',ka_{j'})\leq 2\pi$
(this geometry allows only a ``classical'' 
path from
disk $j$ [under the impact parameter $l/k$] 
to disk $j''$ [under the impact parameter $l''/k$] 
which goes directly {\em through} disk $j'$, see Fig.\ref{fig:conv3}):
\be
 \sum_{l'}\,
{\bf A}^{j j'}_{l l'}{\bf A}^{j' j''}_{l' l''} \!&=&\!
W^{j j''}_{l l''} \left \{\sum_{n=0}^{\infty}
  \oint_{\partial C_+}\!\! {\rm d} {\nu'}
           \frac{\Jb{{\nu'}}{a_{j'}   } }
                    {\Ho{{\nu'}}{a_{j'}   } }  
 \left[
\Ho{l-{\nu'}}{R_{jj'}}  \Ho{{\nu'}-l''}{R_{j'j''}} 
   \e^{\i  {\nu'} ( \Delta\alpha_{j'} +2n\pi-\pi ) } \right. 
\right. \nonumber \\
 &&\qquad\qquad \mbox{} +\left. \left.
  \Ho{l+{\nu'}}{R_{jj'}}  \Ho{-{\nu'}-l''}{R_{j'j''}} 
             \e^{+\i  {\nu'} ( \pi-\Delta\alpha_{j'}+2n\pi ) }  
\right ]
\right.
    \nonumber \\
 &&\left . \qquad \mbox{} - 
\frac{1}{4}  
 \int_{\i\infty(1-\i|\delta|)}^{-\i\infty(1-\i|\delta|)}\!\!\! 
 {\rm d} {\nu'}\, 
      \left\{     \Ho{l-{\nu'}}{R_{jj'}}  \Ho{{\nu'}-l''}{R_{j'j''}} 
             \e^{\i  {\nu'} (-\pi+\Delta\alpha_{j'}) } 
   \right. \right. \nonumber \\
 &&\left.\left. \qquad\qquad\mbox{}   
   +\Ho{l+{\nu'}}{R_{jj'}}  \Ho{-{\nu'}-l''}{R_{j'j''}} 
             \e^{\i  {\nu'} (\pi-\Delta\alpha_{j'}) }\right ]\right\}
  \; .
 \label{alt-3}
\ee
Note that Eq.\equa{alt-3} can also 
be derived from the  ``$+\Delta\alpha_{j'}$ part''
of \equa{alt-1} plus the ``$-\Delta\alpha_{j'}$ part'' of \equa{alt-2}, 
by a rearrangement of the corresponding creeping
and geometrical terms, i.e., by the addition 
of an extra term of smaller creeping length than
the smallest one before and the subtraction of 
the very same piece from the geometrical
terms.

The contour
integrals  of these three alternatives 
are evaluated at the zeros ${\nu'_\ell}$ of the Hankel functions
$\Ho{{\nu'}}{a_{j'}}$, such that
\be
  \oint_{\partial C_+}\!\! {\rm d} {\nu'}\,
           \frac{\Jb{{\nu'}}{a_{j'}   } }
                    {\Ho{{\nu'}}{a_{j'}   } }   Y_{\nu} \e^{\i (2n+1)\nu' \pi} 
  = 2\pi \i \sum_{\ell=1}^{\infty}\,
     \frac{\Jb{{\nu'_\ell}}{a_{j'}   } }
                    {\frac{\partial}{\partial \nu'}
  \Ho{\nu'}{a_{j'}}|_{\nu'=\nu'_\ell }} Y_{\nu_\ell} \e^{\i(2n+1)\nu'_\ell \pi}
 \; . 
 \label{residuum-start}
\ee
Up to this point all expressions are still exact.
The steps introduced so far  just served the 
purpose of generating the three distinct ``classically'' allowed angular
domains and of transforming the original expression \equa{convolution} into
a form ready for the semiclassical approximation. This will be taken next
under the condition ${\rm Re}\, k a_{j'}\gg 1$.
Note that this inequality automatically induces 
${\rm Re}\, k R_{jj'}\gg 1$ and 
${\rm Re}\, k R_{j'j''}\gg 1$.

The contour integral (which, in fact,
is now a sum over the residua) and the straight line integral are now treated
semiclassically in different
ways. 
\subsection{Semiclassical approximation of the straight-line 
integrals\label{app:convol_straightline}}
 
The straight line integrals will be evaluated in the saddle-point
approximation at a saddle $\nu'_s$ where the path crosses the real axis.
For evaluating the saddle-point integral, 
the Debye approximation \equa{H1-Debye}  
will be inserted for the given Hankel functions.
For the first and second alternative, an internal consistency
check on the validity, is the condition $|\nu'_s/k|< a_{j'}$
which in physics terms means that the impact parameter at disk $j'$ has
to be smaller than the disk radius $a_j'$. For the third alternative the
weaker conditions  $| (\nu'_s-l)/k|< R_{jj'}$  and 
$| (\nu'_s-l'')/k|< R_{j'j''}$ are sufficient: The difference in the
impact parameters at successive disks should be smaller than the distance
between the disks. Its validity is guaranteed by the triangular
condition. 

The saddle-point integral is evaluated by expanding the
exponents of the Debye approximates to second order and a successive
integration. 
The reflection angle is determined by the saddle-point condition itself,
the geometrical length of the path can be read off from the total exponent
at zeroth saddle-point order, i.e. from 
the sum of square root terms  of the Debye exponents divided by $k$.
Under the Gauss' integration the second-order fluctuations about
the saddle determine 
the stability factor $1/\sqrt{R_{\rm eff}}$ 
and, together with the already present phases, the overall phase.

The straight line integral of the first two alternatives corresponds 
then to the standard geometrical
path from disk $j$ [under impact parameter $l/k$] to
disk $j''$ [under impact parameter $l''/k$] where there is  
a specular reflection 
from the boundary of  the disk $a_{j'}$ either to the right for the
first alternative (Fig.\ref{fig:conv1}) or to
the left for the second alternative (Fig.\ref{fig:conv2}).
The slope of the  path of this straight line integral, 
which asymptotically is $i(1+\i\delta)$,  has to join smoothly
the slope of the saddle-point path. This condition
determines the sign of the slope. 
The saddle-point integral, which is
of the Fresnel-type, results when the pertinent exponents of
the Debye-approximated Hankel functions 
are expanded to second order around the saddle point
$\nu_s$:  
\be
 \int_{-\e^{-\i\pi/4}\infty}^{\e^{-\i\pi/4}\infty}{\rm d} \delta \nu'
\,\e^{ -\i\half(\delta \nu')^2\left 
(\frac{2}{\sqrt{(ka)^2-{\nu'}_s^2}}
 -\frac{1}{\sqrt{(kR_{jj'})^2-(l-{\nu'}_s)^2}}       
 -\frac{1}{\sqrt{(kR_{j'j''})^2-({\nu'}_s-l'')^2}} \right )} \; .
\ee      
Here $\delta\nu'=\nu-\nu_s$ is the integration variable.
By the substitution $\delta \nu' = \e^{-\i\pi/4}x$
this  integral becomes a standard Gauss' integral
\be
 \e^{-\i\pi/4} \int_{-\infty}^{+\infty}{\rm d} x\,\e^{-x^2 b/2}
 =  \e^{-\i\pi/4}\sqrt{2\pi/b} \label{alt1-gauss}
\ee
with $b={2}\{(ka)^2-{\nu'}_s^2\}^{-1/2}
 -\{(kR_{jj'})^2-(l-{\nu'}_s)^2\}^{-1/2}       
 -\{(kR_{j'j''})^2-({\nu'}_s-l'')^2\}^{-1/2}$ positive as 
$a_{j'} < R_{jj'}-a_{j},R_{j'j''}-a_{j''}$.
The right hand side of Eq.\equa{alt1-gauss} together with the 
prefactors and phases of the Debye-approximated Hankel functions determine
the remaining terms (see below).

Perturbative higher-order $\hbar$-corrections (see 
Refs.\cite{alonso,gasp_hbar,vattay_hbar}) 
result here from higher-order  
terms in 
the Debye approximation
through expansion terms proportional to  $(1/k r)^n$ = $(\hbar/pr)^n$ 
(with $r$ = $a_{j'}$, $R_{jj'}$ or $R_{j'j''}$)
{\em and} from the 
integration of
polynomial second- and higher-order $(\hbar\nu'/p r)^{2n}$ terms under 
the Gauss-type saddle-point integral. The polynomials are generated
by a consistent expansion  of all prefactors and exponential terms  
of the Debye series up to a given order. 
The Debye series reads
\be
   H^{(1/2)}_\nu (z) &\sim& \sqrt{\frac{ 2 }{ \pi z}}
\frac{1}{\left(1-\frac{\nu^2}{z^2}\right)^\fourth}\,
       \e^{\pm \i \sqrt{z^2-\nu^2} \mp \i \nu \arccos(\frac{\nu}{z}) 
\mp \i\frac{\pi}{4}} \nonumber \\
 &&\times
  \left \{ 1 \mp i\left( \frac{1}{ 8z}
                          \frac{1}{\sqrt{1-\frac{\nu^2}{z^2}} }
   +\frac{5}{24z^3} \frac{\nu^2}{\sqrt{1-\frac{\nu^2}{z^2}}^3 }\right )
 \right. 
       \nno \\
  && \mbox{} \quad 
-\left( \frac{9}{128z^2}\frac{1}{1-\frac{\nu^2}{z^2}} 
     +\frac{231}{576z^4} \frac{\nu^2}{(1-\frac{\nu^2}{z^2})^2}
     +\frac{1155}{3456z^6}\frac{\nu^4}{(1-\frac{\nu^2}{z^2})^3 }\right )
 \nonumber \\ &&
\left.\quad \mbox{}+ 
          \order{ \frac{1}{z^{3}}, \frac{\nu^2}{z^5},\cdots} 
    \right \} 
     \label{w:debye}
\ee
where the upper signs apply for the Hankel function of first kind and the
lower ones for the Hankel function of second kind.
Note that this expansion is of asymptotic nature and therefore
induces the asymptotic nature of the $\hbar$-expansion itself. 
Here
we will limit our discussion just to the leading term, such that no
$\hbar$-corrections arise.
 
\subsection{Semiclassical approximation of the residua 
sum\label{app:convol_residua}}
In the contour integral (or residua sum) the Debye approximation is not
justified for the ratio 
\beq
 \frac{\Jb{\nu'_\ell}{a_{j'}}}{\frac{\partial}{\partial \nu'}
 \Ho{\nu'}{a_{j'}}|_{\nu'=\nu'_\ell}}
 \label{residua-ratio}
\eeq
since ${\rm Re}\, \nu'_\ell \simeq ka_{j'}$. It is still valid, however, 
for 
the $R_{jj'}$ and $R_{j'j''}$-dependent 
Hankel functions, since $R_{jj'}> a_{j}+a_{j'}$ and 
$R_{j'j''}>a_{j'}+a_{j''}$. 
Instead, the $a_{j'}$-dependent cylindrical functions are evaluated under the 
Airy approximation.

The latter step is justified as we evaluate the ratio \equa{residua-ratio}
at the zeros $\nu_\ell$ of the Hankel
function $\Ho{\nu}{a}$. In the Sommerfeld representation the contour
of a Hankel function $H^{(1)}_\nu (z)$ has normally two saddles~\cite{franz}.
For
${\rm Re}\, \nu \ll z$ or ${\rm Re}\,\nu \gg z$ 
one of these two saddles dominates 
such that the Hankel function can be approximated by a single
exponential times polynomial corrections (the Debye approximation). In
such a case the Hankel function can obviously not vanish. In order
for it to vanish
the contributions of the two saddles have to be of
the same magnitude. In other words, we have to be in a region of
competing saddles, where the standard saddle-point approximation
(which has been used for the purely geometrical calculation) 
is replaced by
the Airy approximation. This is the case when ${\rm Re}\, \nu \approx
{\rm Re}\, z$. There we have, to leading order 
($\nu=\nu'$ and $a=a_{j'}$), 
\be 
 \Ho{\nu}{a} &\sim&
 \frac{2}{\pi} \e^{-\i\frac{\pi}{3}} \left( \frac{6}{ka} \right)^\third
 A(q^{(1)}) + \order{ (ka)^{-1}}, \label{w:hoairy}\\ \Ht{\nu}{a} &
 \sim& \frac{2}{\pi} \e^{+\i\frac{\pi}{3}} \left( \frac{6}{ka}
 \right)^\third A(q^{(2)}) + \order{(ka)^{-1}}. 
\ee 
Here
\be q^{(1)}
 \equiv \e^{-\i\frac{\pi}{3}} \left( \frac{6}{ka} \right)^\third (\nu
 -ka) + \order{ (ka)^{-1}},\\ q^{(2)} \equiv \e^{+\i\frac{\pi}{3}} \left(
 \frac{6}{ka} \right)^\third (\nu -ka) + \order{ (ka)^{-1} } 
\ee
are
the 
zeros of the Airy integral~\cite{franz} 
\be
A(q)&\equiv& \int_0^\infty {\rm d}\tau\, \cos(q\tau-\tau^3)\nonumber \\
    &=& 3^{-1/3} \pi {\rm Ai}(-3^{-1/3} q) \; ,\label{Airy-integral}
\ee
with ${\rm Ai}(z)$ being the standard Airy function~\cite{Abramowitz};
approximately,  
$q_\ell \approx 6^{1/3} [3\pi(\ell -1/4)]^{2/3}/2$.
Thus Eq.\equa{nu_k} arises.
Note that this is the first term
in an asymptotic expansion where the corrections are of relative
order $\order{ (ka)^{-\twothird}} \sim \order{ \hbar^\twothird}$ as
$k=p/\hbar$. The first correction to the Airy approximation is
therefore more important than the first correction term to the Debye
approximation as the latter  term scales as 
$\order{ \hbar}$.  
Up to order $\order{ (ka)^{-{{5 \over 3}}} }$, the zeros
$\nu_\ell$ read as \cite{FranzGalle} 
\be \nu_\ell\ \sim\ ka&+&
 \e^{\i\frac{\pi}{3}} \left( \frac{ka}{6} \right )^\third q_\ell \ -\
\e^{-\i\frac{\pi}{3}} \left(\frac{6}{ka} \right )^{\third}
\frac{q_\ell^2}{180} \ -\ \frac{1}{70 ka} \left ( 1
-\frac{q_\ell^3}{30} \right ) \nno\\ &+& \e^{\i\frac{\pi}{3} }\left (
\frac{6}{ka} \right )^{{{5 \over 3}}} \frac{1}{3150} \left ( \frac{29
q_\ell}{6^2} - \frac{281 q_\ell^4}{180\cdot 6^3} \right ) \ +\ \cdots.
\label{w:nuzero} 
\ee 
The $\nu$-derivative of the Hankel
function $\Ho{\nu}{a}$ at $\nu=\nu_\ell$ has the form
\cite{FranzGalle} 
\be \frac{\partial}{\partial \nu}
 \Ho{\nu}{a}|_{\nu=\nu_\ell} &\sim& \frac{2 \e^{-\i\frac{2\pi}{3}}}{\pi}
 \left(\frac{6}{ka}\right)^\twothird \!A'(q_\ell) \left \{ 1 -
 \frac{\e^{\i\frac{\pi}{3}}}{5} \left (\frac{6}{ka} \right)^\twothird
 \frac{q_\ell}{6} - \e^{-\i\frac{\pi}{3}} \left ( \frac{6}{ka} \right
 )^{{{4 \over 3}}} \!\frac{37}{630} \left (\frac{q_\ell}{6} \right )^2
 \right. \nno \\ &&\qquad\quad \mbox{} \left.  - \left(\frac{6}{ka}
 \right )^2 \left( \frac{37}{36\cdot 5\cdot 630} - \frac{563}{5\cdot
 630\cdot 9} \left(\frac{q_\ell}{6} \right )^3 \right) + \cdots \right
\} 
\ee 
where $A'(q_\ell)$
is the derivative of the Airy integral $A(q)$ at the position 
$q_\ell$~\cite{franz}.
The Airy approximation to the Bessel function $J_{\nu_\ell}(ka)=\half
\Ht{\nu_\ell}{a}$ reads as 
\be J_{\nu_\ell}(ka) &\sim& \frac{
\e^{\i\frac{\pi}{3}}}{\pi} \left(\frac{6}{ka}\right)^\third
A(q^{(2)}_\ell) \left\{1-\e^{\i\frac{\pi}{3}}
\left(\frac{6}{ka}\right)^\twothird \left(\frac{q_\ell}{45}\right ) -
\e^{-\i\frac{\pi}{3}}\,\frac{29}{14}\left(\frac{6}{ka}\right)^{{{4 \over
3}}} \left(\frac{q_\ell}{45}\right )^2 \right. \nno \\ &&
\qquad\qquad\qquad\qquad\qquad \left.\mbox{} - \left (\frac{6}{ka}
\right)^2 \left(\frac{1}{45^2\cdot 7} -\frac{31}{6}
\left(\frac{q_\ell}{45}\right )^3 \right )+\cdots \right \}\; .
\ee
Applying the Wronsky-relation $A(z)A'(z \e^{\pm 2\pi \i/3}) -A'(z)A( z
\e^{\pm 2\pi \i/3})=-\frac{\pi}{6} \e^{\mp \pi \i/6}$ one gets for
$z=q_\ell^{(1)}$ (with $A(q_\ell^{(1)})=0$) 
\be
 A(q_\ell^{(2)})=\frac{\pi}{6}
 \frac{\e^{-\i\frac{\pi}{6}}}{A'(q_\ell^{(1)} ) }.  
\ee
Thus, under the Airy approximation, each of the residua in
Eq.\equa{residuum-start} becomes
\be
 2\pi \i\frac{\Jb{\nu'_\ell}{a_{j'}}}{\frac{\partial}{\partial \nu'}
 \Ho{\nu'}{a_{j'}}|_{\nu'=\nu'_\ell}}\, Y_{\nu'_\ell} 
 \e^{\i (2n+1)\nu'_\ell \pi} 
  = -\e^{-\i\pi/6} C_\ell\pi^{-1/2} (k a_{j'})^{1/3}\, \frac{\i \pi }{2} 
  Y_{\nu'_\ell} \e^{\i (2n+1)\nu'_\ell \pi} 
 \label{residuum-end}
\ee
with
the coefficient $C_\ell=C_\ell(ka)$
\be
C_\ell(ka)& =& 
2^{-\third} 3^{-{{4\over 3}}} \frac{\pi^\threehalf}{A'(q_\ell)^2}
 \left\{ 1 + \frac{\e^{\i \frac{\pi}{3}}}{18} \left(\frac{q_\ell}{5}\right )
         \left (\frac{6}{ka} \right)^\twothird 
 +\frac{1}{12\cdot 14}  \left(\frac{q_\ell}{5}\right )^2 
               \left (\frac{6}{ka} \right)^{{{4 \over 3}}} \right.\nno \\
 &&\qquad\qquad\qquad\qquad \mbox{} +
\left.\frac{1}{ (ka)^2 } \left ( \frac{29}{9\cdot 25\cdot 14}
            - \frac{281}{6\cdot 81\cdot 14} \left(\frac{q_\ell}{5}\right )^3
  \right ) 
   +\cdots \right \} .
  \label{C_ell} 
\ee
The values of the first zeros $q_\ell$ and the
corresponding coefficients $C_\ell$, truncated at order 
${\cal O}(\{ka\}^{-2/3})$ = ${\cal O}(\hbar^{2/3})$, 
can be found in Ref.\cite{franz}\
and are listed in Table~\ref{tab:franz}.

{\small
\begin{table}[htb]
\centering
\caption[Tab-Franz]{{\small  
The first zeros $q_\ell$ of the Airy Integral $A(q)$ and the
corresponding coefficients $C_\ell$ of the creeping wave under Dirichlet
boundary conditions in the leading Airy approximation.
\label{tab:franz} }}

\vspace{2mm}
\small
\begin{tabular}{ccc}
\hline
$\ell$ & $q_\ell$ & $C_\ell$\\ \hline
 1  & 3.372134 & 0.91072 \\
 2  & 5.895843 & 0.69427 \\
 3  & 7.962025 & 0.59820 \\
 4  & 9.788127 & 0.53974 \\ \hline
\end{tabular}
\end{table}
}
  
We will limit our discussion to the Airy expansion of this leading order, i.e.,
$\nu_\ell'$ as in Eq.\equa{nu_k} and $C_\ell$ as given by the first 
term of Eq.\equa{C_ell}, since all
the higher terms vanish at least as fast as $\hbar^{1/3}$ 
and $\hbar^{2/3}$, respectively, in
the limit $\hbar\to 0$.

Finally, the Debye approximation \equa{H1-Debye} is inserted in 
$Y_{\nu'_\ell}$ for
the $R_{jj'}$ and $R_{j'j''}$ dependent Hankel functions.
The two square root terms in the exponential of the
Debye approximate, e.g.\, $\sqrt{(kR_{jj'})^2 -(l-\nu'_\ell)^2}$, etc., 
under the approximation $\nu'_\ell \simeq ka_{j'}$, give 
the length of the two straight sections of the path times $k$. 
All exponential terms
proportional to $\nu'_\ell$, e.g., $\nu'_\ell\arccos(\cdots)$, 
$n\pi\nu'_\ell$, correspond to 
the creeping sections (of mode number $\ell$) of the path. 
The latter include, of course, the
creeping tunneling suppression factor linked to the imaginary part of
the $\nu_\ell$.
The product of the two Debye prefactors  is just 
the stability of the path times $-\i 2/\pi$. The latter factor cancels the
exposed factor in \equa{residuum-end}.

In summary, the residua of the contour integrals in the Airy approximation 
correspond to those paths from disk $j$ 
[under impact parameter $l/k$] to disk $j''$ [under impact parameter $l''/k$] 
that have straight sections {\em and} circular creeping sections 
of mode number 
$\ell$ which join tangentially 
at the surface of disk $j'$.
For the first term 
of $Y_{\nu_\ell}$, the creeping is 
in the
left hand sense and for the second term in the right hand sense around
disk $j$. The
sum over $n$ counts   $n$ further  {\em complete} 
creeping turns around this disk. 
Note that the smallest creeping angle is less than $2\pi$,
but larger than zero (see Figs.\ref{fig:conv1}--\ref{fig:conv3}). 

\subsection{Resulting Convolutions\label{app:convol_results}}
The first alternative (Fig.\ref{fig:conv1}) reads now
\be
\lefteqn{\sum_{l'}\,
{\bf A}^{j j'}_{l l'}{\bf A}^{j' j''}_{l' l''} } \nonumber \\
&\sim& -W^{j j''}_{l l''} \sum_{\ell=1}^{\infty} \sum_{n=0}^{\infty}\,
 \e^{-\i\pi/6} \pi^{-1/2}C_\ell (ka_{j'})^{1/3} \nonumber \\
&&\quad\mbox{}\times
\left [
 \frac{ \exp{ \i  \sqrt{(kR_{jj'})^2\mbox{$-$}(l\mbox{$-$}{\nu'_\ell})^2}
  }}
  {[(kR_{jj'})^2\mbox{$-$}(l\mbox{$-$}{\nu'_\ell})^2]^{1/4}}\ 
  \frac{\exp{
   \i\sqrt{(kR_{j'j''})^2\mbox{$-$}({\nu'_\ell}\mbox{$-$}l'')^2}
  } }
    { [(kR_{j'j''})^2\mbox{$-$}({\nu'_\ell}\mbox{$-$}l'')^2]^{1/4} } 
  \right . \nonumber \\
 &&\qquad\quad \mbox{} \times
 \e^{ -\i(l-{\nu'_\ell})\arccos[(l-{\nu'_\ell})/kR_{jj'}]
 -\i({\nu'_\ell}-l'')\arccos[({\nu'_\ell}-l'')/kR_{j'j''}] 
 +\i{\nu'_\ell}(\Delta\alpha_{j'}+(2n+1)\pi) } \nonumber \\
&& \quad\quad \mbox{}+ 
 \frac{ 
 \exp{\i  \sqrt{(kR_{jj'})^2\mbox{$-$}(l\mbox{+}{\nu'_\ell})^2}} }
     {[(kR_{jj'})^2\mbox{$-$}(l\mbox{+}{\nu'_\ell})^2]^{1/4}}\ 
 \frac{\exp{\i\sqrt{(kR_{j'j''})^2\mbox{$-$}({\nu'_\ell}\mbox{+}l'')^2}
 }}
{ [(kR_{j'j''})^2\mbox{$-$}({\nu'_\ell}\mbox{+}l'')^2]^{1/4}} \nonumber \\
&& \left. \qquad\quad \mbox{}\times
 \e^{ -\i(l+{\nu'_\ell})\arccos[(l+{\nu'_\ell})/kR_{jj'}]
 +\i({\nu'_\ell}+l'')\arccos[(-{\nu'_\ell}-l'')/kR_{j'j''}]
 +\i{\nu'_\ell}(-\Delta\alpha_{j'}+(2n+1)\pi)\} } \right ]
    \nonumber \\
 && +W^{j j''}_{l l''} \sqrt{\frac{2}{\pi}} \e^{-\i\pi/4}
\,\e^{-\i l\arccos[(l- {\nu'_s})/kR_{jj'}]
             +il''\arccos[({\nu'_s}-l'')/kR_{j'j''}]} \nonumber \\
&& \quad \mbox{}\times 
  \exp\left\{\i\ \sqrt{(kR_{jj'})^2\mbox{$-$}(l\mbox{$-$}{\nu'_s})^2}
             +\i\sqrt{(kR_{jj'})^2\mbox{$-$}({\nu'_s}\mbox{$-$}l'')^2}
             -2\i\sqrt{(ka_{j'})^2\mbox{$-$}{\nu'_s}^2}\right\} \nonumber \\
 && \quad \mbox{}\times    [ (ka_{j'})^2\mbox{$-$}{\nu'_s}^2]^{1/4} \left\{ 
[2(kR_{jj'})^2\mbox{$-$}(l\mbox{$-$}{\nu'_s})^2]^{1/2} 
 [ (kR_{j'j''})^2\mbox{$-$}({\nu'_s}\mbox{$-$}l'')^2]^{1/2}
 \right .\nonumber \\
&& \left.\qquad \mbox{}
 - [(ka_{j'})^2\mbox{$-$}{\nu'_\ell}^2]^{1/2} \{ 
 [(kR_{jj'})^2\mbox{$-$}(l\mbox{$-$}{\nu'_s})^2]^{1/2}+
  [(kR_{j'j''})^2\mbox{$-$}({\nu'_s}\mbox{$-$}l'')^2]^{1/2} 
  \right\}^{-1/2}     
       \; . \nonumber \\ \label{res-alt-1}    
\ee   
Here ${\nu'_\ell}$ is given as in Eq.\equa{nu_k} and $C_\ell$
as in Eq.\equa{C_ell}.
The value of ${\nu'_s}$ follows
from the saddle-point condition
\beq
 \Delta\alpha_{j'}  +2\arccos[{\nu'_s}/ka_{j'}]-\arccos[({\nu'_s}-l)/kR_{jj'}]
      -\arccos[({\nu'_s}-l'')/kR_{j'j''}] = 0 \; .
  \label{cond-alt-1}
\eeq
which fixes the scattering 
angle $\theta_{j'}\equiv\arcsin[{\nu'_s}/ka_{j'}]$
as
\beq
 \theta_{j'}=\Delta \alpha_{j'}+\arcsin[({\nu'_s}-l)/kR_{jj'}]
                        +\arcsin[({\nu'_s}-l'')/kR_{j'j''}] \; .
\eeq
One might wonder why there do not appear  
two different geometrical segments corresponding
to the two terms of the straight line integral in Eq.\equa{alt-1}. The
answer is that the second term of this integral gives the same contribution
as the first one, since the values  of the  pertinent saddles 
just differ by a minus sign. [In fact, it is easy to show with the
help of the transformation laws \equa{hotrafo} and \equa{httrafo} 
that the second term of the straight line
integrals is  identical to the first one.] 
The effective radius $R_{\rm eff}'$ belonging to Eq.\equa{res-alt-1} results
from the prefactors of the Debye-approximated $R_{jj'}$- and 
$R_{j'j''}$-dependent 
Hankel functions, combined with the r.h.s.\ of \equa{alt1-gauss}, and 
reads
\beq
  R_{\rm eff}'=
  \frac{2d_{jj'}d_{j'j''}-\rho_{j'}(d_{jj'}+d_{j'j''})}{\rho_{j'}}
\eeq
with
\be
  d_{jj'}&\equiv&\sqrt{ R_{jj'}^2\mbox{$-$}(\{l\mbox{$-$}{\nu'_s}\}/k)^2} 
\\
  d_{j'j''} &\equiv&
   \sqrt{ R_{j'j''}^2\mbox{$-$}(\{{\nu'_s}\mbox{$-$}l''\}/k)^2}
     \\
  \rho_{j'}&\equiv& \sqrt{a_{j'}^2 - ({\nu'_s}/k)^2} \; .
\ee
This should be compared with effective radius generated by 
the standard evolution of the curvatures in
the corresponding classical problem (see Eqs.\equa{R-eff-creep} 
and \equa{kappa_i})
\beq   
  R_{\rm eff}= L_{0,1} \prod_{i=1}^{m}(1 +\kappa_i L_{i,i+1}) \; .
  \label{Reff-conv}
\eeq
Here
$L_{i,i+1}$ is the length of the leg between the $i^{\rm th}$ 
and the 
$(i+1)^{\rm th}$ reflection.
 The quantity  $\kappa_i$ is the curvature just
after the $i^{\rm th}$ reflection, i.e., 
\beq
 \kappa_i = \frac{1}{\kappa_{i-1}^{-1}+L_{i-1,i}} 
 + \frac{2}{r_i \cos\phi_i} \; ,
\eeq
where, in turn, $r_i$ and  $\phi_i$ are the local radius of curvature and the
deflection angle at the $i^{\rm th}$ reflection. [Note that 
$\kappa_0^{-1}=0$.]\ 
By identifying 
$L_{0,1}=d_{jj'}- \rho_{j'}$, $L_{1,2}=d_{j'j''}-\rho_{j'}$, $r_i=a_{j'}$ and
$\phi_i =\theta_{j'}$ (such that $\rho_{j'}=a_{j'}\cos\theta_{j'}$) 
one can easily show that $R_{\rm eff}'$ and $R_{\rm eff}$
give the same result.

The result of the second alternative (Fig.\ref{fig:conv2}) is as in  
Eqs.\equa{res-alt-1} and \equa{cond-alt-1}  with 
$\Delta\alpha_{j'}$ replaced by $\Delta\alpha_{j'}-2\pi$.
The third alternative (Fig.\ref{fig:conv3}) reads as
\be
\lefteqn{\sum_{l'}\,
{\bf A}^{j j'}_{l l'}{\bf A}^{j' j''}_{l' l''} } \nonumber \\
&\sim& -W^{j j''}_{l l''} \sum_{\ell=1}^{\infty} \sum_{n=0}^{\infty}\,
 \e^{-\i\pi/6} \pi^{-1/2} C_\ell (ka_{j'})^{1/3}\nonumber \\ 
&& \quad \times
\left [
 \frac{ \exp{ \i  \sqrt{(kR_{jj'})^2\mbox{$-$}(l\mbox{$-$}{\nu'_\ell})^2}
  }}
  {[(kR_{jj'})^2\mbox{$-$}(l\mbox{$-$}{\nu'_\ell})^2]^{1/4}}\ 
  \frac{\exp{
   \i\sqrt{(kR_{j'j''})^2\mbox{$-$}({\nu'_\ell}\mbox{$-$}l'')^2}
  } }
    { [(kR_{j'j''})^2\mbox{$-$}({\nu'_\ell}\mbox{$-$}l'')^2]^{1/4} } 
  \right . \nonumber \\
 &&\qquad\quad \mbox{} \times
 \e^{ -\i(l-{\nu'_\ell})\arccos[(l-{\nu'_\ell})/kR_{jj'}]
 -\i({\nu'_\ell}-l'')\arccos[({\nu'_\ell}-l'')/kR_{j'j''}] 
 +\i{\nu'_\ell}(\Delta\alpha_{j'}+(2n-1)\pi) } \nonumber \\
&& \quad\quad \mbox{}+ 
 \frac{ 
 \exp{\i  \sqrt{(kR_{jj'})^2\mbox{$-$}(l\mbox{+}{\nu'_\ell})^2}} }
     {[(kR_{jj'})^2\mbox{$-$}(l\mbox{+}{\nu'_\ell})^2]^{1/4}}\ 
 \frac{\exp{i\sqrt{(kR_{j'j''})^2\mbox{$-$}({\nu'_\ell}\mbox{+}l'')^2}
 }}
{ [(kR_{j'j''})^2\mbox{$-$}({\nu'_\ell}\mbox{+}l'')^2]^{1/4}} \nonumber \\
&& \left. \qquad\quad \mbox{}\times
 \e^{ -\i(l+{\nu'_\ell})\arccos[(l+{\nu'_\ell})/kR_{jj'}]
 +\i({\nu'_\ell}+l'')\arccos[(-{\nu'_\ell}-l'')/kR_{j'j''}]
 +\i{\nu'_\ell}(-\Delta\alpha_{j'}+(2n+1)\pi)\} } \right ]
    \nonumber \\
 && +W^{j j''}_{l l''}
 \sqrt{\frac{2}{\pi}} \e^{-\i\pi/4}
\,\e^{-\i l\arccos[(l-{\nu'_s})/kR_{jj'}]
             +il''\arccos[({\nu'_s}-l'')/kR_{j'j''}]} \nonumber \\
&& \quad \mbox{}\times 
 \frac{ \exp\left\{\i\ \sqrt{(kR_{jj'})^2\mbox{$-$}(l\mbox{$-$}{\nu'_s})^2}
             +\i\sqrt{(kR_{jj'})^2\mbox{$-$}({\nu'_s}\mbox{$-$}l'')^2} 
\right\} }
 {  \sqrt{
 [(kR_{jj'})^2\mbox{$-$}(l\mbox{$-$}{\nu'_s})^2]^{1/2}+
  [(kR_{j'j''})^2\mbox{$-$}({\nu'_s}\mbox{$-$}l'')^2]^{1/2} }}     
       \; .  \label{res-alt-3}    
\ee   
Here ${\nu'_s}$ has to satisfy 
the saddle-point condition
\beq
 \Delta\alpha_{j'}  -\arccos[({\nu'_s}-l)/kR_{jj'}]
      -\arccos[({\nu'_s}-l'')/kR_{j'j''}] = 0 \; .
  \label{cond-alt-3}
\eeq
Again, the two terms in the straight line integral of Eq.\equa{alt-3} give
the same contribution, as the saddle $\nu'_{s_2}$ of the latter term
is $-\nu'_{s_1}$ of the first one.
The minus sign in front of the straight line integral is cancelled by
an additional minus sign [relative to alternative one or two] 
resulting from the positive slope of the straight-line 
section (see Fig.\ref{fig:ghost_path})
and the corresponding changes in the Fresnel integral
\be
 \e^{-2\i\pi/4}\int_{\i\infty(1-\i|\delta|)}^{-\i\infty(1-\i|\delta|)}
 {\rm d} \delta \nu'
\,\e^{ +\i\half(\delta \nu')^2\left 
(
 \frac{1}{\sqrt{(kR_{jj'})^2-(l-\nu'_s)^2}}       
 +\frac{1}{\sqrt{(kR_{j'j''})^2-(\nu'_s-l'')^2}} \right )} \; . 
\ee      
The latter, by the substitution $\delta \nu' = \e^{\i\pi/4}x$, becomes
a negatively transversed Gauss' integral
\be
   \e^{-\i\pi/4}\int_{\infty}^{-\infty}{\rm d}x\, \e^{-x^2 b/2}
 = - \e^{-\i\pi/4}\sqrt{2\pi/b}
\ee
where $b=\{(kR_{jj'})^2-(l-\nu'_s)^2\}^{-1/2}
+\{(kR_{j'j''})^2-(\nu'_s-l'')^2\}^{-1/2}$.
In fact, all dependence of the disk $j'$ is finally 
gone from this expression.
If the third alternative exists, the pertinent straight line integral 
corresponds to  a ``ghost'' segment 
starting at disk $j$ [under the impact parameter $l/k$] and ending
at disk $j''$ [under the impact parameter $l''/k$] 
which is equivalent to the corresponding geometrical segment of
the direct term $A_{ll''}^{jj''}$ ($j''\neq j$). Because of the angular
conditions, specified before Eq.\equa{alt-3}, the ghost path has to cut
disk $j'$, i.e. the modulus of the impact parameter $\nu'_{s}/k$ has
to be smaller than the disk radius $a_{j'}$ (see Fig.\ref{fig:conv3}).

\subsection{Ghost segment\label{app:convol_ghost}}
Let us now discuss the ``ghost'' segment, i.e., the non-creeping terms of
Eq.\equa{res-alt-3}. The ghost cancellation presented here
is, of course, related to Berry's work on the  ghost
cancellation for periodic orbits in the Sinai billiard, 
see Ref.\cite{Berry_KKR}. However, here the calculation is based on
Watson's method which specifies the integration paths, the signs of the
ghost contributions {\em and} encodes the geometries (the choice of the
three alternatives for $\sigma$) into the creeping orbits. 

After restoring $W_{ll''}^{jj''}$
it reads
\be
{\rm ghost}_{jj''}^{\,ll''}({\nu'_s}) &\sim&
(1- \delta_{jj'}) (1- \delta_{jj''})
                  \frac{ a_j   \Jb{l }{a_j   } }
                       { a_{j''}\Ho{l''}{a_{j''}} }
                   (-1)^{l''} 
                  \e^{\i ( l \alpha_{j'' j} - l'' \alpha_{j j''} ) } \nonumber\\
&& \quad \mbox{}\times 
\sqrt{\frac{2}{\pi}} \e^{-\i\pi/4} 
\frac{ \exp\left\{\i\ \sqrt{(kR_{jj'})^2\mbox{$-$}(l\mbox{$-$}{\nu'_s})^2}
             +\i\sqrt{(kR_{jj'})^2\mbox{$-$}({\nu'_s}\mbox{$-$}l'')^2} 
\right\} }
 {  \sqrt{
 [(kR_{jj'})^2\mbox{$-$}(l\mbox{$-$}{\nu'_s})^2]^{1/2}+
  [(kR_{j'j''})^2\mbox{$-$}({\nu'_s}\mbox{$-$}l'')^2]^{1/2} }}   \nonumber \\
        &&\qquad \mbox{}\times
             \e^{+\i  l \, ( \alpha_{j'  j  } -  \alpha_{j'' j  } 
        - \arccos[(l-{\nu'_s})/kR_{jj'}])}\nonumber \\
         &&\qquad \mbox{}\times
             \e^{-\i  l''( \alpha_{j'  j''} -\alpha_{j j''}
 -\arccos[({\nu'_s}-l'')/kR_{j'j''}])} \label{ghost-seg} 
\ee
with 
\beq
 \alpha_{j''j'}-\alpha_{jj'} +\arcsin[({\nu'_s}-l)/kR_{jj'}]
  +\arcsin[({\nu'_s}-l'')/kR_{j'j''}]=\pi  \label{ghost-cond-2}
\eeq
which is equivalent to condition \equa{cond-alt-3}.
As this saddle-point condition implies that the impact parameter ${\nu'_s}/k$
at disk $j'$ lies on the straight line joining the impact parameter $l/k$
at disk $j$, with the impact parameter $l''/k$ at disk $j''$, the following
relation between the lengths of the segments on this line holds
\be
    \sqrt{R_{jj'}^2  -\{(l-{\nu'_s})  /k\}^2}
  + \sqrt{R_{j'j''}^2-\{({\nu'_s}-l'')/k\}^2} 
  = \sqrt{R_{j j''}^2-\{(l-l'')    /k\}^2}\; , \label{alt-3-geo}
\ee
i.e., the length of the straight line 
from the impact parameter $l/k$ to the impact parameter 
$l''/k$ is the sum of the lengths from $l/k$ to ${\nu'}/k$ and from ${\nu'}/k$
to $l''/k$ (see Fig.\ref{fig:conv3}).

The ``ghost'' segment \equa{ghost-seg} 
should be compared with Eq.\equa{A-kernel}, in the semiclassical 
approximation \equa{H1-Debye}, for the Hankel function $\Ho{l-l''}{R_{jj''}}$
\be
 {\bf A}_{ll''}^{jj''} &\sim&
(1- \delta_{jj''}) 
                  \frac{ a_j   \Jb{l }{a_j   } }
                       { a_{j''}\Ho{l''}{a_{j''}} }
                  (-1)^{l''} 
                  \e^{\i ( l \alpha_{j'' j} - l'' \alpha_{j j''} ) } \nonumber \\
  && \mbox{} \times \sqrt{\frac{2}{\pi}} \e^{-\i\pi/4}\,
 \frac{ \exp{\i\sqrt{(kR_{jj'})^2-(l-l')^2}}}
 {[(kR_{jj''})^2-(l-l'')^2]^{1/4}}\, \e^{-\i(l-l'')\arccos[(l-l'')/kR_{jj''}]}
 \; .
 \label{comp-to-ghost}
\ee
Condition \equa{alt-3-geo} implies that the lengths and stabilities
of the ghost segment \equa{ghost-seg} and 
of the direct path \equa{comp-to-ghost}
are the same. The comparison of the phases implies the relations
\be
\pi/2&=& \alpha_{j'j}- \alpha_{j''j}+\arccos[({\nu'_s}-l)/kR_{jj'}]
  +\arcsin[(l''-l)/kR_{jj''}]\\
\pi/2&=&\alpha_{jj''}-\alpha_{j'j''}+\arccos[({\nu'_s}-l'')/kR_{j'j''}]
   +\arcsin[(l-l'')/kR_{jj''}]
\ee
which are valid under the condition \equa{ghost-cond-2}. Thus, we finally
have in the semiclassical approximation 
\be
 {\rm ghost}_{jj''}^{\,ll''}({\nu'_s}) \equiv  
\left({\bf A}^{j \underline{j'}}_{\rm ghost} 
{\bf A}^{\underline{j'}j''}_{\rm ghost}\right)_{ll''}
\simeq  {\bf A}_{ll''}^{jj''} 
\ee
under the condition, of course, that the saddle ${\nu'_s}$ satisfies 
Eq.\equa{ghost-cond-2}. 
\newpage
\newcommand{\AP}[1]{{\em Ann.\ Phys.}\/ {\bf #1}}
\newcommand{\CHAOS}[1]{{\em CHAOS}\/ {\bf #1}}
\newcommand{\CM}[1]{{\em Cont.\ Math.}\/ {\bf #1}}
\newcommand{\CMP}[1]{{\em Commun.\ Math.\ Phys.}\/ {\bf #1}}
\newcommand{\INCB}[1]{{\em Il Nuov.\ Cim.\ B}\/ {\bf #1}}
\newcommand{\JCP}[1]{{\em J.\ Chem.\ Phys.}\/ {\bf #1}}
\newcommand{\JETP}[1]{{\em Sov.\ Phys.\ JETP}\/ {\bf #1}}
\newcommand{\JETPL}[1]{{\em JETP Lett.}\/ {\bf #1}}
\newcommand{\JMP}[1]{{\em J.\ Math.\ Phys.}\/ {\bf #1}}
\newcommand{\JMPA}[1]{{\em J.\ Math.\ Pure Appl.}\/ {\bf #1}}
\newcommand{\JPA}[1]{{\em J.\ Phys.}\/ {\bf A  #1}}
\newcommand{\JPB}[1]{{\em J.\ Phys.}\/ {\bf B  #1}}
\newcommand{\JPC}[1]{{\em J.\ Phys.\ Chem.}\/ {\bf #1}}
\newcommand{\JchemP}[1]{{\em J.\ Chem.\ Phys.}\/ {\bf #1}}
\newcommand{\NPA}[1]{{\em Nucl.\ Phys.}\/ {\bf A #1}}
\newcommand{\NPB}[1]{{\em Nucl.\ Phys.}\/ {\bf B #1}}
\newcommand{\NONLIN}[1]{{\em Nonlinearity}\/ {\bf #1}}
\newcommand{\PLA}[1]{{\em Phys.\ Lett.}\/ {\bf A #1}}
\newcommand{\PLB}[1]{{\em Phys.\ Lett.}\/ {\bf B #1}}
\newcommand{\PRA}[1]{{\em Phys.\ Rev.}\/ {\bf A #1}}
\newcommand{\PRD}[1]{{\em Phys.\ Rev.}\/ {\bf D #1}}
\newcommand{\PRL}[1]{{\em Phys.\ Rev.\ Lett.}\/ {\bf #1}}
\newcommand{\PST}[1]{{\em Phys.\ Scripta}\/ {\bf T #1}}
\newcommand{\RMS}[1]{{\em Russ.\ Math.\ Surv.}\/ {\bf #1}}
\newcommand{\USSR}[1]{{\em Math.\ USSR.\ Sb.}\/ {\bf #1}}
\newcommand{\ZNat}[1]{{\em Z. Naturforschung}\/ {\bf #1}}
\renewcommand{\baselinestretch} {1}

\newpage
\section{Figures of 3-disk resonances 
\label{app:figures}}
\setcounter{equation}{0}
\setcounter{figure}{0}
\setcounter{table}{0}
\noindent\begin{figure}[htb]
\vskip -0.4cm
\centerline{
{\bf(a)} 
\epsfig{file=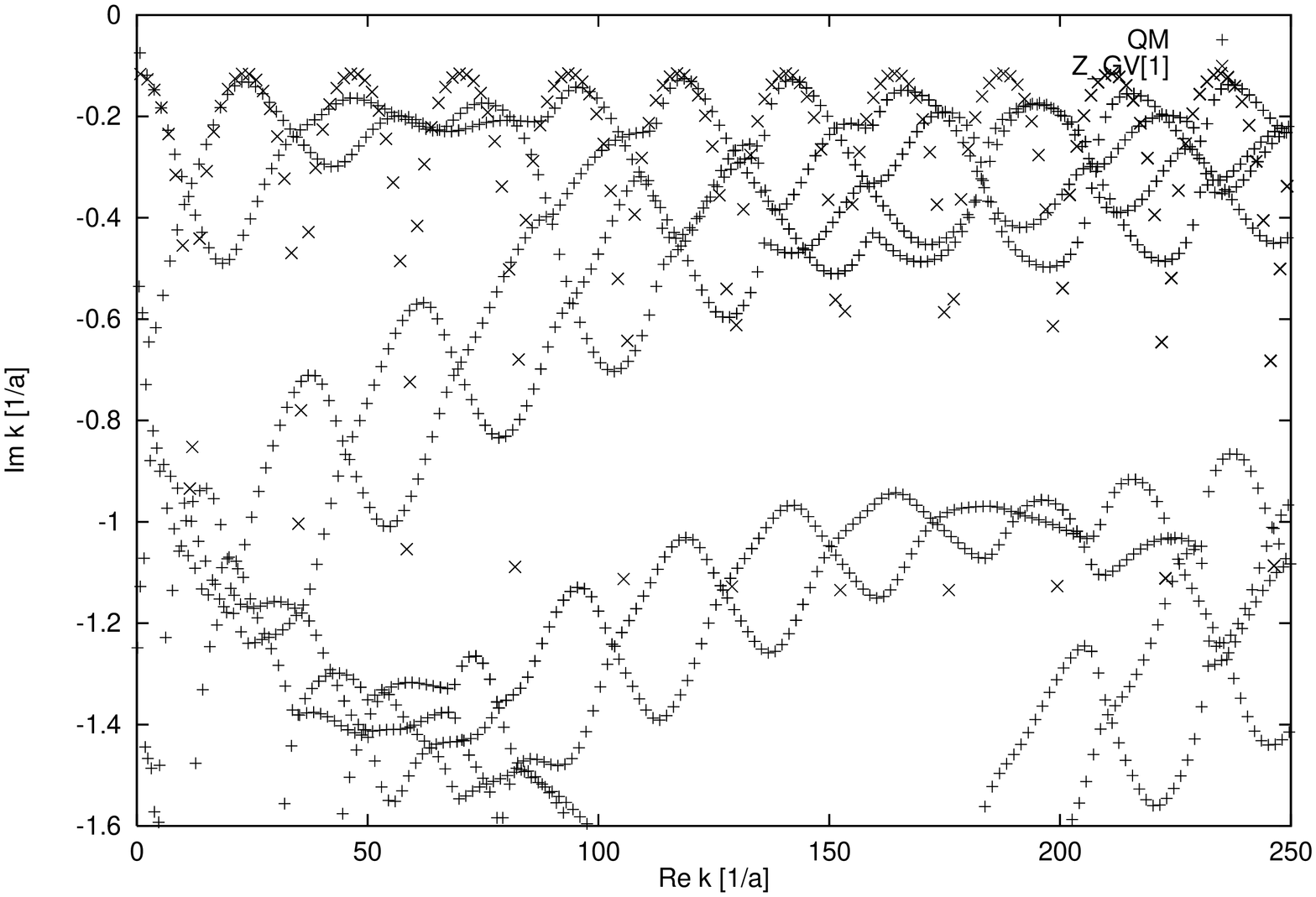,height=7.1cm,angle=-90}}

 \vskip 0.1cm
\centerline{{\bf(b)} 
\epsfig{file=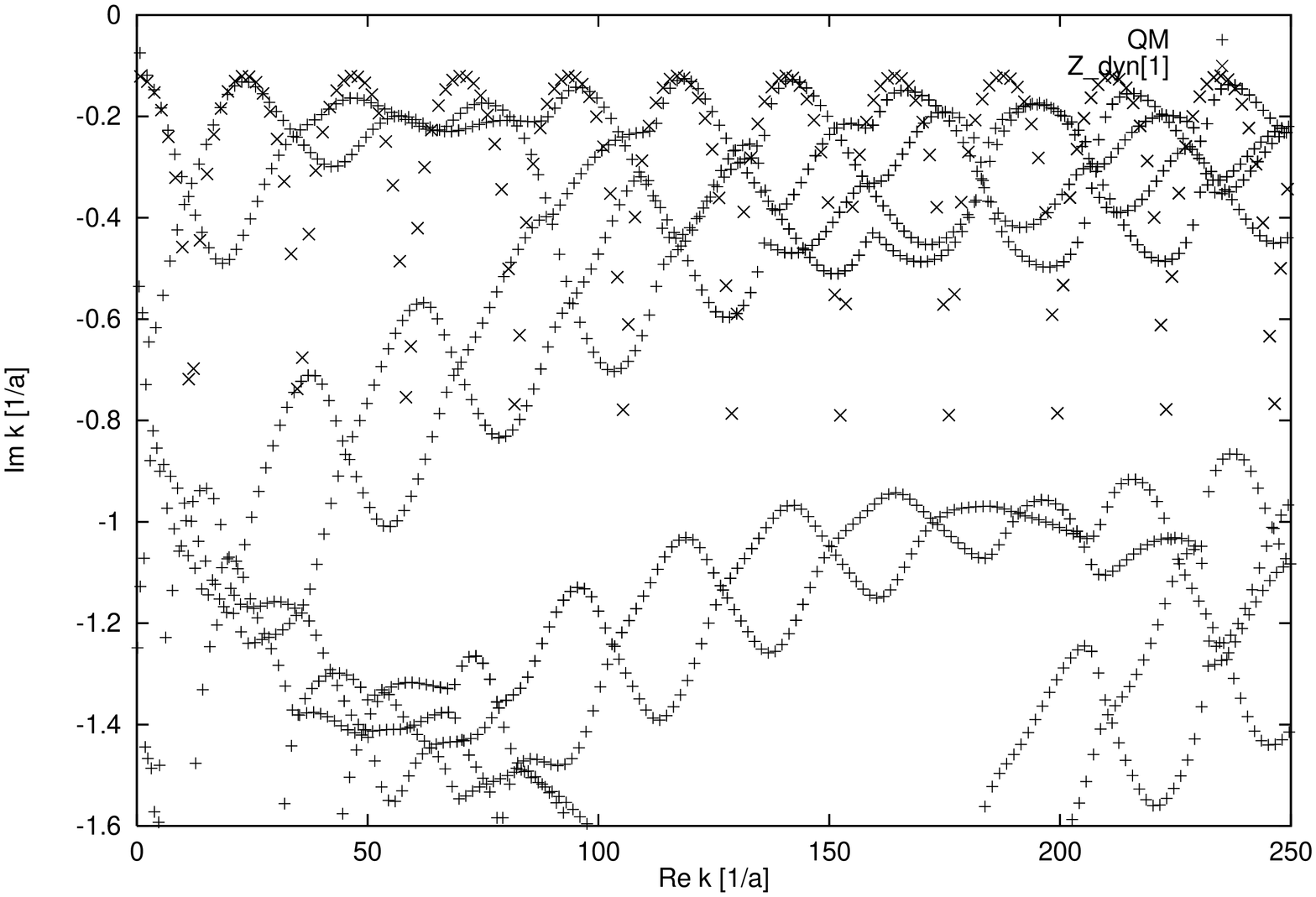,height=7.1cm,angle=-90}}

 \vskip 0.1cm
 \centerline{{\bf(c)} 
\epsfig{file=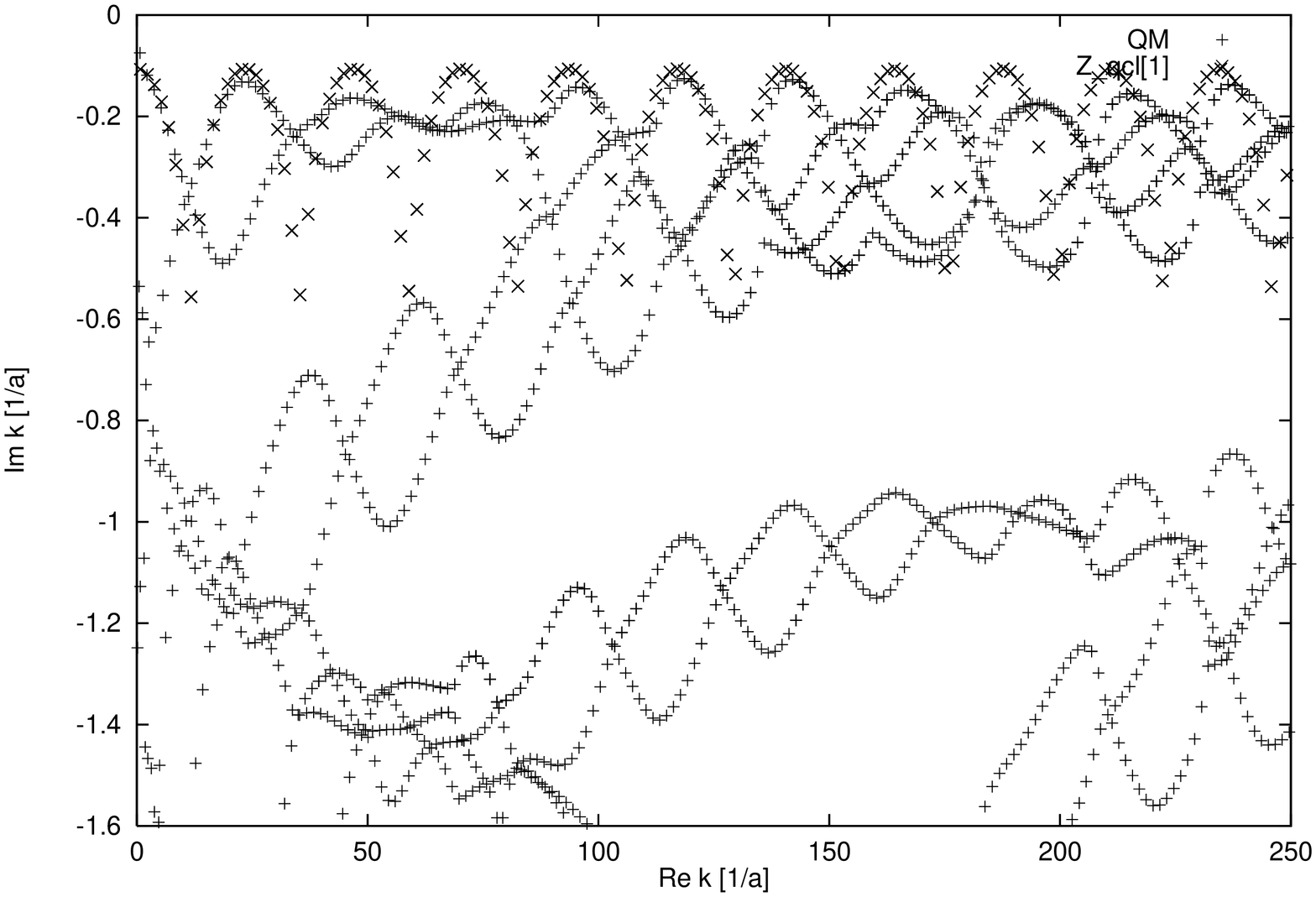,height=7.1cm,angle=-90}}

\caption[fig_n_1]{\small
The ${\rm A}_{\rm 1}$ resonances of the 3-disk system with $R=6a$. The 
exact quantum-mechanical data are denoted by plusses. 
The semiclassical ones are calculated up to   $1^{\,\rm st}$ order
in the curvature expansion and are denoted by crosses: 
{\bf (a)} 
Gutzwiller-Voros 
zeta-function \equa{GV_zeta_app}, {\bf (b)} 
dynamical zeta-function \equa{dyn_zeta_app}, 
{\bf (c)} 
quasiclassical zeta function \equa{qcl_zeta}. 
\label{fig:e_gv1}}
\end{figure}

\noindent\begin{figure}[htb]
\vskip -0.4cm
\centerline{
{\bf(a)} 
\epsfig{file=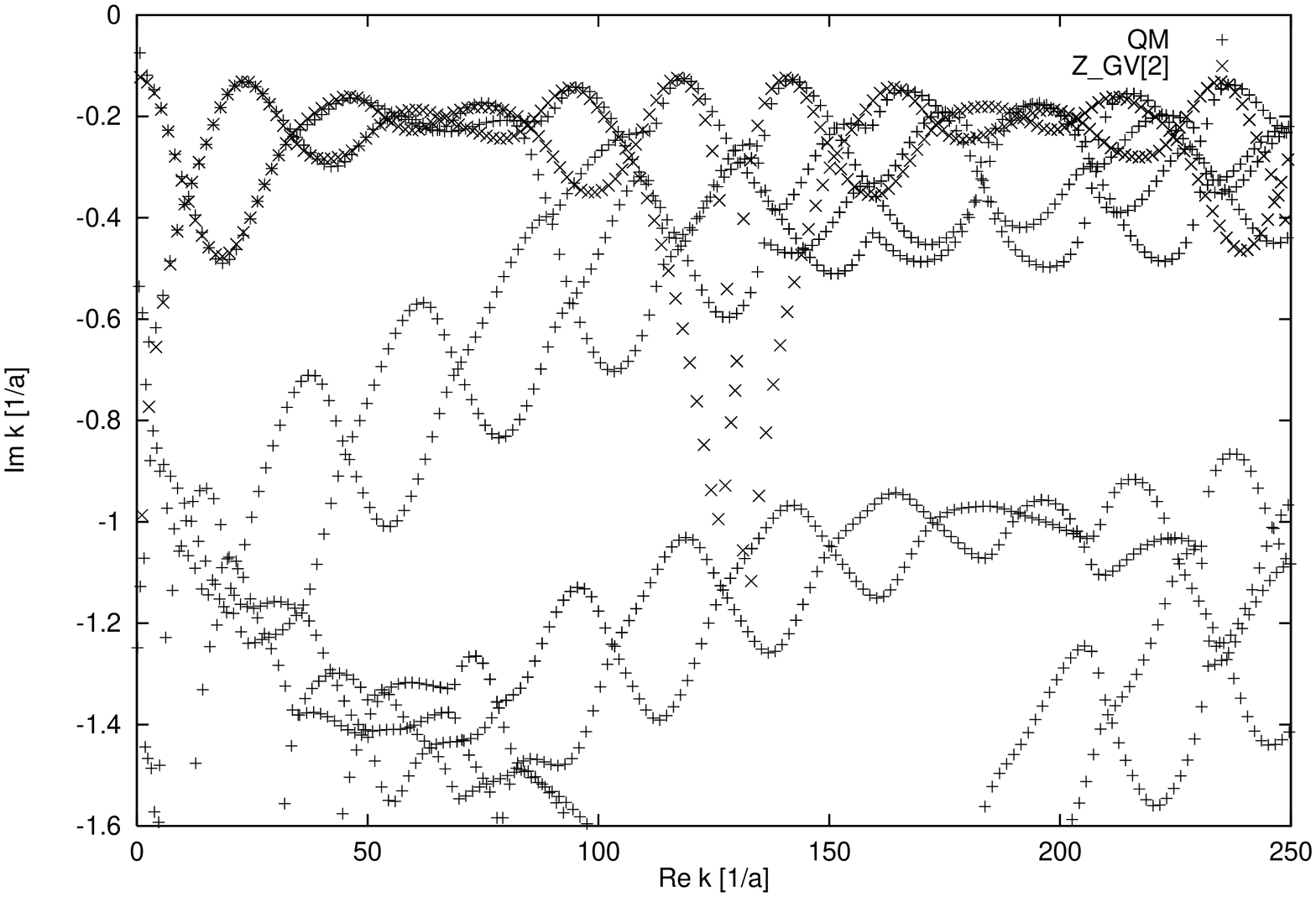,height=7.1cm,angle=-90}}

 \vskip 0.1cm
\centerline{{\bf(b)} \epsfig{file=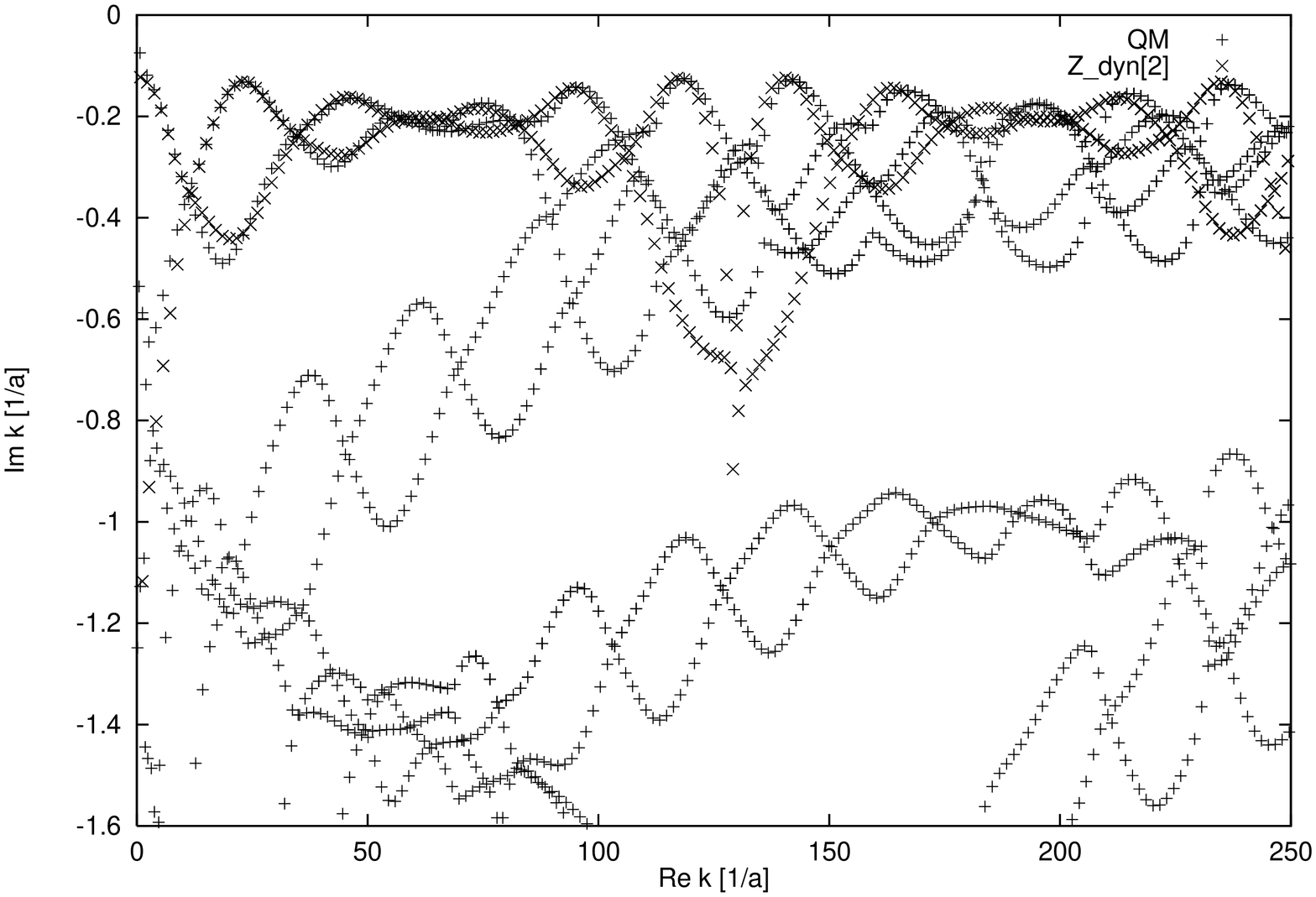,height=7.1cm,angle=-90}}

 \vskip 0.1cm
 \centerline{{\bf(c)} \epsfig{file=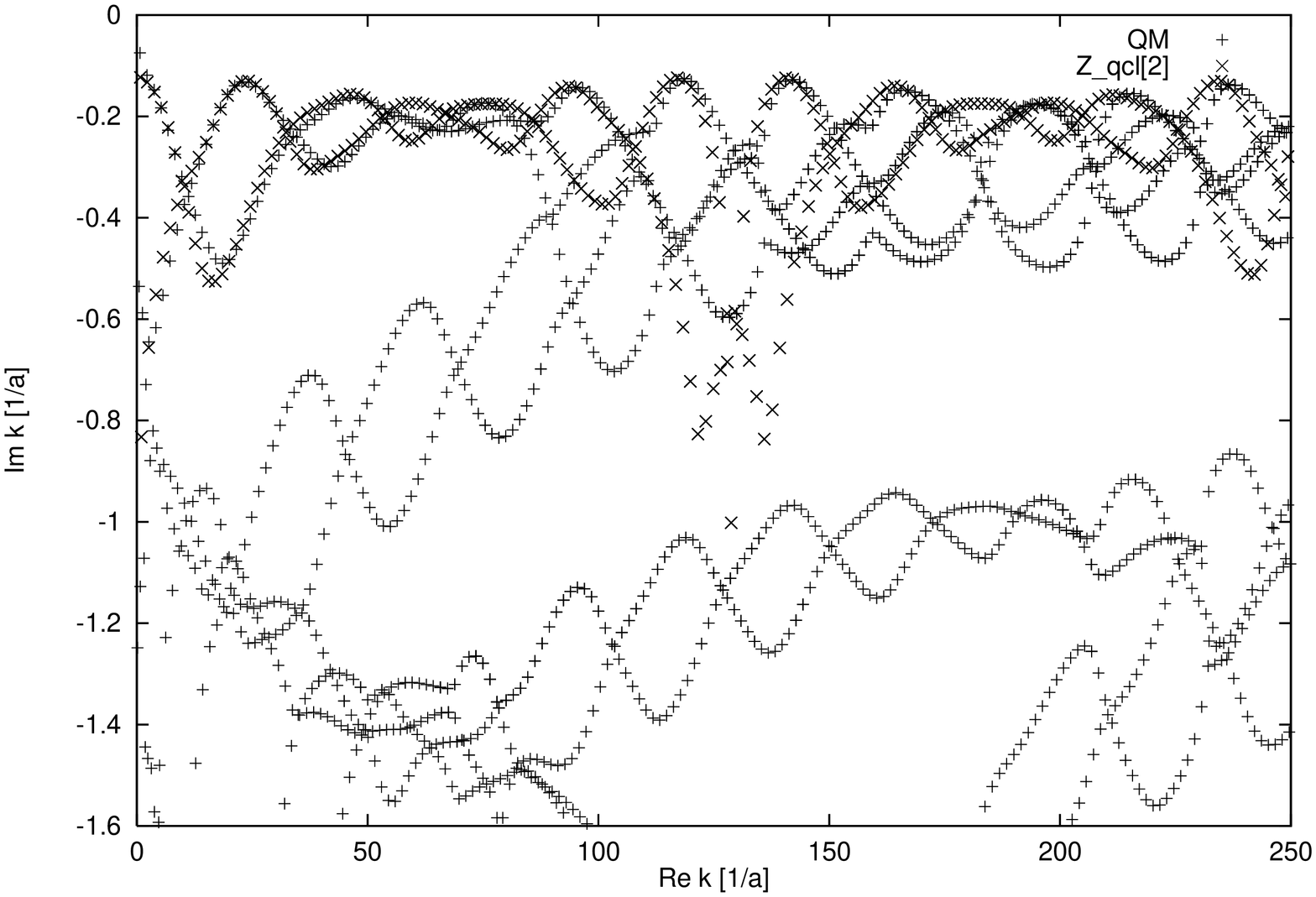,height=7.1cm,angle=-90}}

\caption[fig_n_2]{\small
The ${\rm A}_{\rm 1}$ resonances of the 3-disk system with $R=6a$. The 
exact quantum-mechanical data are denoted by plusses. 
The semiclassical ones  are calculated up to   $2^{\,\rm nd}$ order
in the curvature expansion and are denoted by crosses: 
{\bf (a)} 
Gutzwiller-Voros 
zeta-function \equa{GV_zeta_app}, {\bf (b)} 
dynamical zeta-function \equa{dyn_zeta_app}, 
{\bf (c)} 
quasiclassical zeta function \equa{qcl_zeta}. 
\label{fig:e_gv2}}
\end{figure}

\noindent\begin{figure}[htb]
\vskip -0.4cm
\centerline{
{\bf(a)} 
\epsfig{file=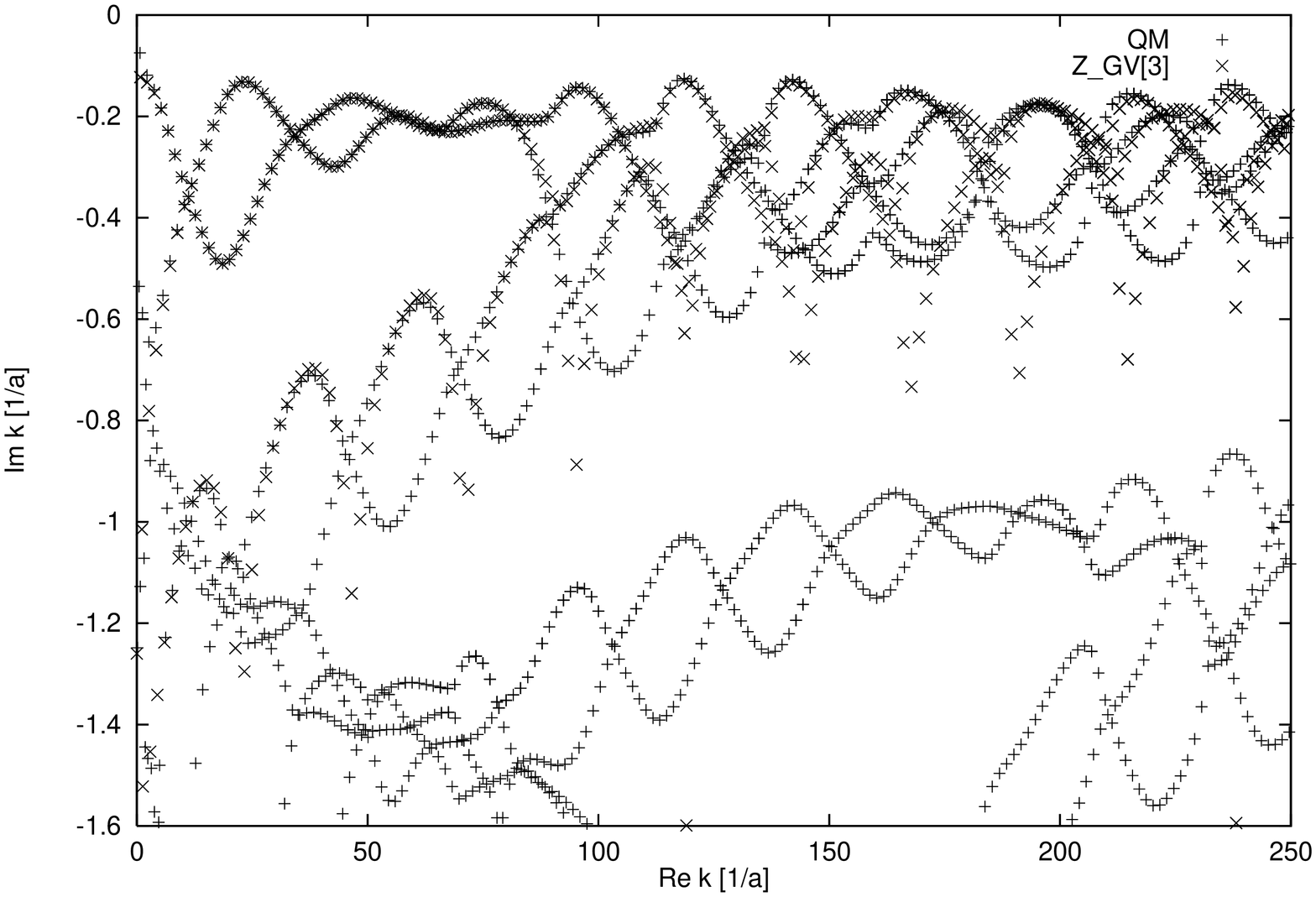,height=7.1cm,angle=-90}}

 \vskip 0.1cm
\centerline{{\bf(b)} \epsfig{file=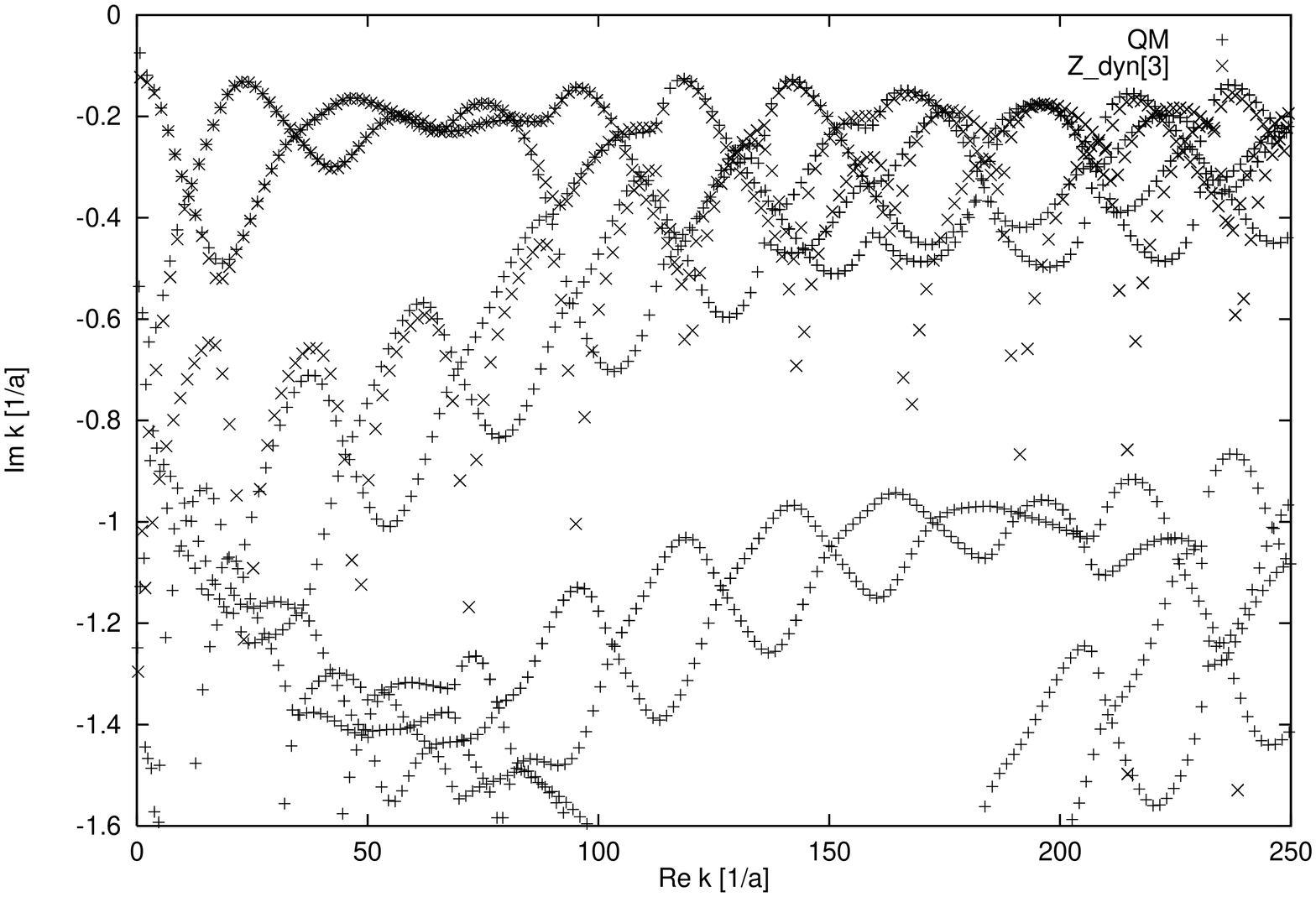,height=7.1cm,angle=-90}}

 \vskip 0.1cm
 \centerline{{\bf(c)} \epsfig{file=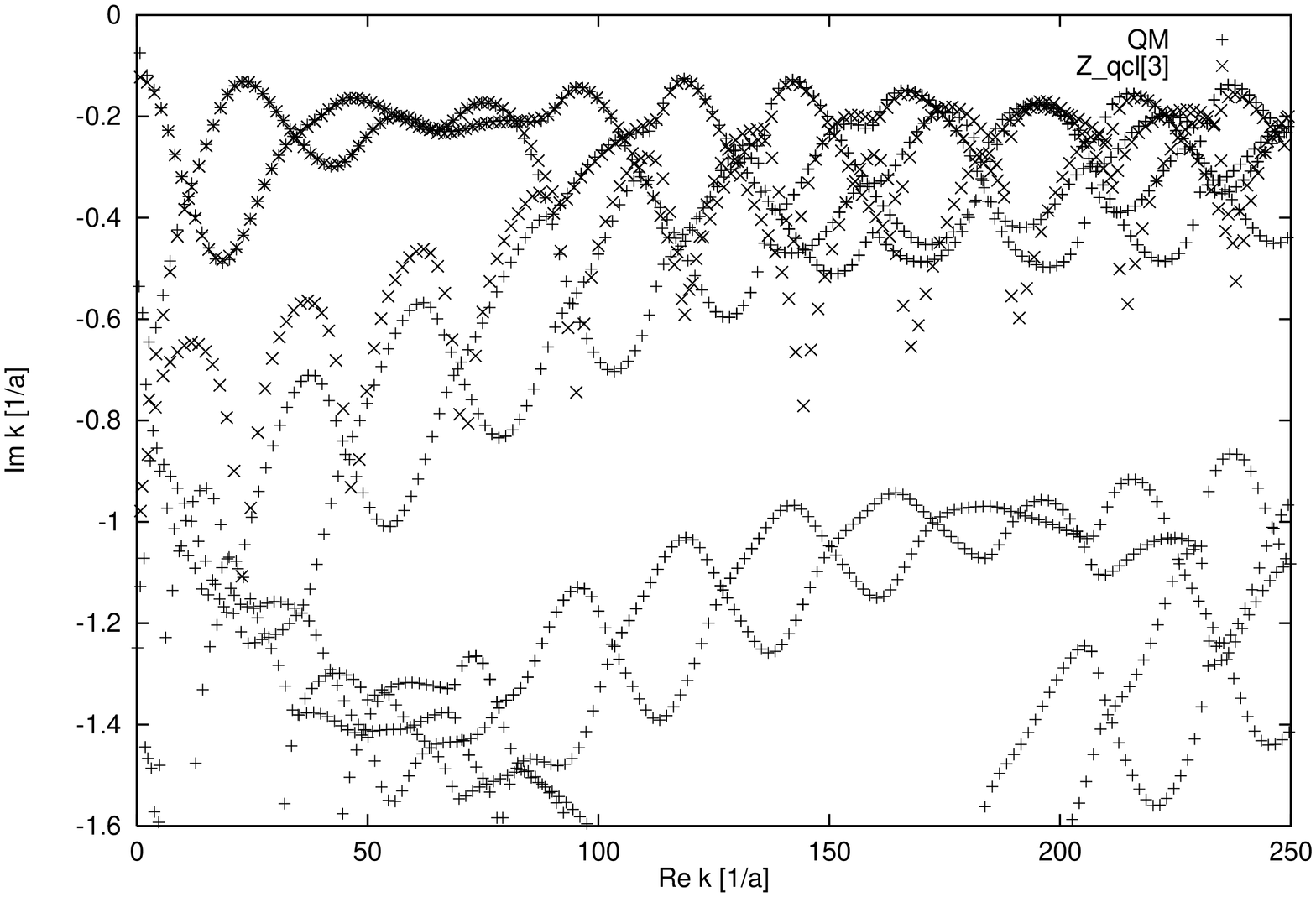,height=7.1cm,angle=-90}}

\caption[fig_n_3]{\small
The ${\rm A}_{\rm 1}$ resonances of the 3-disk system with $R=6a$. The 
exact quantum-mechanical data are denoted by plusses. 
The semiclassical ones are calculated up to   $3^{\,\rm rd}$ order
in the curvature expansion and are denoted by crosses:
{\bf (a)} 
Gutzwiller-Voros 
zeta-function \equa{GV_zeta_app}, {\bf (b)} 
dynamical zeta-function \equa{dyn_zeta_app}, 
{\bf (c)} 
quasiclassical zeta function \equa{qcl_zeta}. 
\label{fig:e_gv3}}
\end{figure}

\noindent\begin{figure}[htb]
\vskip -0.4cm
\centerline{
{\bf(a)} 
\epsfig{file=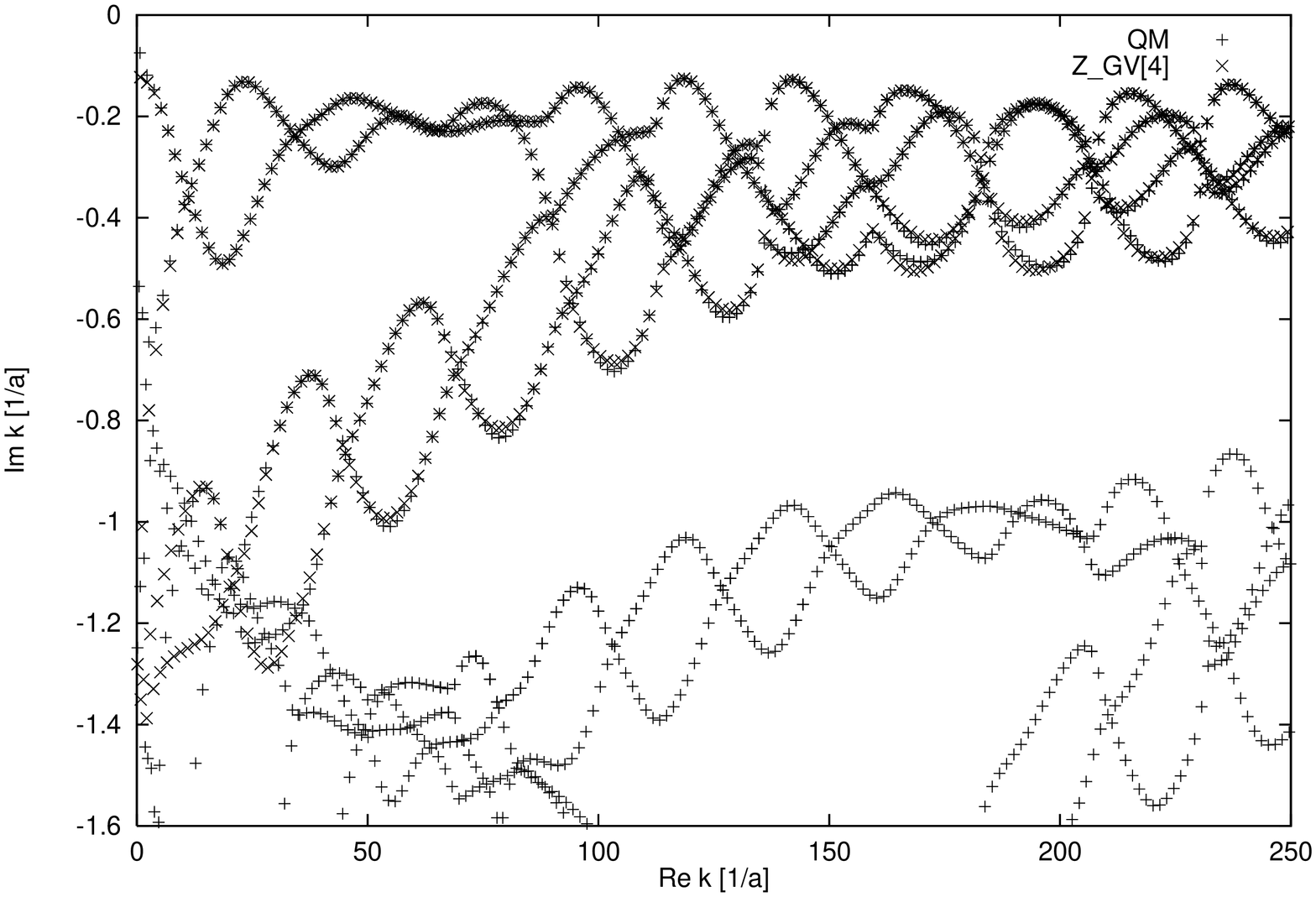,height=7.1cm,angle=-90}}

\vskip 0.1cm
\centerline{{\bf(b)} 
\epsfig{file=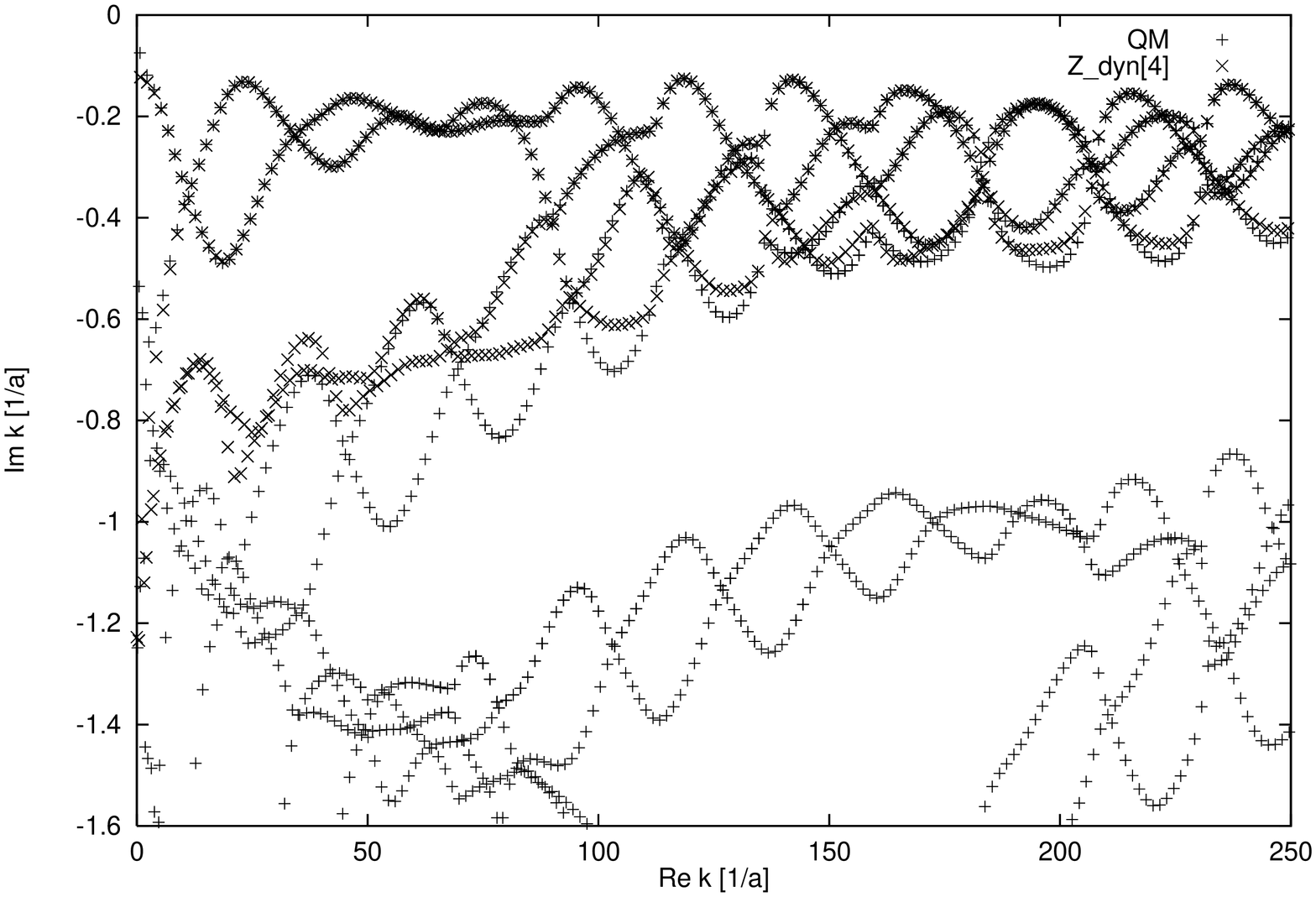,height=7.1cm,angle=-90}}

\vskip 0.1cm
\centerline{{\bf(c)} 
\epsfig{file=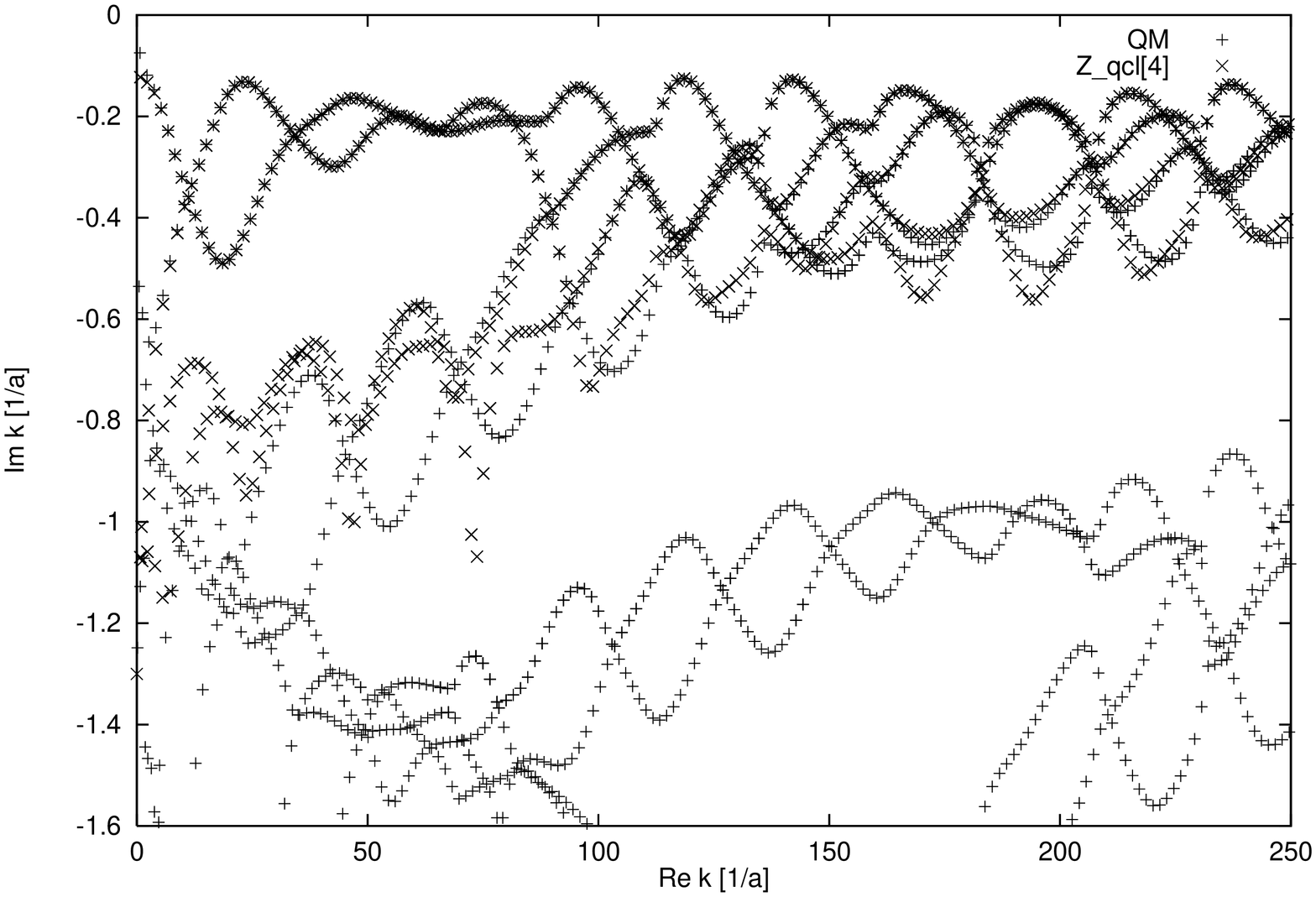,height=7.1cm,angle=-90}}

\caption[fig_n_4]{\small
The ${\rm A}_{\rm 1}$ resonances of the 3-disk system with $R=6a$. The 
exact quantum-mechanical data are denoted by plusses. 
The semiclassical ones are calculated up to   $4^{\,\rm th}$ order
in the curvature expansion and are denoted by crosses:
{\bf (a)} 
Gutzwiller-Voros 
zeta-function \equa{GV_zeta_app}, {\bf (b)} 
dynamical zeta-function \equa{dyn_zeta_app}, 
{\bf (c)} 
quasiclassical zeta function \equa{qcl_zeta}. 
\label{fig:e_gv4}}
\end{figure}

\noindent\begin{figure}[htb]
\vskip -0.4cm
\centerline{
{\bf(a)} 
\epsfig{file=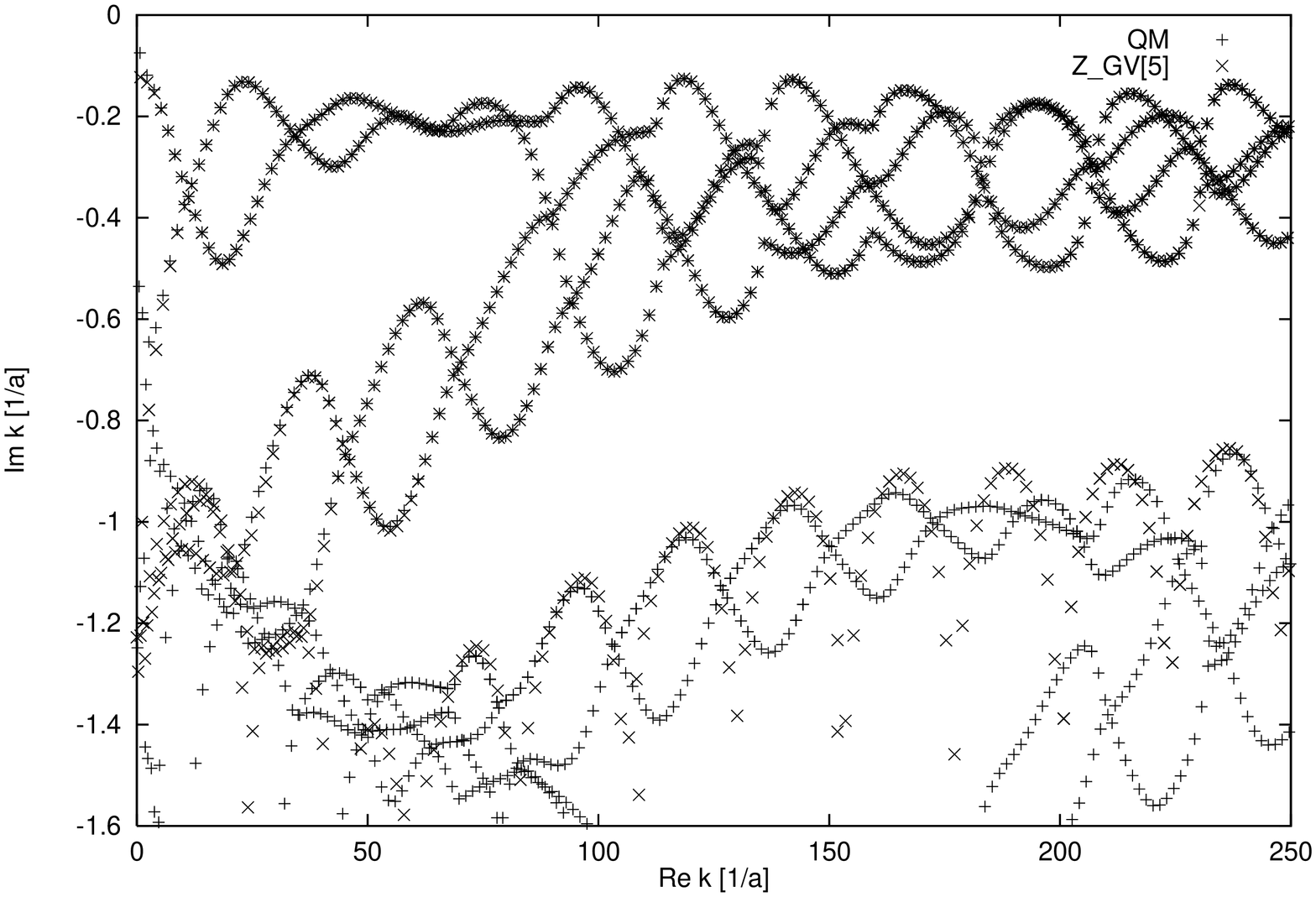,height=7.1cm,angle=-90}}

 \vskip 0.1cm
\centerline{{\bf(b)} \epsfig{file=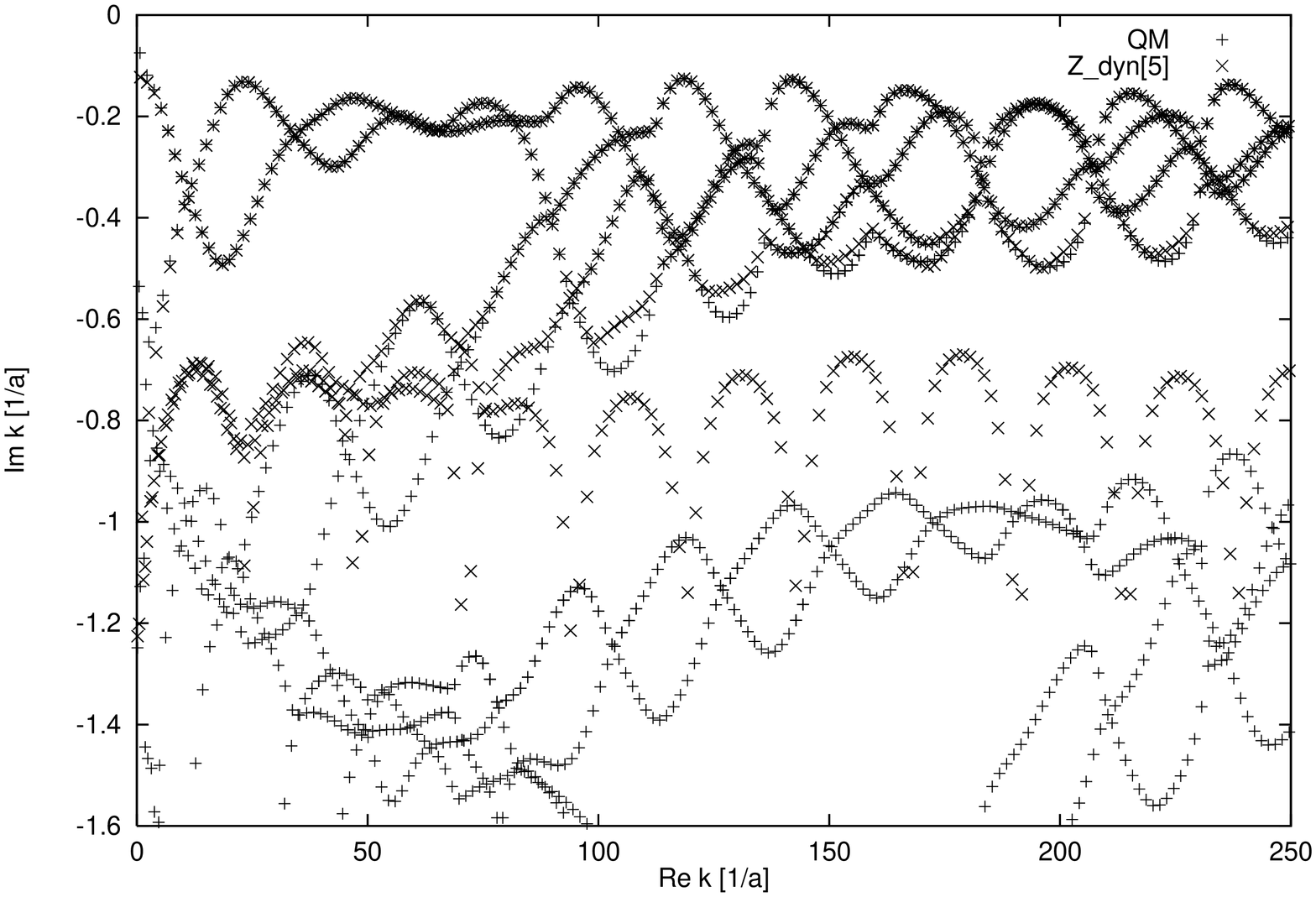,height=7.10cm,angle=-90}}

 \vskip 0.1cm
\centerline{{\bf(c)} \epsfig{file=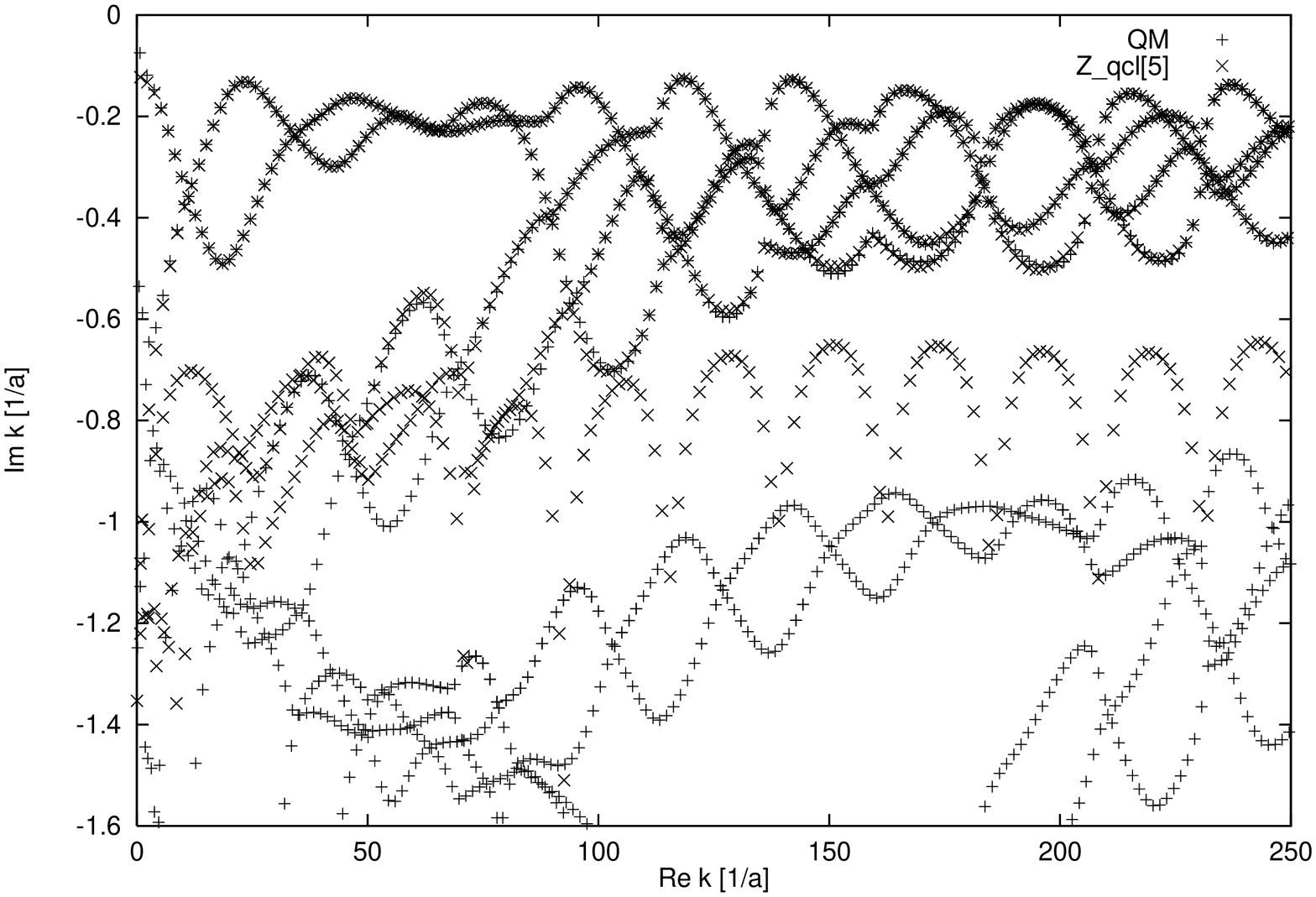,height=7.1cm,angle=-90}}

\caption[fig_n_5]{\small
The ${\rm A}_{\rm 1}$ resonances of the 3-disk system with $R=6a$. The 
exact quantum-mechanical data are denoted by plusses. 
The semiclassical ones are calculated up to  $5^{\,\rm th}$ order
in the curvature expansion and are denoted by crosses: 
{\bf (a)} 
Gutzwiller-Voros 
zeta-function \equa{GV_zeta_app}, {\bf (b)} 
dynamical zeta-function \equa{dyn_zeta_app}, 
{\bf (c)} 
quasiclassical zeta function \equa{qcl_zeta}. 
\label{fig:e_gv5}}
\end{figure}

\noindent\begin{figure}[htb]
\vskip -0.4cm
\centerline{
{\bf(a)} 
\epsfig{file=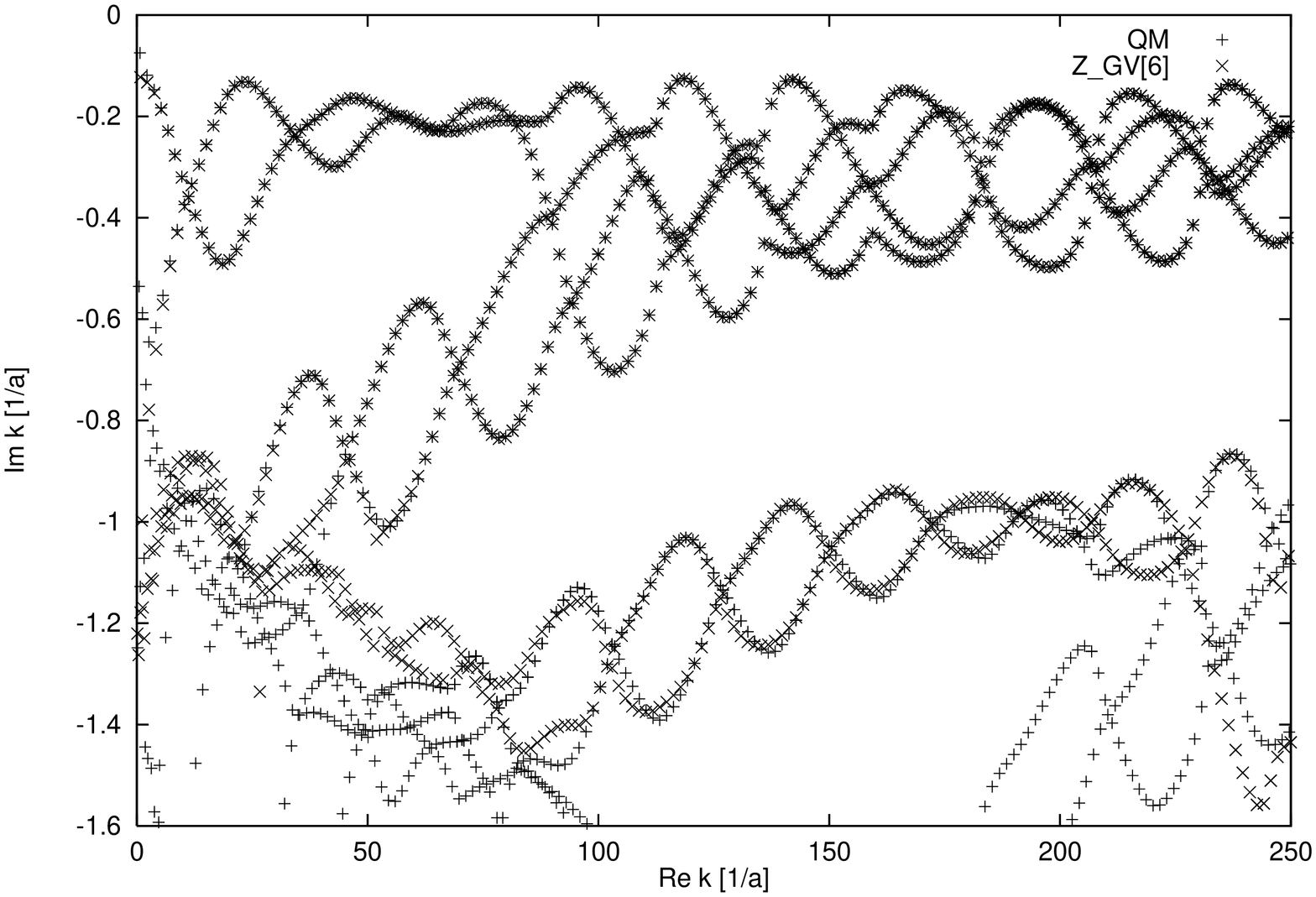,height=7.1cm,angle=-90}}

\vskip 0.1cm
\centerline{{\bf(b)} \epsfig{file=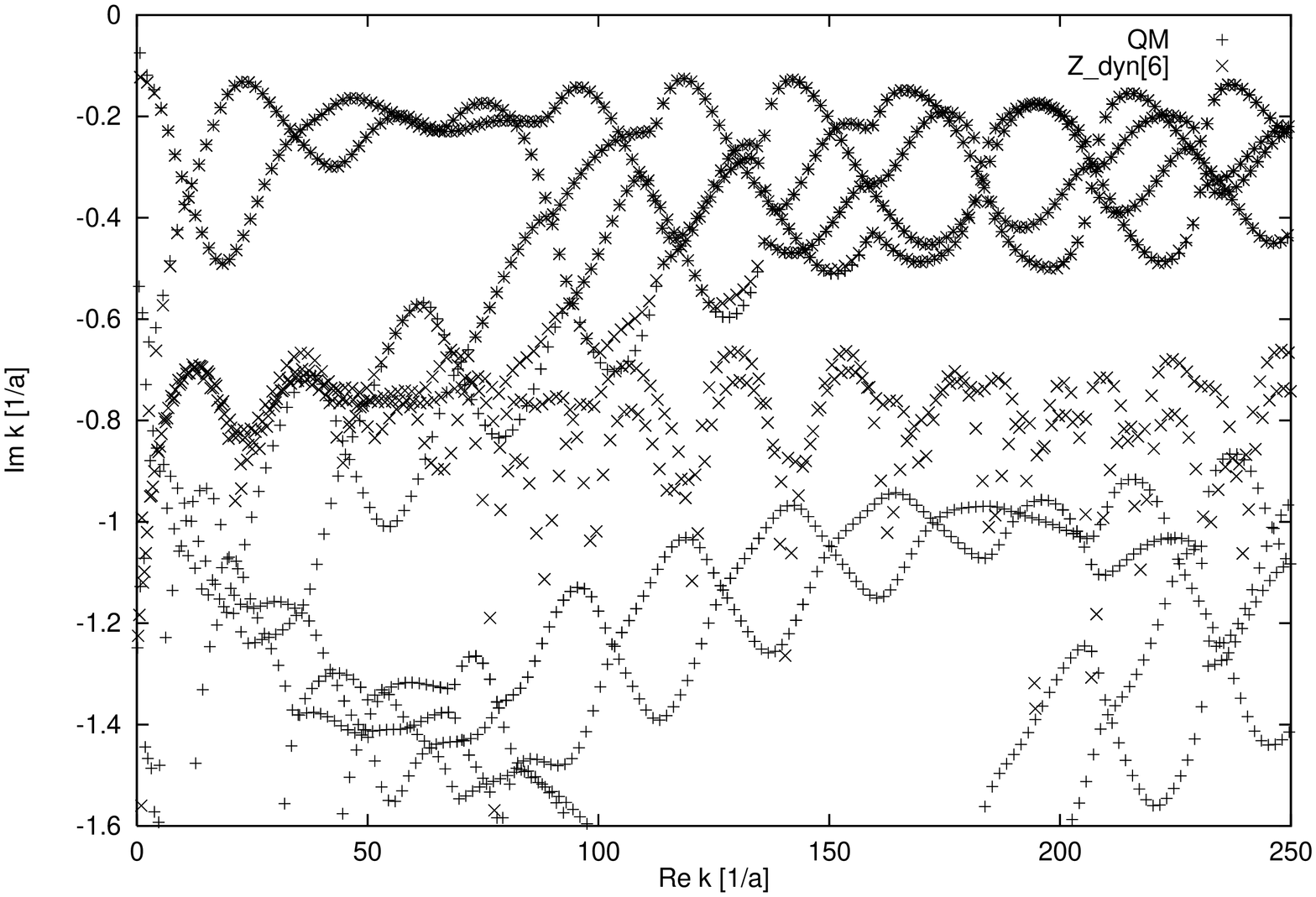,height=7.1cm,angle=-90}}

\vskip 0.1cm
 \centerline{{\bf(c)} \epsfig{file=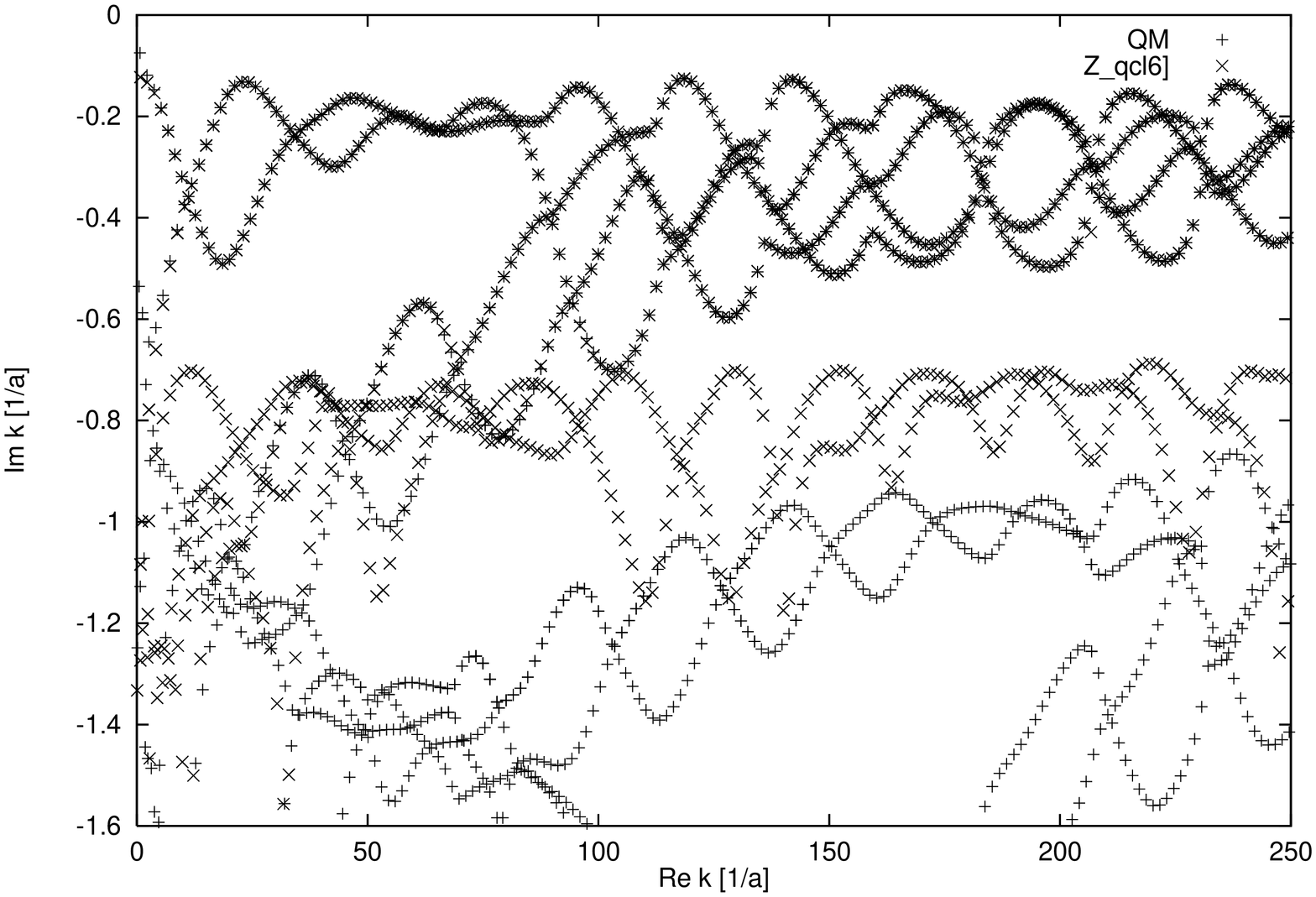,height=7.1cm,angle=-90}}

\caption[fig_n_6]{\small
The ${\rm A}_{\rm 1}$ resonances of the 3-disk system with $R=6a$. The 
exact quantum-mechanical data are denoted by plusses. 
The semiclassical ones are calculated up to  $6^{\,\rm th}$ order
in the curvature expansion and are denoted by crosses:
{\bf (a)} 
Gutzwiller-Voros 
zeta-function \equa{GV_zeta_app}, {\bf (b)} 
dynamical zeta-function \equa{dyn_zeta_app}, 
{\bf (c)} 
quasiclassical zeta function \equa{qcl_zeta}. 
\label{fig:e_gv6}}
\end{figure}

\noindent\begin{figure}[htb]
\vskip -0.4cm
\centerline{
{\bf(a)} 
\epsfig{file=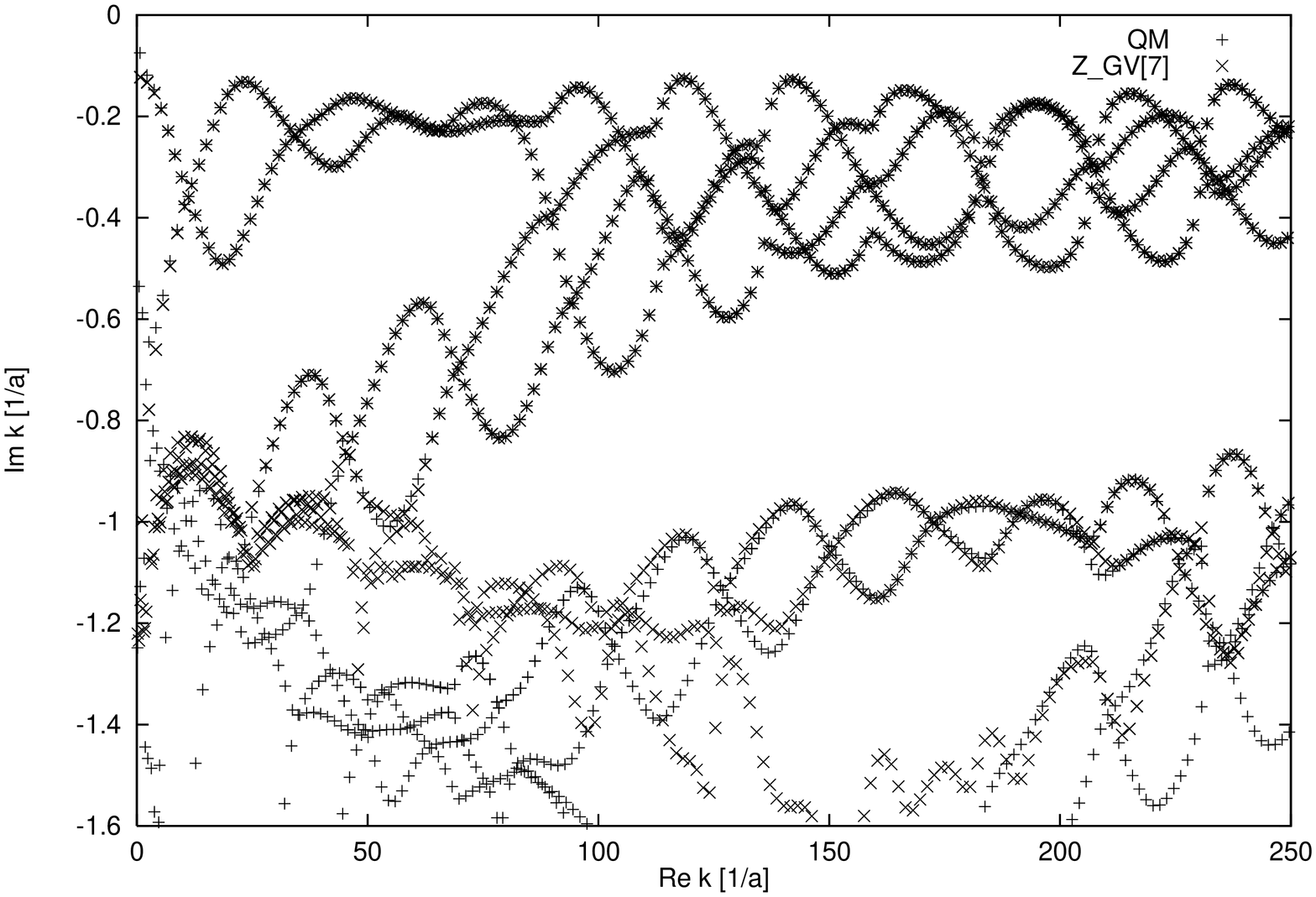,height=7.1cm,angle=-90}}

\vskip 0.1cm
\centerline{{\bf(b)} \epsfig{file=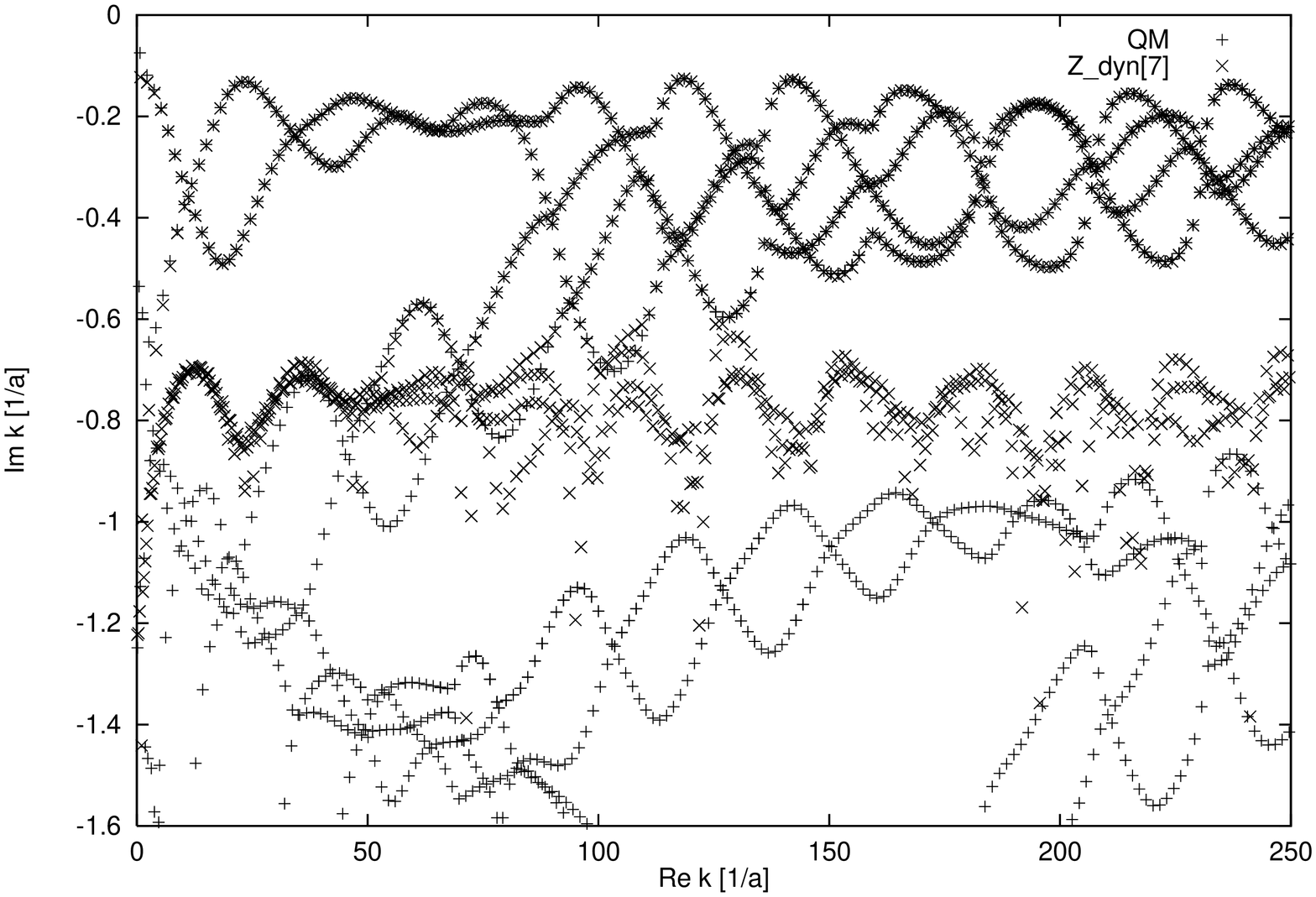,height=7.1cm,angle=-90}}

\vskip 0.1cm
 \centerline{{\bf(c)} \epsfig{file=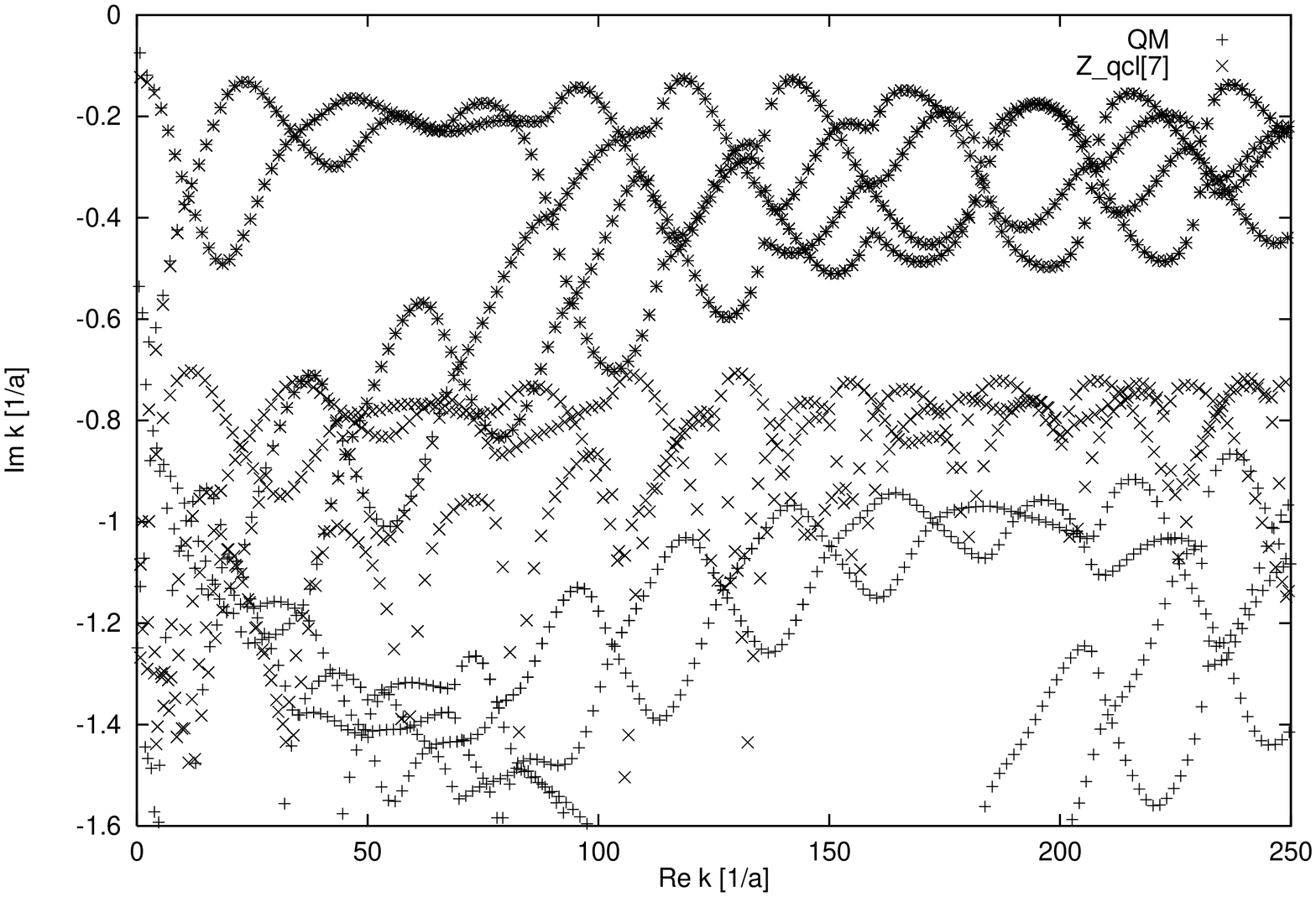,height=7.1cm,angle=-90}}

\caption[fig_n_7]{\small
The ${\rm A}_{\rm 1}$ resonances of the 3-disk system with $R=6a$. The 
exact quantum-mechanical data are denoted by plusses. 
The semiclassical ones are calculated up to  $7^{\,\rm th}$ order
in the curvature expansion and are denoted by crosses:
{\bf (a)} 
Gutzwiller-Voros 
zeta-function \equa{GV_zeta_app}, {\bf (b)} 
dynamical zeta-function \equa{dyn_zeta_app}, 
{\bf (c)} 
quasiclassical zeta function \equa{qcl_zeta}. 
\label{fig:e_gv7}}
\end{figure}

\noindent\begin{figure}[htb]
\vskip -0.4cm
\centerline{
{\bf(a)} 
\epsfig{file=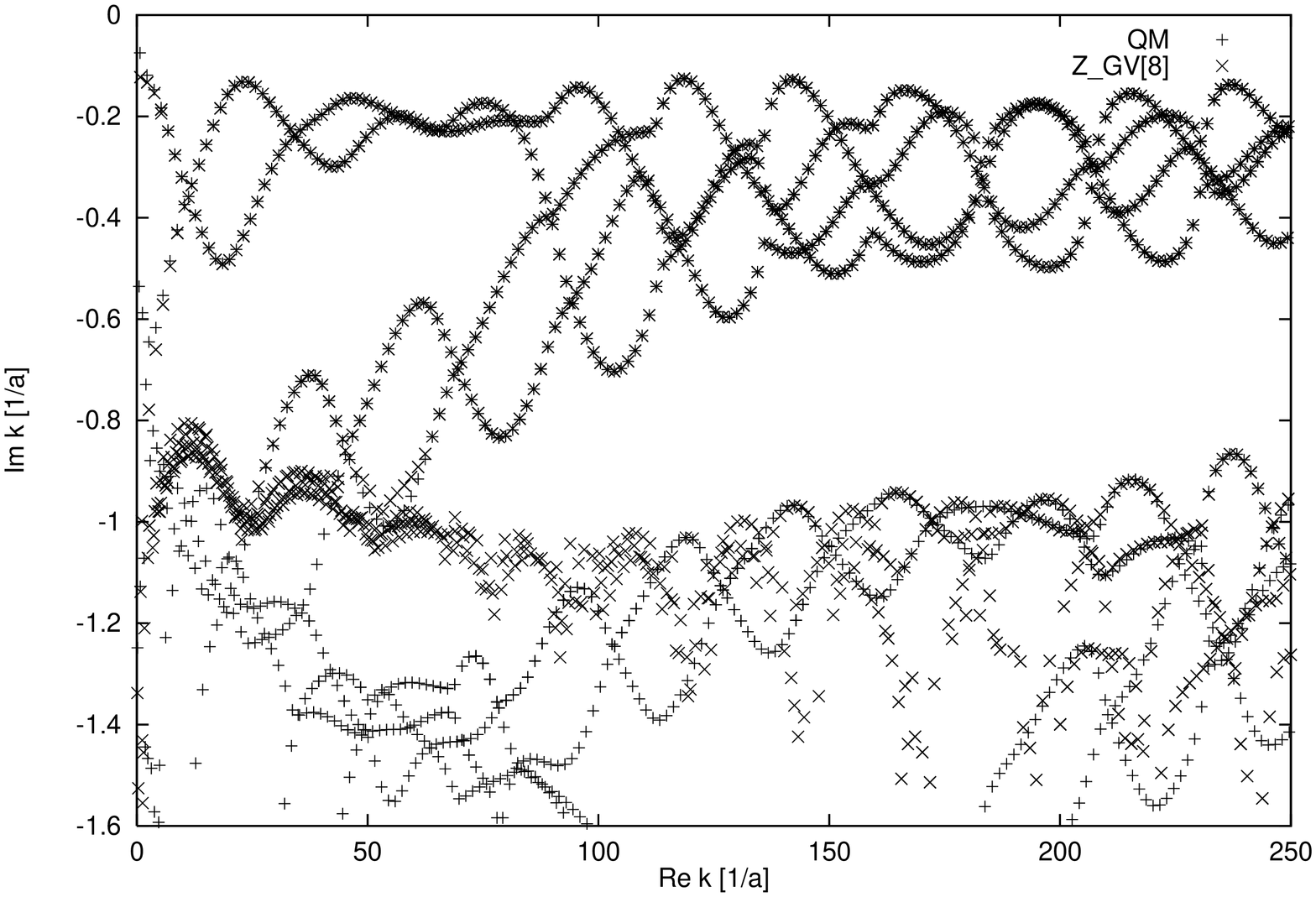,height=7.1cm,angle=-90}}

 \vskip 0.1cm
\centerline{{\bf(b)} \epsfig{file=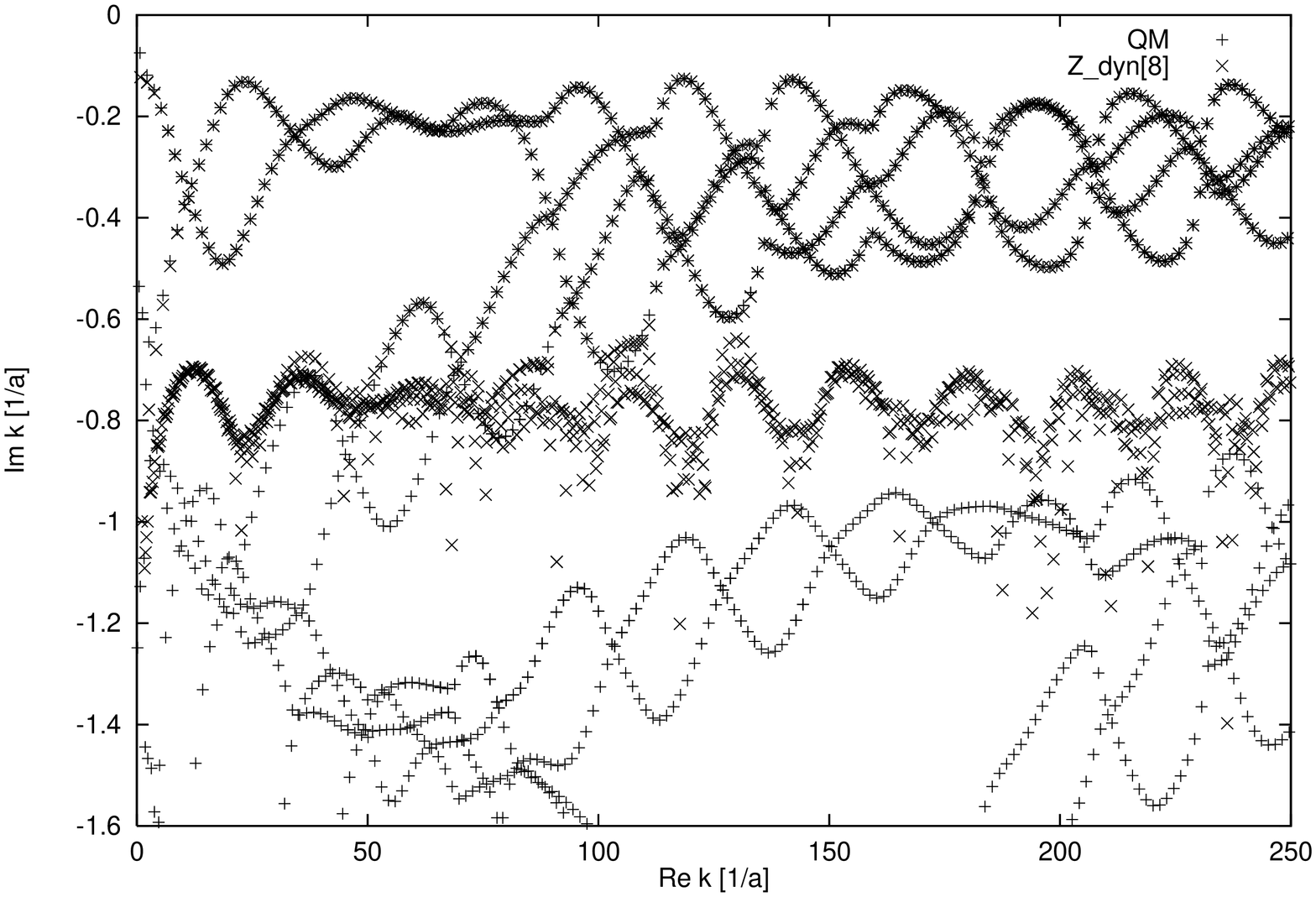,height=7.1cm,angle=-90}}

 \vskip 0.1cm
 \centerline{{\bf(c)} \epsfig{file=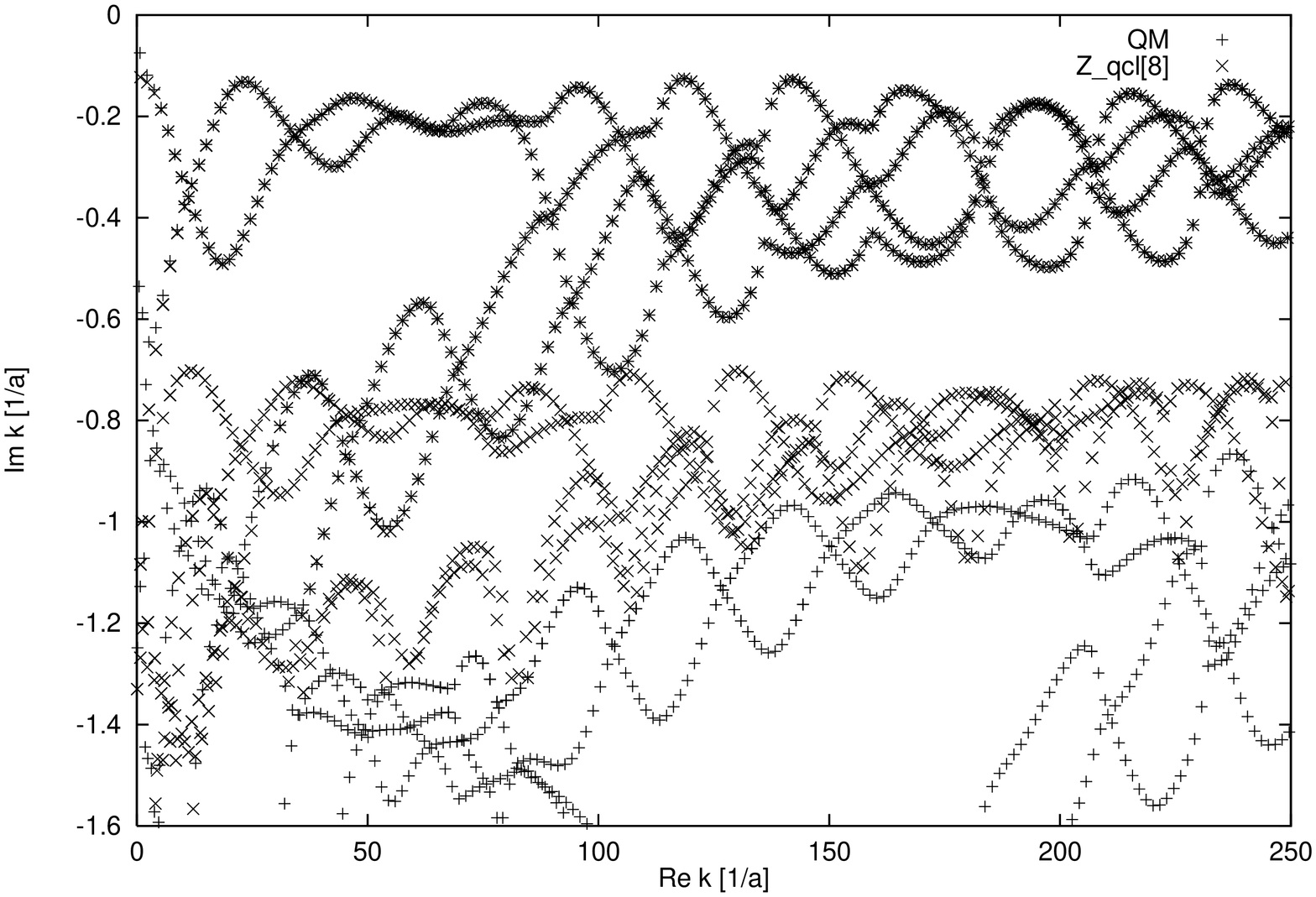,height=7.1cm,angle=-90}}
\caption[fig_n_8]{\small
The ${\rm A}_{\rm 1}$ resonances of the 3-disk system with $R=6a$. The 
exact quantum-mechanical data are denoted by plusses. 
The semiclassical ones are calculated up to $8^{\,\rm th}$ order
in the curvature expansion and are denoted by crosses:
{\bf (a)} 
Gutzwiller-Voros 
zeta-function \equa{GV_zeta_app}, {\bf (b)} 
dynamical zeta-function \equa{dyn_zeta_app}, 
{\bf (c)} 
quasiclassical zeta function \equa{qcl_zeta}. 
\label{fig:e_gv8}}
\end{figure}

\noindent\begin{figure}[htb]
\vskip -0.4cm
\centerline{
{\bf(a)} 
\epsfig{file=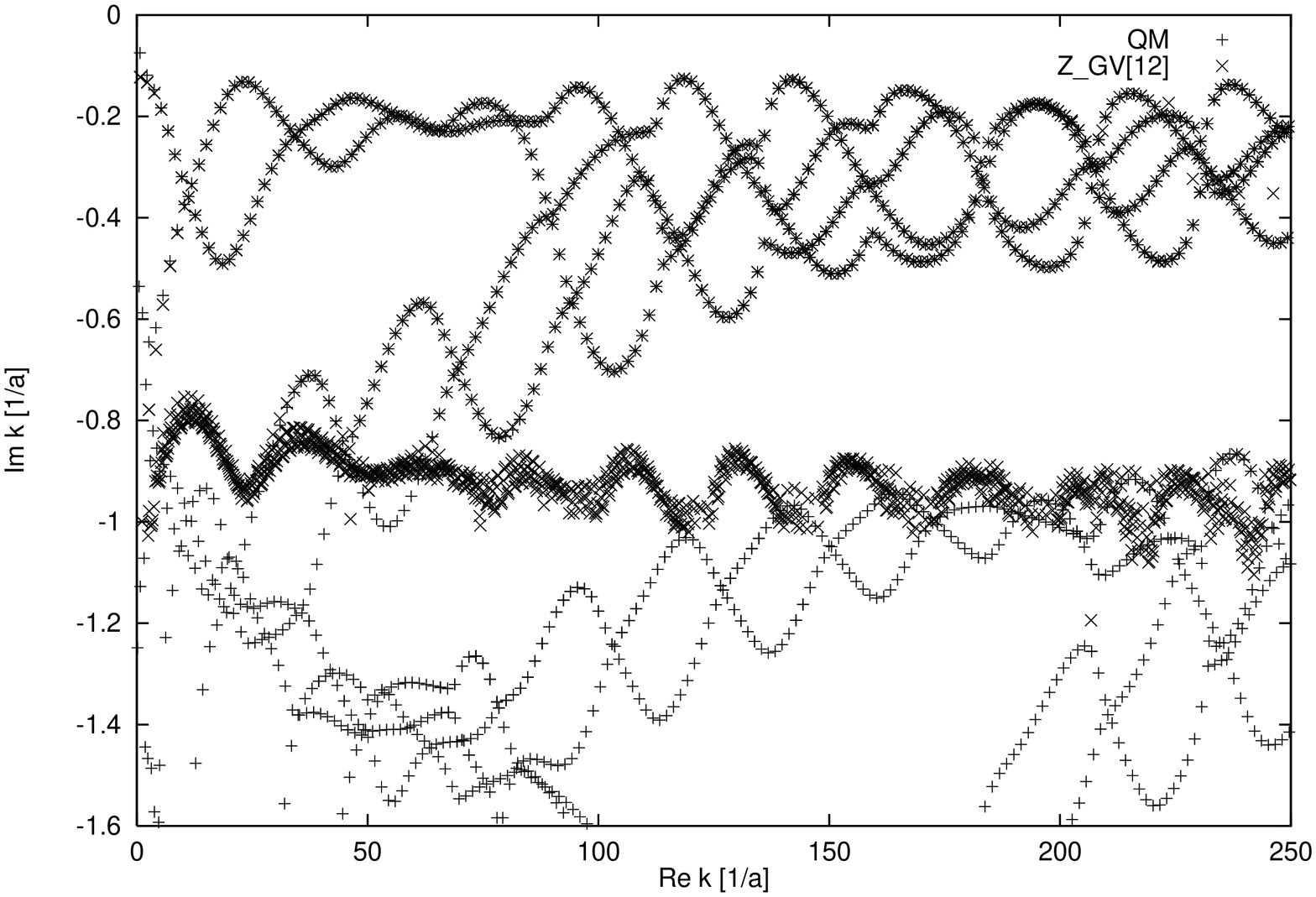,height=7.1cm,angle=-90}}

 \vskip 0.1cm
\centerline{{\bf(b)} \epsfig{file=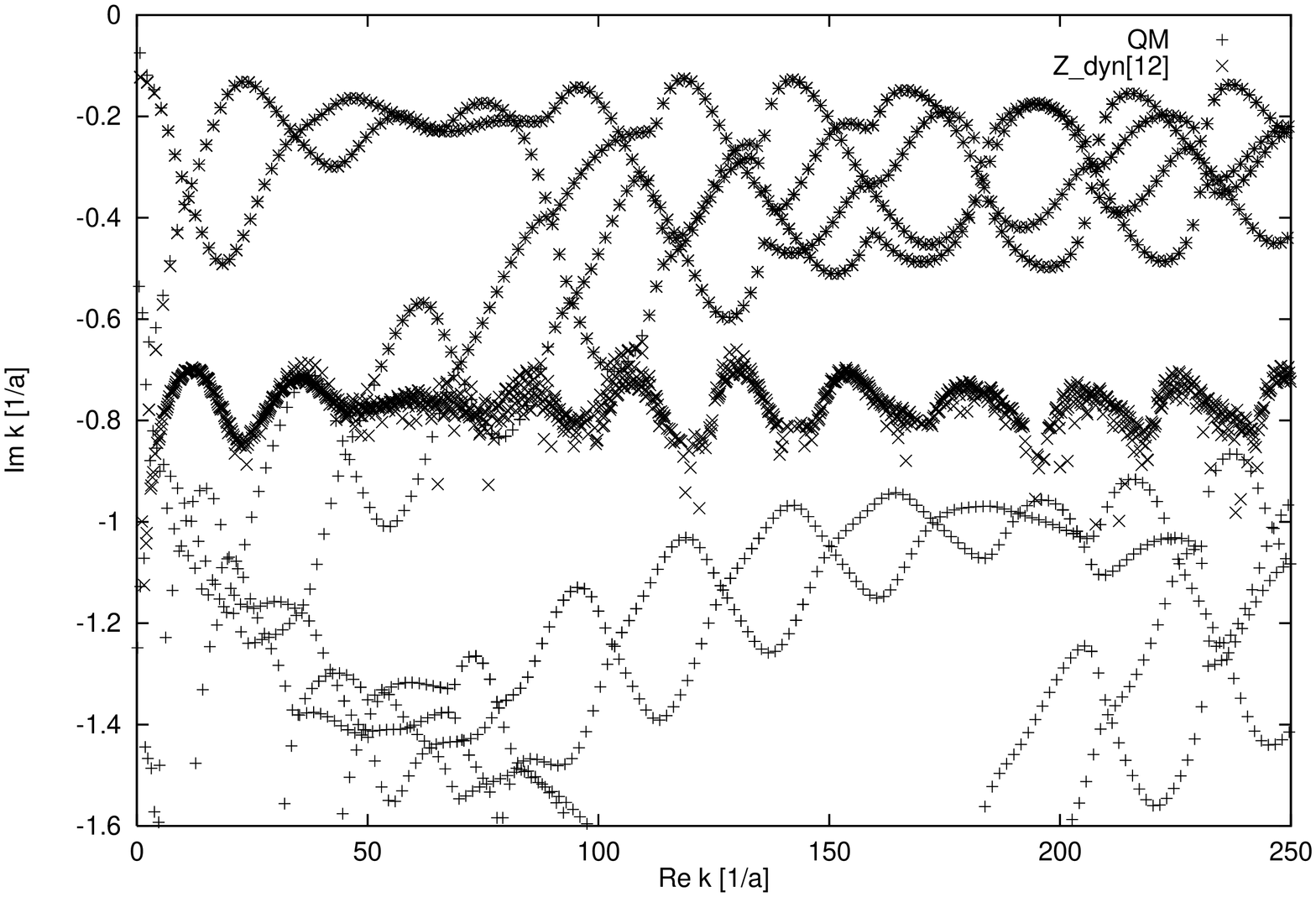,height=7.1cm,angle=-90}}

 \vskip 0.1cm
 \centerline{{\bf(c)} \epsfig{file=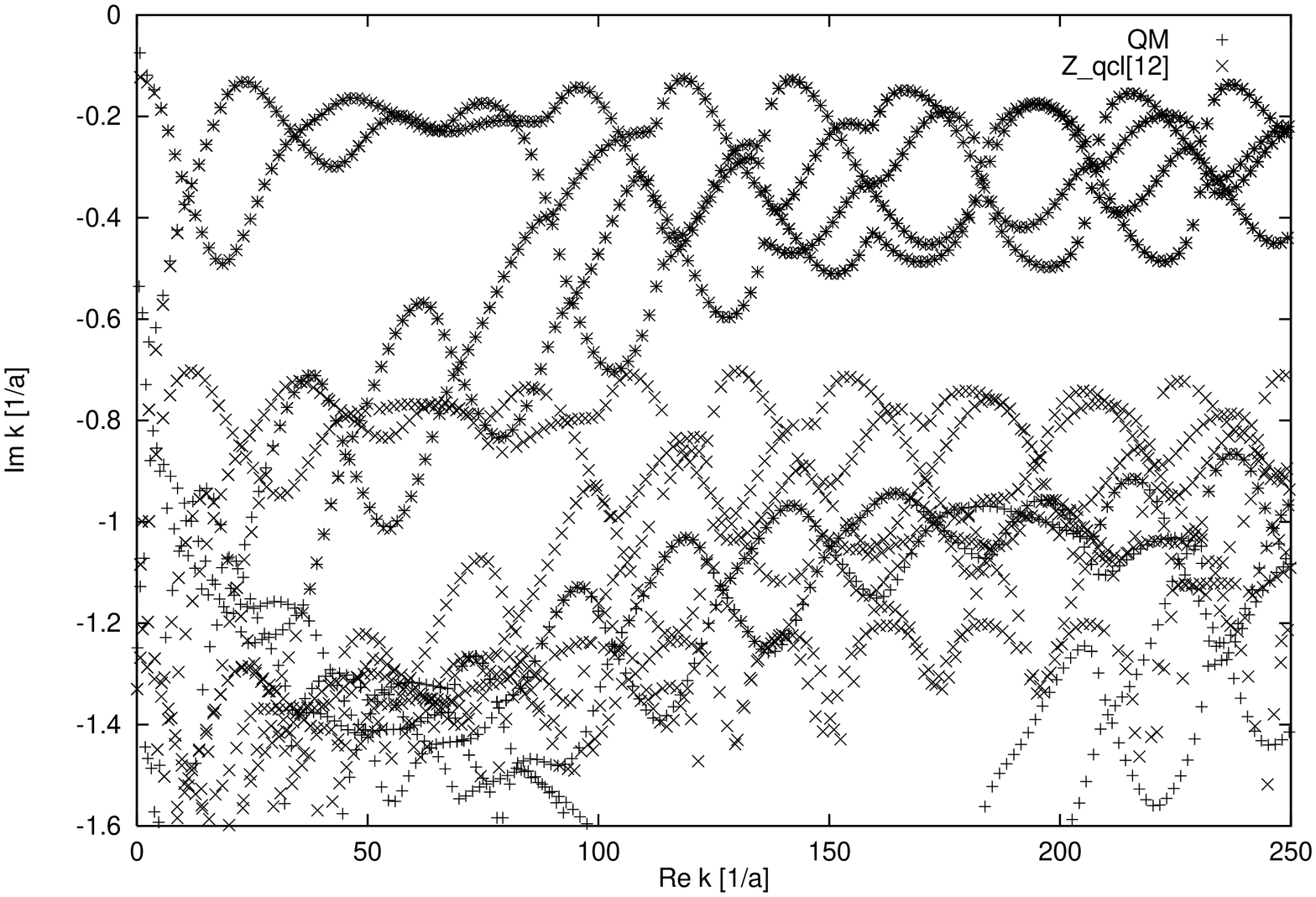,height=7.1cm,angle=-90}}

\caption[fig_n_12]{\small
The ${\rm A}_{\rm 1}$ resonances of the 3-disk system with $R=6a$. The 
exact quantum-mechanical data are denoted by plusses. 
The semiclassical resonances are calculated up to   $12^{\,\rm th}$ order
in the curvature expansion and are denoted by crosses:
{\bf (a)} 
Gutzwiller-Voros 
zeta-function \equa{GV_zeta_app}, {\bf (b)} 
dynamical zeta-function \equa{dyn_zeta_app}, 
{\bf (c)} 
quasiclassical zeta function \equa{qcl_zeta}. 
\label{fig:e_gv12}}
\end{figure}
\end{document}